# Modeling and control of modern wind turbine systems: An introduction†

**Christian Dirscherl, Christoph M. Hackl*** **and Korbinian Schechner**

**Munich School of Engineering, Technische Universität München**

† Revised and translated version of the German book chapter [1] entitled "Modellierung und Regelung von modernen Windkraftanlagen: Eine Einführung", published as Chapter 24 in D. Schröder "Elektrische Antriebe – Regelung von Antriebssystemen" (4. Auflage, 2015), Springer. The authors are most grateful to **Jacob Hamar** for preparing a preliminary draft of this English version.

* Dr.-Ing. Christoph Hackl (corresponding author)
All authors (in alphabetical order) contributed equally to this chapter.

Technische Universität München
Research group "Control of renewable energy systems" (CRES)
Munich School of Engineering

Lichtenbergstraße 4a
85748 Garching

email: christoph.hackl@tum.de
www: www.cres.mse.tum.de

# Contents

# Modeling and control of modern wind turbine systems: An introduction

Christian Dirscherl[†], Christoph M. Hackl[†,⋆] and Korbinian Schechner[†]

**Abstract**

This chapter provides an introduction to the modeling and control of power generation from wind turbine systems. In modeling, the focus is on the electrical components: electrical machine (e.g. permanent-magnet synchronous generators), back-to-back converter (consisting of machine-side and grid-side converter sharing a common DC-link), mains filters and ideal (balanced) power grid. The aerodynamics and the torque generation of the wind turbine are explained in simplified terms using a so-called power coefficient. The overall control system is considered. In particular, the phase-locked loop system for grid-side voltage orientation, the nonlinear speed control system for the generator (and turbine), and the non-minimum phase DC-link voltage control system are discussed in detail; based on a brief derivation of the underlying machine-side and grid-side current control systems. With the help of the power balance of the wind turbine, the operation management and the control of the power flow are explained. Concluding simulation results illustrate the overall system behavior of a controlled wind turbine with a permanent-magnet synchronous generator.

[‡]C. Dirscherl, C.M. Hackl and K. Schechner are with the research group "Control of renewable energy systems" (CRES) at the Munich School of Engineering (MSE), Technische Universität München (TUM), Germany.

[*]Authors are in alphabetical order and contributed equally to the paper. Corresponding author is C.M. Hackl (christoph.hackl@tum.de).

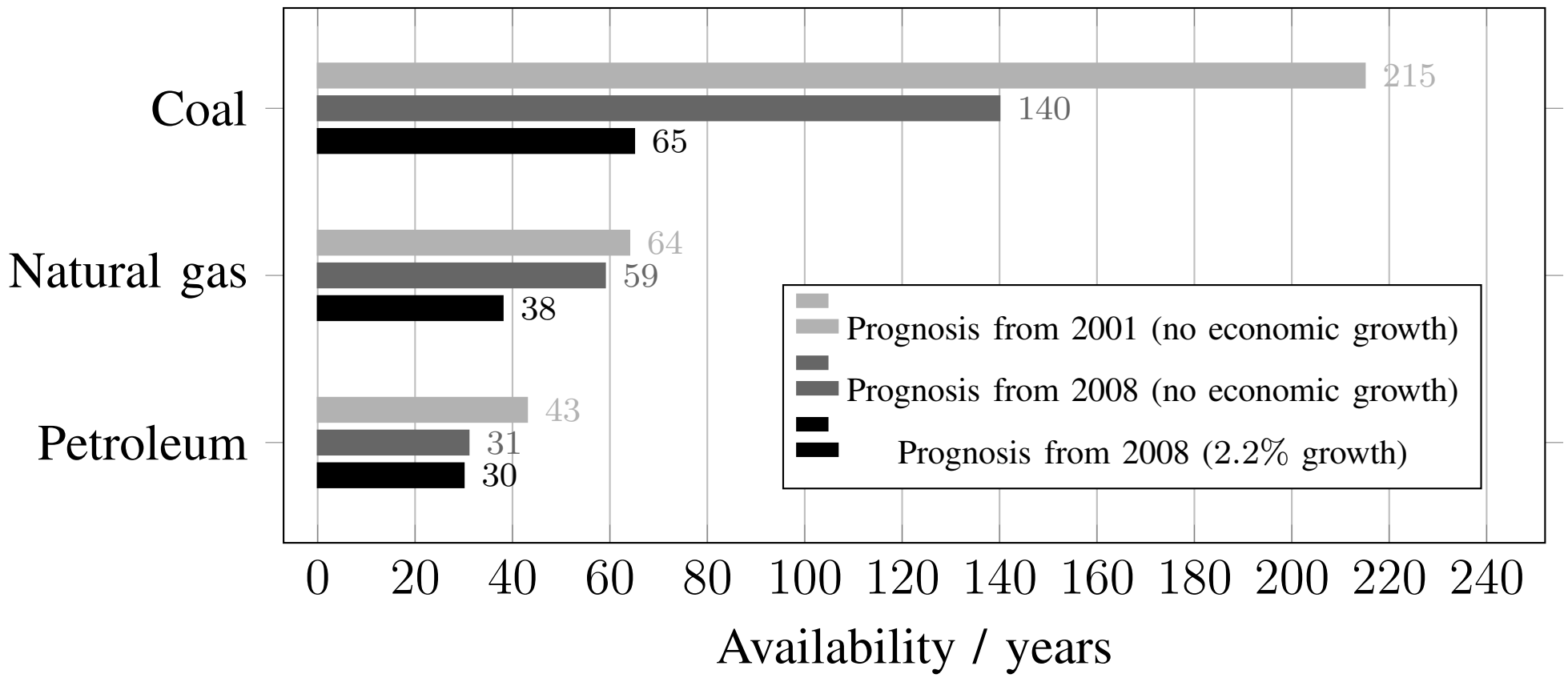

Fig. 1: *Forecasted availability of fossil fuels: Coal, natural gas and petroleum (see [3, Tab. 1.3]).*

# I. Introduction

Recently, because of the introduction of the "Renewable Energy Sources Act[1]" in 2000, a public debate about the future of energy supply was launched: How can a reliable, sustainable and affordable energy supply be achieved in the long term? In the wake of the Fukushima disaster, and the increasingly dramatic sounding reports of the Intergovernmental Panel on Climate Change (IPCC), a new way of thinking seems inevitable: A transition from fossil fuels to "renewable energy sources" such as biomass, solar, hydro and wind.

## *A. Scarcity of resources*

An "anthropogenic climate change" is accepted in the majority of the scientific community. According to [2, p. 17–19], it is extremely probable that human influence is the main cause of the observed warming of the earth since the middle of the 20th century. It is uncertain whether a world-wide rethinking and the paradigm shift from fossil fuels to renewable energy sources can really take place. Most important it is not clear whether a global will to change will persist to significantly and resiliently alter the behavior in the private, economic and political areas.

In the long run, irrespective of all ecological aspects, the *finite* availability of fossil fuels in the world will force humanity to move away from coal, natural gas or petroleum/oil. Based on estimates from the years 2001 and 2008, forecasts on the availability (in years) of coal, natural gas and oil are shown in Fig. 1. For example, if the 2008 forecast were to be based on the assumption of an average global economic growth of $2.2\,\%$, there would be no oil in 2038, no natural gas in 2046, and no coal in 2073. New discoveries and, for example, the new technology of shale gas depletion (fracking) will extend the availability of fossil fuels, but in the end, these "energy sources" remain finite.

## *B. Wind Energy: The beacon of hope of the "Energiewende"*

The "Energiewende" imposes a particular challenge for politics, economics and sciences. The challenge is complex, interdisciplinary and transnational. Apart from, for example, the transformation of the power grid and the development of suitable energy storage technologies, particular focus is on renewable energy sources. The "renewable energies" biomass/biogenic wastes, geothermal energy, photovoltaics (PV), hydro-electric power and wind power (onshore/offshore) are considered as drivers of the "Energiewende" (see [4, p. 1]).

Here, the wind energy plays a special role. It is regarded as the "beacon of hope" (German: "Hoffnungsträger" [5]) of the "Energiewende". In 2012, wind turbines with a total nominal (rated) output power of 31 156 MW were installed in Germany and fed 50 670 GW h into the power grid. This corresponded to almost 8% of the total electricity consumption and 34% of the renewable electricity mix in Germany, respectively (see [8, p. 7–17]). In particular, the growth in the newly installed onshore and offshore wind turbine systems is still high. New onshore wind turbines with a nominal output power of 2 851 MW (growth rate of 9.0 %) and new offshore wind turbines with a nominal output power of 240 MW (growth rate of 80%) were installed in 2013 (see [8, p. 7]).

[1]German: "Erneuerbare Energien Gesetzes" (EEG).

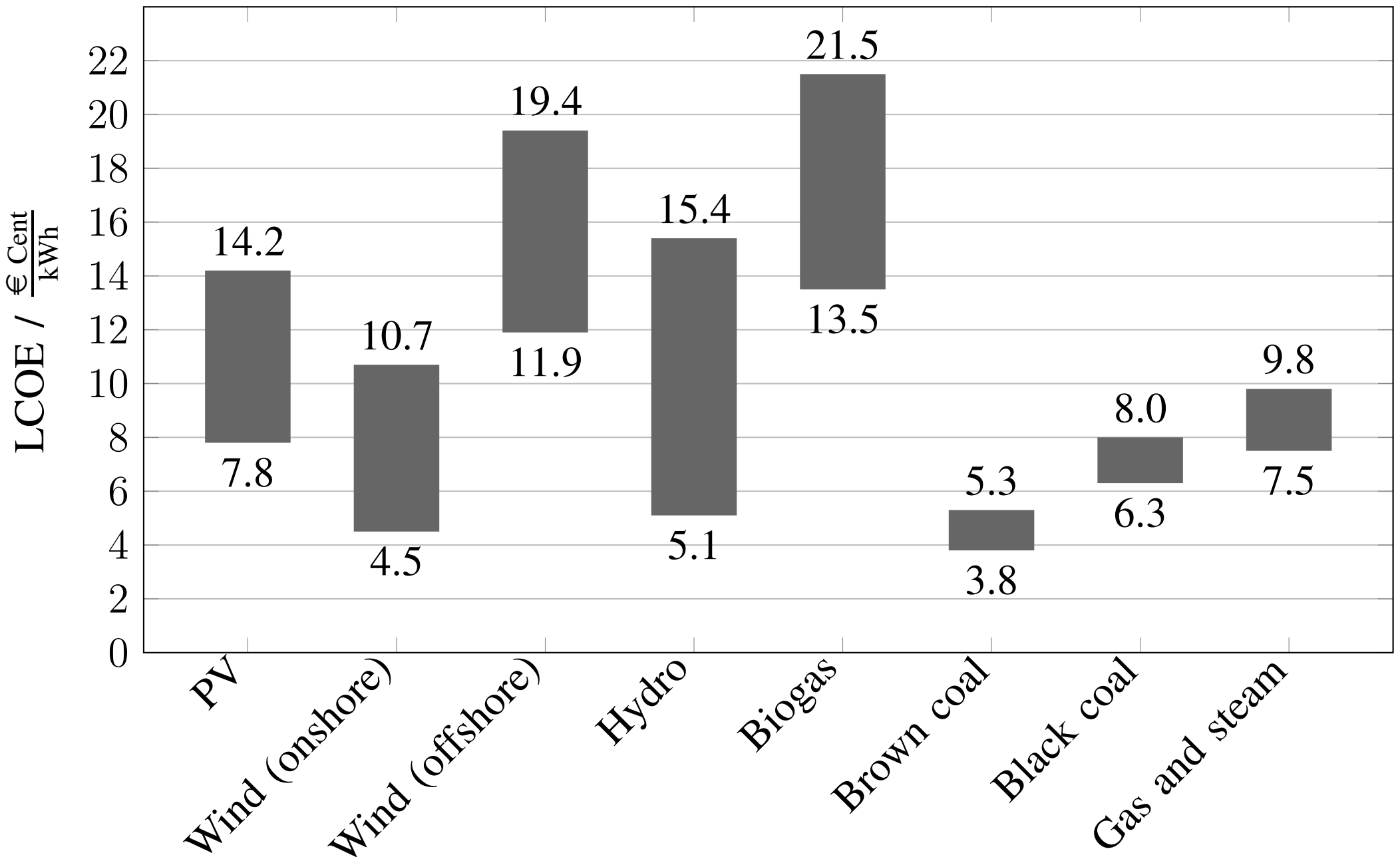

Fig. 2: *Levelized costs of electricity (LCOE) in Germany in 2013: comparison between different technologies (after Fig. 1 in [6] and Tab. 4.12 (from* 500 kW*) in [7]).*

Of particular importance are the already competitive electricity costs of on-shore wind power (see Fig. 2). On-shore large-scale wind turbine systems with *levelized costs of electricity* (LCOE) of 4.5 € Cent/kWh are more favorable than photovoltaics, hydropower, biogas, black coal and gas & steam, and thus represent an economic alternative to most fossil fuels today. Only brown coal power plants, with 3.8 € Cent/kWh, can compete (see [6] and Tab. 4.12 in [7]).

## *C. Development of wind power*

The use of wind energy is hardly a modern development. Wind energy was used as an energy source already in the first centuries before common era (BCE). The following sections provide a brief overview of the development of wind power. A partial time-line is shown in Fig. 3.

*1) Historical overview:* According to historians, the use of wind power started in the 17th century before Christ (BC). In Mesopotamia, the first windmills were used for irrigation of farmland [9, Ch. 2]. The design feature of these historic windmills is considered as a drag[2] with a rotation around the vertical axis. Here the air resistance of the wind sail/area is converted into a drive torque, resulting in a rotational movement about the vertical axis. After a large delay, the same drag rotor technology was used in Persia (7th century anno Domini (AD)), and finally in China (10th century AD).

The use of wind power in Europe began only in the 12th century AD. Wind turbines/mills spread from France and England via Holland and Germany (13th century AD) and as far as Russia (14th century AD). The windmill as a driver of a water wheel became one of the most important technologies of the time [9, Ch. 2]. Unlike drag-based turbines with a vertical rotation axis, these windmills were distinguished by a horizontal axis. The physical principle allowing the rotor blades to rotate is known as the so-called lift principle. Hence, such wind turbines are known as a lift-based turbines[2]. The first lift-based turbines were namely used as grinding mills and it was not until the 15th century in Holland that a new technological construction allowed for the first water pumping application. Thus, it took more than 3 200 years until Europe could be irrigated using wind power.

[2] Drag and lift principles will be further explained in Section I-D4.

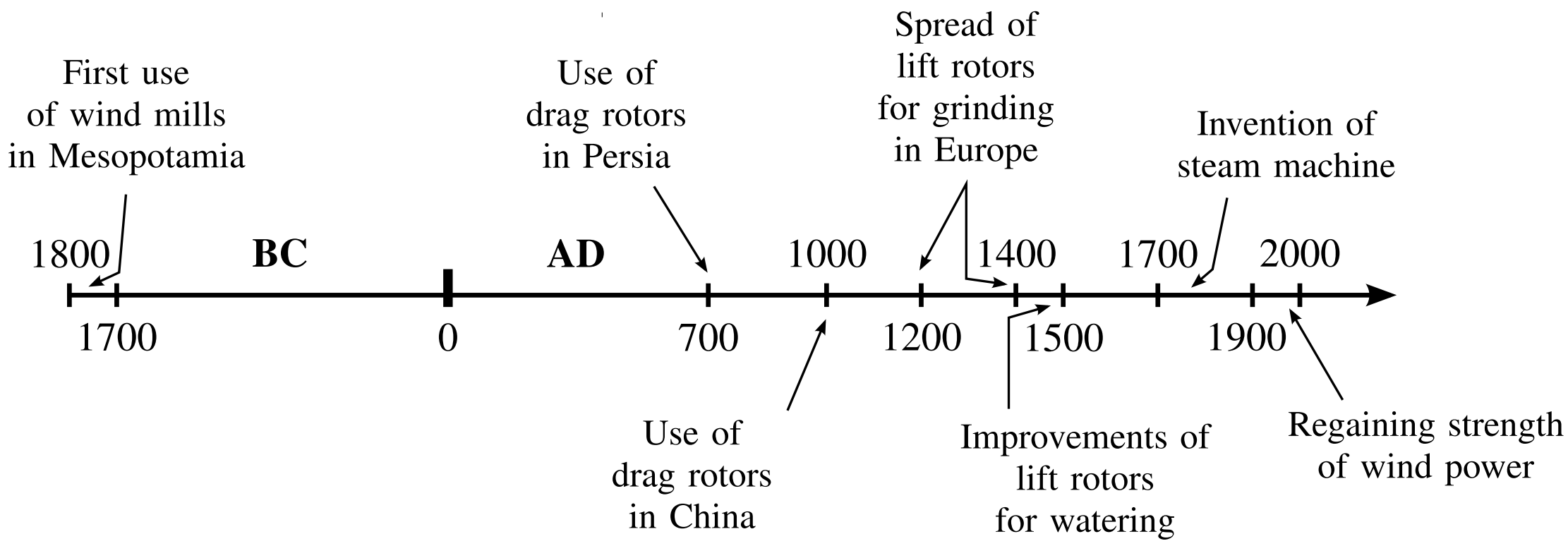

Fig. 3: *Historical development of wind power through civilization.*

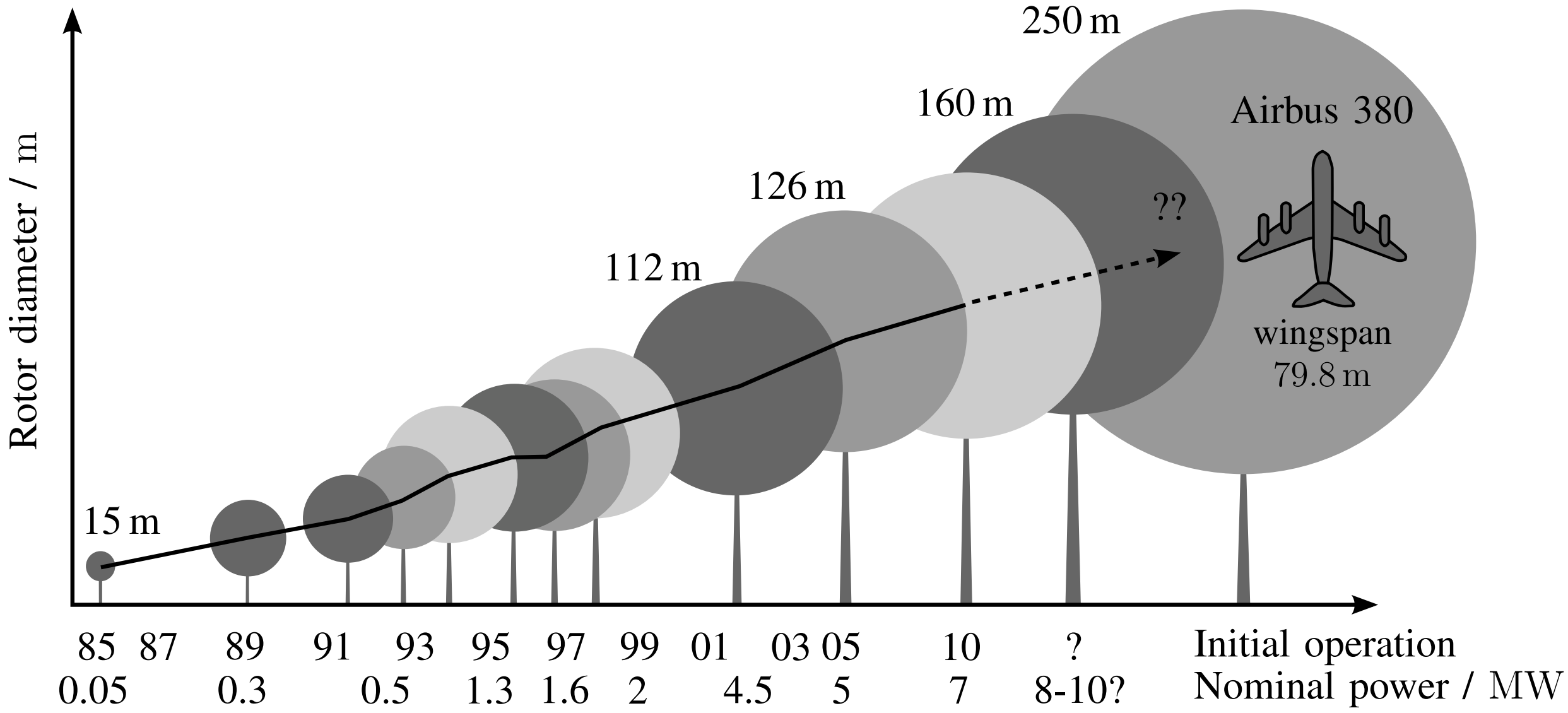

Fig. 4: *Development of rotor blade diameters over the year of commissioning and rated power (see illustration on p. 17 in [10]).*

With the invention of the steam engine, and later the combustion engine, the wind turbines were replaced in the 19th century with these new engines. Only towards the end of the 20th century wind energy did regain interest in (electrical) energy systems. The technical developments of the last 15-20 years in the field of wind turbines has led to todays horizontal axis wind turbines. Its rotor is (almost exclusively) made up of three blades and has a diameter of up to 130 m. Individual wind turbines can harvest nearly 7 MW with rotor diameters of 160 m (see [10, p. 16–19]). In Fig. 4, the past and future wind turbine sizes, rated power and turbine diameter are illustrated.

*2) Recent developments in Germany:* Since the 1990s, the use of wind power in Germany has risen sharply. In Fig 5, the growth of installed capacity of wind turbines in Germany from 1990s to 2013 is shown. While in 1990, only 61 MW were installed, in 2013, that number was already 34 179 MW [8].

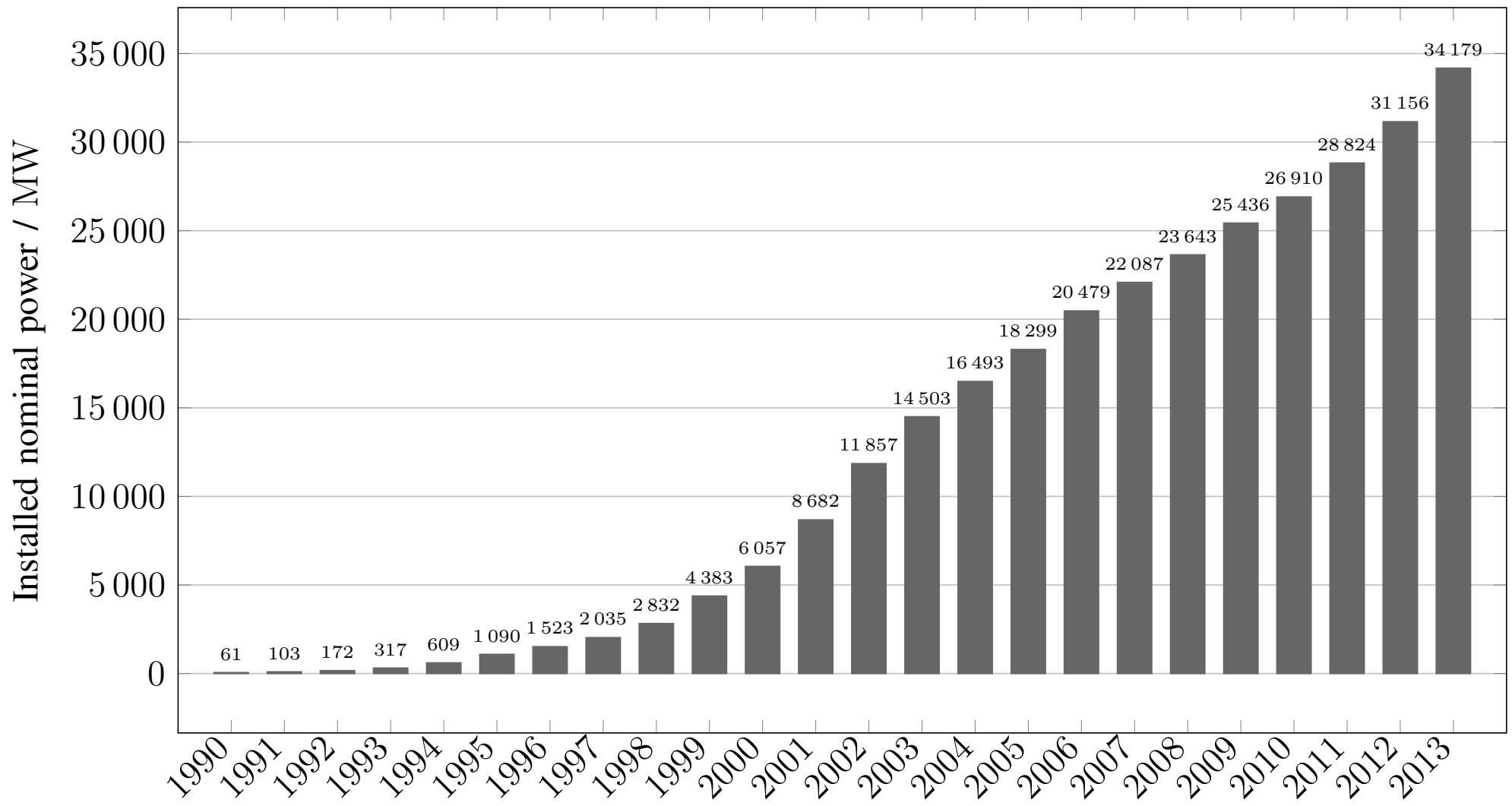

Fig. 5: *Development of the nominal power installation of wind turbines in Germany (from Fig. 4 in [8]).*

This represents an increase of more than $56\,000\,\%$.

The reasons for this sharp increase in Germany in particular can be attributed in large part to the "Stromeinspeisungsgesetz" (from 1991, see [4, p. 30–33]) and the successor, the "Erneuerbare Energien Gesetz (EEG)" (from 2000, see [4, p. 30–33]). The EEG (see [11, 12]) in particular, drove the expansion of renewable energy for power generation. Operators were given financial incentives to support such renewable power generation systems [8]. Research and development in the field of wind was also incentivized. Due to advances in technology and the increase in the number of wind turbines that were produced, electricity costs from wind power significantly decreased (see [13, p. 1]). To compare the costs of energy from various sources, the concept of *Levelized Costs of Electricity (LCOE)*[3]) was introduced (see [6, p. 36–37] and [14, p. 177–178]). In Fig. 2, the levelized costs of electricity of photovoltaic and wind turbine systems are compared with those of conventional power plants such as lignite (brown coal), black coal, gas, hydroelectric and biogas systems. The partly high fluctuations in generation costs are mainly due to the different sizes/capacities of the installed power generation technologies.

Along with hydropower, wind power (onshore) can produce electricity at competitive prices already *today* (see Fig. 2). It is therefore expected that there will be further increases in the installed nominal output power until 2030. In Fig. 6, the projected expansion corridor for wind turbines is shown [15, p. 6]. The reference value of the required maximum electric power in Germany in 2013 is $84\,\mathrm{GW}$ (annual peak load). It is assumed that the new power consumption[4] will decrease by $8\,\%$ in Germany from 2008-2020. This is due to efficiency enhancements. Depending on the power demand and the real future expansion of wind power, there will be periods when the vast majority of the required electric power in Germany is generated by wind turbines. Considering the best-case forecast, it is even possible that more than the required power is produced. The produced wind power surplus could help push Europe (provided adequate

[3]The levelized costs $K_{\mathrm{LCOE}}$ of electricity (e.g. in $\frac{€}{\mathrm{MW\,h}}$) indicate specific costs per unit of electricity. These are averaged over the estimated economic useful life of a total number $n$ (in years) of a power plant. By using this measure, various types of energy can be compared on a per unit basis. The (average discounted) electricity production costs are calculated using the formula

$$K_{\mathrm{LCOE}} = \frac{K_0 + \sum_{i=1}^{n} \frac{K_i}{(1+p/100)^i}}{\sum_{i=1}^{n} \frac{E_{\mathrm{el},i}}{(1+p/100)^i}} \tag{1}$$

with investment costs $K_0$ (in €) at the time of plant commissioning, the estimated total annual cost $K_i$ (in €) (fixed operating costs + variable operating costs + residual value / disposal of the plant) in the year $i \in \{1, \ldots, n\}$, the estimated electricity production $E_{\mathrm{el},i}$ (e.g. in $\mathrm{MW\,h}$) in the $i$-th year and the estimated interest rate $p$ (in $\%$) (see [6, p. 36–37] and [14, p. 177–178]).

[4]"Net electricity consumption $=$ total supply of electricity to consumers $+$ consumption from industry $-$ power plant consumption $-$ pump work $-$ losses from network" [16, p. 499]

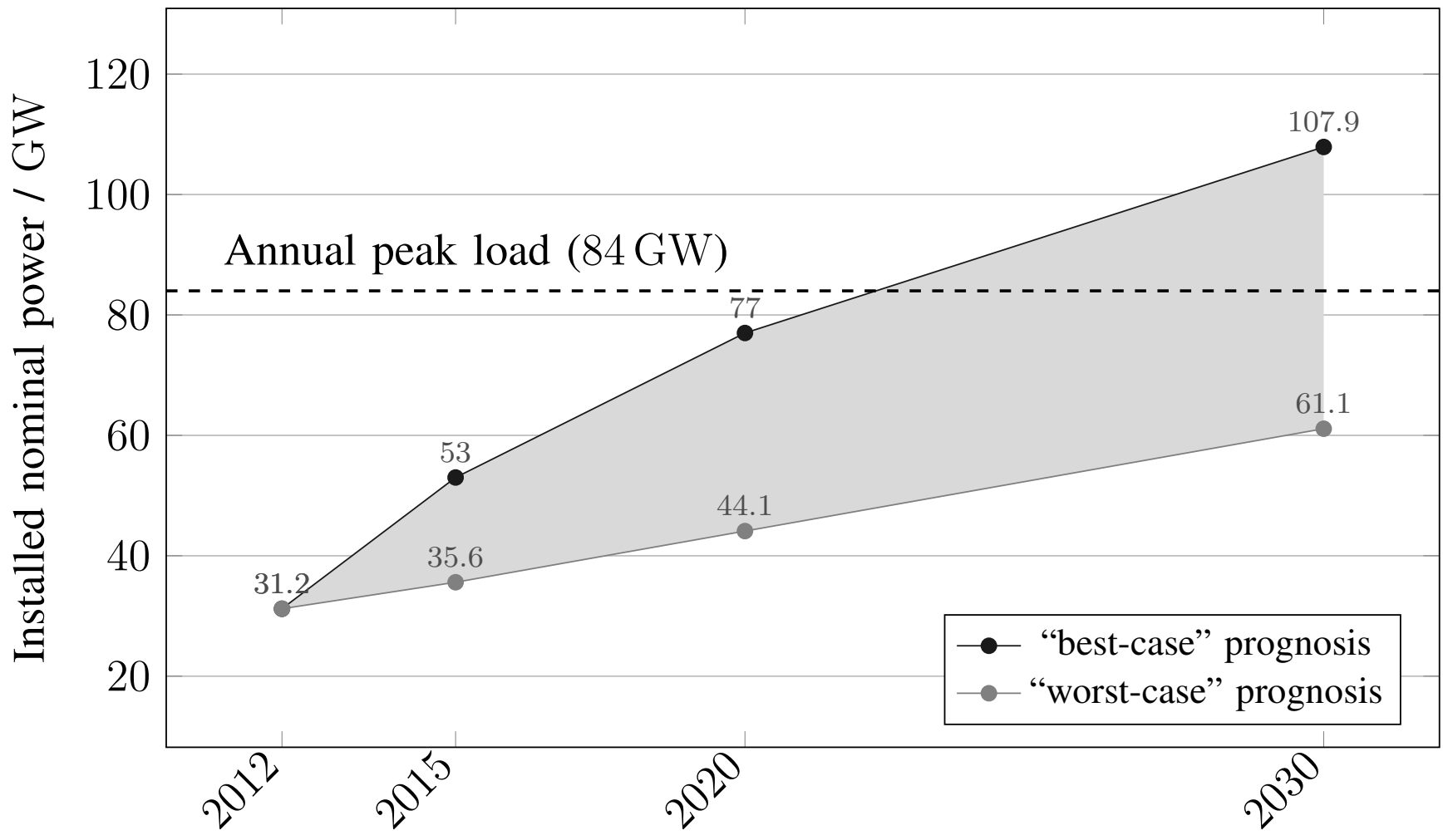

Fig. 6: *Prediction of the rated capacity installation of wind turbines in Germany up to 2030 [15, p. 6]: "Best-case" and "worst-case" prognosis.*

grid installations) towards meeting its power needs, or could be stored (provided adequate storage capacity) for low-wind periods.

### *D. Modern wind turbines*

Modern wind turbines are almost exclusively used for electrical power generation. In the past, various types of turbines have been used: Drag-based or lift-based rotors, with horizontal and vertical axis, and one or more blades. Today, lift-based turbine systems have prevailed on a horizontal axis with three blades. Such a system is shown schematically in Fig. 7. The operating principle of energy conversion from kinetic wind energy into rotational energy is similar across all concepts.

*1) Principle of operation:* In Fig 7, the frontal view of a modern wind turbine is illustrated. The rotor with radius $r_{\mathrm{t}}$ (in $\mathrm{m}$) and rotor area $A_{\mathrm{t}} = \pi r_{\mathrm{t}}^2$ (in $\mathrm{m}^2$) with three rotor blades. Each of the rotor blades can be rotated about the pitch angle $\beta$ (in $^\circ$). To align the rotor perfectly in the wind, modern wind turbines are equipped with a yaw system, which can rotate the entire nacelle about the yaw angle $\gamma$ (in $^\circ$) in (or out of) the wind. In Fig. 7, the wind blows into the image plane. The effective usable area for energy conversion is $A_{\mathrm{t}} = \pi r_{\mathrm{t}}^2 - A_{\mathrm{n}}$ where $A_{\mathrm{n}}$ (in $\mathrm{m}^2$) is the nacelle (and hub) area. In modern wind turbines with large rotor diameters such that $\pi r_{\mathrm{t}}^2 \gg A_{\mathrm{n}}$, it can be assumed that $A_{\mathrm{t}} \approx \pi r_{\mathrm{t}}^2$. The kinetic energy of wind is converted into rotational energy of the rotor, that is in a rotational motion with an angular velocity $\omega_{\mathrm{t}}$ (in $\frac{\mathrm{rad}}{\mathrm{s}}$).

**Remark I.1.** *It is assumed that the yaw system has been optimized, so that the rotor plane of the wind turbine is vertically aligned with the wind direction. The yaw system is not considered further in this chapter.*

*2) Energy and power from wind:* Wind represents a movement of air particles, and the bulk movement possesses kinetic energy. The kinetic energy of a wind volume $V$ (in $\mathrm{m}^3$) with air density $\varrho$ (in $\frac{\mathrm{kg}}{\mathrm{m}^3}$), mass $m$ (in $\mathrm{kg}$) and wind speed $v_{\mathrm{w}}$ (in $\frac{\mathrm{m}}{\mathrm{s}}$) at any time $t \geq 0\,\mathrm{s}$ (see for example [9, Ch. 5], [17, Sec. 6.4] or [18]) is given by

$$\forall t \geq 0\,\mathrm{s}: \qquad E_{\mathrm{w}}(t) := \frac{1}{2} \underbrace{\varrho(t)\, V(t)}_{=:m(t)} v_{\mathrm{w}}(t)^2 \qquad (\text{in } \mathrm{J} = \mathrm{Nm}). \tag{2}$$

Under the following simplifying assumptions

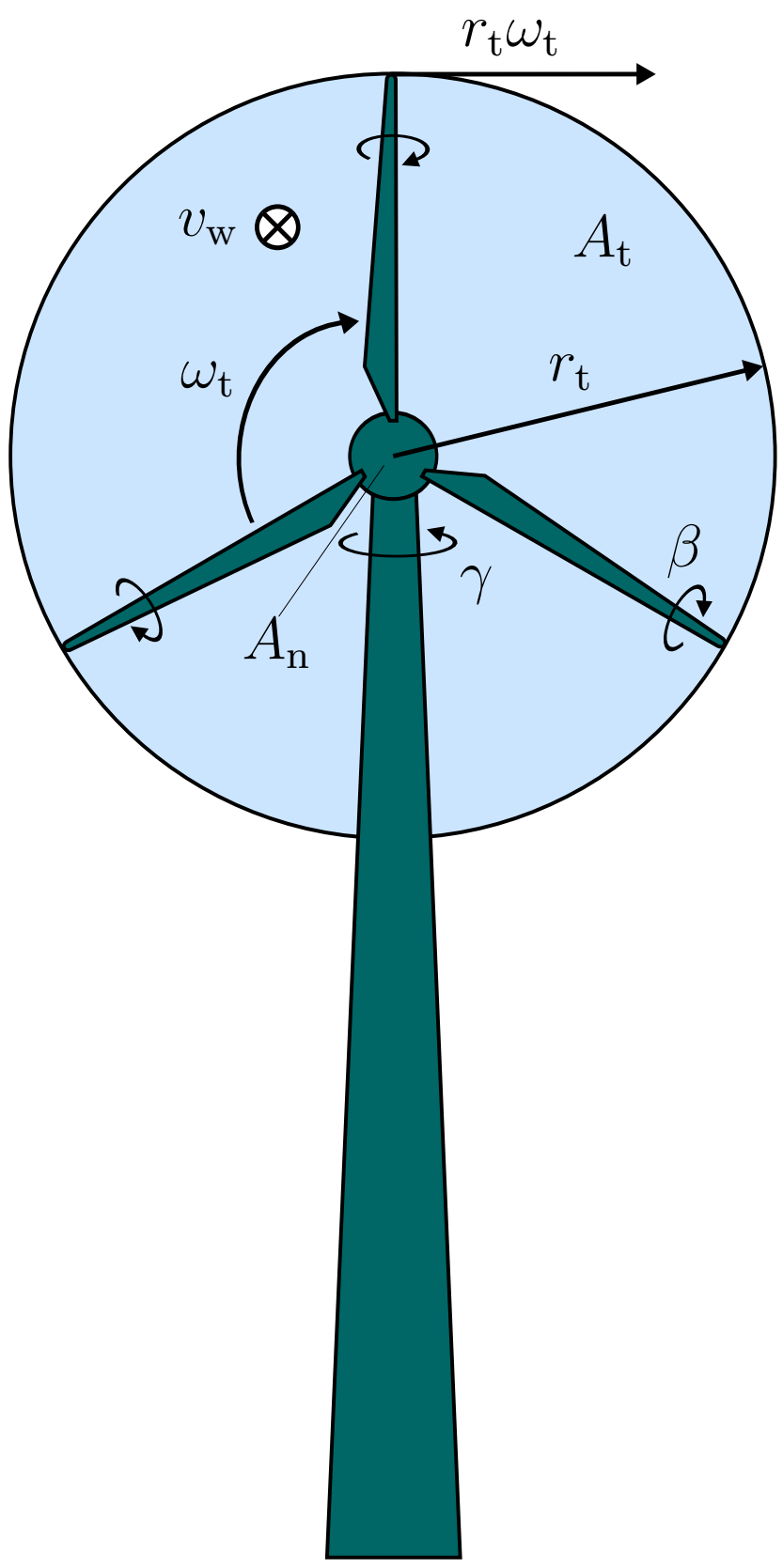

Fig. 7: *Front view of a wind turbine with an ideal air flow (illustration is based on Fig. 6.9 in [17]).*

**Assumption (A.1)** *A* constant *wind speed*[5], *i.e.,* $\frac{\mathrm{d}}{\mathrm{d}t}v_{\mathrm{w}}(t) = 0\,\frac{\mathrm{m}}{\mathrm{s}}$ *for all* $t \geq 0\,\mathrm{s}$, and

**Assumption (A.2)** *A* constant *air density*[6], *i.e.,* $\frac{\mathrm{d}}{\mathrm{d}t}\varrho(t) = 0\,\frac{\mathrm{kg}}{\mathrm{m}^3\,\mathrm{s}}$ *for all* $t \geq 0\,\mathrm{s}$,

the time derivative of the wind energy $E_{\mathrm{w}}(t)$ gives "wind power"

$$\forall t \geq 0\,\mathrm{s}: \qquad p_{\mathrm{w}}(t) := \frac{\mathrm{d}}{\mathrm{d}t}E_{\mathrm{w}}(t) \overset{(2),(A.1)}{=} \frac{1}{2}\,\frac{\mathrm{d}}{\mathrm{d}t}m(t)\,v_{\mathrm{w}}^2 \qquad (\text{in W}). \tag{3}$$

By introducing the mass flow

$$\forall t \geq 0\,\mathrm{s}: \qquad \frac{\mathrm{d}}{\mathrm{d}t}m(t) \overset{(A.2)}{=} \varrho\,\frac{\mathrm{d}}{\mathrm{d}t}V(t) = \varrho\,A\,v_{\mathrm{w}} \quad (\text{in } \frac{\mathrm{kg}}{\mathrm{s}}), \tag{4}$$

with wind speed $v_{\mathrm{w}}$ through an area $A\,(\text{in m}^2)$ of air particles, and by substituting (4) in (3), the wind power

$$\forall t \geq 0\,\mathrm{s}: \qquad p_{\mathrm{w}}(t) \overset{(2),(4)}{=} \frac{1}{2}\varrho\,A\,v_{\mathrm{w}}^3 \geq 0\,\mathrm{W} \tag{5}$$

is obtained. Note that, under the Assumptions (A.1) and (A.2), the wind power (5) is considered *constant* for all time $t \geq 0$ and *non-negative*.

[5] Assumption (A.1) represents an oversimplicifation which is *not* generally valid. The wind speed is highly chaotic and variable.

[6] The air density depends on temperature and pressure. These environmental variables change slowly (see [9, p. 181], [17, p. 248]). Therefore, Assumption (A. 2) is justified in most cases.

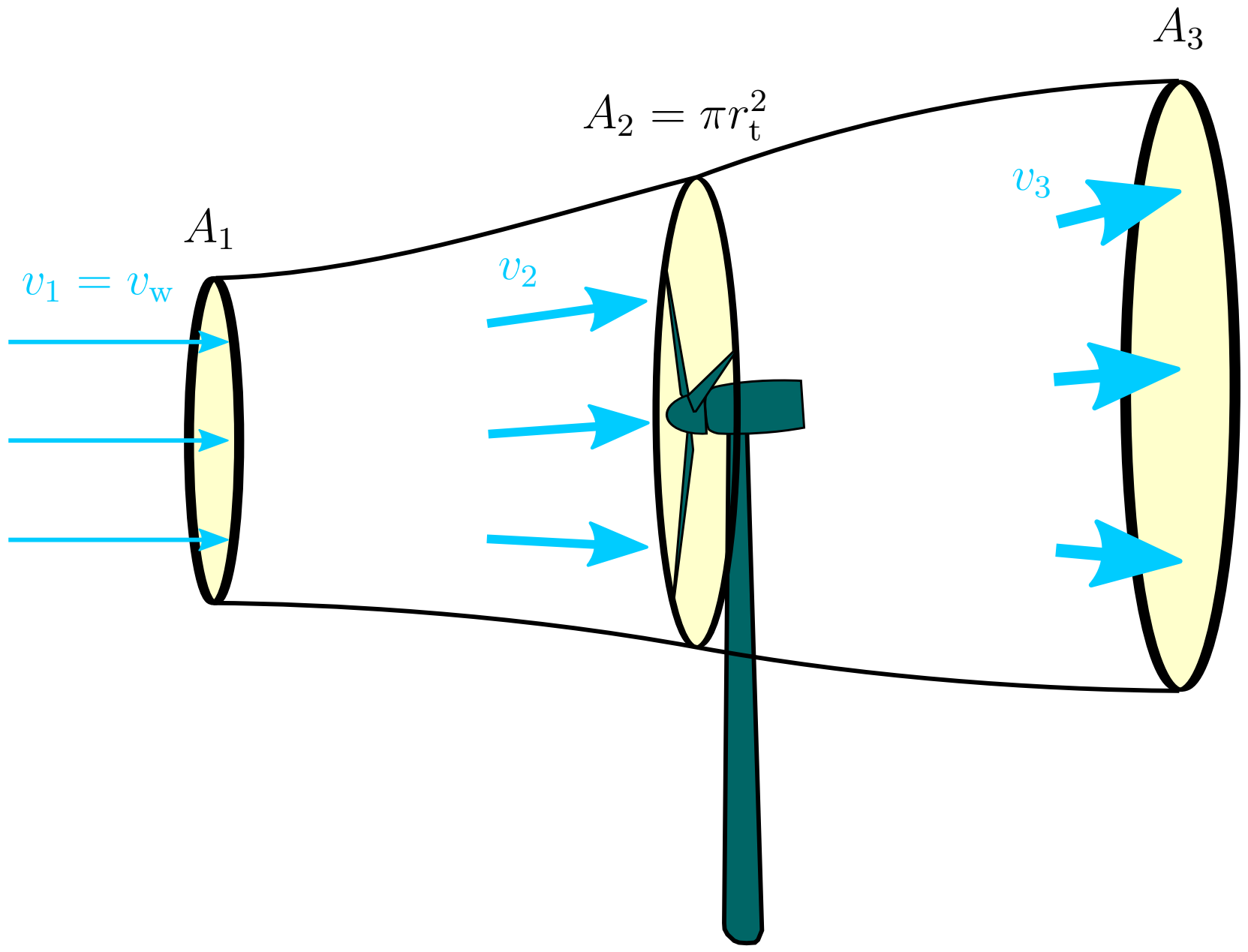

Fig. 8: *Flow profile in a lift-based wind turbine [17, Bild 6.6]: Area $A_2$ is that of the rotor blades.*

*3) Maximum possible power extraction: The Betz limit:* In Fig. 8, the wind flow before, through and after a wind turbine with a turbine radius $r_t$ (in m) is shown. The incoming wind has a velocity of

$$v_1 = v_w \qquad (\text{in } \frac{\text{m}}{\text{s}}) \tag{6}$$

through the first cross section $A_1$ (in $\text{m}^2$) and through the wind channel until reaching the rotor blades, which define the cross section

$$A_2 = \pi r_t^2 \qquad (\text{in } \text{m}^2) \tag{7}$$

with a reduced wind velocity of $v_2$ (in $\frac{\text{m}}{\text{s}}$). As the wind exits the channel at $A_3$ (in $\text{m}^2$), the wind speed is $v_3$ (in $\frac{\text{m}}{\text{s}}$). Because of conservation of mass, a constant mass flow must exist in the flow channel, i.e.,

$$\forall t \geq 0\,\text{s}: \ \tfrac{\text{d}}{\text{d}t} m(t) \overset{(4)}{=} \varrho A_1 v_1 \overset{(6)}{=} \varrho A_1 v_w = \varrho A_2 v_2 \overset{(7)}{=} \varrho \pi r_t^2 v_2 = \varrho A_3 v_3 \geq 0 \, \frac{\text{kg}}{\text{s}}. \tag{8}$$

Therefore,

$$v_1 = v_w > v_2 > v_3 \quad \Longrightarrow \quad A_1 < A_2 = \pi r_t^2 < A_3.$$

From the decrease of wind velocity at the rotor blades (7), a *portion* of the wind power $p_w$ is extracted. This portion can be used as electrical energy. Not all of the wind power can be withdrawn through the turbine as this would require the wind to be decelerated to zero after the turbine at area $A_3$, which would lead to a piling up of air behind the turbine. The turbine extracts energy at $A_2$ which can be written as the energy difference

$$E_t(t) = \frac{1}{2} \underbrace{\varrho V_1(t)}_{=m_1(t)} v_1^2 - \frac{1}{2} \underbrace{\varrho V_3(t)}_{=m_3(t)} v_3^2 \tag{9}$$

where $V_1(t)$ (in $\text{m}^3$), $m_1(t)$ (in kg), $V_3(t)$ (in $\text{m}^3$) and $m_3(t)$ (in kg) correspond to the air volume and the air masses in section $A_1$ and $A_3$, respectively. By taking the time derivative, the turbine power can be computed as follows

$$\forall t \geq 0\,\text{s}: \quad p_t(t) := \tfrac{\text{d}}{\text{d}t} E_t(t) \overset{(9),(A.1)}{=} \frac{1}{2} \left( \tfrac{\text{d}}{\text{d}t} m_1(t)\, v_1^2 - \tfrac{\text{d}}{\text{d}t} m_3(t)\, v_3^2 \right)$$

$$\stackrel{(8)}{=} \frac{1}{2}\frac{\mathrm{d}}{\mathrm{d}t}m(t)\left(v_1^2 - v_3^2\right), \tag{10}$$

where the mass flow rate through the area $A_2$ is given by

$$\frac{\mathrm{d}}{\mathrm{d}t}m(t) \stackrel{(8)}{=} \varrho\, A_2\, v_2 = \varrho\, \pi r_\mathrm{t}^2\, v_2. \tag{11}$$

It has been shown already in 1925 by Albert Betz (see [18]), that the velocity $v_2$ through the turbine plane $A_2$ can be approximated by using the (arithmetic) average of the velocities $v_1 = v_\mathrm{w}$ and $v_3$, i.e.,

$$v_2 = \frac{(v_1 + v_3)}{2} \tag{12}$$

Later, Betz findings were proven in the *Froude-Rankineschem Theorem* [9, pp. 185–186]. Substituting (11) and (12) into (10), the following turbine power equation can be written as

$$\begin{aligned} p_\mathrm{t} &\stackrel{(10)}{=} \frac{1}{2}\frac{\mathrm{d}}{\mathrm{d}t}m\,\left(v_1^2 - v_3^2\right) \stackrel{(11)}{=} \frac{1}{2}\,\varrho\, A_2\, v_2\,\left(v_1^2 - v_3^2\right) \\ &\stackrel{(12)}{=} \frac{1}{2}\,\varrho\, A_2\,\frac{(v_1+v_3)}{2}\,\left(v_1^2 - v_3^2\right) = \frac{1}{2}\,\varrho\, A_2\, v_1^3\left[\frac{1}{2}\,\frac{(v_1+v_3)\,(v_1^2 - v_3^2)}{v_1^3}\right]. \end{aligned}$$

For $v_1 = v_\mathrm{w}$ and $A_2 = \pi\, r_\mathrm{t}^2$, the turbine power equation can be rewritten as follows

$$p_\mathrm{t} = \underbrace{\frac{1}{2}\,\varrho\,\pi\, r_\mathrm{t}^2\, v_\mathrm{w}^3}_{=p_\mathrm{w},\ \text{see (5)}} \underbrace{\left[\frac{1}{2}\left(1 + \frac{v_3}{v_\mathrm{w}}\right)\left(1 - \frac{v_3^2}{v_\mathrm{w}^2}\right)\right]}_{:=c_\mathrm{p}(v_\mathrm{w}, v_3)}. \tag{13}$$

The turbine power $p_\mathrm{t}$ therefore is related to the wind power $p_\mathrm{w}$ and the ratio $\frac{v_3}{v_\mathrm{w}}$. The variant factor $c_\mathrm{p}$ is known as the *power coefficient* and represents a measure of the extractable wind power. For the fraction $\frac{v_3}{v_\mathrm{w}}$, an optimum value can be found which maximizes the turbine power output. From this, the power coefficient is deduced and is given by

$$c_\mathrm{p}(v_\mathrm{w}, v_3) \stackrel{(13)}{=} \frac{1}{2}\left(-\left(\frac{v_3}{v_\mathrm{w}}\right)^3 - \left(\frac{v_3}{v_\mathrm{w}}\right)^2 + \frac{v_3}{v_\mathrm{w}} + 1\right), \tag{14}$$

taking the derivative of this equation with respect to the fraction $\frac{v_3}{v_\mathrm{w}}$ and setting it to zero yields

$$\frac{\mathrm{d}c_\mathrm{p}(v_\mathrm{w}, v_3)}{\mathrm{d}(\frac{v_3}{v_\mathrm{w}})} \stackrel{(14)}{=} \frac{1}{2}\left(-3\left(\frac{v_3}{v_\mathrm{w}}\right)^2 - 2\left(\frac{v_3}{v_\mathrm{w}}\right) + 1\right) \stackrel{!}{=} 0. \tag{15}$$

The resulting quadratic equation has two solutions

$$\boxed{\frac{v_3}{v_\mathrm{w}} = \frac{1}{3}} \quad \text{and} \quad \frac{v_3}{v_\mathrm{w}} = -1. \tag{16}$$

The ratio $\frac{v_3}{v_\mathrm{w}} = -1$ is not considered a meaningful solution of the quadratic equation (15) because it would mean the wind direction in plane $A_3$ would have the opposite sign for $v_3 = -v_\mathrm{w}$. Therefore the solution $\frac{v_3}{v_\mathrm{w}} = \frac{1}{3}$ represents the optimum ratio of incoming and outflowing wind speeds for the maximum turbine power. The wind velocity $v_\mathrm{w}$ in area $A_1$ must then be decelerated through the turbine at area $A_2$ as follows

$$v_2 \stackrel{(16),(12)}{=} \frac{1 + \frac{1}{3}}{2} v_\mathrm{w} = \frac{2}{3} v_\mathrm{w}$$

in order to withdraw the most power from the turbine. Substituting $v_3 = \frac{1}{3}v_\mathrm{w}$ into (14), the maximum power coefficient is

$$\boxed{\text{Betz limit:} \qquad c_\mathrm{p,Betz} := c_\mathrm{p}(v_\mathrm{w}, \tfrac{1}{3}v_\mathrm{w}) \stackrel{(14)}{=} \frac{16}{27} \approx 0.59.} \tag{17}$$

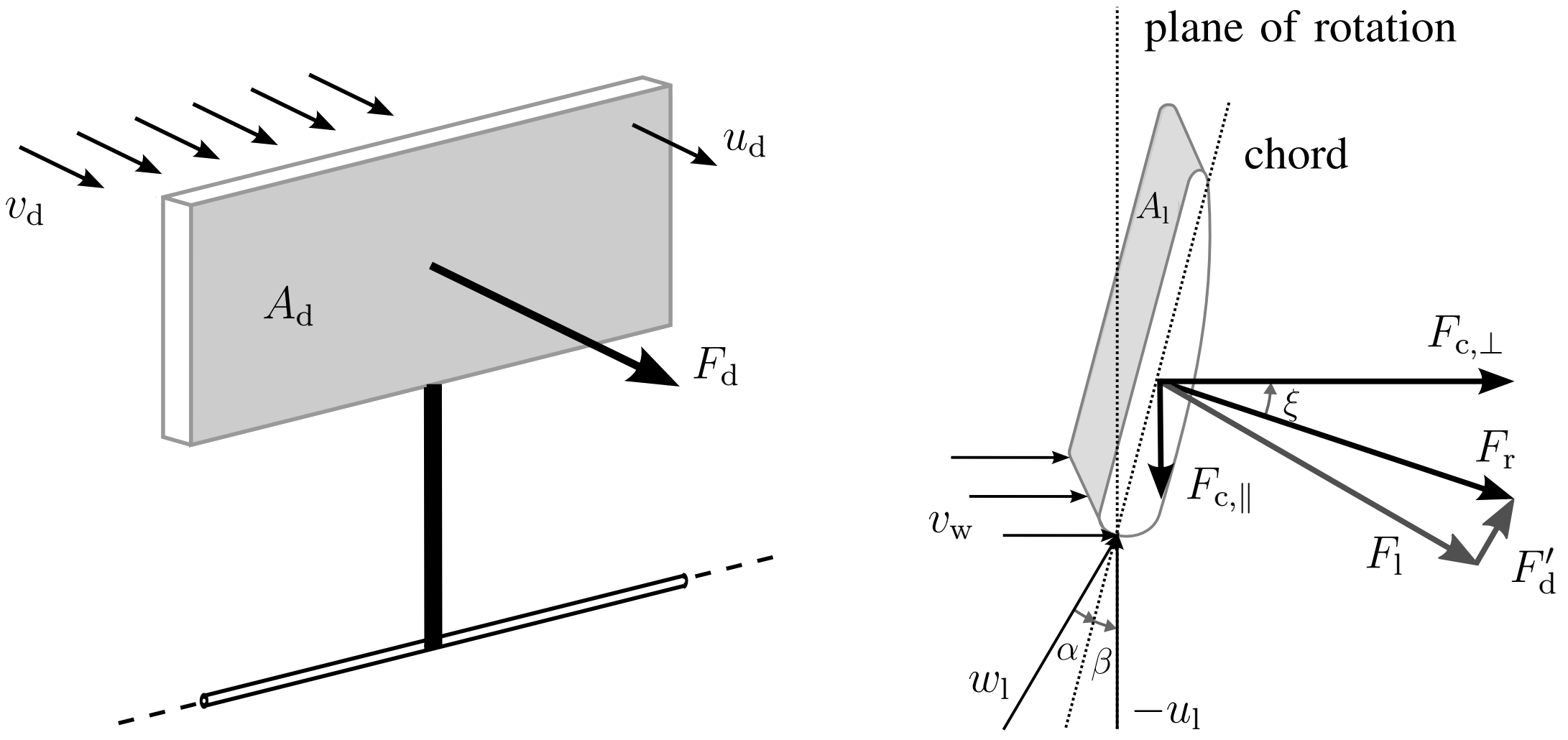

(a) *drag rotor* (b) *lift rotor*

Fig. 9: *Functional principle and power generation for drag and lift rotors [17, Bild 6.10]).*

This general derivation shows that, regardless of which turbine technology is used, no more than $16/27$ of the wind power can be converted to turbine power. Therefore, the following holds

$$p_\mathrm{t} = c_\mathrm{p}(v_\mathrm{w}, v_3)\, p_\mathrm{w} \overset{(5)}{=} c_\mathrm{p}(v_\mathrm{w}, v_3)\frac{1}{2}\varrho \overbrace{\pi r_\mathrm{t}^2}^{=A=A_\mathrm{t}=A_2} v_\mathrm{w}^3 \leq c_\mathrm{p,Betz}\, p_\mathrm{w} \overset{(17)}{=} \frac{16}{27}\, p_\mathrm{w}. \tag{18}$$

**Remark I.2.** *As can be seen from* (18) *there are* three *degrees of freedom to increase the turbine output power:*

- *Increase the rotor radius* $r_\mathrm{t}$*, since* $p_\mathrm{t} \propto r_\mathrm{t}^2$*;*
- *Use of higher wind speeds at higher altitudes* $v_\mathrm{w}$ *(by, for example, higher towers) or in areas with strong winds (e.g. offshore), since* $p_\mathrm{t} \propto v_\mathrm{w}^3$*;*
- *Optimize the rotor blade designs such that* $c_\mathrm{p} \to c_\mathrm{p,Betz}$*, since* $p_\mathrm{t} \propto c_\mathrm{p}$.

*4) Comparison between lift and drag turbines (rotors):* In the previous section, regardless of the turbine type, the optimal deceleration rate and the physical Betz limit $c_\mathrm{p,Betz} = \frac{16}{27}$ were derived. The total amount of energy which can be converted into electrical energy depends on the wind turbine type and its operation principle.

There are two physical principles that allow for the mechanical harvesting of "wind power": The *drag principle* and the *lift principle*. These principles result in a *drag force* $F_\mathrm{d}$ (in N) or a *lift force* $F_\mathrm{l}$ (in N). Today, only *lift -based* turbines are used. These lift-based rotors allow for a significantly higher energy yield than drag-based rotors. This assertion can be proved by a simplified view of the resulting forces on each of the two rotor types.

**Operating principle of the drag rotor (see [9, p. 46]):** In Fig. 9a, the physical operation principle or force diagram is shown for a drag rotor. The wind velocity vector $v_\mathrm{d}$ (in $\frac{\mathrm{m}}{\mathrm{s}}$) is aligned perpendicularly to the plate with surface $A_\mathrm{d}$ (in $\mathrm{m}^2$), which itself rotates with a rotational speed $u_\mathrm{d}$ (in $\frac{\mathrm{m}}{\mathrm{s}}$). The pressure generated by the wind on the plate leads to the drag force $F_\mathrm{d}$. This force drives the rotor and is calculated as follows

$$\left.\begin{aligned} F_\mathrm{d} &= \tfrac{1}{2}\varrho\, c_\mathrm{d}\, A_\mathrm{d}\,(\underbrace{v_\mathrm{d} - u_\mathrm{d}}_{=:w_\mathrm{d}})^2 = \tfrac{1}{2}\varrho\, c_\mathrm{d}\, A_\mathrm{d}\, v_\mathrm{d}^2\Big(1 - \underbrace{\frac{u_\mathrm{d}}{v_\mathrm{d}}}_{=:\lambda_\mathrm{d}}\Big)^2 \\ \text{where}\quad & 0\,\tfrac{\mathrm{m}}{\mathrm{s}} \leq u_\mathrm{d} \leq v_\mathrm{d} \quad\Longrightarrow\quad 0 \leq \lambda_\mathrm{d} \leq 1 \text{ and } 0\,\tfrac{\mathrm{m}}{\mathrm{s}} \leq w_\mathrm{d} \leq v_\mathrm{d}. \end{aligned}\right\} \tag{19}$$

Here, the used quantities are air density $\varrho$ (in $\frac{\mathrm{kg}}{\mathrm{m}^3}$), tip speed ratio $\lambda_\mathrm{d} = \frac{u_\mathrm{d}}{v_\mathrm{d}}$, drag coefficient $c_\mathrm{d}$ and inflow velocity $w_\mathrm{d} = v_\mathrm{d} - u_\mathrm{d}$ (in $\frac{\mathrm{m}}{\mathrm{s}}$) of the rotor. In drag rotors, the rotational speed can be, at most, as large as the incoming wind speed, i.e. $u_\mathrm{d} \leq v_\mathrm{d}$. This means that, tip speed ratio and flow velocity are upper bounded with $\lambda_\mathrm{d} \leq 1$ and $0\,\frac{\mathrm{m}}{\mathrm{s}} \leq w_\mathrm{d} \leq v_\mathrm{d}$.

**Operating principle of lift rotors (see [17, pp. 251–254]):** Figure 9b shows the functional principle and the force distribution of the lift rotor. The wind velocity $v_\mathrm{w}$ strikes the blade with area $A_\mathrm{l}$ (in $\mathrm{m}^2$). The blade is then rotated with rotational velocity of $u_\mathrm{l}$ (in $\frac{\mathrm{m}}{\mathrm{s}}$) (seen below in Fig. 9b). The wind flowing over the flat side of the blade (left in Fig. 9b) has a shorter distance to travel than on the curved opposite side (right in Fig. 9b). The different flow velocities lead to a higher pressure on the upper side (slower flow) and low pressure on the underside (faster flow). The lift force $F_\mathrm{l}$ (in N) on the rotor, resulting from these different air pressures above and below the rotor, acts perpendicularly to the inflow speed $w_\mathrm{l}$ (in $\frac{\mathrm{m}}{\mathrm{s}}$). The lift force can be calculated as follows (see [9, p. 46]):

$$\left.\begin{aligned} &F_\mathrm{l} = \tfrac{1}{2}\varrho\, c_\mathrm{l}\, A_\mathrm{l}\, (\underbrace{v_\mathrm{w}^2 + u_\mathrm{l}^2}_{=:w_\mathrm{l}^2}) = \tfrac{1}{2}\varrho\, c_\mathrm{l}\, A_\mathrm{l}\, v_\mathrm{d}^2\Bigg(1 + \Big(\underbrace{\frac{u_\mathrm{l}}{v_\mathrm{w}}}_{=:\lambda_\mathrm{l}}\Big)^2\Bigg) \\ &\text{where} \quad u_\mathrm{l} \geq 0\,\tfrac{\mathrm{m}}{\mathrm{s}} \quad \Longrightarrow \quad w_\mathrm{l} \geq v_\mathrm{w} \quad \text{and} \quad \lambda_\mathrm{l} \geq 0, \end{aligned}\right\} \tag{20}$$

with tip speed ratio $\lambda_\mathrm{l} = \frac{u_\mathrm{l}}{v_\mathrm{w}}$ and lift coefficient $c_\mathrm{l}$ of the rotor. The rotational velocity $u_\mathrm{l}$ of the rotor can be substantially greater than the wind speed $v_\mathrm{w}$, with tip speed ratios as high as $\lambda_\mathrm{l} = 15$ (see [9, p. 46]).

The direction of the inflow velocity $w_\mathrm{l}$ changes according to the angle of attack $\alpha$ (in °) formed by the chord of the rotor blade. The angle of attack $\alpha$ has a significant impact on the lift coefficient $c_\mathrm{l}$ and therefore the lift force $F_\mathrm{l}$ (see [9, p. 43]). In addition to $F_\mathrm{l}$ is an additional rotor drag force $F'_\mathrm{d}$ (in N) which acts parallel to the wind inflow direction $w_\mathrm{l}$. The drag force $F'_\mathrm{d}$ also depends on the angle of attack. For $\alpha < 15°$, the drag force $F'_\mathrm{d}$ is negligible. The ratio $\frac{F_\mathrm{l}}{F'_\mathrm{d}}$, with small angles of attack, can be as large as 400 (see [17, pp. 251–254]). For $\alpha \geq 15°$, the impact of $F'_\mathrm{d}$ with respect to $F_\mathrm{l}$ rises strongly (see [9, p. 43]).

In modern lift-based wind turbine systems, the blades rotate about their own axis. This pitch system of the wind turbine can align the chord of the blade with the rotation axis of the turbine by adjusting the pitch angle $\beta$ (in °) (see Fig. 9b). The pitch angle $\beta$ allows to vary the angle of attack $\alpha$ and is therefore directly related to the power generated from the rotor blades.

The resultant force $F_\mathrm{r} = F_\mathrm{l} + F'_\mathrm{d}$ (in N) can be divided into a parallel component $F_{\mathrm{c},\|} = F_\mathrm{r} \sin(\xi)$ (in N) and a perpendicular component $F_{\mathrm{c},\perp} = F_\mathrm{r} \cos(\xi)$ (in N). $F_{\mathrm{c},\|}$ is the torque-generating component. $F_{\mathrm{c},\perp}$ is the thrust component and does not contribute to the torque. Figure 9b shows, that drag force $F'_\mathrm{d}$ acts as a braking force on the wind turbine. For high torque outputs $F'_\mathrm{d}$ (i.e. large $F_{\mathrm{c},\|}$) must be as small as possible. Which is – as described above – the case for $\alpha \leq 15\,°$ such that $F_\mathrm{r} \approx F_\mathrm{l}$.

By comparing the two topologies of drag and lift rotors, one can see that, from (19) and (20), $w_\mathrm{l} \geq v_\mathrm{w} \geq w_\mathrm{d} \geq 0\,\frac{\mathrm{m}}{\mathrm{s}}$ and therefore the lift rotor can have (significantly) higher wind inflow velocities than the drag rotor. According to (19) and (20), the drag and lift forces are dependent on the square of the inflow wind velocity, and as such, the difference between the maximum power coefficients $c_{\mathrm{d,max}}$ and $c_{\mathrm{l,max}}$ is low (see [9, pp. 45]). Thus, with approximately the same size of the wind plane area $A_\mathrm{d} \approx A_\mathrm{l}$, the lift force $F_\mathrm{l}$ (and also the torque generating force $F_{\mathrm{c},\|}$) is much greater than the drag force $F_\mathrm{d}$. Concluding, with lift rotors a significantly higher share of the wind energy can be extracted – over 50 % compared to the drag rotor with a maximum of 16 % (see [9, p. 45]).

*5) Core components:* A wind turbine is a complex mechatronic system and consists of a variety of components. For modern wind turbine systems with more than 1 MW output power, the following components can be identified:

- Turbine (lift-based rotor with a horizontal axis and three pitch-controlled rotor blades),
- Gear transmission (can depend on the generator topology),

- Electrical machine (generator, e.g. permanent-magnet synchronous generator or doubly-fed induction generator),
- Back-to-back converter (machine-side and grid-side converter sharing a common DC-link),
- Mains filter (filter to create sinusoidal currents),
- Transformer (to step-up to higher voltage levels, e.g. medium voltage from $1\,\mathrm{kV}$ to $36\,\mathrm{kV}$, see [19, p. 8]) and
- Power grid (modeled as ideal voltage source and power sink with fixed grid frequency $f_{\mathrm{g}} = 50\,\mathrm{Hz}$).

These core components are shown in Fig. 10 and will be further discussed in the indicated sections in the figure. The dashed lines in Fig. 10 indicate the necessary additional wiring/interconnections required for the use of a doubly-fed asynchronous generator. This chapter will focus on the modeling and control of the electrical (sub-)systems. Modeling of the aerodynamics of the rotor is done by the use of an approximation with the so-called power coefficient (or power factor).

*6) Control system, operation management and operation regimes:* The operation of wind turbine systems can be classified according to *four operation regions* depending on the wind speed (see [9, Ch. 12] or [17, Sec. 6.4.2.3]).These four operation regimes are illustrated in Fig. 11[7]. Depending on the wind speed, the operation management system will ensure corresponding operation within that region by passing a reference signal to the control system (see Fig. 10). The control system has three quantities (control variables) it can affect directly for controlling the overall wind turbine system. These variables are the reference pitch angle $\beta_{\mathrm{ref}}$ (in $^\circ$) (neglecting the sub-level position control system) and the machine-side $\boldsymbol{s}_{\mathrm{m}}^{abc}$ and the grid-side $\boldsymbol{s}_{\mathrm{g}}^{abc}$ switching vectors for controlling the back-to-back converter.

The operation status is determined by measuring the wind velocity $v_{\mathrm{w}}$ (in $\frac{\mathrm{m}}{\mathrm{s}}$) (with low accuracy; thus, not usable for speed controller, see Section IV-C), the pitch angle $\beta$ (in $^\circ$), the turbine angular velocity $\omega_{\mathrm{t}}$ (in $\frac{\mathrm{rad}}{\mathrm{s}}$), the machine angular velocity $\omega_{\mathrm{m}}$ (in $\frac{\mathrm{rad}}{\mathrm{s}}$), the machine-side (stator) phase currents $\boldsymbol{i}_{\mathrm{s}}^{abc}$ (in $\mathrm{A})^3$ and the grid-side (filter) phase currents $\boldsymbol{i}_{\mathrm{f}}^{abc}$ (in $\mathrm{A})^3$, the DC-link voltage $u_{\mathrm{dc}}$ (in V) and the (transformed) grid voltages $\boldsymbol{u}_{\mathrm{g}}^{abc}$ (in $\mathrm{V})^3$ at the Point of Common Coupling (PCC). At the PCC, the instantaneous (active) power $p_{\mathrm{pcc}}$ (in W) and the reactive power $Q_{\mathrm{pcc}}$ (in var) are fed into or exchanged with the grid, respectively. The four operation regimes of a wind turbine are:

- **Operation regime I**: The wind speed $v_{\mathrm{w}} \in [0,\, v_{\mathrm{cut-in}})$ is below the necessary (minimum) cut-in wind speed $v_{\mathrm{cut-in}}$ (in $\frac{\mathrm{m}}{\mathrm{s}}$) to ensure efficient operation. The turbine is at stand-still or coasting. Since the minimum active power $p_{\mathrm{pcc,min}} > 0$ (in W) can not be produced, the generator is not connect to the grid. Thus the active power $p_{\mathrm{pcc}}$ (in W) fed into the grid is zero. The speed control of the generator is inactive. According to the reference pitch angle $\beta_{\mathrm{ref}}$ from the operation management system, the turbine angular velocity $\omega_{\mathrm{t}}$ of the turbine is controlled (stand still or controlled coasting). The DC-link voltage controller is active (or can be active) to allow for set-point tracking of the voltage to the constant reference value $u_{\mathrm{dc,ref}} > 0$ (in V) such that, should the cut-in wind speed $v_{\mathrm{cut-in}}$ be exceeded, the operation management system is ready to move directly into operation regime II.
- **Operation regime II**: The wind velocity $v_{\mathrm{w}} \in [v_{\mathrm{cut-in}},\, v_{\mathrm{nom}})$ is between the cut-in wind speed $v_{\mathrm{cut-in}}$ and the rated wind speed $v_{\mathrm{nom}}$ (in $\frac{\mathrm{m}}{\mathrm{s}}$) of the turbine. The wind turbine system is then operated at variable speed. The turbine power $p_{\mathrm{t}}$ (in W) varies between zero and the turbine rated power. The goal is to operate at a maximum power point with an optimum tip speed ratio $\lambda^\star$, which is enabled via the speed controller according to the "Maximum Power Point Tracking (MPPT)". The pitch control system maintains the pitch angle at zero, i.e. $\beta = \beta_{\mathrm{ref}} = 0$ (in $^\circ$). Depending on the reactive power reverence signal $q_{\mathrm{pcc,ref}}$ (in var), the desired $q_{\mathrm{pcc}}$ is fed into the grid using the current reference $i_{\mathrm{f,ref}}^{q}$ (in A). The instantaneous power $p_{\mathrm{pcc}}$ fed into to grid (or active power $P_{\mathrm{pcc}}$ (in W)) depends on the turbine power $p_{\mathrm{t}}$ and varies between the minimum power $p_{\mathrm{pcc,min}}$ and the nominal power $p_{\mathrm{pcc,nom}}$ (in W), i.e. $p_{\mathrm{pcc,min}} \le p_{\mathrm{pcc}} < p_{\mathrm{pcc,nom}} < p_{\mathrm{t}}$ in *steady-state* (see Fig. 11).

[7] Other operation conditions (for example, due to faults in the wind turbine or the network) and start-up/shut-down routines are not considered. A more detailed description of the overall operation management can be found in Chapter 5.6 in [20].

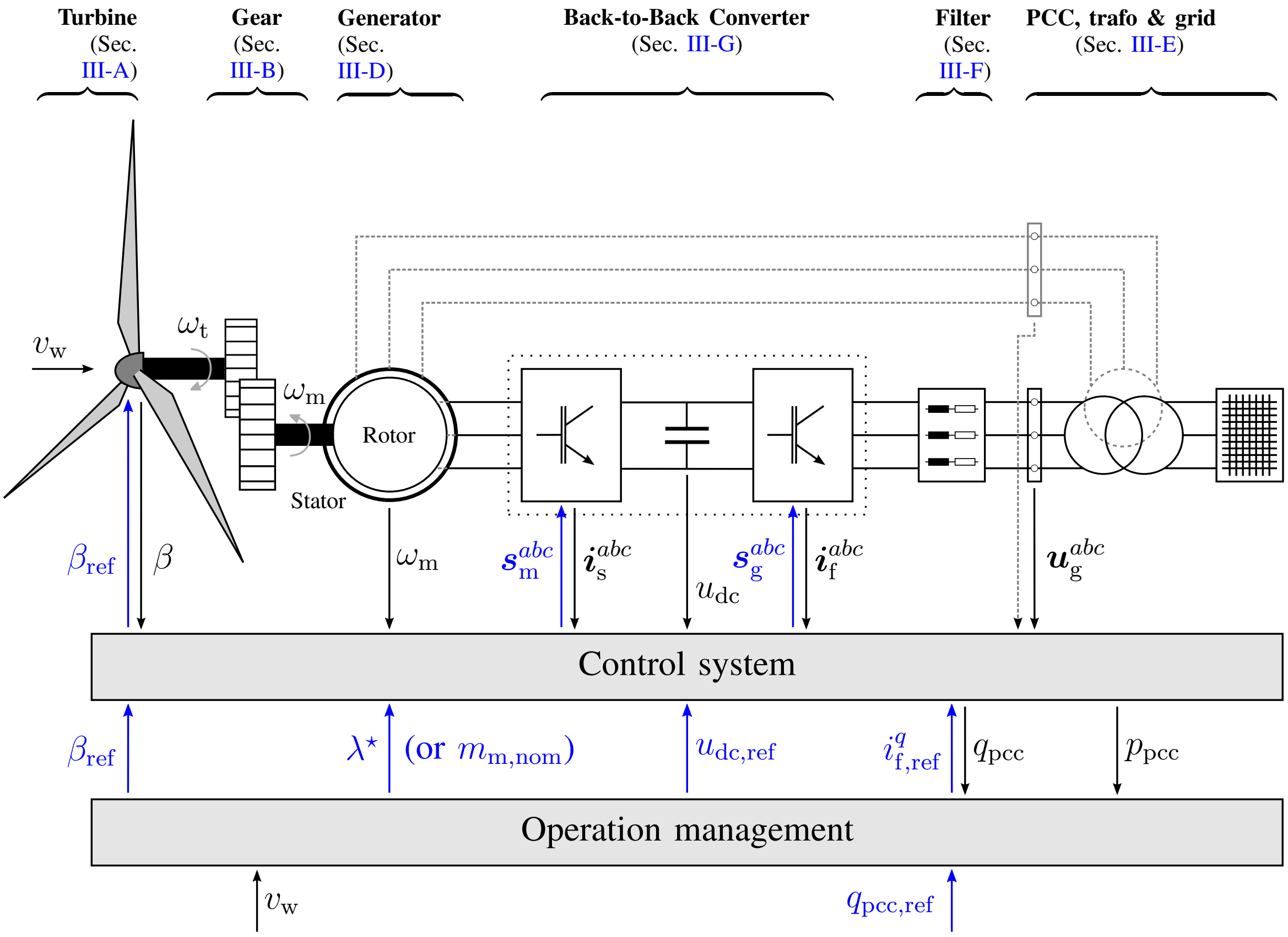

Fig. 10: *Core components, control and operation of a wind turbine with guidance, positioning and reference variables.*

- **Operation regime III**: The wind velocity $v_\mathrm{w} \in [v_\mathrm{nom}, v_\mathrm{cut-out})$ is between rated wind speed $v_\mathrm{nom}$ and cut-out wind speed $v_\mathrm{cut-out}$ (in $\frac{\mathrm{m}}{\mathrm{s}}$). The wind turbine is running at its rated power. The turbine power $p_\mathrm{t}$ is also operating at its rated value. The generator is producing its nominal torque $m_\mathrm{m,nom}$ (in $\mathrm{N\,m}$). The turbine angular velocity $\omega_\mathrm{t}$ is controlled via the pitch control system, to maintain steady power generation in the nominal range. Control of the reactive power is again maintained using the reference signal $i^q_\mathrm{f,ref}$ to maintain the target reactive power $q_\mathrm{pcc,ref}$. The new output power $p_\mathrm{pcc}$ corresponds to the nominal output power $p_\mathrm{pcc,nom}$, i.e. $p_\mathrm{pcc} = p_\mathrm{pcc,nom} < p_\mathrm{t}$ in *steady-state*.
- **Operation regime IV**: For wind speeds $v_\mathrm{w} > v_\mathrm{cut-out}$, a safe operation of the wind turbine can no longer be ensured because there is too much power in the wind (safety shutdown). The turbine remains active until the rotor is slowed to a complete stop (i.e. by brakes on the rotor blade tips or by turning the blades out of the wind). Turbine power and output power become zero, i.e. $p_\mathrm{t} = p_\mathrm{pcc} = 0\,\mathrm{W}$.

The particular values of the wind speeds $v_\mathrm{cut-in}$, $v_\mathrm{nom}$ and $v_\mathrm{cut-out}$ are system specific and may vary with the type and design of the wind turbine.

### E. Future challenges for wind turbines

The afore mentioned increase in the share of wind power in the electricity supply in Germany requires a strengthening of the grid code requirements for wind turbines [15]. Future wind turbines must be able to help to maintain a stable grid operation. For example, they must be capable to support and stabilize grid frequency, voltage and respond to power outages by adequate grid re-initialization procedures [15].

In order to comply with these grid code requirements, wind turbines of the future must have flexible operating options. This robust and flexible operation management must be ensured by a dynamic, and above all reliable control system. The control system of wind turbines is not (any more) just about the control of conventional drives. The complex coupling and interaction of various control systems necessitates a holistic approach when considering the system controller design.

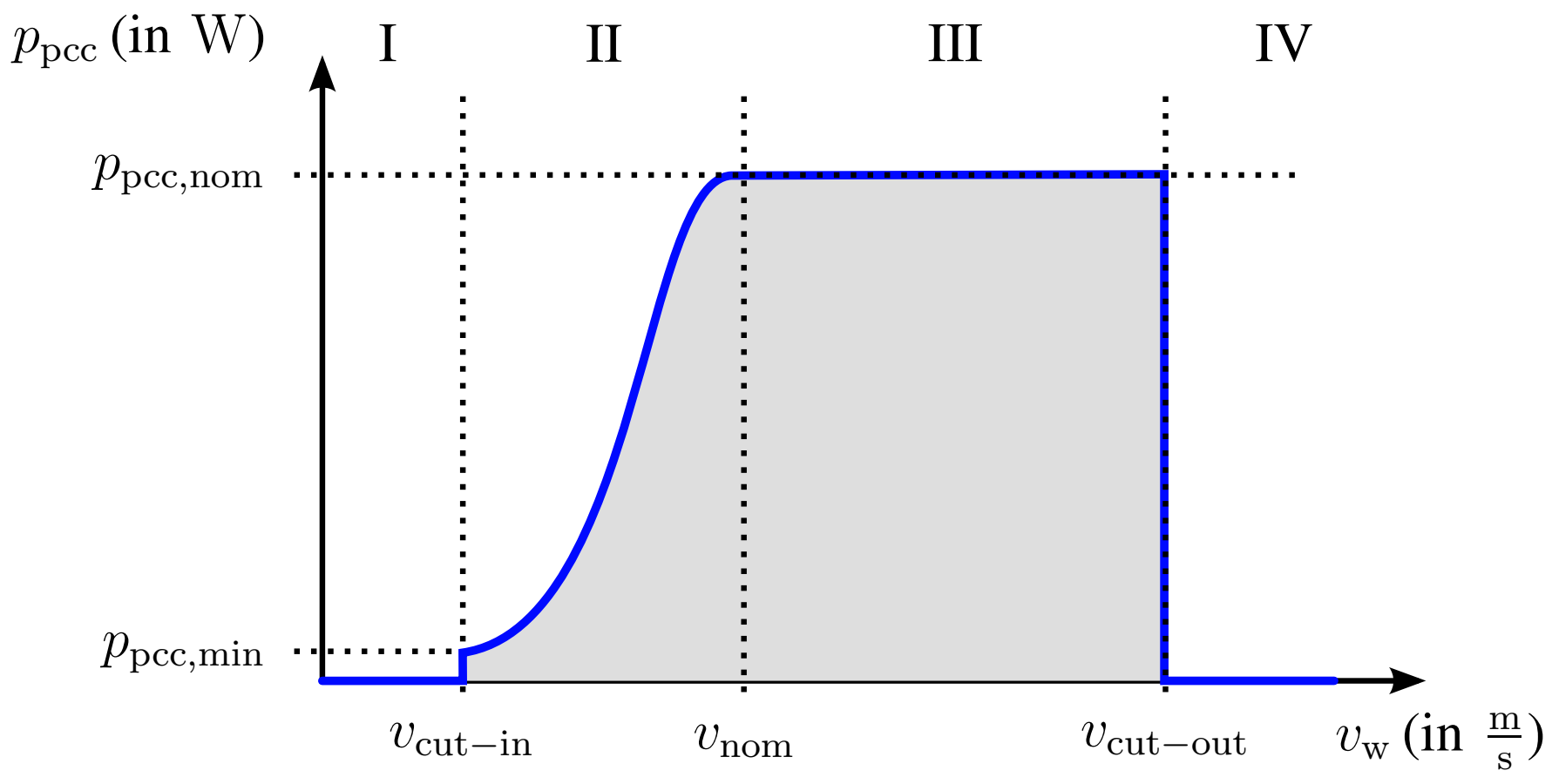

Fig. 11: *The four operation regimes of a wind turbine (Fig. based on [9, Abb. 12-2]).*

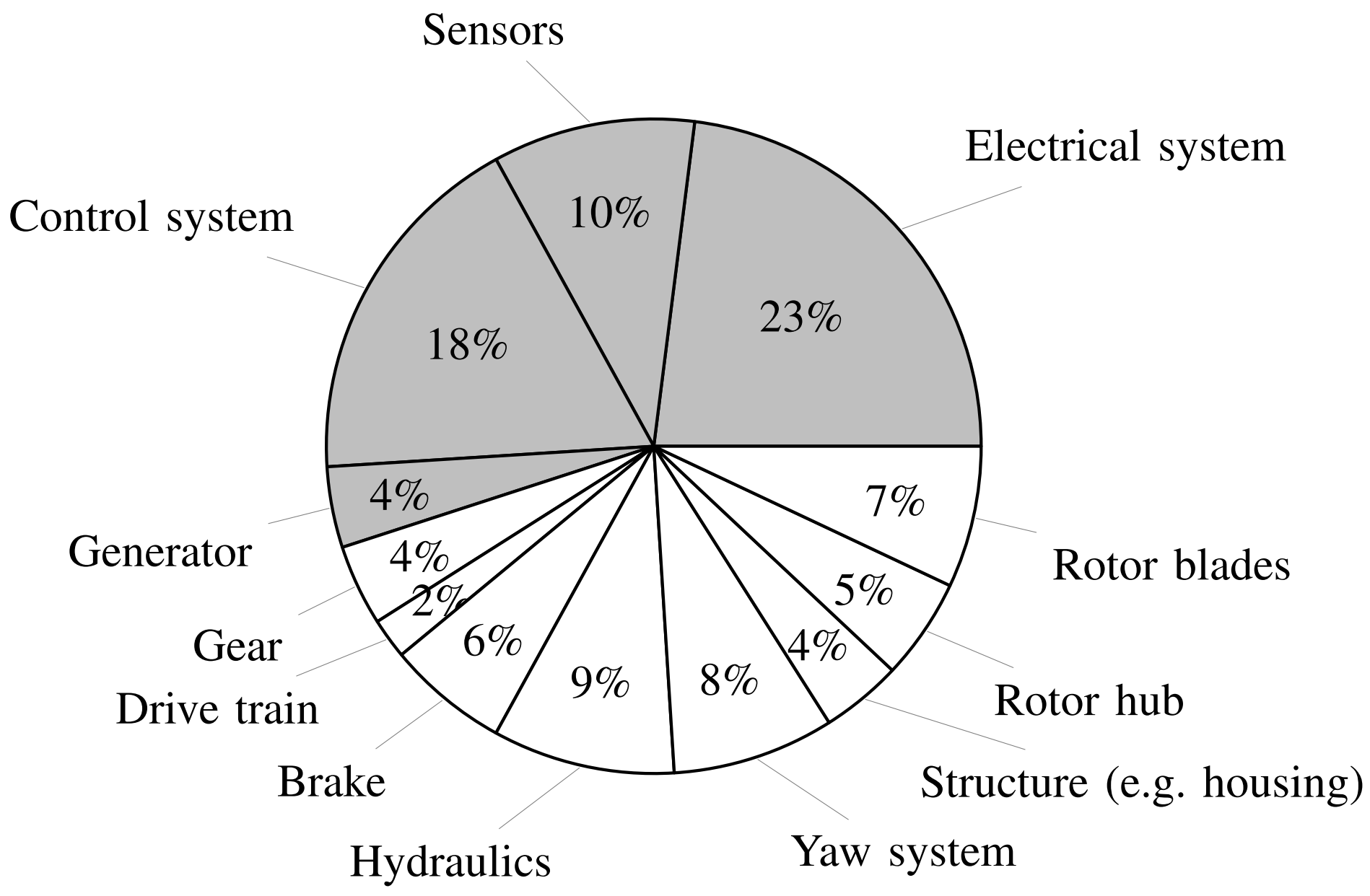

Fig. 12: *Percentage distribution of the failure factors of a wind turbine system according to different sub-components (based on Fig. 2 in [21]).*

The need for a special focus on the control of wind turbines is also supported by long-term studies on downtime duration and probability of a fault by individual system components. Modern wind turbines are designed for a service life of 20 years. With regular maintenance of onshore wind turbines, an average annual availability of 98%, with a total downtime of 7 days can be expected (see [21] and [22, Sec. 2.1]). In the pie chart in Fig. 12, the reasons for failures of a wind turbine system are broken down as a percentage of different subcomponents. Functional faults in the overall system are mainly due to faults in the electrical system (23%), the control system (18%) and the sensors (10%), whereas the failures in the mechanical components (i.e. transmission, rotor or housing) are less likely (see [21], [23] and references therein). In addition to the likelihood of a breakdown caused by a single system or component, the average downtime over the lifetime of the wind turbine is of particular interest. The failure of a single component usually results in a system shutdown and thus reduced power yield and production losses. Figure 13 shows a bar graph of the failure probability and downtime (in days) of individual system components per failure. If the failure probability and time for the average life span of 20 years are estimated, the following average total failure times of selected components are:

- Electrical systems: $0.57 \cdot 1.53 \, \frac{1}{\text{turbine year}} \cdot 20 \, \text{years} = \mathbf{17.4} \, \frac{\text{days}}{\text{turbine}}$,
- Control system: $0.43 \cdot 1.59 \, \frac{1}{\text{turbine year}} \cdot 20 \, \text{years} = \mathbf{13.7} \, \frac{\text{days}}{\text{turbine}}$, and
- Gear: $0.1 \cdot 6.21 \, \frac{1}{\text{turbine year}} \cdot 20 \, \text{years} = \mathbf{12.4} \, \frac{\text{days}}{\text{turbine}}$.

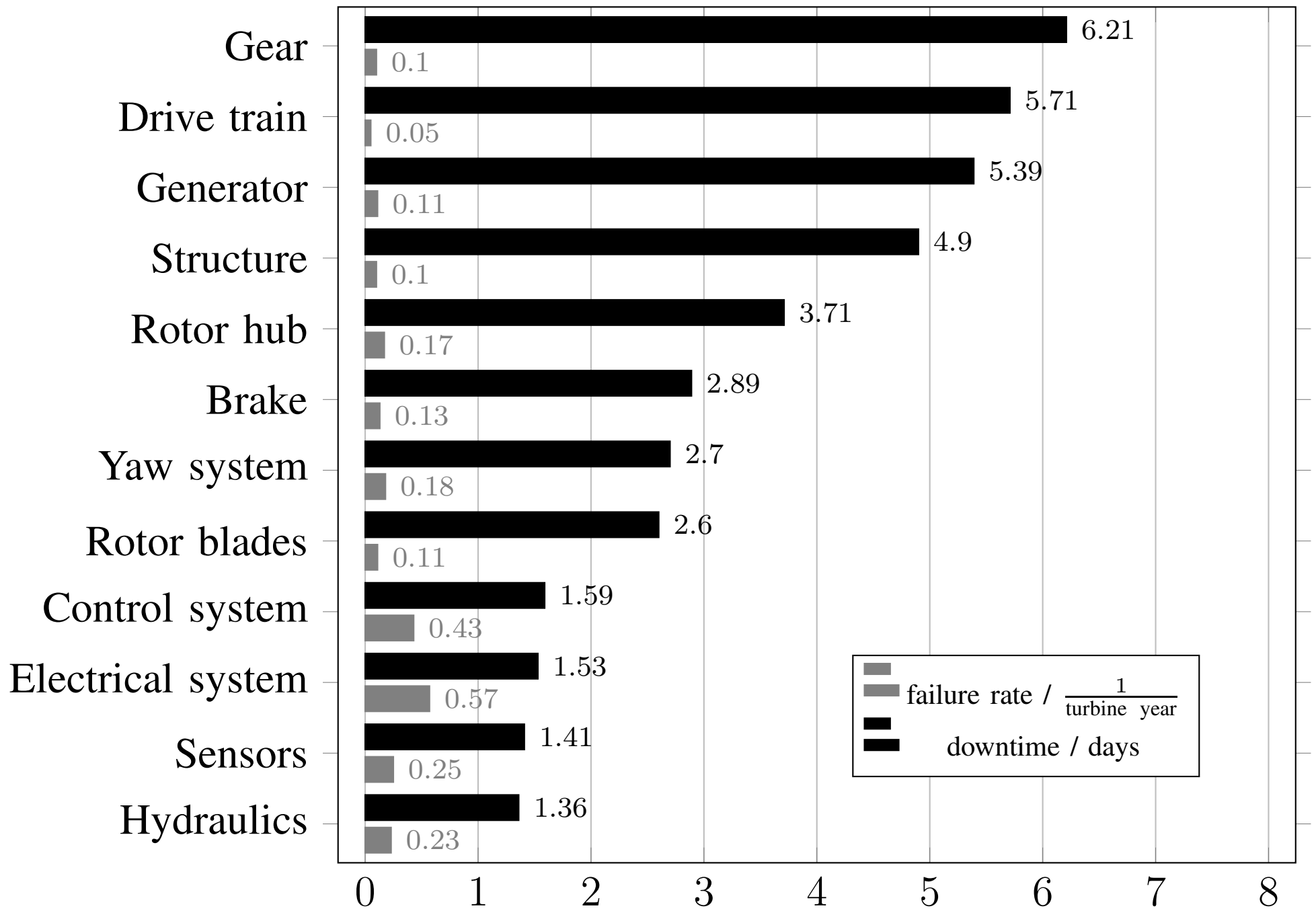

Fig. 13: *Average failure rate and average downtime per failure of several different components (based on Fig. 2 in [24]).*

As can be seen by the items above, special attention should be paid to the electrical system (i.e generator and power electronics) and control system. In particular, for offshore wind turbine systems, an increase in availability and reliability is essential for economic viability (see [23]; study is based on data of 11 years from Denmark and Germany).

## II. Nomenclature and preliminaries

In this section, the utilized nomenclature is introduced and the preliminaries for electrical three-phase systems are laid out. In contrast to the usually employed complex representation/notation, in this chapter, the space vectors are introduced in *vector-/matrix* notation and the calculation of *instantaneous* active, reactive and apparent power is done according to the "Instantaneous Power Theory" (see [25, App. B]).

### A. Electrical three-phase systems

Vector variables are used to describe three-phase electrical systems. For this reason, the following (signal) vector

$$\boldsymbol{x}^{abc} \colon \mathbb{R}_{>0} \to \mathbb{R}^3, \quad t \mapsto \boldsymbol{x}^{abc}(t) := \begin{pmatrix} x^a(t) \\ x^b(t) \\ x^c(t) \end{pmatrix} := \begin{pmatrix} \hat{x}^a(t) \, \cos(\phi_x^a(t)) \\ \hat{x}^b(t) \, \cos(\phi_x^b(t)) \\ \hat{x}^c(t) \, \cos(\phi_x^c(t)) \end{pmatrix}$$

with sinusoidal components is defined. The vector is time dependent and exists for all $t \geq 0$ (in s). The individual phase quantities $x^a(t)$, $x^b(t)$, $x^c(t)$ have amplitudes $\hat{x}^a(t)$, $\hat{x}^b(t)$, $\hat{x}^c(t) > 0$ and phase angles $\phi_x^a(t), \phi_x^b(t), \phi_x^c(t) \in \mathbb{R}$ (in rad), respectively. The variable $\boldsymbol{x}$ (or $x$) corresponds to the phase current vector $\boldsymbol{i}$ (in A)$^3$ (or phase current $i$ (in A)), phase voltage vector $\boldsymbol{u}$ (in V)$^3$ (or phase voltage $u$ (in V)) or the flux linkage vector $\boldsymbol{\psi}$ (in V s)[3] (or, the flux linkage $\psi$ (in V s)), i.e., $\boldsymbol{x} \in \{\boldsymbol{i}, \boldsymbol{u}, \boldsymbol{\psi}\}$ (or $x \in \{i,\, u,\, \psi\}$). Thus, for example, the phase currents can be written compactly as

$$\boldsymbol{i}^{abc}(t) = \begin{pmatrix} i^a(t) \\ i^b(t) \\ i^c(t) \end{pmatrix} = \begin{pmatrix} \hat{\imath}^a(t)\, \cos(\phi_i^a(t)) \\ \hat{\imath}^b(t)\, \cos(\phi_i^b(t)) \\ \hat{\imath}^c(t)\, \cos(\phi_i^c(t)) \end{pmatrix}$$

for the phases $a, b, c$ at time $t \geq 0\,\text{s}$.

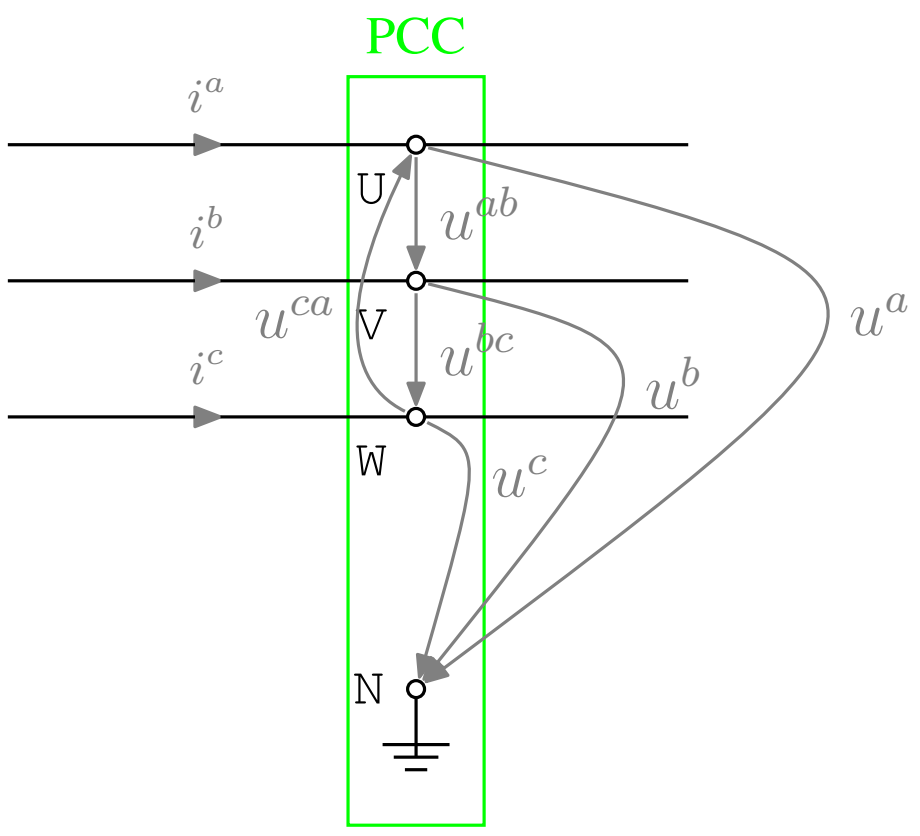

Fig. 14: *Signals at the Point of Common Coupling (PCC) with (neutral) reference point* N*, and terminals* U*,* V *and* W*: phase currents* $\boldsymbol{i}^{abc} = (i^a, i^b, i^c)^\top$*, phase voltages* $\boldsymbol{u}^{abc} = (u^a, u^b, u^c)^\top$ *and line-to-line voltages* $\boldsymbol{u}^{ltl} = (u^{ab}, u^{bc}, u^{ca})^\top$*.*

*1) Balanced (or symmetrical) three-phase systems:* In this chapter, mainly *balanced* (or *symmetrical* three-phase systems (alternating current systems) are considered, and as such, the following assumptions can be imposed on the phase currents and phase voltages:

**Assumption (A.3)** *All phases have the same amplitude for all time, i.e.*

$$\forall\, t \geq 0\,\text{s}: \qquad \hat{x}(t) := \hat{x}^a(t) = \hat{x}^b(t) = \hat{x}^c(t). \tag{21}$$

**Assumption (A.4)** *The phase angles are offset from one another by* $\frac{2}{3}\pi$ *for all time, i.e.*

$$\forall\, t \geq 0\,\text{s}: \qquad \phi_x(t) := \phi_x^a(t) = \phi_x^b(t) + \tfrac{2}{3}\pi = \phi_x^c(t) + \tfrac{4}{3}\pi. \tag{22}$$

Due to the trigonometric identity (see [26, p. 124])

$$\forall\, \alpha, \beta \in \mathbb{R}: \;\; \cos(\alpha) + \cos(\beta) = 2\cos\left(\tfrac{\alpha+\beta}{2}\right)\cos\left(\tfrac{\alpha-\beta}{2}\right) \tag{23}$$

and the symmetry properties (21) and (22), the following property can be derived: The sum of the balanced (symmetrical) three-phase quantities is zero, since

$$\begin{aligned} \forall\, t \geq 0\,\text{s}\, x^a(t) + x^b(t) + x^c(t) &= \\ &\overset{(21),(22)}{=} \hat{x}(t)\big[\cos\left(\phi_x(t)\right) + \cos\left(\phi_x(t) - \tfrac{2}{3}\pi\right) + \cos\left(\phi_x(t) - \tfrac{4}{3}\pi\right)\big] \\ &\overset{(23)}{=} \hat{x}(t)\big[2\cos\left(\phi_x(t) - \tfrac{1}{3}\pi\right)\underbrace{\cos\left(\tfrac{\pi}{3}\right)}_{=\frac{1}{2}} + \cos\left(\phi_x(t) - \tfrac{4}{3}\pi\right)\big] \\ &\overset{(23)}{=} \hat{x}(t)\big[\cos\left(\phi_x(t)\right)\underbrace{\cos\left(\tfrac{\pi}{2}\right)}_{=0}\big] = 0. \end{aligned} \tag{24}$$

**Remark II.1.** *For three-phase systems in star connection with freely floating star point, the relationship* $i^a(t) + i^b(t) + i^c(t) = 0\,\text{A}$ *can be directly derived from Kirchhoff's Law.*

*2) Relationship between phases and line-to-line quantities:* A general three-phase system is shown in Fig. 14. In such a three-phase system, if the neutral point $\mathtt{N}$ is not accessible, the *line-to-line* voltages are the only voltages which can directly be measured. Applying Kirchhoff's voltage law in Fig. 14 leads to the following line-to-line voltage vector

$$\forall t \geq 0\,\mathrm{s}: \qquad \boldsymbol{u}^{ltl}(t) := \begin{pmatrix} u^{ab}(t) \\ u^{bc}(t) \\ u^{ca}(t) \end{pmatrix} = \begin{pmatrix} u^{a}(t) - u^{b}(t) \\ u^{b}(t) - u^{c}(t) \\ u^{c}(t) - u^{a}(t) \end{pmatrix}. \tag{25}$$

Under certain assumptions, the line-to-line voltages $\boldsymbol{u}^{ltl}(t)$ can be expressed through the phase voltages

$$\boldsymbol{u}^{abc}(t) := \big(u^{a}(t), \quad u^{b}(t), \quad u^{c}(t)\big)^{\top}. \tag{26}$$

To show this, rewrite (25) in matrix notation as follows

$$\boldsymbol{u}^{ltl}(t) = \underbrace{\begin{bmatrix} 1 & -1 & 0 \\ 0 & 1 & -1 \\ -1 & 0 & 1 \end{bmatrix}}_{=:\, \boldsymbol{T}_{\mathrm{v}}^{\star}} \boldsymbol{u}^{abc}(t). \tag{27}$$

Note that the matrix $\boldsymbol{T}_{\mathrm{v}}^{\star}$ is singular[8]. Thus, $\boldsymbol{u}^{abc}(t)$ can *not* be solved. But for balanced (symmetrical) three-phase systems (i.e. (21) and (22) hold true) a linearly independent basis can be found: According to (24), for example $u^{a}(t) = -u^{b}(t) - u^{c}(t)$ holds true for all $t \geq 0$. Substituting this fact into the last row of $\boldsymbol{T}_{\mathrm{v}}^{\star}$ yields

$$u^{ca}(t) = \begin{pmatrix} -1 & 0 & 1 \end{pmatrix} \boldsymbol{u}^{abc}(t) = \begin{pmatrix} 0 & 1 & 2 \end{pmatrix} \boldsymbol{u}^{abc}(t) \tag{28}$$

and one obtains

$$\boldsymbol{u}^{ltl}(t) = \underbrace{\begin{bmatrix} 1 & -1 & 0 \\ 0 & 1 & -1 \\ 0 & 1 & 2 \end{bmatrix}}_{=:\, \boldsymbol{T}_{\mathrm{v}}} \boldsymbol{u}^{abc}(t) \quad \overset{\det(\boldsymbol{T}_{\mathrm{v}})=3}{\Longleftrightarrow} \quad \boldsymbol{u}^{abc}(t) = \underbrace{\frac{1}{3}\begin{bmatrix} 3 & 2 & 1 \\ 0 & 2 & 1 \\ 0 & -1 & 1 \end{bmatrix}}_{=\, \boldsymbol{T}_{\mathrm{v}}^{-1}} \boldsymbol{u}^{ltl}(t). \tag{29}$$

Clearly now, the altered matrix $\boldsymbol{T}_{\mathrm{v}} \in \mathbb{R}^{3\times 3}$ is invertible, and therefore, a direct relationship between the phase voltage vector $\boldsymbol{u}^{abc}(t)$ and the line-to-line voltage vector $\boldsymbol{u}^{ltl}(t)$ is found.

*3) Space vector representation in vector/matrix notation:* The use of space vectors to represent the three-phase systems is well established in practice (see [27, p. 288–296] or [28, Sec. 13.1]). As an alternative to the widespread *complex* representation of the space vector, they can be represented by vectors and adequate transformation and rotation matrices as described in this section. For this explanation, consider the sketch of the electrical machine in Fig. 15. The points $\mathtt{U}$, $\mathtt{V}$, and $\mathtt{W}$ represent the connection terminals. The machine consists of a stator and a rotor. Both are connected in star-configuration, therefore the stator phase windings $a, b, c$ and the rotor phase windings $a_r, b_r, c_r$ are connected with their respective star (or neutral) point. For simplicity, only a machine with a single pole pair is considered. Thus, the windings in the stator phases $a, b, c$ (or in the rotor phases $a_r, b_r, c_r$) are *spatially* offset from each other by $120^\circ = \frac{2}{3}\pi$. The phases $a, b, c$ (and $a_r, b_r, c_r$, resp.) are plotted in the right-hand side of Fig. 15 as coordinate axes of respective reference frames (coordinate systems). To plot a vector in the *coordinate plane* one does not need *three* axes. One orthogonal coordinate system (CoSy) or reference frame is sufficient. In Fig. 15, three coordinate systems are shown:

(i) The stator-fixed $s = (\alpha, \beta)$-reference frame[9] with the axes $\alpha, \beta$ and the vector $\boldsymbol{x}^s = (x^\alpha, x^\beta)^\top$;
(ii) The rotor-fixed $r = (d', q')$-reference frame[10] with the axis $d', q'$ and the vector $\boldsymbol{x}^r = (x^{d'}, x^{q'})^\top$; and

[8] The rows are linearly dependent. For example: Row 1 + Row 2 = − Row 3.
[9] Convention: The $\alpha$-axis of the $s$-coordinate system/reference frame lies on the $a$-axis of the 3-phase coordinate system $(a, b, c)$.
[10] Convention: The $d'$-axis of the $r$-coordinate system/reference frame lies on the $a_r$-axis of the 3-phase rotor-coordinate system $(a_r, b_r, c_r)$.

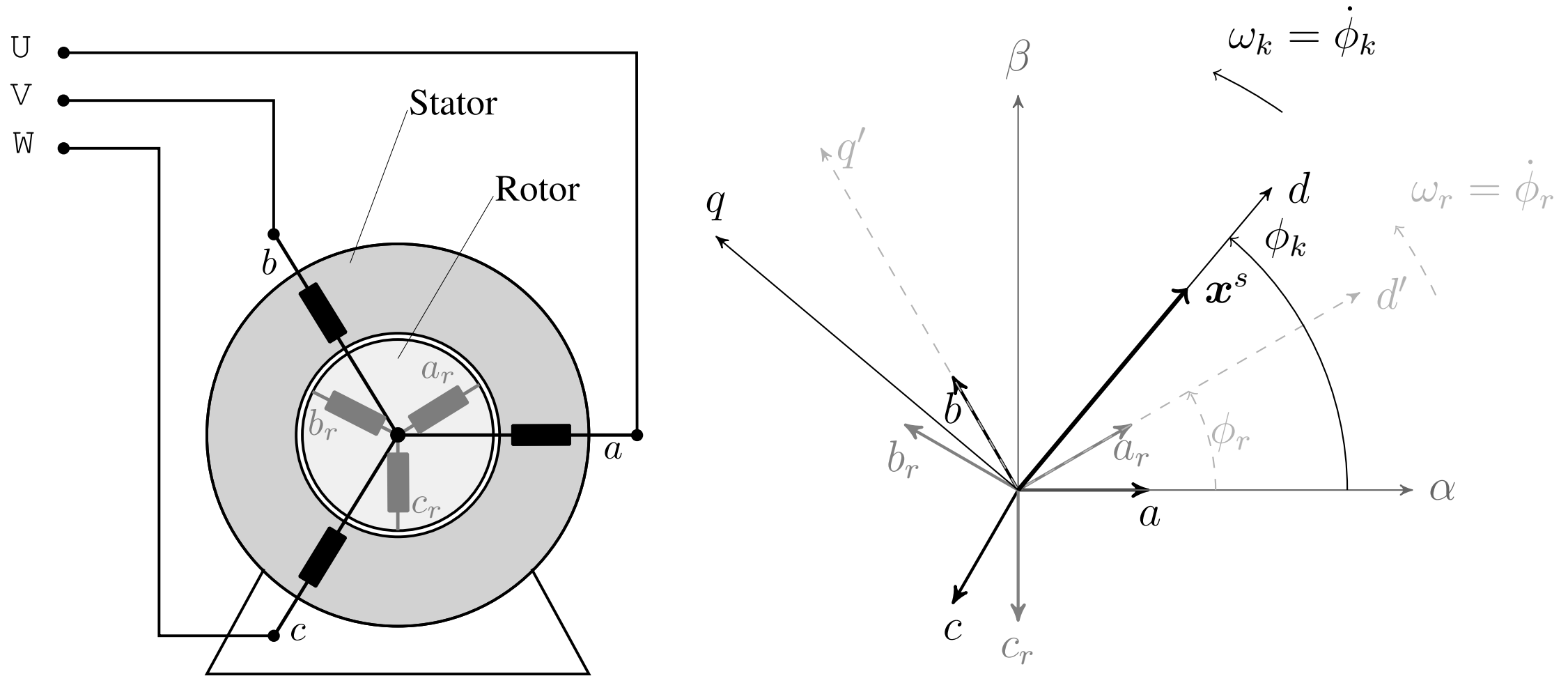

Fig. 15: *Space vector theory: A machine with the terminals* `U`*,* `V`*,* `W`*, stator windings a, b, c and rotor windings $a_r$, $b_r$, $c_r$ (left) and various coordinate systems (or reference frames, right): three-phase $(a, b, c)$-reference frame, three-phase (rotor) $(a_r, b_r, c_r)$-reference frame, stator-fixed $s = (\alpha, \beta)$-reference frame, rotor-fixed $r = (d', q')$-reference frame and (arbitrarily) rotating $k = (d, q)$-reference frame. Signal vectors $\boldsymbol{x}^s = (x^\alpha,\, x^\beta)^\top$ in the stator-fixed s-reference frame with length $\|\boldsymbol{x}^s\| = \sqrt{(x^\alpha)^2 + (x^\beta)^2}$.*

(iii) The (arbitrarily) rotating $k = (d, q)$-reference frame with the axes $d$, $q$ and vector $\boldsymbol{x}^k = (x^d, x^q)^\top$.

In the following section, the representation of an electrical quantity of a three-phase system as a (vectorized) space vector and the conversion between the individual reference frames will be discussed.

*Clarke transformation: $(a, b, c) \leftrightarrow (\alpha, \beta)$:* The signal vector $\boldsymbol{x}^{abc}(t) = (x^a(t),\, x^b(t), x^c(t))^\top \in \mathbb{R}^3$ of a three-phase system can be transformed to the stator-fixed $s$-reference frame by invoking the *Clarke transformation* and can be brought back again with the inverse Clarke transformation. The signal vector in the $s$-reference frame is denoted with

$$\boldsymbol{x}^s = (x^\alpha, x^\beta)^\top \in \mathbb{R}^2.$$

Next, the general relationships should be considered. In addition to the components $x^\alpha$ and $x^\beta$, the general Clarke transformation also takes into account a zero component $x^0$ (see [29]) and is given by

$$\begin{pmatrix} x^\alpha(t) \\ x^\beta(t) \\ x^0(t) \end{pmatrix} = \underbrace{\kappa \begin{bmatrix} 1 & -\frac{1}{2} & -\frac{1}{2} \\ 0 & \frac{\sqrt{3}}{2} & -\frac{\sqrt{3}}{2} \\ \frac{1}{\sqrt{2}} & \frac{1}{\sqrt{2}} & \frac{1}{\sqrt{2}} \end{bmatrix}}_{=:\boldsymbol{T}_\mathrm{c}^\star \in \mathbb{R}^{3\times 3}} \boldsymbol{x}^{abc}(t) \Leftrightarrow \boldsymbol{x}^{abc}(t) = \underbrace{\frac{1}{\kappa} \begin{bmatrix} \frac{2}{3} & 0 & \frac{\sqrt{2}}{3} \\ -\frac{1}{3} & \frac{1}{\sqrt{3}} & \frac{\sqrt{2}}{3} \\ -\frac{1}{3} & -\frac{1}{\sqrt{3}} & \frac{\sqrt{2}}{3} \end{bmatrix}}_{=(\boldsymbol{T}_\mathrm{c}^\star)^{-1}} \begin{pmatrix} x^\alpha(t) \\ x^\beta(t) \\ x^0(t) \end{pmatrix}. \tag{30}$$

Here, $\boldsymbol{T}_\mathrm{c}^\star$ and $(\boldsymbol{T}_\mathrm{c}^\star)^{-1}$ are the general Clarke and the general inverse Clarke transformation matrix, respectively, with coefficient $\kappa \in \{\frac{2}{3}, \sqrt{\frac{2}{3}}\}$. It then follows that $\boldsymbol{T}_\mathrm{c}^\star(\boldsymbol{T}_\mathrm{c}^\star)^{-1} = (\boldsymbol{T}_\mathrm{c}^\star)^{-1}\boldsymbol{T}_\mathrm{c}^\star = \boldsymbol{I}_3$[11]. For balanced or symmetric three-phase systems (see Assumption (A.3) and (A.4)), the transformation can be simplified. The zero component can be neglected, since

$$\forall\, t \geq 0: \qquad x^0(t) = \frac{\kappa}{\sqrt{2}}\big(x^a(t) + x^b(t) + x^c(t)\big) \overset{(24)}{=} 0.$$

[11]For $n \in \mathbb{N}$, $\boldsymbol{I}_n = \begin{bmatrix} 1 & & \\ & \ddots & \\ & & 1 \end{bmatrix} \in \mathbb{R}^{n\times n}$ is the $n$-dimensional unit matrix.

Thus, in $\boldsymbol{T}_\mathrm{c}^\star$ and $(\boldsymbol{T}_\mathrm{c}^\star)^{-1}$, the last row and column, respectively, can be neglected. This yields the *simplified Clarke transformation* given by

$$\boldsymbol{x}^s(t) = \underbrace{\kappa \begin{bmatrix} 1 & -\frac{1}{2} & -\frac{1}{2} \\ 0 & \frac{\sqrt{3}}{2} & -\frac{\sqrt{3}}{2} \end{bmatrix}}_{=:\boldsymbol{T}_\mathrm{c}\in\mathbb{R}^{2\times 3}} \boldsymbol{x}^{abc}(t) \iff \boldsymbol{x}^{abc}(t) = \underbrace{\frac{1}{\kappa}\begin{bmatrix} \frac{2}{3} & 0 \\ -\frac{1}{3} & \frac{1}{\sqrt{3}} \\ -\frac{1}{3} & -\frac{1}{\sqrt{3}} \end{bmatrix}}_{=:\boldsymbol{T}_\mathrm{c}^{-1}\in\mathbb{R}^{3\times 2}} \boldsymbol{x}^s(t). \tag{31}$$

Here, $\boldsymbol{T}_\mathrm{c}$ corresponds to the *simplified* Clarke transformation matrix and $\boldsymbol{T}_\mathrm{c}^{-1}$ to the *simplified* inverse Clarke transformation matrix. It follows then that $\boldsymbol{T}_\mathrm{c}\boldsymbol{T}_\mathrm{c}^{-1} = \boldsymbol{I}_2$, but $\boldsymbol{T}_\mathrm{c}^{-1}\boldsymbol{T}_\mathrm{c} \neq \boldsymbol{I}_3$.

**Remark II.2.** *The coefficient $\kappa$ allows the following:*

- *for $\kappa = \frac{2}{3}$, it yields an amplitude-correct transformation, i.e., $|x^\alpha(t)| = |x^a(t)|$ for all $t \geq 0\,\mathrm{s}$*
- *for $\kappa = \sqrt{\frac{2}{3}}$, it gives a power-correct transformation, i.e., $\boldsymbol{u}^{abc}(t)^\top \boldsymbol{i}^{abc}(t) = \boldsymbol{u}^s(t)^\top \boldsymbol{i}^s(t)$ for all $t \geq 0\,\mathrm{s}$.*

**Remark II.3.** *For a symmetrical three-phase system, one can directly calculate the phase quantities from the line-to-line quantities. The following holds*

$$\boldsymbol{x}^s(t) = \boldsymbol{T}_\mathrm{c}\boldsymbol{x}^{abc}(t) \overset{(29)}{=} \boldsymbol{T}_\mathrm{c}\boldsymbol{T}_\mathrm{v}^{-1}\boldsymbol{x}^{ltl}(t).$$

*In this case, the matrix $\boldsymbol{T}_\mathrm{c}\boldsymbol{T}_\mathrm{v}^{-1}$ is* not *unique, since $\boldsymbol{T}_\mathrm{v}$ in (29) would also have full rank if other rows (or columns) were replaced.*

*Park transformation: $(\alpha,\beta) \leftrightarrow (d,q)$ (or $(\alpha,\beta) \leftrightarrow (d',q')$):* By using the *Park transformation* stator-fixed quantities $\boldsymbol{x}^s(t) = (x^\alpha, x^\beta)^\top$ can be represented in the the (arbitrarily) rotating $k = (d,q)$-reference frame. The transformed signal vector in the $k$-reference frame is denoted with

$$\boldsymbol{x}^k(t) := (x^d(t),\, x^q(t))^\top \in \mathbb{R}^2.$$

The Park transformation represents a rotation of the vector $\boldsymbol{x}^s(t)$ counterclockwise by the (time-variant) angle $\phi_\mathrm{k}\colon \mathbb{R}_{>0} \to \mathbb{R}$ (in rad). The conversion between stator-fixed quantities and quantities in the $k$-reference frame is given by

$$\boldsymbol{x}^k(t) = \underbrace{\begin{bmatrix} \cos(\phi_\mathrm{k}(t)) & \sin(\phi_\mathrm{k}(t)) \\ -\sin(\phi_\mathrm{k}(t)) & \cos(\phi_\mathrm{k}(t)) \end{bmatrix}}_{=:\boldsymbol{T}_\mathrm{p}^{-1}(\phi_\mathrm{k}(t))\in\mathbb{R}^{2\times 2}} \boldsymbol{x}^s(t) \Leftrightarrow \boldsymbol{x}^s(t) = \underbrace{\begin{bmatrix} \cos(\phi_\mathrm{k}(t)) & -\sin(\phi_\mathrm{k}(t)) \\ \sin(\phi_\mathrm{k}(t)) & \cos(\phi_\mathrm{k}(t)) \end{bmatrix}}_{=\boldsymbol{T}_\mathrm{p}(\phi_\mathrm{k}(t))\in\mathbb{R}^{2\times 2}} \boldsymbol{x}^k(t). \tag{32}$$

Where $\boldsymbol{T}_\mathrm{p}(\phi_\mathrm{k}(t))$ is the Park transformation matrix and $\boldsymbol{T}_\mathrm{p}(\phi_\mathrm{k}(t))^{-1}$ is the inverse Park transformation matrix. Both are rotational matrices which satisfy

$$\forall\, t \geq 0\,\mathrm{s}\colon \quad \boldsymbol{T}_\mathrm{p}(\phi_\mathrm{k}(t))^{-1}\boldsymbol{T}_\mathrm{p}(\phi_\mathrm{k}(t)) = \boldsymbol{T}_\mathrm{p}(\phi_\mathrm{k}(t))\boldsymbol{T}_\mathrm{p}(\phi_\mathrm{k}(t))^{-1} = \boldsymbol{I}_2.$$

If $\phi_\mathrm{k}(t) = \phi_\mathrm{r}(t)$ (see Fig. 15), the transformation into the $r$-reference frame is obtained, i.e.,

$$\boldsymbol{x}^r(t) = (x^{d'}(t), x^{q'}(t))^\top = \boldsymbol{T}_\mathrm{p}^{-1}(\phi_\mathrm{r}(t))\boldsymbol{x}^s(t) \iff \boldsymbol{x}^s(t) = \boldsymbol{T}_\mathrm{p}(\phi_\mathrm{r}(t))\boldsymbol{x}^r(t).$$

**Remark II.4.** *The Park transformation matrix $\boldsymbol{T}_\mathrm{p}(\cdot)$ in (32) has some important properties which are required later for the modeling and control sections. The following properties apply*

$$\forall \alpha \in \mathbb{R}\colon \qquad \boldsymbol{T}_\mathrm{p}(\alpha)^{-1} = \begin{bmatrix} \cos(\alpha) & \sin(\alpha) \\ -\sin(\alpha) & \cos(\alpha) \end{bmatrix} = \boldsymbol{T}_\mathrm{p}(\alpha)^\top = \boldsymbol{T}_\mathrm{p}(-\alpha). \tag{33}$$

*and*

$$\boldsymbol{J} := \begin{bmatrix} 0 & -1 \\ 1 & 0 \end{bmatrix} = \boldsymbol{T}_\mathrm{p}\left(\tfrac{\pi}{2}\right). \tag{34}$$

*From which it follows*

$$\forall \alpha \in \mathbb{R}: \qquad \boldsymbol{J}\boldsymbol{T}_{\rm p}(\alpha) = \begin{bmatrix} -\sin(\alpha) & -\cos(\alpha) \\ \cos(\alpha) & -\sin(\alpha) \end{bmatrix} = \boldsymbol{T}_{\rm p}(\alpha)\boldsymbol{J}. \tag{35}$$

*Furthermore, using the trigonometric identities (see [26, p. 124])*

$$\forall\, \alpha, \beta \in \mathbb{R}: \qquad \sin(\alpha \pm \beta) = \sin(\alpha)\cos(\beta) \pm \cos(\alpha)\sin(\beta) \tag{36}$$

*and*

$$\forall\, \alpha, \beta \in \mathbb{R}: \qquad \cos(\alpha \pm \beta) = \cos(\alpha)\cos(\beta) \mp \sin(\alpha)\sin(\beta) \tag{37}$$

*one obtains*

$$\forall\, \alpha, \beta \in \mathbb{R}: \qquad \boldsymbol{T}_{\rm p}(\alpha \pm \beta) = \boldsymbol{T}_{\rm p}(\alpha)\boldsymbol{T}_{\rm p}(\pm\beta) = \boldsymbol{T}_{\rm p}(\pm\beta)\boldsymbol{T}_{\rm p}(\alpha). \tag{38}$$

*For* $\phi(t) = \int_0^t \omega(\tau)\,\mathrm{d}\tau + \phi_{\rm g}$ *with* $\phi_{\rm g} \in \mathbb{R}$ *(in* rad*),* $\frac{\mathrm{d}}{\mathrm{d}t}\phi(t) = \omega(t)$ *for all* $t \geq 0\,\mathrm{s}$ *holds true and thus the following identities can be derived*

$$\forall t \geq 0\,\mathrm{s}: \;\; \tfrac{\mathrm{d}}{\mathrm{d}t}\boldsymbol{T}_{\rm p}(\phi(t)) =: \dot{\boldsymbol{T}}_{\rm p}(\phi(t)) = \omega(t)\boldsymbol{J}\boldsymbol{T}_{\rm p}(\phi(t)) = \omega(t)\boldsymbol{T}_{\rm p}(\phi(t))\boldsymbol{J} \tag{39}$$

*and*

$$\forall t \geq 0\,\mathrm{s}: \;\; \tfrac{\mathrm{d}}{\mathrm{d}t}\boldsymbol{T}_{\rm p}(\phi(t))^{-1} =: \dot{\boldsymbol{T}}_{\rm p}(\phi(t))^{-1} = -\omega(t)\boldsymbol{J}\boldsymbol{T}_{\rm p}(\phi(t))^{-1} = -\omega(t)\boldsymbol{T}_{\rm p}(\phi(t))^{-1}\boldsymbol{J}. \tag{40}$$

### *B. Power calculation*

The *instantaneous* active, reactive and apparent power at the *Point of Common Coupling (PCC)* (see Fig. 14) are important to consider now. For that assume that the phase voltages and phase currents are given by

$$\begin{aligned} \boldsymbol{u}^{abc}(t) &= \hat{u}(t)\begin{pmatrix} \cos(\phi(t) + \varphi_u(t)) \\ \cos\left(\phi(t) - \frac{2}{3}\pi + \varphi_u(t)\right) \\ \cos\left(\phi(t) - \frac{4}{3}\pi + \varphi_u(t)\right) \end{pmatrix} \quad \text{and} \\ \boldsymbol{i}^{abc}(t) &= \hat{\imath}(t)\begin{pmatrix} \cos(\phi(t) + \varphi_i(t)) \\ \cos\left(\phi(t) - \frac{2}{3}\pi + \varphi_i(t)\right) \\ \cos\left(\phi(t) - \frac{4}{3}\pi + \varphi_i(t)\right) \end{pmatrix} \end{aligned} \tag{41}$$

with phase angle $\phi(t)$ (in rad) and phase offset $\varphi_u(t)$ (in rad) or $\varphi_i(t)$ (in rad) of voltage and current, respectively. Thus, in addition to the assumptions made in (A.3) and (A. 4) of a symmetrical system (see (21) and (22)), the following assumption shall hold as well:

**Assumption (A.5)** *Voltage and current phase angles in* (22) *can be written as the sum of* $\phi(t)$ *and* $\varphi_u(t)$ *respectively* $\varphi_i(t)$*, i.e.,*

$$\begin{aligned} \forall\, t \geq 0\,\mathrm{s}: \;\; & \phi_u^a(t) = \phi(t) + \varphi_u(t) = \phi_u^b(t) + \tfrac{2}{3}\pi = \phi_u^c(t) + \tfrac{4}{3}\pi \qquad \textit{and} \\ & \phi_i^a(t) = \phi(t) + \varphi_i(t) = \phi_i^b(t) + \tfrac{2}{3}\pi = \phi_i^c(t) + \tfrac{4}{3}\pi. \end{aligned} \tag{42}$$

The phase voltages and currents in (41) correspond to the time-varying *phase (displacement) angel* or *phase shift*

$$\forall\, t \geq 0\,\mathrm{s}: \qquad \varphi(t) := \varphi_u(t) - \varphi_i(t) \in \mathbb{R}. \tag{43}$$

*1) Instantaneous power:* By using the trigonometric identity (see [26, p. 124])

$$\forall\,\alpha,\beta \in \mathbb{R}\colon\quad \cos(\alpha)\cos(\beta) = \frac{1}{2}\big(\cos(\alpha-\beta)+\cos(\alpha+\beta)\big), \tag{44}$$

the *instantaneous power* $p_{3\sim}(t)$ (in W) can be defined for all $t \geq 0\,\mathrm{s}$:

$$\begin{aligned}
p_{3\sim}(t) \quad &:= \quad \boldsymbol{u}^{abc}(t)^\top \boldsymbol{i}^{abc}(t) = u^a(t)i^a(t) + u^b(t)i^b(t) + u^c(t)i^c(t) \\
&\overset{(41)}{=} \quad \hat{u}(t)\,\hat{\imath}(t)\Big[\cos\big(\phi(t)+\varphi_u(t)\big)\cos\big(\phi(t)+\varphi_i(t)\big) \\
&\qquad + \cos\big(\phi(t)-\tfrac{2}{3}\pi+\varphi_u(t)\big)\cos\big(\phi(t)-\tfrac{2}{3}\pi+\varphi_i(t)\big) \\
&\qquad + \cos\big(\phi(t)-\tfrac{4}{3}\pi+\varphi_u(t)\big)\cos\big(\phi(t)-\tfrac{4}{3}\pi+\varphi_i(t)\big)\Big] \\
&\overset{(44),(43)}{=} \quad \tfrac{3}{2}\hat{u}(t)\,\hat{\imath}(t)\cos\big(\varphi(t)\big) + \\
&\underbrace{+\tfrac{1}{2}\hat{u}(t)\,\hat{\imath}(t)\Big[\cos(2\phi(t)+\varphi(t)) + \cos(2\phi(t)-\tfrac{4}{3}\pi+\varphi(t)) + \cos(2\phi(t)-\overbrace{\tfrac{8}{3}\pi}^{=\frac{2}{3}\pi}+\varphi(t))\Big]}_{\overset{(24)}{=}0\,\mathrm{W}} \\
&= \quad \tfrac{3}{2}\,\hat{u}(t)\,\hat{\imath}(t)\cos\big(\varphi(t)\big).
\end{aligned} \tag{45}$$

In summary, the instantaneous power can be written as follows

$$\left.\begin{aligned}
\forall\,t\geq 0\,\mathrm{s}\colon\quad p_{3\sim}(t) := \boldsymbol{u}^{abc}(t)^\top \boldsymbol{i}^{abc}(t) &\overset{(31)}{=} \tfrac{3}{2}\boldsymbol{u}^s(t)^\top \boldsymbol{i}^s(t) \overset{(32)}{=} \tfrac{3}{2}\boldsymbol{u}^k(t)^\top \boldsymbol{i}^k(t) \\
&\overset{(45)}{=} \tfrac{3}{2}\,\hat{u}(t)\,\hat{\imath}(t)\,\cos\big(\varphi(t)\big).
\end{aligned}\right\} \tag{46}$$

**Remark II.5.** *The instantaneous power $p_{3\sim}(t)$ can also be calculated from the line-to-line voltages $\boldsymbol{u}^{ltl}(t)$ and currents $\boldsymbol{i}^{ltl}(t)$ as follows*

$$\begin{aligned}
\boldsymbol{u}^{ltl}(t)^\top \boldsymbol{i}^{ltl}(t) &= u^{ab}(t)i^{ab}(t) + u^{bc}(t)i^{bc}(t) + u^{ca}(t)i^{ca}(t) \\
&= u^a(t)i^a(t) + u^b(t)i^b(t) - (u^a(t)i^b(t) + u^b(t)i^a(t)) \\
&\quad + u^b(t)i^b(t) + u^c(t)i^c(t) - (u^b(t)i^c(t) + u^c(t)i^b(t)) \\
&\quad + u^c(t)i^c(t) + u^a(t)i^a(t) - (u^c(t)i^a(t) + u^a(t)i^c(t)) \\
&= 2u^a(t)i^a(t) + 2u^b(t)i^b(t) + 2u^c(t)i^c(t) \\
&\quad - u^a(t)\underbrace{(i^b(t)+i^c(t))}_{=-i^a(t)} - u^b(t)\underbrace{(i^a(t)+i^c(t))}_{=-i^b(t)} - u^c(t)\underbrace{(i^a(t)+i^b(t))}_{=-i^c(t)} \\
&= 3\boldsymbol{u}^{abc}(t)^\top \boldsymbol{i}^{abc}(t).
\end{aligned}$$

*From which it can be concluded that*

$$\forall\,t\geq 0\,\mathrm{s}\colon\qquad p_{3\sim}(t) = \boldsymbol{u}^{abc}(t)^\top \boldsymbol{i}^{abc}(t) = \frac{1}{3}\boldsymbol{u}^{ltl}(t)^\top \boldsymbol{i}^{ltl}(t). \tag{47}$$

*2) Active, reactive and apparent power:* For the definition of (averaged) active, reactive and apparent power, three additional assumptions are required:

**Assumption (A.6)** *The amplitudes $\hat{u}(t)$ and $\hat{\imath}(t)$ of phase voltages $\boldsymbol{u}^{abc}(t)$ and phase currents $\boldsymbol{i}^{abc}(t)$, respectively, are constant, i.e.,*

$$\forall\,t\geq 0\,\mathrm{s}\colon\qquad \hat{u}(t) = \hat{u} > 0\,\mathrm{V} \qquad \textit{and} \qquad \hat{\imath}(t) = \hat{\imath} > 0\,\mathrm{A}. \tag{48}$$

**Assumption (A.7)** *All phase quantities have a constant angular frequency* $\omega = \frac{2\pi}{T}$ *(in* $\frac{\mathrm{rad}}{\mathrm{s}}$*) over the time period* $T$ *(in* $\mathrm{s}$*), therefore, for quantities in* (41) *it follows that:*

$$\forall\, t \geq 0\,\mathrm{s}: \qquad \phi(t) = \int_0^t \omega\,\mathrm{d}\tau \;= \omega\, t = \frac{2\pi}{T}\, t. \tag{49}$$

**Assumption (A.8)** *The phase angle between voltage and current remains constant, therefore* $\varphi(t)$ *in* (43) *is such that*

$$\forall\, t \geq 0\,\mathrm{s}: \qquad \varphi(t) = \varphi_0 := \varphi_u - \varphi_i \in \mathbb{R}. \tag{50}$$

The active power $P(t)$ (in $\mathrm{W}$) is defined as the value of the instantaneous power $p_{3\sim}(t)$ averaged over a certain period $T$ (see example [30, p. 15–17]), i.e.,

$$\boxed{\forall\, t \geq T: \quad P(t) \overset{(49)}{:=} \frac{1}{T}\int_{t-T}^{t} p_{3\sim}(\tau)\,\mathrm{d}\tau \overset{(46),(48),(50)}{=} \frac{3}{2}\,\hat{u}\,\hat{\imath}\,\cos(\varphi_0)\,.} \tag{51}$$

The reactive power $Q(t)$ (in $\mathrm{var}$) is defined as the scalar product of the voltage $\boldsymbol{u}^{abc}(t)$ and the *orthogonal* current $\boldsymbol{i}^{abc}(t-\frac{T}{4}) = \boldsymbol{i}^{abc}(t-\frac{\pi}{2\omega})\;\; \boldsymbol{i}^{abc}(t)$ (as in (41)) averaged over a period $T$, i.e.,

$$\boxed{\forall\, t \geq T: \quad Q(t) \overset{(49)}{:=} -\frac{1}{T}\int_{t-T}^{t} \boldsymbol{u}^{abc}(\tau)^\top \boldsymbol{i}^{abc}(\tau - \tfrac{T}{4})\,\mathrm{d}\tau \overset{(45),(48),(50)}{=} \frac{3}{2}\hat{u}\hat{\imath}\,\sin(\varphi_0)}. \tag{52}$$

Finally, the apparent power $S(t)$ (in $\mathrm{V\,A}$) is defined as the square mean value of the active power $P(t)$ (in $\mathrm{W}$) and the reactive power $Q(t)$ (in $\mathrm{var}$), i.e.,

$$\boxed{\forall\, t \geq T: \quad S(t) := \sqrt{P(t)^2 + Q(t)^2} \overset{(51),(52)}{=} \frac{3}{2}\,\hat{u}\,\hat{\imath}.} \tag{53}$$

**Remark II.6.** *Under the assumptions (A.6), (A.7) and (A.8), the (averaged) active power* $P(t)$ *corresponds to the instantaneous active power* $p_{3\sim}(t)$*, since*

$$\forall\, t \geq T: \qquad P(t) \overset{(50),(51)}{=} p_{3\sim}(t). \tag{54}$$

*3) Active, reactive and apparent power in space vectors:* Particularly for Section IV, the understanding of space vectors is important. In the following section, it is shown that active, reactive and apparent power can be represented using the voltage and current components of the $s$ or $k$-reference frames. Then, the computation using three-phase quantities becomes obsolete for balanced three-phase systems.

In Fig. 16, the voltage space vector $\boldsymbol{u}^s(t) = \boldsymbol{T}_\mathrm{c}\boldsymbol{u}^{abc}(t)$ and current space vector $\boldsymbol{i}^s(t) = \boldsymbol{T}_\mathrm{c}\boldsymbol{i}^{abc}(t)$ of the phase variables $\boldsymbol{u}^{abc}(t)$ and $\boldsymbol{i}^{abc}(t)$, respectively, from (41) are displayed in the $s$-reference frame for the time instant $t \geq 0\,\mathrm{s}$. The Assumptions (A.3)–(A.8) apply. In the case of the amplitude-correct Clarke transformation (31), i.e., $\kappa = \frac{2}{3}$ (see Section II-A3), the amplitude of voltage and current space vectors correspond to the peak value of the sinusoidal phase voltage $\boldsymbol{u}^{abc}(t)$ and $\boldsymbol{i}^{abc}(t)$, respectively, from (41), i.e., $\|\boldsymbol{u}^s(t)\| = \hat{u}$ and $\|\boldsymbol{i}^s(t)\| = \hat{\imath}$, respectively, for all $t \geq 0\,\mathrm{s}$. Due to the Clarke transformation, the phase shift or phase offset $\varphi(t) = \varphi_0 = \varphi_u - \varphi_i$ as in (50) – between the phase voltages $\boldsymbol{u}^{abc}(t)$ and phase currents $\boldsymbol{i}^{abc}(t)$ as in (41) becomes a spatial offset between the corresponding space vectors $\boldsymbol{u}^s(t)$ and $\boldsymbol{i}^s(t)$ with the angle $\varphi_0$.

Now, a simple calculation shows that, for all $t \geq 0\,\mathrm{s}$ and $\phi_\mathrm{k}\colon \mathbb{R}_{>0} \to \mathbb{R}$, the following holds

$$\begin{aligned}
\boldsymbol{u}^{abc}(t)^\top \boldsymbol{i}^{abc}(t) &\overset{(31)}{=} \boldsymbol{u}^s(t)^\top (\boldsymbol{T}_\mathrm{c}^{-1})^\top \boldsymbol{T}_\mathrm{c}^{-1} \boldsymbol{i}^s(t) = \frac{3}{2}\boldsymbol{u}^s(t)^\top \boldsymbol{i}^s(t) \\
&\overset{(32)}{=} \frac{3}{2}\boldsymbol{u}^k(t)^\top \underbrace{\boldsymbol{T}_\mathrm{p}(\phi_\mathrm{k}(t))^\top \boldsymbol{T}_\mathrm{p}(\phi_\mathrm{k}(t))}_{=\boldsymbol{I}_2} \boldsymbol{i}^k(t)
\end{aligned}$$

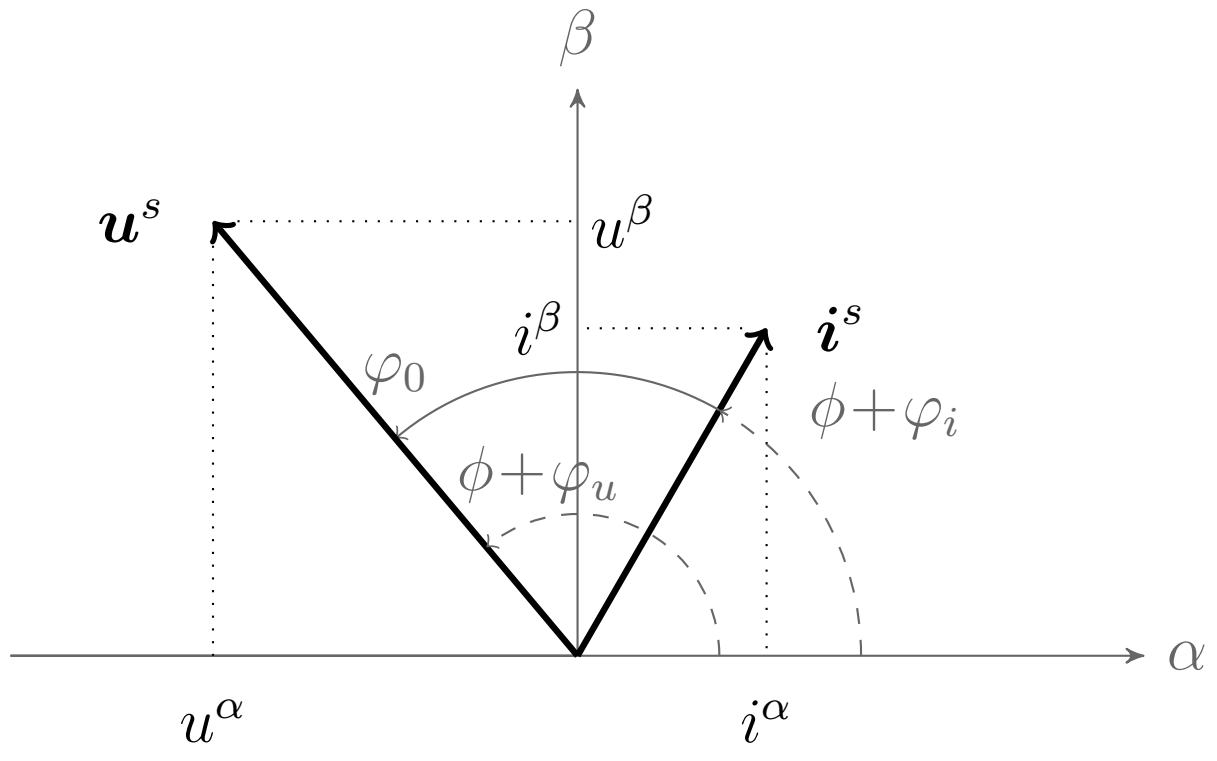

Fig. 16: *Voltage space vector* $\boldsymbol{u}^s = (u^\alpha,\, u^\beta)^\top$ *and current space vector* $\boldsymbol{i}^s = (i^\alpha,\, i^\beta)^\top$ *in the s-reference frame with amplitude* $\|\boldsymbol{u}^s\| = \hat{u}$ *and* $\|\boldsymbol{i}^s\| = \hat{\imath}$*, respectively.*

$$= \frac{3}{2}\hat{u}\,\hat{\imath}\cos(\varphi_0) \overset{(51)}{=} P(t). \tag{55}$$

This relation corresponds to the scalar product of the voltage and current space vectors (see [26, p. 78]) in the $s$- or $k$-reference frame, respectively, and thus equals to the active power $P(t)$ as in (51).

Furthermore, it can be concluded with the help of Fig. 16, that for all $t \geq 0$ and $\phi_\mathrm{k} \colon \mathbb{R}_{>0} \to \mathbb{R}$, the following holds

$$\begin{aligned}
\frac{3}{2}\boldsymbol{u}^s(t)^\top \boldsymbol{J}\boldsymbol{i}^s(t) &\overset{(32),(35)}{=} \frac{3}{2}\boldsymbol{u}^k(t)^\top \underbrace{\boldsymbol{T}_\mathrm{p}(\phi_\mathrm{k}(t))^\top \boldsymbol{T}_\mathrm{p}(\phi_\mathrm{k}(t))}_{\overset{(33)}{=}\boldsymbol{I}_2} \boldsymbol{J}\boldsymbol{i}^k(t) \\
&\overset{(32)}{=} \frac{3}{2}\left(\boldsymbol{T}_\mathrm{p}(\phi(t)+\varphi_u)\begin{pmatrix}\hat{u}\\0\end{pmatrix}\right)^\top \boldsymbol{J}\left(\boldsymbol{T}_\mathrm{p}(\phi(t)+\varphi_i)\begin{pmatrix}\hat{\imath}\\0\end{pmatrix}\right) \\
&\overset{(33),(35)}{=} \frac{3}{2}\begin{pmatrix}\hat{u}\\0\end{pmatrix}^\top \boldsymbol{T}_\mathrm{p}(-\phi(t)-\varphi_u)\boldsymbol{T}_\mathrm{p}(\phi(t)+\varphi_i)\boldsymbol{J}\begin{pmatrix}\hat{\imath}\\0\end{pmatrix} \\
&\overset{(38),(50)}{=} \frac{3}{2}\begin{pmatrix}\hat{u}\\0\end{pmatrix}^\top \boldsymbol{T}_\mathrm{p}(-\varphi_0)\boldsymbol{J}\begin{pmatrix}\hat{\imath}\\0\end{pmatrix} \\
&\overset{(33),(34)}{=} \frac{3}{2}\begin{pmatrix}\hat{u}\\0\end{pmatrix}^\top \begin{bmatrix}\cos(\varphi_0) & \sin(\varphi_0)\\ -\sin(\varphi_0) & \cos(\varphi_0)\end{bmatrix}\begin{pmatrix}0\\\hat{\imath}\end{pmatrix} \\
&= \frac{3}{2}\hat{u}\,\hat{\imath}\,\sin(\varphi_0) \overset{(52)}{=} Q(t).
\end{aligned} \tag{56}$$

Thus, as shown, there is a relationship between voltage and current space vectors in the $s$- and $k$-reference frame. In conclusion, with the Assumptions (A.3)–(A.8), for the phase voltages and phase currents from (41), the following definitions can be summarized in the *space vector notation*: One obtains the following expressions for the active power

$$\boxed{\begin{aligned}\forall t \geq T\colon \qquad P(t) &= \tfrac{3}{2}\boldsymbol{u}^s(t)^\top \boldsymbol{i}^s(t) = \tfrac{3}{2}\big(u^\alpha(t)i^\alpha(t) + u^\beta(t)i^\beta(t)\big) \\ &= \tfrac{3}{2}\boldsymbol{u}^k(t)^\top \boldsymbol{i}^k(t) = \tfrac{3}{2}\big(u^d(t)i^d(t) + u^q(t)i^q(t)\big),\end{aligned}}, \tag{57}$$

for the reactive power (see, e.g., [31, p. 50])

$$\boxed{\begin{aligned}\forall t \geq T\colon \qquad Q(t) &= \tfrac{3}{2}\boldsymbol{u}^s(t)^\top \boldsymbol{J}\boldsymbol{i}^s(t) = \tfrac{3}{2}\big(u^\beta(t)i^\alpha(t) - u^\alpha(t)i^\beta\big) \\ &= \tfrac{3}{2}\boldsymbol{u}^k(t)^\top \boldsymbol{J}\boldsymbol{i}^k(t) = \tfrac{3}{2}\big(u^q(t)i^d(t) - u^d(t)i^q(t)\big)\end{aligned}}, \tag{58}$$

and for the apparent power

$$\boxed{\forall t \geq T: \qquad S(t) = \frac{3}{2}\hat{u}\,\hat{\imath} = \frac{3}{2}\|\boldsymbol{u}^s(t)\|\|\boldsymbol{i}^s(t)\| = \frac{3}{2}\|\boldsymbol{u}^k(t)\|\|\boldsymbol{i}^k(t)\|.} \tag{59}$$

**Remark II.7.** *The definitions of active, reactive and apparent power in* (57)*,* (58) *and* (59)*, respectively, applies for all* $t \geq T$ *and a constant period* $T > 0\,\mathrm{s}$*. In contrast to these classical definitions, instantaneous expressions are possible for the active, reactive and apparent power even for three-wire or four-wire systems [25, Appendix B].*

### *C. Unit definitions, energy units and conversion factors*

Energy quantities are often not uniformly expressed in terms of a one to one comparison. This can lead to uncertainty among researchers and readers and skew perceptions and comparisons. For this reason, various unit conversions (as, for example, $\mu\mathrm{m} = 10^{-6}\mathrm{m}$), and corresponding energy sources/quantities and different energy units (e.g. in $\mathrm{kW\,h}$) are listed in the following Tables I, III and II.

TABLE I: *Prefixes, symbols and factors*

| Prefix name | Symbol | Value | English term |
|---|---|---|---|
| Yotta | Y | $10^{24}$ | (septillion) |
| Zetta | Z | $10^{21}$ | (sextillion) |
| Exa | E | $10^{18}$ | (quintillion) |
| Peta | P | $10^{15}$ | (quadrillion) |
| Tera | T | $10^{12}$ | (trillion) |
| Giga | G | $10^{9}$ | (billion) |
| Mega | M | $10^{6}$ | (million) |
| Kilo | k | $10^{3}$ | (thousand) |
| Hekto | h | $10^{2}$ | (hundred) |
| Deka | da | $10^{1}$ | (ten) |
| – | – | $10^{0}$ | (one) |
| Dezi | d | $10^{-1}$ | (tenth) |
| Zenti | c | $10^{-2}$ | (hundredth) |
| Milli | m | $10^{-3}$ | (thousandth) |
| Mikro | $\mu$ | $10^{-6}$ | (millionth) |
| Nano | n | $10^{-9}$ | (billionth) |
| Piko | p | $10^{-12}$ | (trillionth) |
| Femto | f | $10^{-15}$ | (quadrillionth) |
| Atto | a | $10^{-18}$ | (quintillionth) |
| Zepto | z | $10^{-21}$ | (sextillionth) |
| Yokto | y | $10^{-24}$ | (septillionth) |

TABLE II: *Conversion factors between different energy units (see [17, Tab. 1.1]) with the abbreviations* $\mathrm{kJ}$*: kilo-Joule,* $\mathrm{W\,s}$*: watt-second,* $\mathrm{kcal}$*: kilo-calorie,* $\mathrm{kW\,h}$*: kilo-watthour,* $\mathrm{kg}$*: kilo-gram,* SKE*: Steinkohleeinheit (black coal-unit),* toe*: tonne of oil equivalent,* $\mathrm{m}^3$*: cubic meter (volume), and nat. gas: natural gas.*

| | kJ | kcal | kW h | kg SKE | kg toe | $\mathrm{m}^3$ nat. gas |
|---|---|---|---|---|---|---|
| 1 kJ | **1** | 0.2388 | $\frac{1}{3\,600}$ | $3.4 \cdot 10^{-5}$ | $2.4 \cdot 10^{-5}$ | $3.2 \cdot 10^{-5}$ |
| 1 kcal | 4.1868 | **1** | $1.163 \cdot 10^{-3}$ | $1.43 \cdot 10^{-4}$ | $1 \cdot 10^{-4}$ | $1.3 \cdot 10^{-4}$ |
| 1 kW h | 3 600 | 860 | **1** | 0.123 | 0.086 | 0.113 |
| 1 kg SKE | 29 308 | 7 000 | 8.14 | **1** | 0.7 | 0.923 |
| 1 kg toe | 41 868 | 10 000 | 11.63 | 1.428 | **1** | 1.319 |
| 1 $\mathrm{m}^3$ nat. gas | 31 736 | 7 580 | 8.816 | 1.083 | 0.758 | **1** |

TABLE III: *Conversion factors between different energy quantities (see [3, Tab. 1.2]).*

| Energy source | Energy stored | Remark |
|---|---|---|
| $1\,\mathrm{kg}$ Black coal | $8.14\,\mathrm{kW\,h}$ | – |
| $1\,\mathrm{kg}$ Crude oil | $11.63\,\mathrm{kW\,h}$ | Gasoline: $8.7\,\frac{\mathrm{kW\,h}}{\mathrm{l}}$; Diesel: $9.8\,\frac{\mathrm{kW\,h}}{\mathrm{l}}$ |
| $1\,\mathrm{m}^3$ Natural gas | $8.82\,\mathrm{kW\,h}$ | – |
| $1\,\mathrm{kg}$ Wood | $4.3\,\mathrm{kW\,h}$ | (for $15\%$ humidity) |

## III. Modeling of wind turbine systems

This section deals with the modeling of a wind turbine system with permanent-magnet synchronous generator. The presented models of the turbine (aerodynamical torque generation), gearbox, permanent-magnet synchronous generator, back-to-back converter, grid filter and the (ideal) grid can be used directly for the construction of a simplified simulation model.

### A. Turbine

The turbine (rotor with three blades) converts part of the kinetic wind energy into rotational energy, which is then converted into electrical energy via the generator. Fig. 7 shows a frontal view of a wind turbine. The wind velocity $v_{\mathrm{w}}$ is directed into the plane. The turbine rotors have radius $r_{\mathrm{t}}$ (in $\mathrm{m}$) and the encircled area is described by

$$A_{\mathrm{t}} = \pi\, r_{\mathrm{t}}^2 \quad (\text{in } \mathrm{m}^2) \tag{60}$$

(neglecting the area $A_{\mathrm{n}}$ of the nacelle, see Section I-D1). By changing the turbine angular velocity $\omega_{\mathrm{t}}$ (in $\frac{\mathrm{rad}}{\mathrm{s}}$), the rotational velocity $r_{\mathrm{t}}\omega_{\mathrm{t}}$ and therefore the tip speed ratio

$$\lambda := \frac{r_{\mathrm{t}}\,\omega_{\mathrm{t}}}{v_{\mathrm{w}}} \tag{61}$$

of the wind turbine can also be changed. The tip speed ratio $\lambda$ indicates the ratio between the speed at the extreme tip of the rotor blades and the incoming wind speed. For three-bladed wind turbines, the optimal tip speed ratio varies between 7...8 [17, p. 259]. The tip speed ratio $\lambda$ is an important control variable for the control of wind turbine systems. It allows for the operation at an optimum, for which the maximum possible power can be extracted from the wind. In addition to the tip speed ratio $\lambda$, the second most important control variable is the pitch angle $\beta$ (in $^\circ$). By changing the pitch angle, the inflow of the wind power from the rotor blades can be increased or decreased. Thus, for example, an increase in the pitch angle reduces the turbine power.

Both control variables have a direct influence on the amount of power the wind turbine can remove from the wind. In the next sections, these relationships with regard to the turbine power, power coefficient and turbine torque will be explained in more detail.

*1) Turbine power:* In Section I-D3, it was derived that all of the power can not be completely extracted from the wind. The removable fraction of the wind power is limited by the Betz limit $c_{\mathrm{p,Betz}} = 16/27$ and the turbine power is given by

$$\forall t \geq 0\,\mathrm{s}: \qquad p_{\mathrm{t}}(t) = c_{\mathrm{p}} \underbrace{\frac{1}{2}\varrho\pi r_{\mathrm{t}}^2 v_{\mathrm{w}}(t)^3}_{=p_{\mathrm{w}}(t)} \leq c_{\mathrm{p,Betz}}\, p_{\mathrm{w}}(t).$$

The power coefficient $c_{\mathrm{p}}$ must be determined for each wind turbine system and is a function of the pitch angle $\beta$ and tip speed ratio $\lambda$, i.e.,

$$c_{\mathrm{p}} : \mathbb{R}_{\geq 0} \times \mathbb{R}_{\geq 0} \to \mathbb{R}_{>0}, \qquad (\beta,\, \lambda) \mapsto c_{\mathrm{p}} := c_{\mathrm{p}}(\beta,\, \lambda).$$

In summary, the following expression describes the turbine power output

$$p_{\mathrm{t}}(t, \beta, \lambda) = c_{\mathrm{p}}(\beta,\, \lambda)\, p_{\mathrm{w}}(t) \overset{(5)}{=} c_{\mathrm{p}}(\beta,\, \lambda)\, \frac{1}{2}\varrho\pi r_{\mathrm{t}}^2 v_{\mathrm{w}}(t)^3 \leq c_{\mathrm{p,Betz}}\, p_{\mathrm{w}}(t). \tag{62}$$

TABLE IV: *Exemplary parameterizations of the power coefficient approximation in* (63) *for two distinct* 2 MW *wind turbines (see [32, p. 9],[33], [34]).*

| | $c_{\rm p,1}$ (without pitch system) | $c_{\rm p,2}$ (with pitch system) |
|---|---|---|
| $c_1$ | 1 | 0.73 |
| $c_2$ | 46.6 | 151 |
| $c_3$ | 0 | 0.58 |
| $c_4$ | 0 | 0.002 |
| $c_5$ | 2.0 | 13.2 |
| $c_6$ | 15.6 | 18.4 |
| $f(\beta,\lambda)$ | $\frac{1}{\lambda} - 0.01$ | $\frac{1}{\lambda-0.02\beta} - \frac{0.003}{\beta^3+1}$ |
| $x$ | – | 2.14 |

*2) Approximation of the power coefficient:* An approximation of the power coefficient $c_{\rm p}(\beta,\lambda)$ as a function of the pitch angle $\beta$ (in °) and tip speed ratio $\lambda$ is used for modeling. In [20, (2.38)], the following function is proposed

$$c_{\rm p}\colon \overline{\mathcal{D}} \to \mathbb{R}_{>0},\ (\beta,\lambda) \mapsto c_{\rm p}(\beta,\lambda) := c_1\big[c_2\, f(\beta,\lambda) - c_3\beta - c_4\beta^x - c_5\big]e^{-c_6\, f(\beta,\lambda)}$$
$$\text{where } \mathcal{D} := \{\ (\beta,\lambda) \in \mathbb{R}_{>0} \times \mathbb{R}_{>0} \mid c_{\rm p}(\beta,\lambda) > 0\ \} \quad (63)$$

to approximate the power coefficient of wind turbine systems. The constants $c_1,\dots,c_6 > 0$, the exponent $x \geq 0$ and the continuously differentiable function $f\colon \overline{\mathcal{D}} \to \mathbb{R}$ can be determined from the measurement data of a real wind turbine system or by aerodynamic simulation tools. The approximation function $c_{\rm p}(\cdot,\cdot)$ in (63) of the power coefficient is defined for the range $\overline{\mathcal{D}} := \mathcal{D} \cap \partial\mathcal{D}$ (with the boundary $\partial\mathcal{D}$ of the set $\mathcal{D}$) and has the following **properties**:

**(P$_1$)** $c_{\rm p}(\cdot,\cdot)$ is continuous and not negative,
**(P$_2$)** $c_{\rm p}(\cdot,\cdot)$ is continuously differentiable, and
**(P$_3$)** $c_{\rm p}(\cdot,\cdot)$ has a *unique* maximum $c_{\rm p}(\beta_0,\lambda^\star_{\beta_0})$ for every $\beta_0$ and optimal tip speed ratio $\lambda^\star_{\beta_0}$, i.e.,

$$\forall(\beta_0,\lambda) \in \mathcal{D}\ \exists\, \lambda^\star_{\beta_0} > 0\colon \quad c_{\rm p}(\beta_0,\lambda) \leq c_{\rm p}(\beta_0,\lambda^\star_{\beta_0}) \quad \wedge$$
$$\left.\frac{\partial c_{\rm p}(\beta_0,\lambda)}{\partial\lambda}\right|_{\lambda^\star_{\beta_0}} = 0 \quad \wedge \quad \left.\frac{\partial^2 c_{\rm p}(\beta_0,\lambda)}{\partial\lambda^2}\right|_{\lambda^\star_{\beta_0}} < 0. \quad (64)$$

**(P$_4$)** $c_{\rm p}(0,\cdot)$ is proportional to $e^{-\frac{1}{\cdot}}$, i.e.,

$$\forall\lambda \geq 0\colon \qquad c_{\rm p}(0,\lambda) \propto e^{-\frac{1}{\lambda}}. \quad (65)$$

Examples for the parameterization of the power coefficient approximation (63) are summarized in Tab. IV for actual power coefficient curves of two different 2 MW wind turbines: $c_{\rm p,1}(\cdot)$ *without* pitch control system and $c_{\rm p,2}(\cdot,\cdot)$ *with* pitch control system. If the parameters from Tab. IV are inserted into (63), the following explicit approximations of the power coefficient for the 2 MW wind turbine systems are given:

- Power coefficient $c_{\rm p,1}(\cdot)$ *without* pitch control system (i.e., $\beta = \beta_0 = 0°$):

$$c_{\rm p,1}\colon\ \overline{\mathcal{D}} \to \mathbb{R}_{>0}, \qquad (0,\lambda) \mapsto c_{\rm p,1}(0,\lambda) :=$$
$$c_{\rm p,1}(\lambda) := \left[46.4 \cdot \left(\frac{1}{\lambda} - 0.01\right) - 2.0\right] e^{-15.6\left(\frac{1}{\lambda}-0.01\right)}. \quad (66)$$

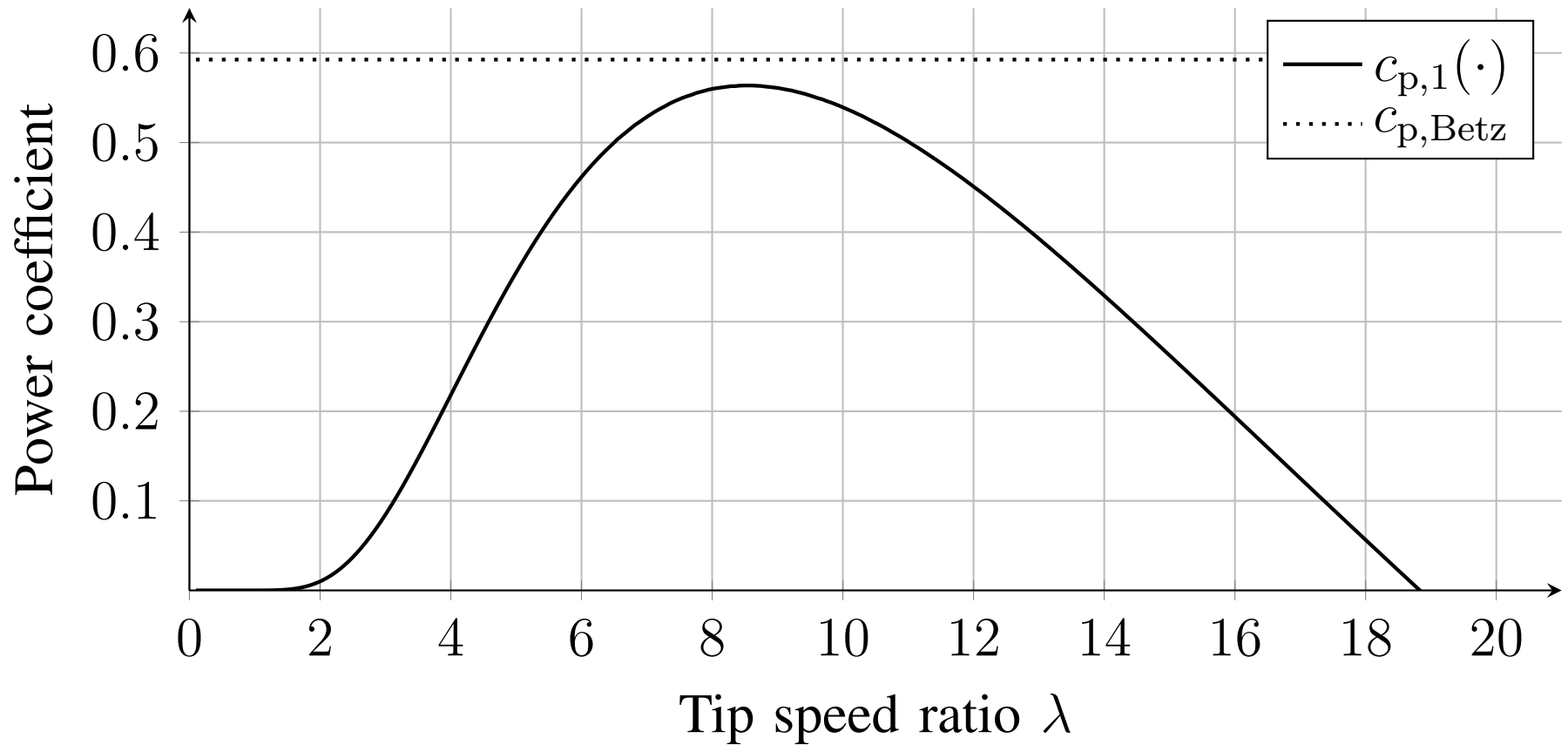

Fig. 17: *Power coefficient for a* $2\,\mathrm{MW}$ *wind turbine* without *pitch control system (i.e.* $\beta = 0$*) as in* (66).

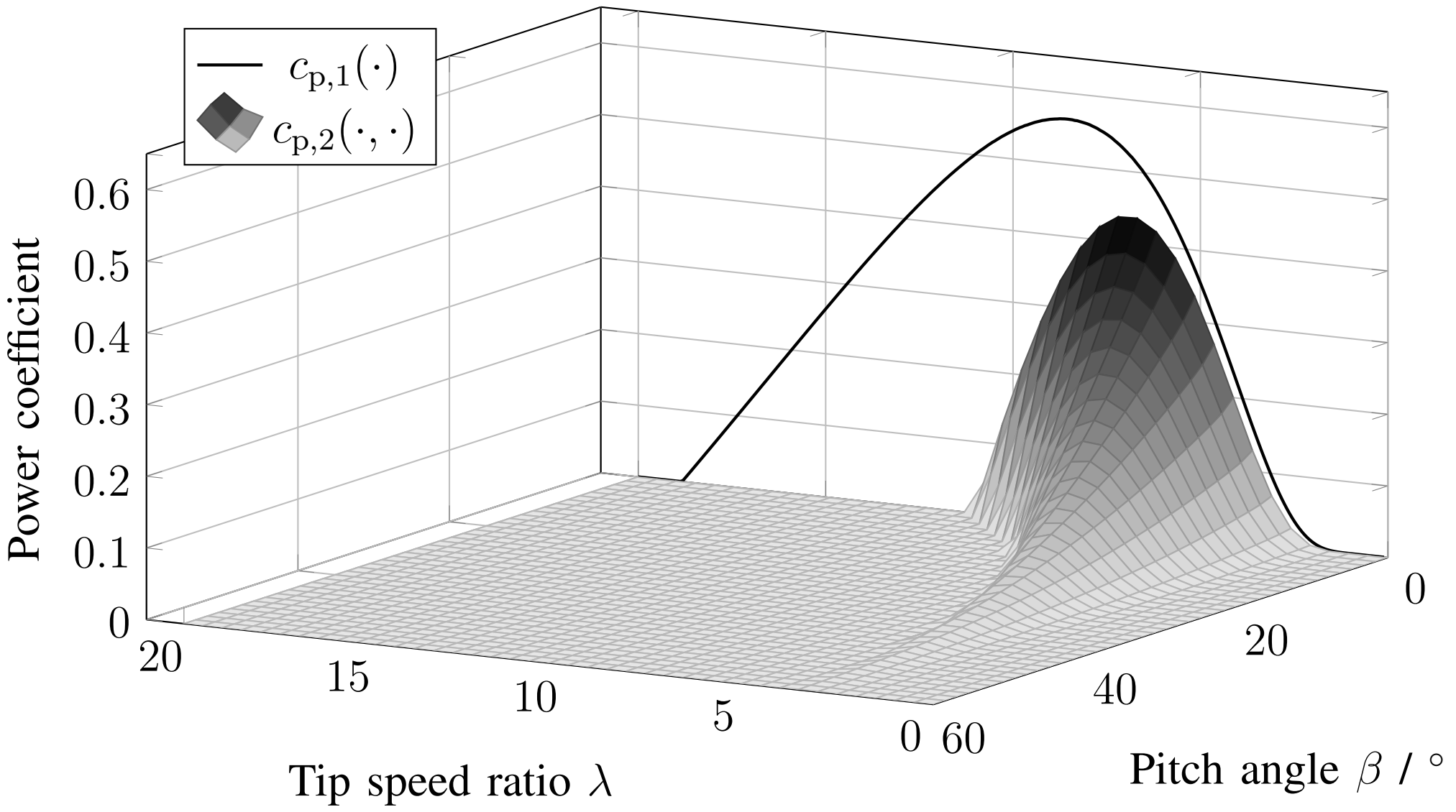

Fig. 18: *Power coefficient for a* $2\,\mathrm{MW}$ *wind turbine with* variable *pitch angle, expressed in* (67).

$c_{\mathrm{p},1}(\cdot)$ has a global maximum at $\lambda^\star = \left(\frac{\frac{46.4}{15.6}+2.0}{46.4} + 0.01\right)^{-1} \approx 8.53$ with $c^\star_{\mathrm{p},1} := c_{\mathrm{p},1}(\lambda^\star) = 0.564$.

- Power coefficient $c_{\mathrm{p},2}(\cdot,\cdot)$ *with* pitch control system:

$$c_{\mathrm{p},2}\colon\ \overline{\mathcal{D}} \to \mathbb{R}_{>0}, \qquad (\beta,\lambda) \mapsto$$
$$c_{\mathrm{p},2}(\beta,\lambda) := 0.73\left[151\left(\frac{1}{\lambda - 0.02\beta} - \frac{0.003}{\beta^3+1}\right) - 0.58\beta - 0.002\beta^{2.14} - 13.2\right]\cdot$$
$$\cdot \exp\left(-18.4\left(\frac{1}{\lambda - 0.02\beta} - \frac{0.003}{\beta^3+1}\right)\right). \tag{67}$$

For $\beta_0 = 0°$, $c_{\mathrm{p},2}(\beta_0,\cdot)$ has a maximum at $\lambda^\star := \lambda^\star_{\beta_0} = \left(\frac{\frac{151}{18.4}+13.2}{151} + 0.003\right)^{-1} \approx 6.91$ with $c^\star_{\mathrm{p},2} := c_{\mathrm{p},2}(0,\lambda^\star) = 0.441$.

The pitch angle $\beta$ is always in degrees (i.e., °) in (67). The graphs of $c_{\mathrm{p},1}(\cdot)$ and $c_{\mathrm{p},2}(\cdot,\cdot)$ are shown in Fig. 17 or 18, respectively. Both power coefficients are below the possible Betz limit of $c_{\mathrm{p,Betz}} = 16/27$. The maximum value of $c_{\mathrm{p},1}(\cdot)$ is above $c_{\mathrm{p},2}(\cdot,\cdot)$. This does not hold in general for all wind turbines, but seems to be a characteristic feature of the wind turbines considered in [32, p. 9], [33], [34].

*3) Aerodynamic turbine torque:* The turbine converts kinetic wind energy (translational energy) into rotational energy. The turbine thus exerts a torque on the drive train, which leads to an acceleration and rotation of the generator. If friction losses are neglected then, from turbine power $p_\mathrm{t}$, as in (62), turbine angular velocity $\omega_\mathrm{t}$ (in $\frac{\mathrm{rad}}{\mathrm{s}}$) and power coefficient $c_\mathrm{p}(\beta,\lambda)$ as in (63), the turbine torque $m_\mathrm{t}$ (in $\mathrm{N\,m}$) can be directly derived (see [35, (M 7.29)]) as follows

$$p_\mathrm{t} = m_\mathrm{t}\,\omega_\mathrm{t} \quad\Longrightarrow\quad m_\mathrm{t} \overset{(62),(63)}{=} \underbrace{\frac{1}{2}\,\varrho\,\pi\,r_\mathrm{t}^2\,v_\mathrm{w}^3}_{=p_\mathrm{w}}\,\frac{c_\mathrm{p}(\beta,\lambda)}{\omega_\mathrm{t}}. \tag{68}$$

With the definition of the tip speed ratio $\lambda = \lambda(v_\mathrm{w},\omega_\mathrm{t}) = \frac{r_\mathrm{t}\,\omega_\mathrm{t}}{v_\mathrm{w}}$ in (61) the turbine torque can be rewritten as

$$m_\mathrm{t}(v_\mathrm{w},\beta,\omega_\mathrm{t}) \overset{(68),(61)}{=} \frac{1}{2}\,\varrho\,\pi\,r_\mathrm{t}^2\,v_\mathrm{w}^3\,\frac{c_\mathrm{p}(v_\mathrm{w},\beta,\omega_\mathrm{t})}{\omega_\mathrm{t}} \tag{69}$$

or, with $\omega_\mathrm{t} = \frac{v_\mathrm{w}\lambda}{r_\mathrm{t}}$, as

$$m_\mathrm{t}(v_\mathrm{w},\beta,\lambda) \overset{(68),(61)}{=} \frac{1}{2}\,\varrho\,\pi\,r_\mathrm{t}^3\,v_\mathrm{w}^2\,\frac{c_\mathrm{p}(\beta,\lambda)}{\lambda}. \tag{70}$$

The turbine torque $m_\mathrm{t}(v_\mathrm{w},\beta,\lambda) = m_\mathrm{t}(v_\mathrm{w},\beta,\omega_\mathrm{t})$ is a *nonlinear function* of pitch angle $\beta$, wind speed $v_\mathrm{w}$ and tip speed ratio $\lambda$ or turbine angular velocity $\omega_\mathrm{t}$, respectively.

**Remark III.1.** *The approximation of the power coefficient $c_\mathrm{p}(\cdot,\cdot)$ according to* (63) *does* not *allow for the simulation of the start-up behavior of a wind turbine system. For start-up, $v_\mathrm{w} \geq v_\mathrm{cut-in} > 0\,\frac{\mathrm{m}}{\mathrm{s}}$ and $\beta = 0°$ (operation regime II) hold true, and the turbine is initially at a standstill, i.e., $\omega_\mathrm{t} = 0\,\frac{\mathrm{rad}}{\mathrm{s}}$* and *$\lambda = \frac{r_\mathrm{t}\omega_\mathrm{t}}{v_\mathrm{w}} = 0$. Considering the limit $\omega_\mathrm{t} \to 0\,\frac{\mathrm{rad}}{\mathrm{s}}$, the turbine torque becomes*

$$\lim_{\omega_\mathrm{t}\to 0\,\frac{\mathrm{rad}}{\mathrm{s}}} m_\mathrm{t}(v_\mathrm{w},0,\omega_\mathrm{t}) \overset{(69)}{=} \lim_{\omega_\mathrm{t}\to 0\,\frac{\mathrm{rad}}{\mathrm{s}}} \underbrace{\frac{1}{2}\varrho r_\mathrm{t}^2 v_\mathrm{w}^3}_{>0\,\frac{\mathrm{kg\,m^2}}{\mathrm{s^3}}} \frac{c_\mathrm{p}(0,\lambda)}{\omega_\mathrm{t}} \overset{(70)}{=} \lim_{\lambda\to 0} \underbrace{\frac{1}{2}\varrho r_\mathrm{t}^3 v_\mathrm{w}^2}_{>0\,\frac{\mathrm{kg\,m^2}}{\mathrm{s^3}}} \frac{c_\mathrm{p}(0,\lambda)}{\lambda}$$

$$\overset{(E_4)}{\propto} \lim_{\lambda\to 0} \frac{e^{-\frac{1}{\lambda}}}{\lambda} \overset{\text{[36, Satz III.6.5(iii)]}}{=} \lim_{x\to\infty} \frac{x}{e^x} = 0.$$

*Therefore, there is* no *accelerating torque at standstill; which shows the limitations of using of the power coefficient approximation* (63) *in system modeling. The approximation* (63) *only yields physically meaningful results for $\lambda > 0$.*

## *B. Gearbox and gear transmission*

A gearbox transmits the mechanical power of the turbine shaft to the rotor of the generator. In wind turbines, the angular velocity $\omega_\mathrm{t}$ (in $\frac{\mathrm{rad}}{\mathrm{s}}$) of the turbine is significantly slower than the (nominal) angular velocity $\omega_\mathrm{m}$ (in $\frac{\mathrm{rad}}{\mathrm{s}}$) of the machine (generator). Therefore, the use of an step-up gearbox with ratio $g_\mathrm{r} > 1$ is necessary (exceptions are wind turbine systems with a "Direct Drive"[12] configuration). Based on the explanation in [37, Sec. 1.1], the modeling of a gearbox is now briefly presented. First, the constant gear ratio

$$\forall\, t \geq 0\,\mathrm{s}: \qquad g_\mathrm{r} := \frac{\omega_\mathrm{m}(t)}{\omega_\mathrm{t}(t)} > 1 \tag{71}$$

is defined which relates turbine angular velocity $\omega_\mathrm{t}$ and machine angular velocity $\omega_\mathrm{m}$. If the turbine shaft rotates with $\omega_\mathrm{t}$ this causes the machine shaft to rotate with $\omega_\mathrm{m} = g_\mathrm{r}\omega_\mathrm{t}$. Figure 19 (see [37, p. 13–15]) shows a simplified example of the construction of a rotatory/rotatory gearbox. In addition to the angular velocities

[12] In "Direct Drive" turbines, the generator is connected directly and rigidly to the rotor of the wind turbine. Thus, the generator and turbine rotate at the same angular velocity $\omega_\mathrm{t}(t) = \omega_\mathrm{m}(t)$ for all $t \geq 0\,\mathrm{s}$.

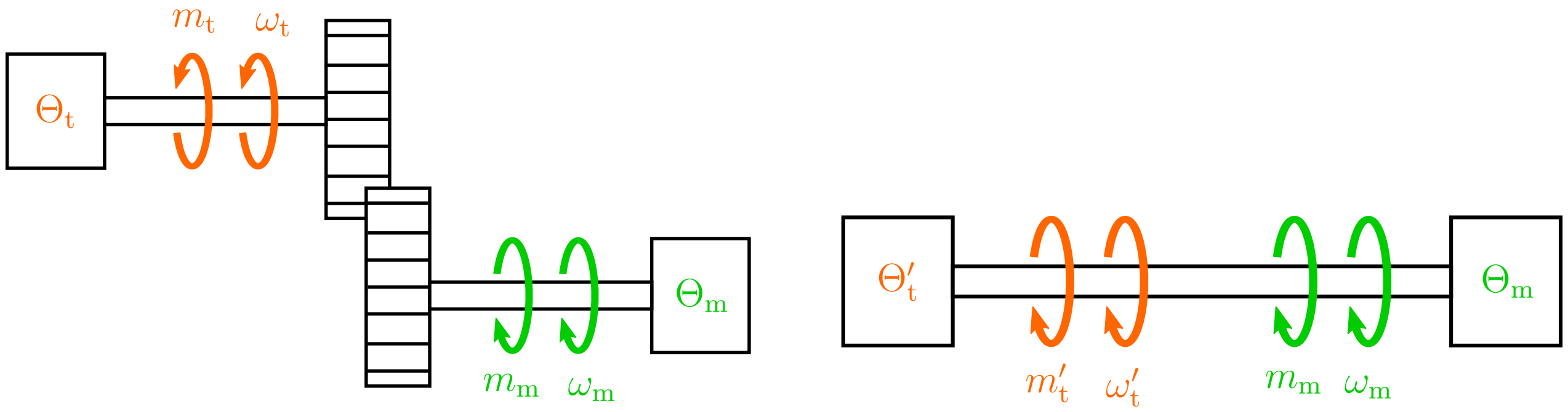

(a) *Drive train with gearbox.*

(b) *Drive train with turbine-side quantities related to the machine-side.*

Fig. 19: *Drive train of a wind turbine with gearbox.*

$\omega_\mathrm{t}$ and $\omega_\mathrm{m}$, the turbine torque $m_\mathrm{t}$ (in $\mathrm{N\,m}$) (as in (69) or (70)) and the machine torque $m_\mathrm{m}$ (in $\mathrm{N\,m}$), as well as, the turbine inertia $\Theta_\mathrm{t}$ (in $\mathrm{kg\,m^2}$) and the machine inertia $\Theta_\mathrm{m}$ (in $\mathrm{kg\,m^2}$) are shown in Fig. 19. Friction[13], gearbox inertia and gear play[14] are neglected for simplicity.

For the modeling of the electrical machine as described in III-D, the turbine variables $m_\mathrm{t}$ and $\omega_\mathrm{t}$ are converted to the machine side in order to take into account their effect on the machine-side shaft. Figure 19 should therefore be replaced by Fig. 19b. The quantities related to the machine shaft are indicated by $'$. The machine quantities do not have to be adjusted. It follows

$$\forall t \geq 0\,\mathrm{s}\colon\; \omega'_\mathrm{m}(t) = \omega_\mathrm{m}(t), \quad m'_\mathrm{m}(t) = m_\mathrm{m}(t) \quad \text{and} \quad \Theta'_\mathrm{m} = \Theta_\mathrm{m}. \tag{72}$$

The relationship for the turbine variables $\omega'_\mathrm{t}$, $\Theta'_\mathrm{t}$ and $m'_\mathrm{t}$ which are converted to the machine side are now derived. The starting point for this is the conversation of energy in the drive train (i.e., rotational energy without friction loss)

$$\forall t \geq 0\,\mathrm{s}\colon\; \frac{1}{2}\left(\Theta_\mathrm{t}\omega_\mathrm{t}(t)^2 + \Theta_\mathrm{m}\omega_\mathrm{m}(t)^2\right) = \frac{1}{2}\left(\Theta'_\mathrm{t}\omega'_\mathrm{t}(t)^2 + \Theta_\mathrm{m}\omega_\mathrm{m}(t)^2\right) \tag{73}$$

and the conservation of power

$$\forall t \geq 0\,\mathrm{s}\colon\; m_\mathrm{t}(t)\omega_\mathrm{t}(t) = \omega'_\mathrm{t}(t)m'_\mathrm{t}(t). \tag{74}$$

According to Fig. 19b, the identity $\omega'_\mathrm{t}(t) = \omega_\mathrm{m}(t)$ must hold for all $t \geq 0\,\mathrm{s}$. It then follows that

$$\forall t \geq 0\,\mathrm{s}\colon\; \omega'_\mathrm{t}(t) = \omega_\mathrm{m}(t) \overset{(71)}{=} g_\mathrm{r}\omega_\mathrm{t}(t) \implies m'_\mathrm{t}(t) \overset{(74)}{=} \frac{m_\mathrm{t}(t)}{g_\mathrm{r}}. \tag{75}$$

It can be inferred from (73) that

$$\forall t \geq 0\,\mathrm{s}\colon \qquad \Theta'_\mathrm{t} \overset{(73)}{=} \Theta_\mathrm{t}\frac{\omega_\mathrm{t}(t)^2}{\omega'_\mathrm{t}(t)^2} \overset{(75)}{=} \Theta_\mathrm{t}\frac{\cancel{\omega_\mathrm{t}(t)^2}}{g_\mathrm{r}^2\cancel{\omega_\mathrm{t}(t)^2}} = \frac{\Theta_\mathrm{t}}{g_\mathrm{r}^2}. \tag{76}$$

This allows for the total inertia—relative to the machine side—to be

$$\Theta := \Theta'_\mathrm{t} + \Theta_\mathrm{m} \overset{(76)}{=} \frac{\Theta_\mathrm{t}}{g_\mathrm{r}^2} + \Theta_\mathrm{m}. \tag{77}$$

In summary, the dynamics of the *machine-side* mechanics can be expressed as follows

$$\tfrac{\mathrm{d}}{\mathrm{d}t}\omega_\mathrm{m}(t) = \frac{1}{\Theta}\big(m'_\mathrm{t}(t) + m_\mathrm{m}(t)\big) \overset{(75)}{=} \frac{1}{\Theta}\left(\frac{m_\mathrm{t}(t)}{g_\mathrm{r}} + m_\mathrm{m}(t)\right), \quad \omega_\mathrm{m}(0\,\mathrm{s}) = \omega_\mathrm{m}^0. \tag{78}$$

This completes the modeling of the mechanical components (turbine and gear transmission). The following sections describe the electrical components of the wind turbine system.

[13] A detailed discussion of on the (controlled) system can be found in [38] or Chapter [39].

[14] Gear play (or backlash) and its compensation are dealt with in [40, Sec. 6.4].

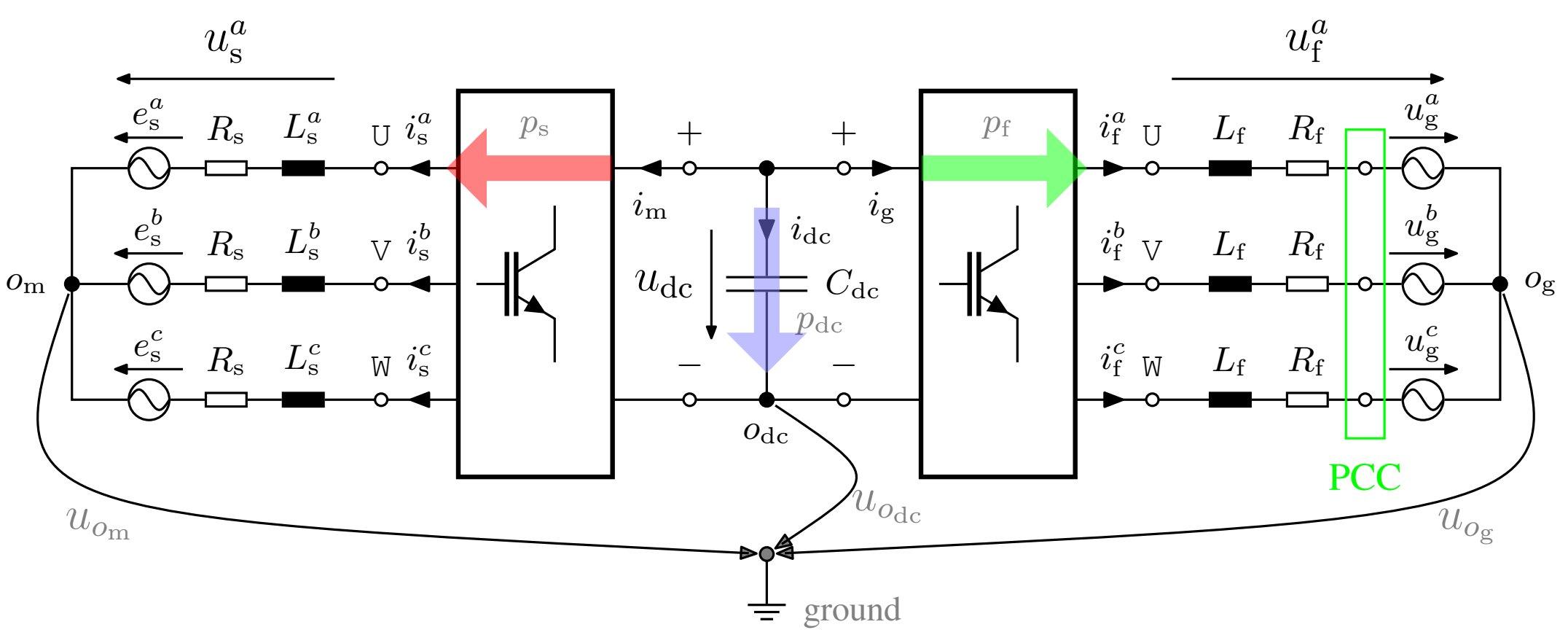

Fig. 20: *Electrical network of a rotating field generator (left) with a back-to-back converter, grid-side filter, Point of Common Coupling (PCC) and an ideal grid (right, neglecting the transformer).*

## C. *Electrical network*

For the modeling the electrical components, the (simplified) electrical network of a wind turbine with permanent-magnet synchronous generator is considered as shown in Fig. 20. The machine-side network (left) shows stator phase voltages $\boldsymbol{u}_{\mathrm{s}}^{abc} = (u_{\mathrm{s}}^{a}, u_{\mathrm{s}}^{b}, u_{\mathrm{s}}^{c})^{\top}$ (in V)$^3$, stator phase current $\boldsymbol{i}_{\mathrm{s}}^{abc} = (i_{\mathrm{s}}^{a}, i_{\mathrm{s}}^{b}, i_{\mathrm{s}}^{c})^{\top}$ (in A)$^3$, stator phase resistance $R_{\mathrm{s}}$ (in $\frac{\mathrm{V}}{\mathrm{A}}$), phase inductances $L_{\mathrm{s}}^{a}$, $L_{\mathrm{s}}^{b}$, $L_{\mathrm{s}}^{c}$ (in $\frac{\mathrm{V\,s}}{\mathrm{A}}$) and induced phase voltages $\boldsymbol{e}_{\mathrm{s}}^{abc} = (e_{\mathrm{s}}^{a}, e_{\mathrm{s}}^{b}, e_{\mathrm{s}}^{c})^{\top}$ (in V)$^3$. The grid-side network (right) is comprised of filter phase voltages $\boldsymbol{u}_{\mathrm{f}}^{abc} = (u_{\mathrm{f}}^{a}, u_{\mathrm{f}}^{b}, u_{\mathrm{f}}^{c})^{\top}$ (in V)$^3$, filter phase currents $\boldsymbol{i}_{\mathrm{f}}^{abc} = (i_{\mathrm{f}}^{a}, i_{\mathrm{f}}^{b}, i_{\mathrm{f}}^{c})^{\top}$ (in A)$^3$, filter resistance $R_{\mathrm{f}}$ (in $\frac{\mathrm{V}}{\mathrm{A}}$), filter inductance $L_{\mathrm{f}}$ (in $\frac{\mathrm{V\,s}}{\mathrm{A}}$) and the grid phase voltages $\boldsymbol{u}_{\mathrm{g}}^{abc} = (u_{\mathrm{g}}^{a}, u_{\mathrm{g}}^{b}, u_{\mathrm{g}}^{c})^{\top}$ (in V)$^3$. The (transformed) grid voltage $\boldsymbol{u}_{\mathrm{g}}^{abc}$ is measured at the Point of Common Coupling (PCC). The transmission ratio of the transformer (not shown in Fig. 10) is not explicitly modeled.

The back-to-back converter consists of two fully-controlled converters which share a common DC-link. Machine-side and grid-side converters exchange the machine-side power $p_{\mathrm{m}}$ (in W) and the grid-side power $p_{\mathrm{f}}$ (in W) with the DC-link. In the continuous operation of the wind turbine, the DC-link capacitor $C_{\mathrm{dc}}$ (in $\frac{\mathrm{A\,s}}{\mathrm{V}}$) on average is *not* charged or discharged. On average, no DC-link power, $p_{\mathrm{dc}}$ (in W), is exchanged within the circuit and the DC-link voltage $u_{\mathrm{dc}}$ (in V) stays (nearly) constant.

The reference points $o_{\mathrm{m}}$ (machine-side star point), $o_{\mathrm{g}}$ (grid-side star point) and $o_{\mathrm{dc}}$ (negative potential of the DC-link circuit) are not connected to ground. The reference potentials $u_{o_{\mathrm{m}}}$ (in V), $u_{o_{\mathrm{g}}}$ (in V) and $u_{o_{\mathrm{dc}}}$ (in V) are thus freely floating and can vary with respect to ground.

## D. *Electrical machine (generator)*

In this chapter, only a permanent-magnet synchronous generator (PMSG) is considered. Using Kirchhoff's laws, the electrical circuits in Fig. 20 can be analyzed and described as follows

$$\boldsymbol{u}_{\mathrm{s}}^{abc}(t) = R_{\mathrm{s}}\boldsymbol{i}_{\mathrm{s}}^{abc}(t) + \underbrace{\frac{\mathrm{d}}{\mathrm{d}t}\begin{pmatrix} L_{\mathrm{s}}^{a} i_{\mathrm{s}}^{a}(t) \\ L_{\mathrm{s}}^{b} i_{\mathrm{s}}^{b}(t) \\ L_{\mathrm{s}}^{c} i_{\mathrm{s}}^{c}(t) \end{pmatrix} + \begin{pmatrix} e_{\mathrm{s}}^{a}(t) \\ e_{\mathrm{s}}^{b}(t) \\ e_{\mathrm{s}}^{c}(t) \end{pmatrix}}_{=:\frac{\mathrm{d}}{\mathrm{d}t}\boldsymbol{\psi}_{\mathrm{s}}^{abc}(t)}, \quad \boldsymbol{i}_{\mathrm{s}}^{abc}(0\,\mathrm{s}) = \boldsymbol{i}_{\mathrm{s}}^{abc,0}\ (\text{in A})^3. \tag{79}$$

Applying the Clarke transformation (31) leads to the fundamental model of a permanent-magnet synchronous generator (see [28, Sec. 16.6]):

$$\left.\begin{aligned}
&\text{Stator:} && \boldsymbol{u}_\mathrm{s}^s(t) = R_\mathrm{s}\boldsymbol{i}_\mathrm{s}^s(t) + \tfrac{\mathrm{d}}{\mathrm{d}t}\boldsymbol{\psi}_\mathrm{s}^s(t) && , \boldsymbol{\psi}_\mathrm{s}^s(0\,\mathrm{s}) = \boldsymbol{\psi}_\mathrm{s}^{s,0}\,(\text{in V s})^2 \\
&\text{Flux:} && \boldsymbol{\psi}_\mathrm{s}^k(t) = \underbrace{\begin{bmatrix} L_\mathrm{s}^d & 0\,\frac{\mathrm{V\,s}}{\mathrm{A}} \\ 0\,\frac{\mathrm{V\,s}}{\mathrm{A}} & L_\mathrm{s}^q \end{bmatrix}}_{=:\boldsymbol{L}_\mathrm{s}^k\in\mathbb{R}^{2\times 2}} \boldsymbol{i}_\mathrm{s}^k(t) + \underbrace{\begin{pmatrix} \psi_\mathrm{pm} \\ 0\,\mathrm{V\,s} \end{pmatrix}}_{=:\boldsymbol{\psi}_\mathrm{pm}^k} && \\
&\text{Mechan.:} && \tfrac{\mathrm{d}}{\mathrm{d}t}\,\omega_\mathrm{m}(t) \overset{(78)}{=} \tfrac{1}{\Theta}\left(\tfrac{m_\mathrm{t}(v_\mathrm{w},\beta,\omega_\mathrm{t})}{g_\mathrm{r}} + m_\mathrm{m}(t)\right) && , \omega_\mathrm{m}(0\,\mathrm{s}) = \omega_\mathrm{m}^0\,(\text{in } \tfrac{\mathrm{rad}}{\mathrm{s}}) \\
&\text{Torque:} && m_\mathrm{m}(t) = \tfrac{3}{2}\,n_\mathrm{p}\,\boldsymbol{i}_\mathrm{s}^s(t)^\top \boldsymbol{J}\boldsymbol{\psi}_\mathrm{s}^s(t). &&
\end{aligned}\right\} \tag{80}$$

Here, the variables are stator voltage $\boldsymbol{u}_\mathrm{s}^s$ (in V)$^2$, stator current $\boldsymbol{i}_\mathrm{s}^s$ (in A)$^2$, stator resistance $R_\mathrm{s}$ (in $\Omega$), stator inductance matrix $\boldsymbol{L}_\mathrm{s}^k$ (in $\frac{\mathrm{V\,s}}{\mathrm{A}})^{2\times 2}$ (with inductances $L_\mathrm{s}^d$, $L_\mathrm{s}^q > 0\,\frac{\mathrm{V\,s}}{\mathrm{A}}$), flux linkage of the permanent magnet $\boldsymbol{\psi}_\mathrm{pm}^k = (\psi_\mathrm{pm}, 0)^\top$ (in V s)$^2$ (with $\psi_\mathrm{pm} > 0\,$(in V s)), number of pole pairs $n_\mathrm{p}$, machine angular velocity $\omega_\mathrm{m}$ (in $\frac{\mathrm{rad}}{\mathrm{s}}$), gear ratio $g_\mathrm{r} \geq 1$, total inertia of the wind turbine (referring to the machine side) $\Theta = \Theta_\mathrm{m} + \Theta_\mathrm{t}/g_\mathrm{r}^2$ (in kg m$^2$), machine torque is $m_\mathrm{m}$ (in N m) and turbine torque $m_\mathrm{t}$ (in N m) as in (69). In view of (75), it follows that $\omega_\mathrm{m} = g_\mathrm{r}\omega_\mathrm{t}$. The electrical angular velocity

$$\omega_\mathrm{r}(t) = n_\mathrm{p}\,\omega_\mathrm{m}(t) \quad (\text{in } \frac{\mathrm{rad}}{\mathrm{s}}) \tag{81}$$

depends on the mechanical angular velocity $\omega_\mathrm{m}$ and the number of pole pairs $n_\mathrm{p}$. The stator currents $\boldsymbol{i}_\mathrm{s}^{abc}(t)$ and the mechanical angular velocity $\omega_\mathrm{m}(t)$ are available as measured variables for feedback control.

### *E. Point of Common Coupling (PCC), transformer and grid*

At the grid-side feed-in point or *Point of Common Coupling (PCC)*, the generated power is fed into the grid (see Fig. 10). The grid voltages $\boldsymbol{u}_\mathrm{g}^{abc}(t) = \big(u_\mathrm{g}^a(t),\;\, u_\mathrm{g}^b(t),\;\, u_\mathrm{g}^c(t)\big)^\top$ (in V)$^3$ are measured just before (or after) the transformer. The transformer steps up the voltage to a higher voltage level according to the voltage level at the PCC (for example, to the medium voltage level of the power grid). The measured grid voltages

$$\boldsymbol{u}_\mathrm{g}^{abc}(t) := \begin{pmatrix} u_\mathrm{g}^a(t) \\ u_\mathrm{g}^b(t) \\ u_\mathrm{g}^c(t) \end{pmatrix} = \hat{u}_\mathrm{g} \begin{pmatrix} \cos(\omega_\mathrm{g}t + \alpha_0) \\ \cos(\omega_\mathrm{g}t - 2/3\pi + \alpha_0) \\ \cos(\omega_\mathrm{g}t - 4/3\pi + \alpha_0) \end{pmatrix} \tag{82}$$

are assumed to be balanced (ideal) and have constant amplitude $\hat{u}_\mathrm{g}$ (in V), constant grid frequency $\omega_\mathrm{g}$ (in $\frac{\mathrm{rad}}{\mathrm{s}}$) and phase offset $\alpha_0$ (in rad). In the European power grid, the grid frequency is strictly defined around $f_\mathrm{g} = 50\,\mathrm{Hz} \pm 0.5\,\mathrm{Hz}$ to ensure frequency stability of the grid (see [19, p. 13,20,27]). In Fig. 21, balanced grid voltages $\boldsymbol{u}_\mathrm{g}^{abc}(\cdot) = (u_\mathrm{g}^a(\cdot),\, u_\mathrm{g}^b(\cdot),\, u_\mathrm{g}^c(\cdot))^\top$ with angular frequency $\omega_\mathrm{g} = 2\pi f_\mathrm{g} = 100\pi\,\frac{\mathrm{rad}}{\mathrm{s}}$, voltage amplitude $\hat{u}_\mathrm{g} = 2\,700\,\mathrm{V}$ (at PCC) and initial phase offset $\alpha_0 = 0\,\mathrm{rad}$ are shown.

### *F. Filter*

In order to be able to induce sinusoidal phase currents into the network, a line filter must be used. A simple $RL$-filter (in each phase) with filter inductance $L_\mathrm{f}$ (in $\frac{\mathrm{V\,s}}{\mathrm{A}}$) and filter resistance $R_\mathrm{f}$ (in $\frac{\mathrm{V}}{\mathrm{A}}$) will be discussed. The grid-side converter generates the (filter) voltages $\boldsymbol{u}_\mathrm{f}^{abc} = (u_\mathrm{f}^a, u_\mathrm{f}^b, u_\mathrm{f}^c)^\top$ (in V)$^3$ (or $\boldsymbol{u}_\mathrm{f}^{ltl}$ (in V)$^3$, respectively) which are applied to the filter and, due to the inductance $L_\mathrm{f}$, lead to (approximately) sinusoidal phase (filter) currents $\boldsymbol{i}_\mathrm{f}^{abc} = (i_\mathrm{f}^a, i_\mathrm{f}^b, i_\mathrm{f}^c)^\top$ (in A)$^3$. At the filter resistance $R_\mathrm{f}$, the copper losses $R_\mathrm{f}\|\boldsymbol{i}_\mathrm{f}^{abc}\|^2$ (in W) are dissipated and converted into heat. The grid-side electrical network with an ideal grid voltage $\boldsymbol{u}_\mathrm{g}^{abc}$ as in (82) is shown in Fig. 20. According to Kirchoff's voltage law, the following grid-side dynamics are derived

$$\boldsymbol{u}_\mathrm{f}^{abc}(t) = R_\mathrm{f}\boldsymbol{i}_\mathrm{f}^{abc}(t) + L_\mathrm{f}\tfrac{\mathrm{d}}{\mathrm{d}t}\boldsymbol{i}_\mathrm{f}^{abc}(t) + \boldsymbol{u}_\mathrm{g}^{abc}(t), \quad \boldsymbol{i}_\mathrm{f}^{abc}(0\,\mathrm{s}) = \boldsymbol{i}_\mathrm{f}^{abc,0}\,(\text{in A})^3. \tag{83}$$

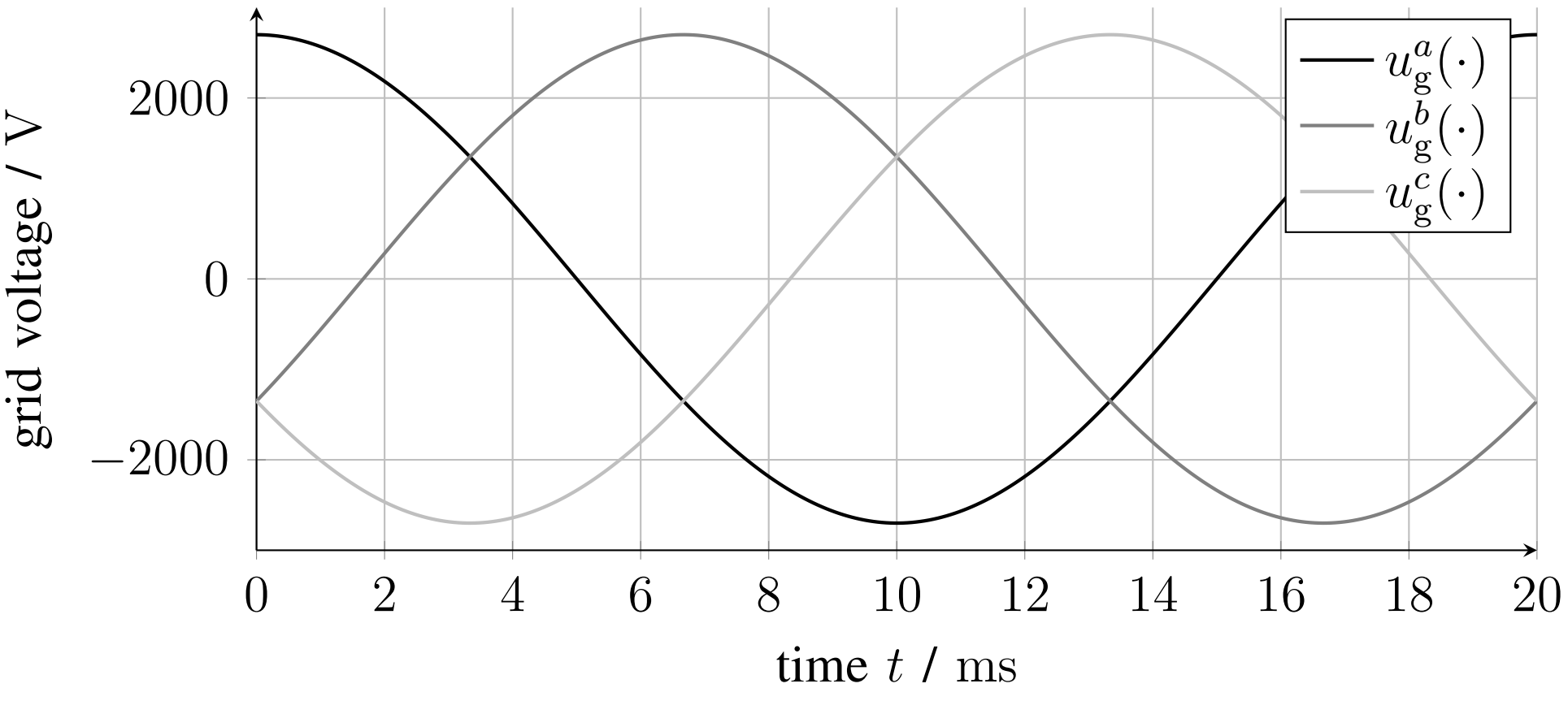

Fig. 21: *Balanced (ideal) grid voltage vector* $\boldsymbol{u}_\mathrm{g}^{abc}(\cdot) = (u_\mathrm{g}^a(\cdot),\, u_\mathrm{g}^b(\cdot),\, u_\mathrm{g}^c(\cdot))^\top$ *with* $\omega_\mathrm{g} = 2\pi f_\mathrm{g} = 100\pi\,\frac{\mathrm{rad}}{\mathrm{s}}$, $\hat{u}_\mathrm{g} = 2\,700\,\mathrm{V}$ *and* $\alpha_0 = 0\,\mathrm{rad}$.

Using the Clarke transformation, the representation (83) in the three-phase $abc$-reference frame is transformed to the two-phase $s$-reference frame with the transformed quantities $\boldsymbol{u}_\mathrm{f}^s = (u_\mathrm{f}^\alpha, u_\mathrm{f}^\beta)^\top$ (in V)$^2$, $\boldsymbol{i}_\mathrm{f}^s = (i_\mathrm{f}^\alpha, i_\mathrm{f}^\beta)^\top$ (in A)$^2$ and $\boldsymbol{u}_\mathrm{g}^s = (u_\mathrm{g}^\alpha, u_\mathrm{g}^\beta)^\top$ (in V)$^2$. One obtains

$$\begin{aligned}\boldsymbol{u}_\mathrm{f}^s(t) &\overset{(31)}{=} \boldsymbol{T}_\mathrm{c}\boldsymbol{u}_\mathrm{f}^{abc}(t) = R_\mathrm{f}\boldsymbol{T}_\mathrm{c}\boldsymbol{i}_\mathrm{f}^{abc}(t) + L_\mathrm{f}\tfrac{\mathrm{d}}{\mathrm{d}t}\boldsymbol{T}_\mathrm{c}\boldsymbol{i}_\mathrm{f}^{abc}(t) + \boldsymbol{T}_\mathrm{c}\boldsymbol{u}_\mathrm{g}^{abc}(t)\\ &= R_\mathrm{f}\boldsymbol{i}_\mathrm{f}^s(t) + L_\mathrm{f}\tfrac{\mathrm{d}}{\mathrm{d}t}\boldsymbol{i}_\mathrm{f}^s(t) + \boldsymbol{u}_\mathrm{g}^s(t), \qquad \boldsymbol{i}_\mathrm{f}^s(0\,\mathrm{s}) = \boldsymbol{T}_\mathrm{c}\boldsymbol{i}_\mathrm{f}^{s,0} \text{ (in A)}^2.\end{aligned} \tag{84}$$

**Remark III.2.** *In wind turbine systems, also $LCL$-filters could used instead of $RL$-filters. The design of an $LCL$-filter allows for smaller inductances. Thus, an $LCL$-filter can be made smaller than an $RL$-filter. In [25, Ch. 11], there is a detailed discussion of the design of an $LCL$-filter. The control of grid-side power converters connected to the grid via $LCL$-filters is discussed in e.g. [41, 42].*

## *G. Back-to-back converter*

Although the multi-level converters for the regulation of wind power plants are likely to be used in the future due to the ever-increasing power ratings [43], the widely used two-level back-to-back converter should still be understood and will be explained briefly in the following sections. A detailed discussion can be found in [44, Sec. 8.3.4].

A back-to-back converter (or inverter with Active Front End) consists of two converters which share a common DC-link voltage (see Fig. 22). The machine-side converter feeds the electrical machine (generator) with the line-to-line stator voltages $\boldsymbol{u}_\mathrm{s}^{ltl} = (u_\mathrm{s}^{ab},\, u_\mathrm{s}^{bc},\, u_\mathrm{s}^{ca})^\top$ (in V)$^3$, while the grid-side converter applies the line-to-line voltages $\boldsymbol{u}_\mathrm{f}^{ltl} = (u_\mathrm{f}^{ab},\, u_\mathrm{f}^{bc},\, u_\mathrm{f}^{ca})^\top$ (in V)$^3$ to the grid filter. The task of the back-to-back converter is to generate the phase voltages $\boldsymbol{u}_\mathrm{s}^{abc}$ (in V)$^3$ in the machine and the phase voltages $\boldsymbol{u}_\mathrm{f}^{abc}$ (in V)$^3$ in the filter in accordance with the predetermined reference voltages $\boldsymbol{u}_\mathrm{s,ref}^{abc}$ and $\boldsymbol{u}_\mathrm{f,ref}^{abc}$, respectively. The output voltages of the machine-side or grid-side converter depend on the DC-link voltage $u_\mathrm{dc}$ (in V) across the DC-link capacitance $C_\mathrm{dc}$ (in $\frac{\mathrm{A\,s}}{\mathrm{V}}$). The machine-side and grid-side converters are controlled by applying adequate switching vectors $\boldsymbol{s}_\mathrm{m}^{abc} = (s_\mathrm{m}^a,\, s_\mathrm{m}^b,\, s_\mathrm{m}^c)^\top$ or $\boldsymbol{s}_\mathrm{g}^{abc} = (s_\mathrm{g}^a,\, s_\mathrm{g}^b,\, s_\mathrm{g}^c)^\top$ (e.g. coming from a modulator), respectively.

*1) Switching vectors, voltage vectors and voltage hexagon:* In Fig. 23, the eight possible voltage vectors $\boldsymbol{u}_{000}^s$, $\boldsymbol{u}_{100}^s$, …, and $\boldsymbol{u}_{111}^s$ of a two-level inverter are illustrated in the $s$-reference frame for the eight possible switching vectors $(\boldsymbol{s}^{abc})^\top \in \{\mathtt{000}, \mathtt{100}, \ldots, \mathtt{111}\}$. The eight voltage vectors are:

$$\begin{aligned}(\boldsymbol{s}^{abc})^\top = \mathtt{000} \vee \mathtt{111} \quad &\Longrightarrow \quad \boldsymbol{u}_{000}^s = \boldsymbol{u}_{111}^s = (0\,\mathrm{V}, 0\,\mathrm{V})^\top,\\ (\boldsymbol{s}^{abc})^\top = \mathtt{100} \quad &\Longrightarrow \quad \boldsymbol{u}_{100}^s = -\boldsymbol{u}_{011}^s = \left(\tfrac{2}{3}u_\mathrm{dc},\ \ 0\,\mathrm{V}\right)^\top,\\ (\boldsymbol{s}^{abc})^\top = \mathtt{110} \quad &\Longrightarrow \quad \boldsymbol{u}_{110}^s = -\boldsymbol{u}_{001}^s = \left(\tfrac{1}{3}u_\mathrm{dc},\ \tfrac{1}{\sqrt{3}}u_\mathrm{dc}\right)^\top \text{ and}\end{aligned}$$

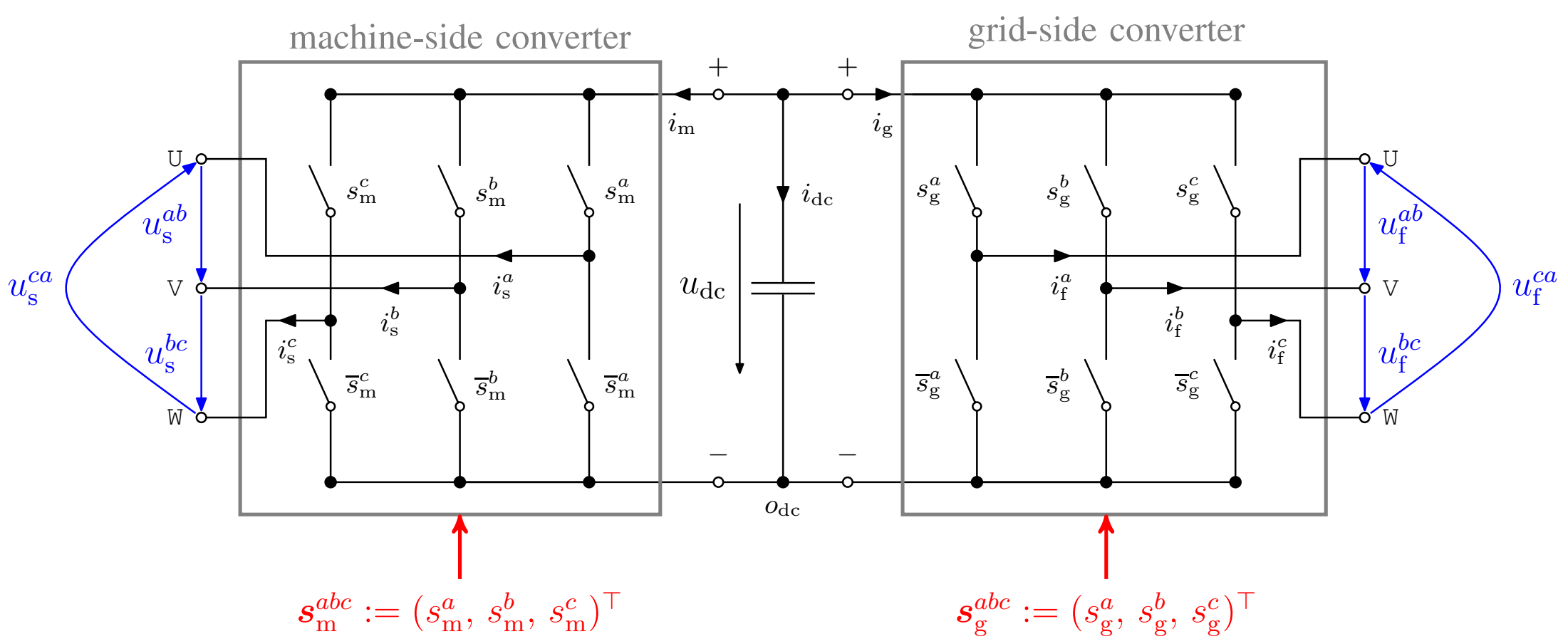

Fig. 22: *Back-to-back converter: Machine-side and grid-side converter (with ideal switches, without free-wheel diodes) with common DC-link.*

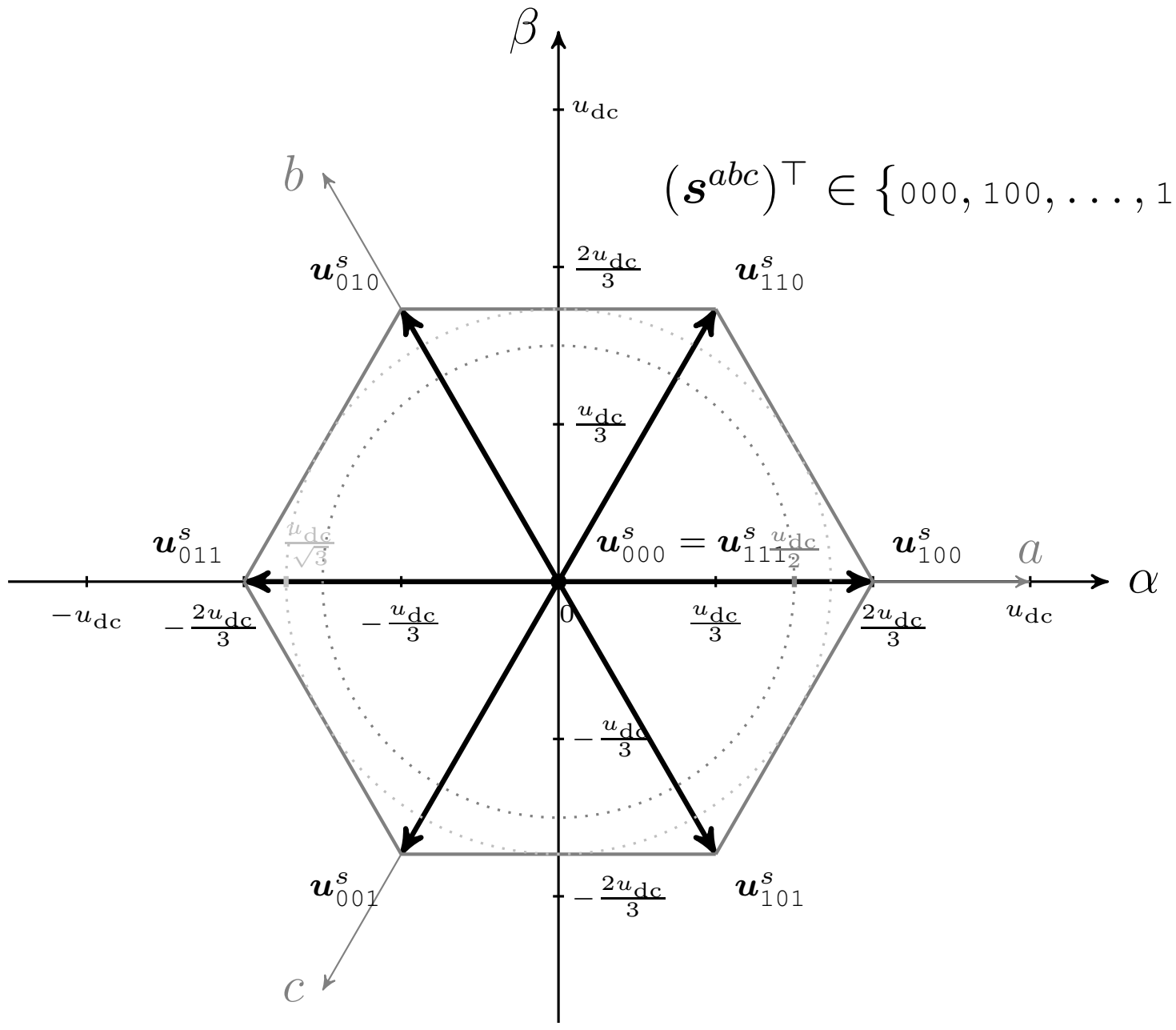

Fig. 23: *Voltage hexagon: The* eight *possible switching vectors* $\boldsymbol{s}^{abc}$ *and the resulting voltage vectors* $\boldsymbol{u}^s_{000}, \boldsymbol{u}^s_{100}, \dots, \boldsymbol{u}^s_{111}$ *of a two-level voltage source converter.*

$$(\boldsymbol{s}^{abc})^\top = 010 \quad \Longrightarrow \quad \boldsymbol{u}^s_{010} = -\boldsymbol{u}^s_{101} = \left(-\tfrac{1}{3}u_{\rm dc},\, \tfrac{1}{\sqrt{3}}u_{\rm dc}\right)^\top.$$

The six *active* voltage vectors have the same amplitude, i.e.,

$$\forall\, t \geq 0\,\mathrm{s}: \; \|\boldsymbol{u}^s_{100}(t)\| = \|\boldsymbol{u}^s_{010}(t)\| = \cdots = \|\boldsymbol{u}^s_{011}(t)\| = \frac{2}{3}u_{\rm dc}(t),$$

and thus depend on the (possibly time-varying) DC-link voltage $u_{\rm dc}(t)$; whereas the *zero* (inactive) vectors have the amplitude $\|\boldsymbol{u}^s_{000}(t)\| = \|\boldsymbol{u}^s_{111}(t)\| = 0\,\mathrm{V}$.

**Remark III.3.** *The "circular areas" in Fig. 23 with radius* $u_{\rm dc}/2$ *and* $u_{\rm dc}/\sqrt{3}$ *(see [44, p. 658–720] and [45, p. 132–136]) within the voltage hexagon can be reproduced by* pulse width modulation *(PWM) or* space vector modulation *(SVM), respectively.*

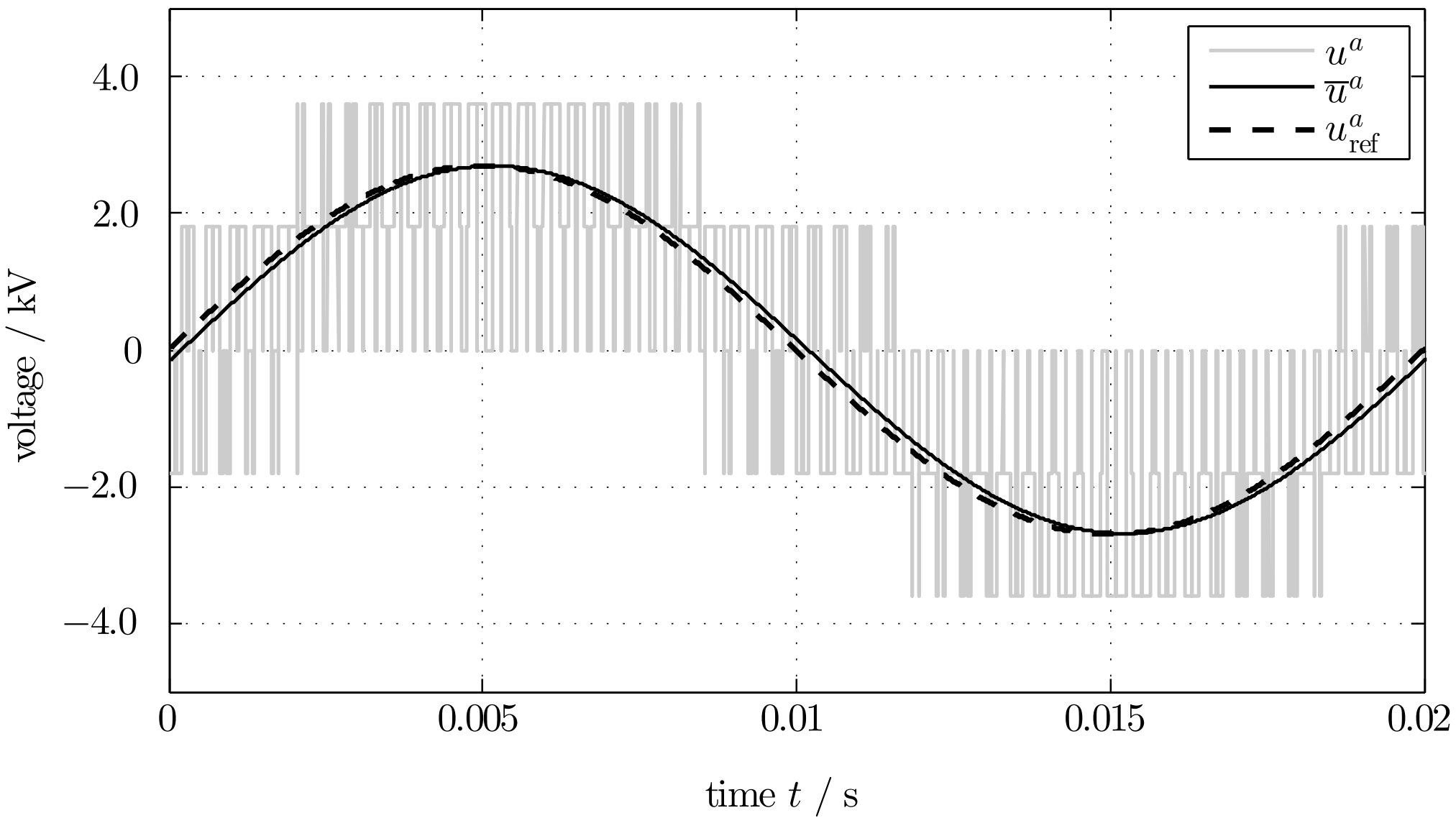

Fig. 24: *Example delay time of a voltage source converter with a switching frequency of* $2.5\,\mathrm{kHz}$*.*

*2) DC-link circuit:* The DC-link capacitor $C_{\rm dc}$ (in $\frac{\rm A\,s}{\rm V}$) is charged or discharged via the DC-link current $i_{\rm dc} = -i_{\rm m} - i_{\rm g}$ (in A) which depends on the machine-side and grid-side currents, $i_{\rm m}$ (in A) and $i_{\rm g}$ (in A), respectively (see Figure 20). Therefore the following relationship applies to the dynamics of the DC-link voltage:

$$\tfrac{\rm d}{{\rm d}t}\, u_{\rm dc}(t) = \frac{1}{C_{\rm dc}} i_{\rm dc}(t) = \frac{1}{C_{\rm dc}}\Big( - i_{\rm g}(t) - i_{\rm m}(t)\Big), \qquad u_{\rm dc}(0\,{\rm s}) = u^0_{\rm dc} > 0\,(\text{in V}). \tag{85}$$

*3) Line-to-line voltages and currents in the DC-link:* Depending on the DC-link voltage $u_{\rm dc}(t)$ and the actual switching vector $\boldsymbol{s}^{abc}(t)$, the line-to-line output voltages $\boldsymbol{u}^{ltl}_{\rm s/f}$ (neglecting losses) and the DC-link currents $i^{+}_{\rm m/g}$ of the machine-side and grid-side converter, as shown in Fig. 22, can be derived. The line-to-line voltages at the `U`, `V`, and `W` terminals are given by

$$\forall\, t \geq 0\,{\rm s}: \quad \boldsymbol{u}^{ltl}_{\rm s/f}(t) = u_{\rm dc}(t) \begin{bmatrix} 1 & -1 & 0 \\ 0 & 1 & -1 \\ -1 & 0 & 1 \end{bmatrix} \boldsymbol{s}^{abc}_{\rm m/g}(t) \overset{(27)}{=} u_{\rm dc}(t) \boldsymbol{T}^{\star}_{\rm v} \boldsymbol{s}^{abc}_{\rm m/g}(t), \tag{86}$$

whereas the DC-link currents, flowing from the DC-link into the respective converter, are given by

$$\forall\, t \geq 0\,{\rm s}: \qquad i_{\rm m/g}(t) = \boldsymbol{i}^{abc}_{\rm s/f}(t)^{\top} \boldsymbol{s}^{abc}_{\rm m/g}(t). \tag{87}$$

**Remark III.4.** *If stator voltages $\boldsymbol{u}^{abc}_{\rm s}$ and filter voltages $\boldsymbol{u}^{abc}_{\rm f}$ are balanced (though this is* not *correct in general), the following holds*

$$\forall t \geq 0\,{\rm s}: \quad u^a_{\rm s}(t) + u^b_{\rm s}(t) + u^c_{\rm s}(t) = u^a_{\rm f}(t) + u^b_{\rm f}(t) + u^c_{\rm f}(t) = 0\,{\rm V}, \tag{88}$$

*and the phase voltages of the machine and of the filter can be calculated directly from the switching vectors $\boldsymbol{s}^{abc}_{\rm m}$ or $\boldsymbol{s}^{abc}_{\rm g}$ and the DC-link voltage $u_{\rm dc}$ as follows*

$$\begin{aligned} \forall t \geq 0\,{\rm s}: \quad \boldsymbol{u}^{abc}_{\rm s/f}(t) &\overset{(29)}{=} \boldsymbol{T}^{-1}_{\rm v} \boldsymbol{u}^{ltl}(t) \overset{(86)}{=} u_{\rm dc}(t) \boldsymbol{T}^{-1}_{\rm v} \boldsymbol{T}^{\star}_{\rm v} \boldsymbol{s}^{abc}_{\rm s/f}(t) \\ &= \frac{u_{\rm dc}(t)}{3} \begin{bmatrix} 2 & -1 & -1 \\ -1 & 2 & -1 \\ -1 & -1 & 2 \end{bmatrix} \boldsymbol{s}^{abc}_{\rm s/f}(t). \end{aligned} \tag{89}$$

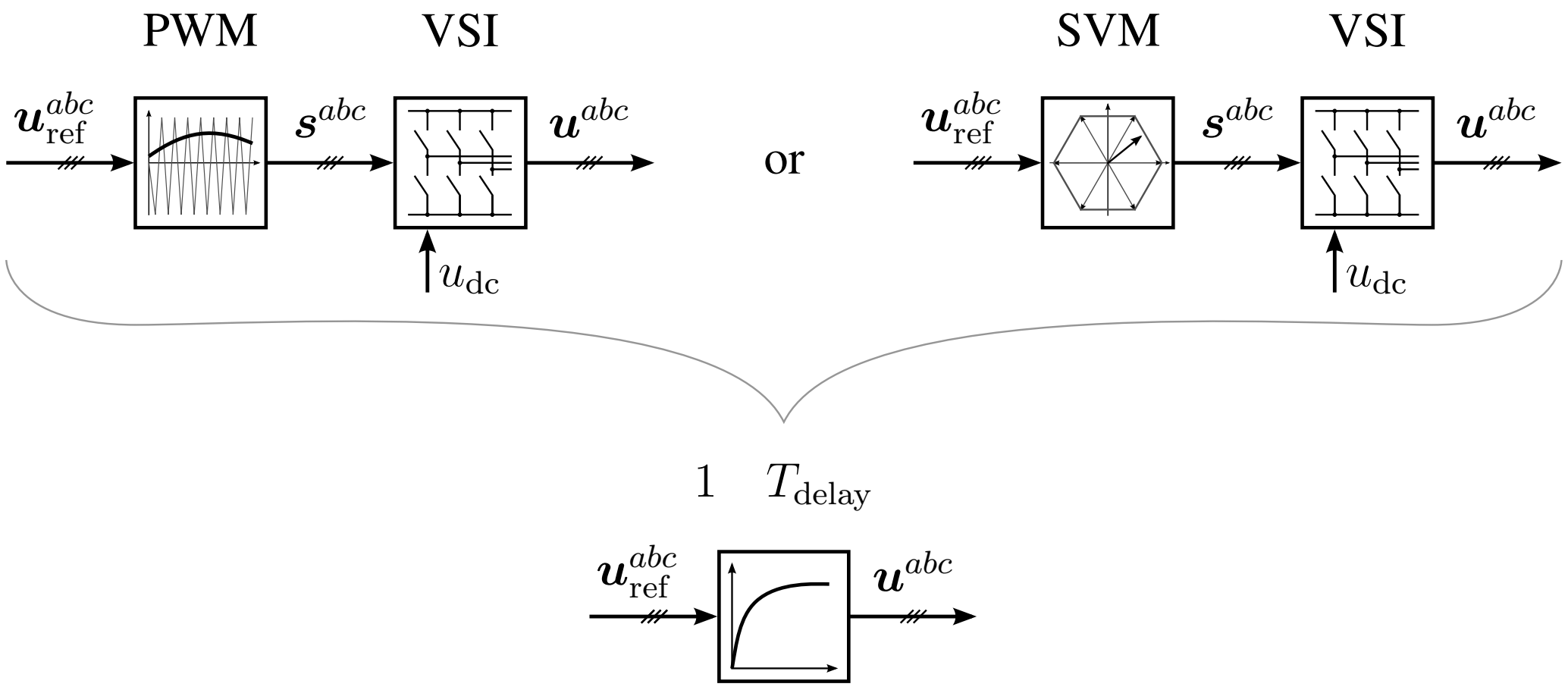

Fig. 25: *Signal flow diagram of the voltage generation in a converter/inverter by means of pulse width modulation (PWM) or space vector modulation (SVM) and signal flow diagram of the approximated dynamics of the voltage generation.*

*4) Delay in the voltage generation:* The desired reference voltage $\boldsymbol{u}_{s/f,ref}^{abc}$ (or $\boldsymbol{u}_{s/f,ref}^{s}$ or $\boldsymbol{u}_{s/f,ref}^{k}$) can not be instantaneously generated by the converter (see Fig. 24). The reference voltage can only be reproduced on average over a switching period by pulse width modulation (PWM, [28, Sec. 14.3]) or space vector modulation (SVM, [28, Sec. 14.4]). The switching period is indirectly proportional to the switching frequency $f_{sw}$ (in Hz) of the converter. The generation of the reference voltage thus leads to a delay with a lag time of $T_{delay} \propto 1/f_{sw} > 0$ (in s). This delaying behavior of the converter is illustrated in Fig. 24 for the phase reference $u_{ref}^{a}(\cdot)$ and for a symmetrical, regularly-sampled PWM-pulse pattern. The phase voltage $u^{a}(\cdot)$ represents the typical PWM voltage pulse pattern. The shape of the averaged fundamental wave $\bar{u}^{a}(\cdot)$ of the phase voltage $u^{a}$ is delayed by $T_{delay} = \frac{1}{2 \cdot 2.5\,\mathrm{kHz}} = 0.2\,\mathrm{ms}$ in relation to the reference voltage $u_{ref}^{a}(\cdot)$.

For the later controller design, the converter delay on the machine and grid side is simplified by means of a first-order lag system, so the dynamic relationship between reference and actual voltage is approximated through the following transfer function

$$
\begin{aligned}
F_{VSI}(s) = F_{VSI}(s) &= \frac{u^{\alpha/\beta}(s)}{u_{ref}^{\alpha/\beta}(s)} = \frac{u^{a/b/c}(s)}{u_{ref}^{a/b/c}(s)} \\
&= e^{-sT_{delay}} = \frac{1}{1 + \frac{sT_{delay}}{1!} + \frac{s^2 T_{delay}^2}{2!} + \ldots} \\
&\approx \frac{1}{1 + sT_{delay}} \quad \text{for} \quad 1 \gg T_{delay} > 0 \quad (\text{in s}).
\end{aligned} \tag{90}
$$

This simplified idea of the dynamics of voltage generation in a converter is illustrated in Fig. 25.

*5) Relationship between line-to-line voltages and phase voltages in the stator-fixed reference frame:* The models of the machine-side and grid-side electrical network were derived in the stator-fixed reference frame (see (80) and (84)), therefore, the input voltage vectors (for the models) should be $\boldsymbol{u}_s^s = (u_s^\alpha, u_s^\beta)^\top$ and $\boldsymbol{u}_f^s = (u_f^\alpha, u_f^\beta)^\top$ in the $s$-reference frame, respectively.

Invoking the general Clarke transformation (30) (with $\kappa = \frac{2}{3}$) yields

$$
\begin{pmatrix} u_{s/f}^{\alpha}(t) \\ u_{s/f}^{\beta}(t) \\ u_{s/f}^{0}(t) \end{pmatrix} = \boldsymbol{T}_c^\star \boldsymbol{u}_{s/f}^{abc}(t) \overset{(30)}{=} \frac{2}{3} \begin{pmatrix} \frac{1}{2}\big(u_{s/f}^{a}(t) - u_{s/f}^{b}(t)\big) - \frac{1}{2}\big(u_{s/f}^{c}(t) - u_{s/f}^{a}(t)\big) \\ \frac{\sqrt{3}}{2}\big(u_{s/f}^{b}(t) - u_{s/f}^{c}(t)\big) \\ \frac{1}{\sqrt{2}}\big(u_{s/f}^{a}(t) + u_{s/f}^{b}(t) + u_{s/f}^{c}(t)\big) \end{pmatrix}
$$

$$= \frac{2}{3}\begin{bmatrix} \frac{1}{2} & 0 & -\frac{1}{2} \\ 0 & \frac{\sqrt{3}}{2} & 0 \\ 0 & 0 & 0 \end{bmatrix} \boldsymbol{u}^{ltl}_{\mathrm{s/f}}(t) + \frac{2}{3}\begin{bmatrix} 0 & 0 & 0 \\ 0 & 0 & 0 \\ \frac{1}{\sqrt{2}} & \frac{1}{\sqrt{2}} & \frac{1}{\sqrt{2}} \end{bmatrix} \boldsymbol{u}^{abc}_{\mathrm{s/f}}(t). \tag{91}$$

Due to the star connection in the generator and filter, the zero-sequence components of the currents are zero (because $i^a_{\mathrm{s}} + i^b_{\mathrm{s}} + i^c_{\mathrm{s}} = i^a_{\mathrm{f}} + i^b_{\mathrm{f}} + i^c_{\mathrm{f}} = 0\,\mathrm{A}$) and do not contribute to the generation of torque or power. Therefore, the consideration of these zero-sequence voltage components $u^0_{\mathrm{s}}$ and $u^0_{\mathrm{f}}$ in the modeling is therefore not necessary and, by invoking the line-to-line voltages $\boldsymbol{u}^{ltl}_{\mathrm{s}}(t)$ and $\boldsymbol{u}^{ltl}_{\mathrm{f}}(t)$, the phase voltages in the stator-fixed $s$-reference frame can be derived as follows

$$\boldsymbol{u}^{s}_{\mathrm{s/f}}(t) = \begin{pmatrix} u^{\alpha}_{\mathrm{s/f}}(t) \\ u^{\beta}_{\mathrm{s/f}}(t) \end{pmatrix} = \underbrace{\frac{2}{3}\begin{bmatrix} \frac{1}{2} & 0 & -\frac{1}{2} \\ 0 & \frac{\sqrt{3}}{2} & 0 \end{bmatrix}}_{=:\boldsymbol{T}^{ltl}_{\mathrm{c}} \in \mathbb{R}^{2\times 3}} \boldsymbol{u}^{ltl}_{\mathrm{s/f}}(t). \tag{92}$$

**Remark III.5.** *For balanced (symmetrical) grid voltages with star-point connection, it follows that*

$$\begin{aligned} \forall t \geq 0\,\mathrm{s}\colon u^a_{\mathrm{g}}(t) + u^b_{\mathrm{g}}(t) + u^c_{\mathrm{g}}(t) &= 0\,\mathrm{V} \\ i^a_{\mathrm{f}}(t) + i^b_{\mathrm{f}}(t) + i^c_{\mathrm{f}}(t) &= 0\,\mathrm{A} \\ \tfrac{\mathrm{d}}{\mathrm{d}t} i^a_{\mathrm{f}}(t) + \tfrac{\mathrm{d}}{\mathrm{d}t} i^b_{\mathrm{f}}(t) + \tfrac{\mathrm{d}}{\mathrm{d}t} i^c_{\mathrm{f}}(t) &= 0\,\frac{\mathrm{A}}{\mathrm{s}}, \end{aligned} \tag{93}$$

*and it can be concluded that, for the grid-side dynamics, the following holds*

$$\begin{aligned} \forall\, t \geq 0\,\mathrm{s}\colon \quad u^a_{\mathrm{f}}(t) + u^b_{\mathrm{f}}(t) + u^c_{\mathrm{f}}(t) &\overset{(83)}{=} R_{\mathrm{f}}\big(i^a_{\mathrm{f}}(t) + i^b_{\mathrm{f}}(t) + i^c_{\mathrm{f}}(t)\big) \\ &\quad + L_{\mathrm{f}}\Big(\tfrac{\mathrm{d}}{\mathrm{d}t}\, i^a_{\mathrm{f}}(t) + \tfrac{\mathrm{d}}{\mathrm{d}t}\, i^b_{\mathrm{f}}(t) + \tfrac{\mathrm{d}}{\mathrm{d}t}\, i^c_{\mathrm{f}}(t)\Big) \\ &\quad + u^a_{\mathrm{g}}(t) + u^b_{\mathrm{g}}(t) + u^c_{\mathrm{g}}(t) \\ &\overset{(93)}{=} 0\,\mathrm{V}. \end{aligned} \tag{94}$$

*Thus, the zero-sequence component $u^0_{\mathrm{f}}$ in (91) is always zero and can be neglected, and the phase voltages*

$$\boldsymbol{u}^{abc}_{\mathrm{f}}(t) \overset{(91)}{=} (\boldsymbol{T}^{\star}_{\mathrm{c}})^{-1} \frac{2}{3}\begin{bmatrix} \frac{1}{2} & 0 & -\frac{1}{2} \\ 0 & \frac{\sqrt{3}}{2} & 0 \\ 0 & 0 & 0 \end{bmatrix} \boldsymbol{u}^{ltl}_{\mathrm{f}}(t) = \begin{bmatrix} \frac{1}{3} & 0 & -\frac{1}{3} \\ -\frac{1}{6} & \frac{1}{2} & \frac{1}{6} \\ -\frac{1}{6} & -\frac{1}{2} & \frac{1}{6} \end{bmatrix} \boldsymbol{u}^{ltl}_{\mathrm{f}}(t). \tag{95}$$

*can be calculated directly from the line-to-line filter voltages $\boldsymbol{u}^{ltl}_{\mathrm{f}}$.*

## IV. Control of wind turbine systems

In this section, based on the models described in Section III, the control of the entire system is discussed in more detail. A particular focus is on the control of the wind turbine system in the operating regime II (see Fig. 11). In this operating regime, the maximum power from wind should be "harvested". For this "Maximum Power Point Tracking (MPPT)", a sufficiently accurate speed feedback control system of the generator is required. The torque control is made possible by the machine-side current control system. On the grid side, the underlying current control system forms the basis for the superimposed feedforward control of the reactive power at the PCC. Finally, the control of the power flow within the wind turbine by a proper DC-link control system and the effects of the operation management on the power balance are discussed.

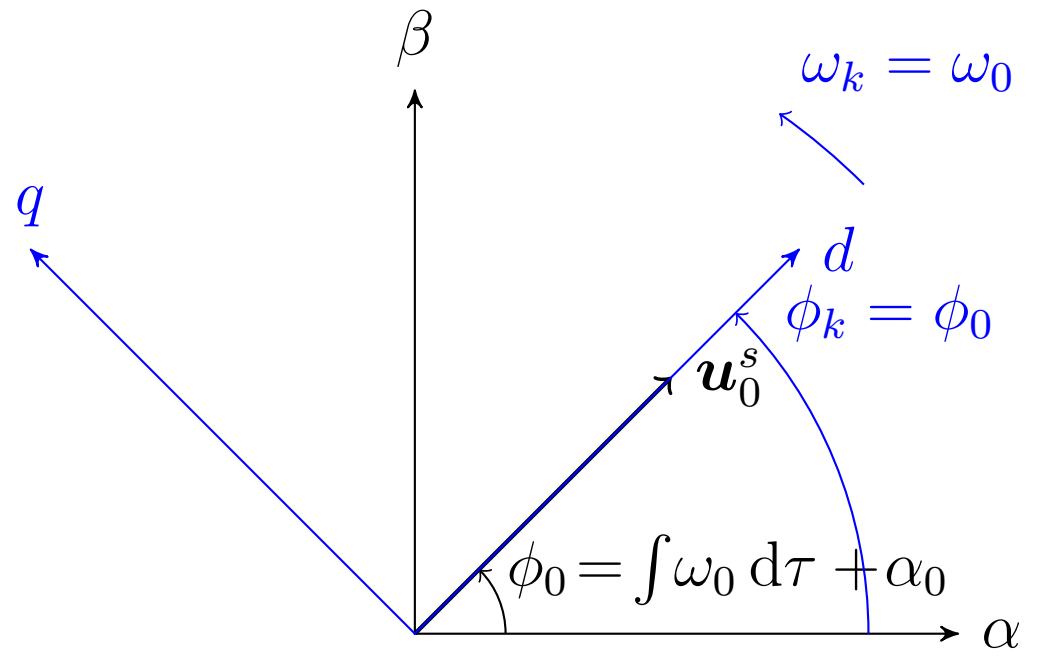

Fig. 26: *Vector diagram of grid voltage orientation: Orientation of the $d$-axis of the $k$-reference frame on the grid voltage $\boldsymbol{u}_g^s$.*

### A. Control of the grid-side currents

The grid-side current control is tasked with adjusting the grid-side filter currents $\boldsymbol{i}_f^{abc} = (i_f^a,\, i_f^b,\, i_f^c)^\top$ (in A)$^3$ (see Fig. 20) such that an independent supply of active and reactive power into the power grid can be achieved. For this purpose, the phase angle (and the amplitude) of the currents $\boldsymbol{i}_f^{abc}$ in the filter, having resistance $R_f$ (in $\frac{V}{A}$) and inductance $L_f$ (in $\frac{Vs}{A}$), must be controlled with respect to the grid voltages $\boldsymbol{u}_g^{abc} = (u_g^a,\, u_g^b,\, u_g^c)^\top$ (in V)$^3$ with amplitude $\hat{u}_g > 0$ (in V). In order to achieve a desired phase angle between the grid voltage vector $\boldsymbol{u}_g^{abc}$ and the filter current vector $\boldsymbol{i}_f^{abc}$, the phase angle of the grid voltage vector must be dynamically determined. This is usually achieved by a phase-locked loop (PLL). The detected grid voltage phase angle is used for *grid voltage orientation* of the $k$-reference frame on the grid side. This grid voltage orientation simplifies the design of the grid-side current controllers and permits an (almost) decoupled active and reactive power output to the grid.

*1) Voltage orientation through phase-locked loop (PLL):* In order to achieve a grid voltage orientation online, a phase-locked loop (PLL) is most commonly used (see e.g. [46] or [25, Sec. 4.2.2]). The aim is to dynamically align the $d$-axis of the $k$-reference frame onto the grid voltage vector

$$\forall t \geq 0\,\mathrm{s}: \qquad \boldsymbol{u}_g^s(t) = \boldsymbol{T}_c \boldsymbol{u}_g^{abc}(t) \overset{(82)}{=} \hat{u}_g \left(\cos(\phi_g(t)),\; \sin(\phi_g(t))\right)^\top \tag{96}$$

with voltage angle $\phi_g(t) = \int_0^t \omega_g(\tau)\, d\tau + \alpha_0$ (see Fig. 26). For this orientation, the angle $\phi_g$ must be estimated using the PLL. In the following, the estimated value is denoted by $\widetilde{\phi}_g$. In the ideal case, $\widetilde{\phi}_g = \phi_g$ holds (exact estimation) and one obtains

$$\forall t \geq 0\,\mathrm{s}: \qquad \boldsymbol{u}_g^k(t) = \left(u_g^d(t),\, u_g^d(t)\right)^\top = \boldsymbol{T}_p(\widetilde{\phi}_g(t))^{-1} \boldsymbol{u}_g^s(t) = \left(\hat{u}_g,\, 0\,\mathrm{V}\right)^\top. \tag{97}$$

Therefore, the grid voltage vector $\boldsymbol{u}_g^k$ has only a $d$-component (the $q$-component is zero). To achieve this, the $k$-reference frame must be initialized with initial angle $\alpha_0$ and then must rotate with the grid angular frequency $\omega_g = 2\pi f_g$ (where $f_g = 50\,\mathrm{Hz}$). Thus, in the ideal case, a dynamic rotation about

$$\forall\, t \geq 0\,\mathrm{s}: \;\; \phi_k(t) = \widetilde{\phi}_g(t) = \phi_g(t) = \int_0^t \omega_g(\tau)\, d\tau + \alpha_0 \;\; \text{with} \;\; \omega_k(t) = \omega_g(t) \tag{98}$$

can be obtained. The implementation of a phase-locked loop is illustrated in Fig. 27 as a signal flow diagram.

The measured three-phase grid voltages $\boldsymbol{u}_g^{abc}$ are transformed into the $s$-reference frame via the Clarke transformation. The transformed voltage vector $\boldsymbol{u}_g^s$ is finally mapped to the $k$-reference frame by rotating around the estimated angle $\widetilde{\phi}_g$ with the Park transformation. Then, for the estimated grid voltage vector $\widetilde{\boldsymbol{u}}_g^k$ in $k$-reference frame, the following holds

$$\widetilde{\boldsymbol{u}}_g^k(t) = \boldsymbol{T}_p(\widetilde{\phi}_g(t))^{-1} \boldsymbol{u}_g^s(t) \overset{(96)}{=} \hat{u}_g \begin{pmatrix} \cos(\phi_g(t)) \cos(\widetilde{\phi}_g(t)) + \sin(\phi_g(t)) \sin(\widetilde{\phi}_g(t)) \\ \sin(\phi_g(t)) \cos(\widetilde{\phi}_g(t)) - \cos(\phi_g(t)) \sin(\widetilde{\phi}_g(t)) \end{pmatrix}. \tag{99}$$

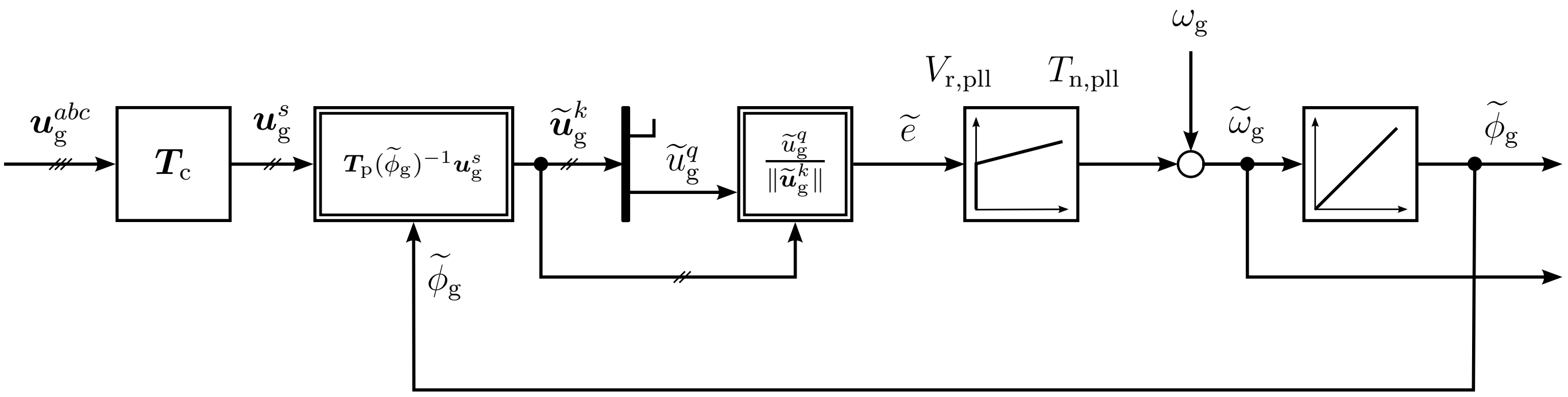

Fig. 27: *Phase-locked loop (PLL) for grid voltage orientation.*

With the help of the trigonometric formula (36), $\tilde{u}_g^q$ and thus $\tilde{e}$ (see Fig. 27) can be represented as follows

$$\forall t \geq 0\,\text{s}: \quad \tilde{e}(t) = \frac{\tilde{u}_g^q(t)}{\|\tilde{u}_g^k(t)\|} = \frac{\tilde{u}_g^q(t)}{\hat{u}_g} \overset{(99),(36)}{=} \sin\big(\phi_g(t) - \tilde{\phi}_g(t)\big). \tag{100}$$

For small angle differences $|\phi_g - \tilde{\phi}_g| \ll 1\,\text{rad}$, a common small-signal approximation of (100) is

$$\tilde{e}(t) = \sin\big(\phi_g(t) - \tilde{\phi}_g(t)\big) \overset{\phi_g - \tilde{\phi}_g \ll 1\,\text{rad}}{\approx} \phi_g(t) - \tilde{\phi}_g(t). \tag{101}$$

In this case, $\tilde{e}$ corresponds to the "control error" between the rotation angle $\tilde{\phi}_g$ (actual value) from the Park transformation and the desired grid angle $\phi_g$ (reference). By the use of a PI controller

$$F_{C,\text{pll}}(s) = V_{r,\text{pll}} \frac{1 + sT_{n,\text{pll}}}{sT_{n,\text{pll}}} \tag{102}$$

and the integrator in Fig. 27, this error is caused to asymptotically approach zero. If the small-signal approximation in (101) is valid, the analysis of the closed-loop system of the phase-locked loop is considerably simplified: The nonlinear signal flow diagram in Fig. 27 can be converted into the linear control loop depicted in Fig. 28 having the following transfer function

$$F_{CL,\text{pll}}(s) := \frac{\tilde{\phi}_g(s)}{\phi_g(s)} = \frac{V_{r,\text{pll}} \frac{1+sT_{n,\text{pll}}}{sT_{n,\text{pll}}} \frac{1}{s}}{1 + V_{r,\text{pll}} \frac{1+sT_{n,\text{pll}}}{sT_{n,\text{pll}}} \frac{1}{s}} = \frac{1 + sT_{n,\text{pll}}}{1 + sT_{n,\text{pll}} + s^2 \frac{T_{n,\text{pll}}}{V_{r,\text{pll}}}}. \tag{103}$$

A controller design which guarantees aperiodic damping and asymptotic accuracy can be achieved, for example, by the correct assignment of the closed-loop poles (see [28, Sec. 5.5.5]). The following parameterization of the PI controller (102) is derived from pole placement by comparing the coefficients of the denominator of the closed-loop transfer function (103) and the desired polynomial $1 + s2T_p + s^2T_p^2$ (where $\frac{1}{f_g} \gg T_p > 0\,\text{s}$):

$$1 + sT_{n,\text{pll}} + s^2 \frac{T_{n,\text{pll}}}{V_{r,\text{pll}}} \overset{!}{=} 1 + s2T_p + s^2T_p^2 \quad \text{with} \quad \frac{1}{f_g} = \frac{2\pi}{\omega_g} \gg T_p > 0\,\text{s}$$
$$\Longrightarrow \quad T_{n,\text{pll}} = 2T_p \quad \text{and} \quad V_{r,\text{pll}} = \frac{2}{T_p}. \tag{104}$$

In order to ensure a sufficiently fast settling time, the pole placement time constant $T_p$ (in s) should be chosen to be significantly smaller than the grid period $\frac{1}{f_g} = \frac{2\pi}{\omega_g} = 0.02\,\text{s}$. The angular frequency $\omega_g$ can be used as feedforward control signal to improve the dynamic behavior (see Fig. 27 and 28).

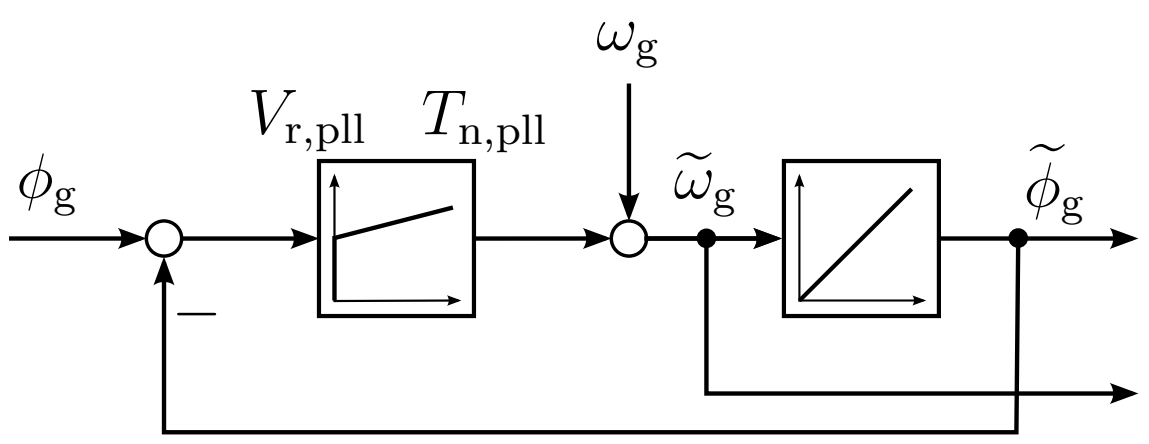

Fig. 28: *Small-signal approximation of the PLL for grid voltage orientation.*

2) *Grid-side control system in grid voltage orientation:* In Section III-F, the model (84) of the grid-side electrical network was derived and illustrated in the $s$-reference frame. This model can be further simplified by considering grid voltage orientation. Here, the model (84) can be transformed by the inverse Park transformation (32) with $\phi_k = \phi_g$ (grid phase angle) in the $k$-reference frame, such that the $d$-axis lies on the vector $\boldsymbol{u}_g^s$ of the grid voltage (see Fig. 26). In the ideal case, (97) holds true. For a constant grid angular frequency $\omega_g = \frac{d}{dt}\phi_g(t)$, one obtains the following dynamical model

$$\begin{aligned}
\boldsymbol{u}_f^k(t) &\overset{(32)}{=} \boldsymbol{T}_p(\phi_g(t))^{-1}\boldsymbol{u}_f^s(t) \\
&\overset{(84)}{=} R_f \underbrace{\boldsymbol{T}_p(\phi_g(t))^{-1}\boldsymbol{i}_f^s(t)}_{=\boldsymbol{i}_f^k(t)} + \underbrace{\boldsymbol{T}_p(\phi_g(t))^{-1}\boldsymbol{u}_g^s(t)}_{=\boldsymbol{u}_g^k(t)} \\
&\quad + L_f \boldsymbol{T}_p(\phi_g(t))^{-1}\Big[\underbrace{\big(\tfrac{d}{dt}\boldsymbol{T}_p(\phi_g(t))\big)\boldsymbol{i}_f^k(t) + \boldsymbol{T}_p(\phi_g(t))\tfrac{d}{dt}\boldsymbol{i}_f^k(t)}_{=\frac{d}{dt}\boldsymbol{i}_f^s(t)}\Big] \\
&\overset{(39)}{=} R_f \boldsymbol{i}_f^k(t) + \boldsymbol{u}_g^k(t) + \omega_g L_f \boldsymbol{J}\boldsymbol{i}_f^k(t) + L_f \tfrac{d}{dt}\boldsymbol{i}_f^k(t)
\end{aligned} \tag{105}$$

of the grid-side electrical network in voltage orientation. For the grid-side current control system, the most interesting factor is the current dynamics. Rewriting (105) yields the differential equation

$$\frac{d}{dt}\boldsymbol{i}_f^k(t) = \frac{1}{L_f}\Big(\boldsymbol{u}_f^k(t) - R_f\boldsymbol{i}_f^k(t) \underbrace{-\omega_g L_f \boldsymbol{J}\boldsymbol{i}_f^k(t) - \boldsymbol{u}_g^k(t)}_{=:\boldsymbol{u}_{f,\text{dist}}^k(t)\ \text{(disturbance terms)}}\Big) \tag{106}$$

of the grid-side current dynamics, where $\omega_g = 2\pi f_g$ (in $\frac{\text{rad}}{\text{s}}$) is the (constant) grid angular frequency, $R_f$ (in $\frac{\text{V}}{\text{A}}$) is the filter resistance, $L_f$ (in $\frac{\text{Vs}}{\text{A}}$) is the filter inductance, $\boldsymbol{u}_g^k = \big(u_g^d, u_g^q\big)^\top = (\hat{u}_g, 0)^\top$ (in V)$^2$ is the grid voltage vector in the $k$-reference frame (having amplitude $\hat{u}_g$ (in V)) and $\boldsymbol{u}_f^k = \big(u_f^d, u_f^q\big)^\top$ (in V)$^2$ are the applied filter phase voltages by the grid-side converter. The compact model (106) in voltage orientation can be decomposed into the corresponding $d$- and $q$-components for the controller design as follows

$$\left.\begin{aligned}
\tfrac{d}{dt}\, i_f^d(t) &= \tfrac{1}{L_f}\Big(u_f^d(t) - R_f i_f^d(t) \overbrace{+\omega_g L_f i_f^q(t) - \hat{u}_g}^{=:u_{f,\text{dist}}^d(t)}\Big) \\
\tfrac{d}{dt}\, i_f^q(t) &= \tfrac{1}{L_f}\Big(u_f^q(t) - R_f i_f^q(t) \underbrace{-\omega_g L_f i_f^d(t)}_{=:u_{f,\text{dist}}^q(t)}\Big).
\end{aligned}\right\} \tag{107}$$

Figure 29 shows the signal flow diagram of the coupled relationships in (106) and (107) of the grid-side electrical network with the approximated converter dynamics, and the PI controllers (90). The current dynamics are (linearly) coupled to one another by the “disturbances” $u_{f,\text{dist}}^d(t) = u_{f,\text{dist}}^d(i_f^q(t))$ and $u_{f,\text{dist}}^q(t) = u_{f,\text{dist}}^q(i_f^d(t))$.

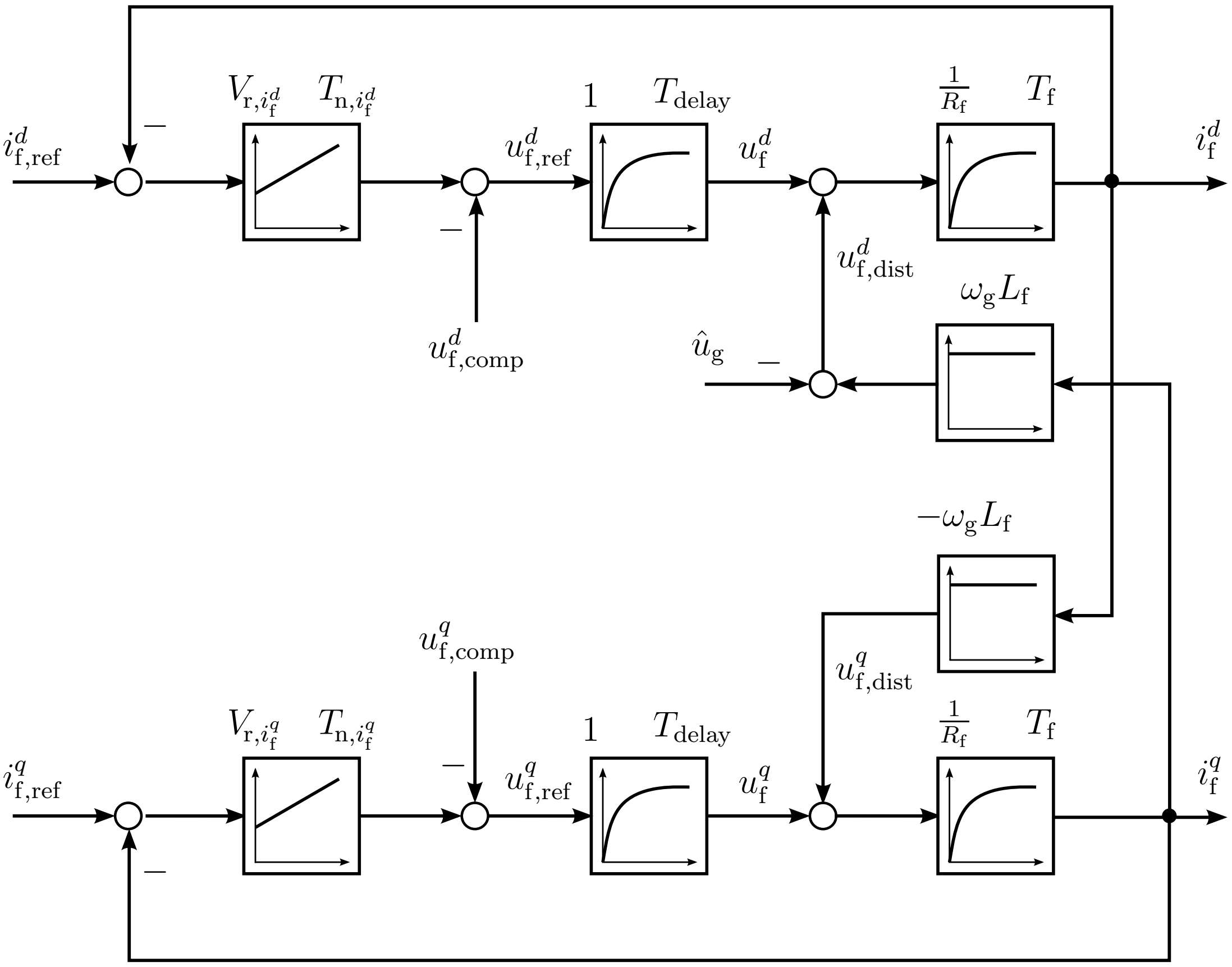

Fig. 29: *Coupled grid-side current control circuit in the $k$-reference frame (*voltage orientation*) with PI-controllers* (113)*, disturbance compensation* (108)*, approximated converter dynamics* (90) *and network dynamics* (107)*.*

*3) Compensation of grid-side disturbances:* For controller design, a (reasonably) simplified and decoupled system behavior should be available. For this purpose, a dynamic compensation of the disturbances in (106) and (107) is beneficial. To achieve this, the compensation terms must be transferred to the converter via the transfer function[15]

$$\boldsymbol{u}^k_{\mathrm{f,comp}}(t) = \begin{pmatrix} u^d_{\mathrm{f,comp}}(t) \\ u^q_{\mathrm{f,comp}}(t) \end{pmatrix} = \mathscr{L}^{-1}\left\{F_K(s)\right\} \begin{pmatrix} \omega_\mathrm{g} L_\mathrm{f} i^q_\mathrm{f}(t) - \hat{u}_\mathrm{g} \\ -\omega_\mathrm{g} L_\mathrm{f} i^d_\mathrm{f}(t) \end{pmatrix} \tag{108}$$

in the form of a (dynamic) *disturbance feedforward network* (see Fig. 30).

An *exact* (ideal) compensation can only be achieved for $F_K(s) = 1 + s\,T_\mathrm{delay} = F_\mathrm{VSI}(s)^{-1}$ (which is *not* causal!) and exactly known parameters $\hat{u}_\mathrm{g}$, $\omega_\mathrm{g}$ and $L_\mathrm{f}$ and accurate measurements from $i^d_\mathrm{f}$ and $i^q_\mathrm{f}$. Due to the delay in the converter (see (90)), measurement errors and/or noise and parameter uncertainties, an ideal compensation is not feasible. Typically the following compensation transfer function

$$F_K(s) = \begin{cases} V_K & \text{, static compensation with } V_K \leq 1 \\ V_K \frac{1+sT_\mathrm{delay}}{1+sT_h} & \text{, dynamic compensation with } T_h \ll T_\mathrm{delay},\; V_K \leq 1 \end{cases} \tag{109}$$

is implemented (for more details see [28, Sec. 7.1.1.1]). The selection of the compensation gain $V_K \leq 1$ avoids an overcompensation in the case of actuator saturation (input constraints), whereas the choice of $T_h \ll T_\mathrm{delay}$ is a causal approximation of the proportional-derivative term $1 + s\,T_\mathrm{delay}$ (inverse converter dynamics).

[15] The Laplace Transformation of a function $f(\cdot) \in \mathcal{L}^1_\mathrm{loc}(\mathbb{R}_{>0};\mathbb{R})$ is given by $f(s) := \mathscr{L}\{f(t)\} := \int_0^\infty f(t)\exp(-st)\,\mathrm{d}t$ or $f(s)$ •—○ $f(t)$ with $\Re(s) \geq \alpha$, provided $\alpha \in \mathbb{R}$ exists, such that $[t \mapsto \exp(-\alpha t) f(t)] \in \mathcal{L}^1(\mathbb{R}_{>0};\mathbb{R})$ [47, p. 742]. The inverse Laplace transformation is written as $f(t) = \mathscr{L}^{-1}\{f(s)\}$ or $f(t)$ ○—• $f(s)$ where $\mathcal{L}^p_{(\mathrm{loc})}(I;Y)$ is the space of measurable, (locally) $p$-integrable functions mapping $I \to Y$ and $\Re(s)$ to the real part of the complex number $s \in \mathbb{C}$.

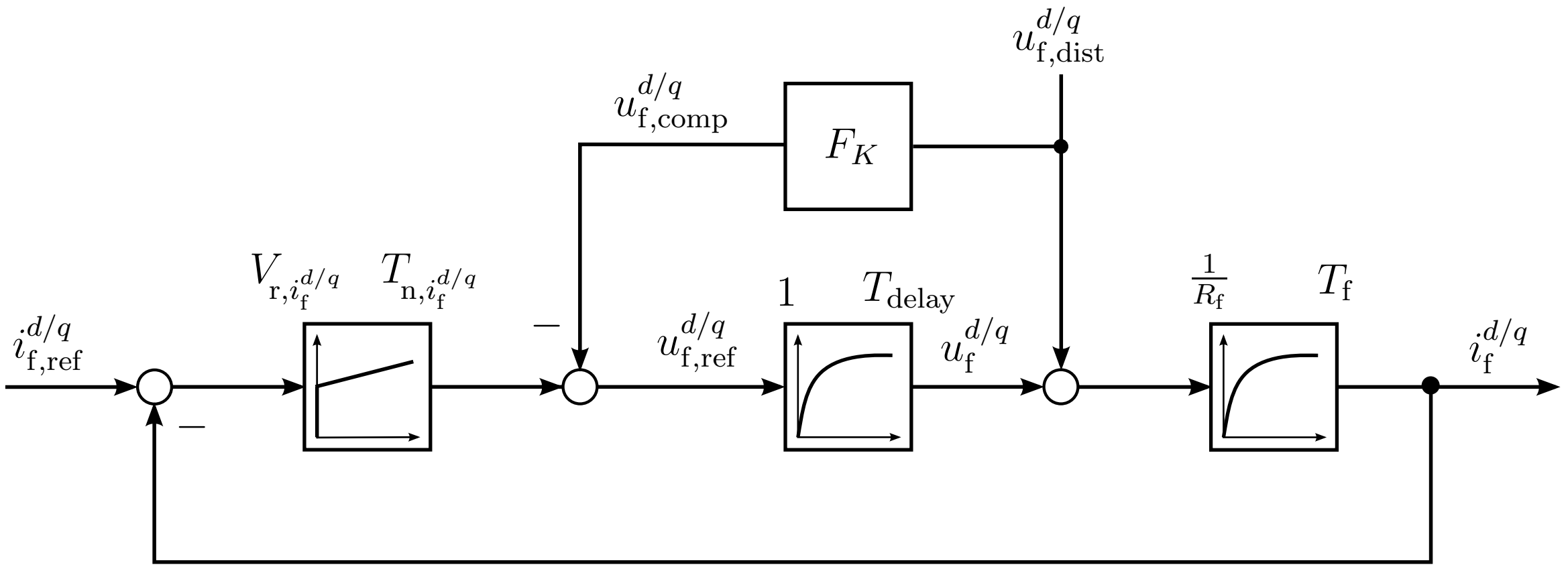

Fig. 30: *Signal flow diagram of the line-side current control loop with disturbance compensation (ideally, $F_K(s) = 1+sT_{\mathrm{delay}}$).*

*4) Design of the grid-side current controllers:* Under the assumption

**Assumption (A.9)** *For an* ideal *compensation of the grid-side disturbances in* (106) *or* (107)*, it follows that*

$$\forall t \geq 0\,\mathrm{s}: \qquad u_{\mathrm{f,comp}}^{d/q}(t) = \mathscr{L}^{-1}\left\{1 + s\,T_{\mathrm{delay}}\right\} u_{\mathrm{f,dist}}^{d/q}(t). \tag{110}$$

Thus, the $d$- and $q$-components can be regarded as decoupled and the current dynamics of the $d$- and $q$-components simplify to the following transfer functions

$$\left.\begin{array}{rclcl} \frac{\mathrm{d}}{\mathrm{d}t}\, i_{\mathrm{f}}^{d}(t) &=& \frac{1}{L_{\mathrm{f}}}\Big(u_{\mathrm{f}}^{d}(t) - R_{\mathrm{f}} i_{\mathrm{f}}^{d}(t)\Big) & \circ\!\!-\!\!\bullet & \frac{i_{\mathrm{f}}^{d}(s)}{u_{\mathrm{s}}^{d}(s)} = \frac{\frac{1}{R_{\mathrm{f}}}}{1+s\,\frac{L_{\mathrm{f}}}{R_{\mathrm{f}}}} \\ \frac{\mathrm{d}}{\mathrm{d}t}\, i_{\mathrm{f}}^{q}(t) &=& \frac{1}{L_{\mathrm{f}}}\Big(u_{\mathrm{f}}^{q}(t) - R_{\mathrm{f}} i_{\mathrm{f}}^{q}(t)\Big) & \circ\!\!-\!\!\bullet & \frac{i_{\mathrm{f}}^{q}(s)}{u_{\mathrm{s}}^{q}(s)} = \frac{\frac{1}{R_{\mathrm{f}}}}{1+s\,\frac{L_{\mathrm{f}}}{R_{\mathrm{f}}}}. \end{array}\right\} \tag{111}$$

Considering the approximated converter dynamics (90), the overall system transfer functions

$$\left.\begin{array}{rcl} F_{S,i_{\mathrm{f}}^{d}}(s) = \frac{i_{\mathrm{f}}^{d}(s)}{u_{\mathrm{f,ref}}^{d}(s)} &=& \frac{\frac{1}{R_{\mathrm{f}}}}{(1+sT_{\mathrm{f}})(1+sT_{\mathrm{delay}})} \\ F_{S,i_{\mathrm{f}}^{q}}(s) = \frac{i_{\mathrm{f}}^{q}(s)}{u_{\mathrm{f,ref}}^{q}(s)} &=& \frac{\frac{1}{R_{\mathrm{f}}}}{(1+sT_{\mathrm{f}})(1+sT_{\mathrm{delay}})} \end{array}\right\} \text{ with filter time constant } T_{\mathrm{f}} := \frac{L_{\mathrm{f}}}{R_{\mathrm{f}}}. \tag{112}$$

can be obtained for the current controller design. Note that, on the grid side, $F_{S,i_{\mathrm{f}}^{d}}(s) = F_{S,i_{\mathrm{f}}^{q}}(s)$. Due to the ideal disturbance compensation, *no* disturbances act on both subsystems (112) which represent two second-order delay systems. Therefore, in order to achieve a good and fast tracking response, the PI controllers are tuned according to the Magnitude Optimum (german: Betragsoptimum (BO)) with the help of the tuning rules summarized in the optimization table in [28, p. 81/82]. For both current controllers, one obtains the identical tuning:

$$\left.\begin{array}{l} F_{C,i_{\mathrm{f}}^{d}}(s) = V_{\mathrm{r},i_{\mathrm{f}}^{d}}\frac{1+sT_{\mathrm{n},i_{\mathrm{f}}^{d}}}{sT_{\mathrm{n},i_{\mathrm{f}}^{d}}} \text{ with } V_{\mathrm{r},i_{\mathrm{f}}^{d}} = \frac{T_{\mathrm{f}}}{2\cdot\frac{1}{R_{\mathrm{f}}}\cdot T_{\mathrm{delay}}} \text{ and } T_{\mathrm{n},i_{\mathrm{f}}^{d}} = T_{\mathrm{f}}, \\ F_{C,i_{\mathrm{f}}^{q}}(s) = V_{\mathrm{r},i_{\mathrm{f}}^{q}}\frac{1+sT_{\mathrm{n},i_{\mathrm{f}}^{q}}}{sT_{\mathrm{n},i_{\mathrm{f}}^{q}}} \text{ with } V_{\mathrm{r},i_{\mathrm{f}}^{q}} = \frac{T_{\mathrm{f}}}{2\cdot\frac{1}{R_{\mathrm{f}}}\cdot T_{\mathrm{delay}}} \text{ and } T_{\mathrm{n},i_{\mathrm{f}}^{q}} = T_{\mathrm{f}}. \end{array}\right\} \tag{113}$$

The choice of the integrator time constant $T_{\mathrm{n},i_{\mathrm{f}}^{d/q}} = T_{\mathrm{f}}$ of the PI controllers results in the elimination of the filter time constant $T_{\mathrm{f}}$ (which is, $T_{\mathrm{delay}} \ll T_{\mathrm{f}}$, cf. [28, Sec. 3.1]). Due to the disturbance feedforward compensation, both current components are (almost perfectly) decoupled and can be controlled independently from one another. For the tuning (113), the closed-loop transfer functions for both components become

$$\begin{aligned} F_{CL,i_\mathrm{f}^{d/q}}(s) = \frac{i_\mathrm{f}^{d/q}(s)}{i_{\mathrm{f,ref}}^{d/q}(s)} &= \frac{V_{\mathrm{r},i_\mathrm{f}^{d/q}} \frac{1+sT_{\mathrm{n},i_\mathrm{f}^{d/q}}}{sT_{\mathrm{n},i_\mathrm{f}^{d/q}}} \frac{1}{R_\mathrm{f}} \frac{1}{(1+sT_\mathrm{f})(1+sT_\mathrm{delay})}}{1 + V_{\mathrm{r},i_\mathrm{f}^{d/q}} \frac{1+sT_{\mathrm{n},i_\mathrm{f}^{d/q}}}{sT_{\mathrm{n},i_\mathrm{f}^{d/q}}} \frac{1}{R_\mathrm{f}} \frac{1}{(1+sT_\mathrm{f})(1+sT_\mathrm{delay})}} \\ &= \frac{\frac{1}{s2T_\mathrm{delay}(1+sT_\mathrm{delay})}}{1+\frac{1}{s2T_\mathrm{delay}(1+sT_\mathrm{delay})}} = \frac{1}{s2T_\mathrm{delay}(1+sT_\mathrm{delay})+1} \\ &= \frac{1}{1+s\,2T_\mathrm{delay}+s^2\,2T_\mathrm{delay}^2} \overset{T_\mathrm{delay}^2 \approx 0\,\mathrm{s}}{\approx} \frac{1}{1+s\,2T_\mathrm{delay}}. \end{aligned} \tag{114}$$

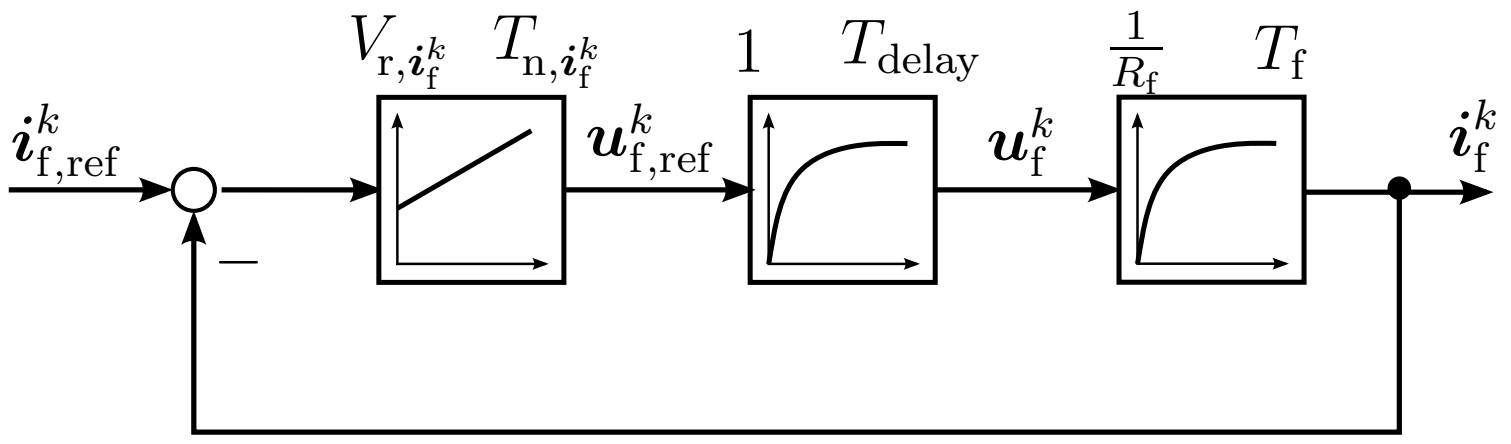

Fig. 31: *Simplified grid-side control circuit of the $d$ or $q$-axis current component with ideal disturbance compensation.*

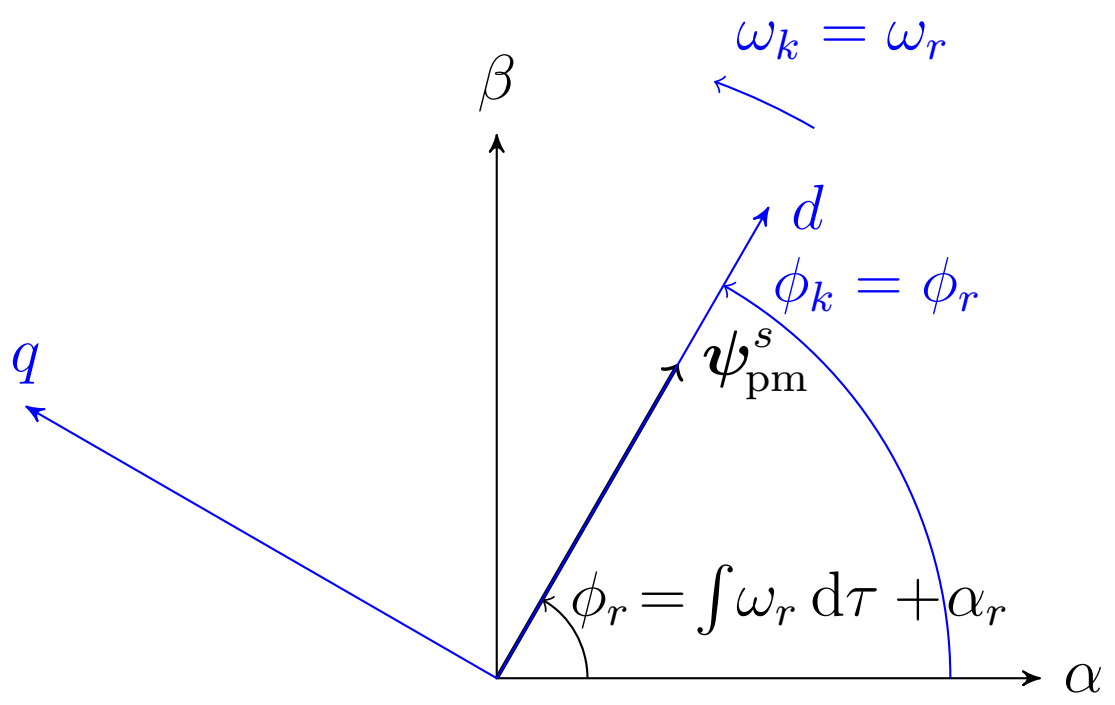

Fig. 32: *Space vector diagram for permanent magnet flux linkage orientation: $d$-axis of the $k$-reference frame is aligned with $\boldsymbol{\psi}_{\mathrm{pm}}^s$.*

The converter delay $T_\mathrm{delay}$ is usually very small, so the above approximation can be made and the current control loop dynamics simplify to a first-order lag system

$$F_{CL,i_\mathrm{f}^{d/q}}(s) = \frac{i_\mathrm{f}^{d/q}(s)}{i_{\mathrm{f,ref}}^{d/q}(s)} \approx \frac{1}{1+s\,T_{\mathrm{app},\boldsymbol{i}_\mathrm{f}^k}}$$
$$\text{with approximated time constant} \quad T_{\mathrm{app},\boldsymbol{i}_\mathrm{f}^k} := 2\,T_\mathrm{delay} > 0\,\mathrm{s}. \tag{115}$$

### *B. Control of the machine-side currents and torque generation*

In the following section, the machine-side current control of the permanent-magnet synchronous generator is discussed as a basis for the torque generation and for the speed regulation of the turbine. It will be shown that the illustrated current control can be designed in the same way as the current regulation of a permanent magnet synchronous motor.

*1) Permanent-magnet flux linkage orientation:* Similar to the grid voltage orientation of the grid-side control system, a simplified model of the generator shall be derived for the machine-side control system by using the model representation in permanent-magnet flux linkage orientation. The aim is to align the $d$-axis of the $k$-reference frame dynamically along the space vector of the permanent-magnet flux linkage $\boldsymbol{\psi}_{\mathrm{pm}}^{s}$ (see Fig. 32), i.e.,

$$\forall t \geq 0\,\mathrm{s}\colon\ \boldsymbol{\psi}_{\mathrm{pm}}^{k}(t) = \left(\psi_{\mathrm{pm}}^{d}(t),\, \psi_{\mathrm{pm}}^{q}(t)\right)^{\top} = \boldsymbol{T}_{\mathrm{p}}(\phi_{\mathrm{k}}(t))^{-1}\boldsymbol{\psi}_{\mathrm{pm}}^{s}(t) = \left(\psi_{\mathrm{pm}},\, 0\,\mathrm{V\,s}\right)^{\top}. \tag{116}$$

After a correct initialization of the initial phase angle $\alpha_r$ of the magnet in the rotor, the $k$-reference frame must only rotate with the electrical angular rotor velocity $\omega_{\mathrm{r}}(t) = n_{\mathrm{p}}\omega_{\mathrm{m}}(t)$. This allows a rotation about

$$\forall\, t \geq 0\,\mathrm{s}\colon\ \phi_{\mathrm{k}}(t) = \phi_{\mathrm{r}}(t) = \int_0^t \omega_{\mathrm{r}}(\tau)\,\mathrm{d}\tau\ + \alpha_r \text{ with } \omega_{\mathrm{k}}(t) = \omega_{\mathrm{r}}(t) = n_{\mathrm{p}}\,\omega_{\mathrm{m}}(t) \tag{117}$$

dynamically. Since the mechanical angular velocity $\omega_{\mathrm{m}}$ is a measured quantity (e.g. obtained from an encoder) and the number of pole pairs $n_{\mathrm{p}}$ can be assumed to be known, only the initial angle $\alpha_r$ must be determined. This can be achieved in a simple manner by applying a voltage $\boldsymbol{u}_{\mathrm{s}}^{abc}(0\,\mathrm{s}) = \boldsymbol{T}_{\mathrm{c}}^{-1}(u_{\mathrm{s}}^{\alpha},\, 0\,\mathrm{V})^{\top}$ in the direction of the $\alpha$-axis. The voltage $u_{\mathrm{s}}^{\alpha} > 0\,\mathrm{V}$ leads to a current flow $\boldsymbol{i}_{\mathrm{s}}^{s}$ and therefore a machine torque aligning the permanent magnet in the rotor in the direction of the $\alpha$-axis of the $s$-reference frame. When the voltage $u_{\mathrm{s}}^{\alpha}$ is specified, care must be taken that the maximum permissible phase currents are not exceeded.

*2) Machine-side control system in permanent-magnet flux linkage orientation:* The model (80) derived in Section III-D of the generator can also be represented in permanent-magnet flux linkage orientation. For this purpose, the inverse Park transformation (32) with $\phi_{\mathrm{k}} = \phi_{\mathrm{r}}$ is applied to (80). Thus, the generator model in the $k$-reference frame is obtained and the $d$-axis aligned with the flux linkage vector $\boldsymbol{\psi}_{\mathrm{pm}}^{s}$ of the permanent magnet (see Fig. 32), i.e. (116) holds. With $\omega_{\mathrm{k}}(t) := \frac{\mathrm{d}}{\mathrm{d}t}\phi_{\mathrm{k}}(t) = \frac{\mathrm{d}}{\mathrm{d}t}\phi_{\mathrm{r}}(t)$, the generator model

$$\begin{aligned}
\boldsymbol{u}_{\mathrm{s}}^{k}(t) &\overset{(32)}{=} \boldsymbol{T}_{\mathrm{p}}(\phi_{\mathrm{k}}(t))^{-1}\boldsymbol{u}_{\mathrm{s}}^{s}(t) \overset{(80)}{=} R_{\mathrm{s}}\underbrace{\boldsymbol{T}_{\mathrm{p}}(\phi_{\mathrm{k}}(t))^{-1}\boldsymbol{i}_{\mathrm{s}}^{s}(t)}_{=\boldsymbol{i}_{\mathrm{s}}^{k}(t)} + \boldsymbol{T}_{\mathrm{p}}(\phi_{\mathrm{k}}(t))^{-1}\tfrac{\mathrm{d}}{\mathrm{d}t}\underbrace{\boldsymbol{\psi}_{\mathrm{s}}^{s}(t)}_{=\boldsymbol{T}_{\mathrm{p}}(\phi_{\mathrm{k}}(t))\boldsymbol{\psi}_{\mathrm{s}}^{k}(t)} \\
&= R_{\mathrm{s}}\boldsymbol{i}_{\mathrm{s}}^{k}(t) + \boldsymbol{T}_{\mathrm{p}}(\phi_{\mathrm{k}}(t))^{-1}\left(\tfrac{\mathrm{d}}{\mathrm{d}t}\boldsymbol{T}_{\mathrm{p}}(\phi_{\mathrm{k}}(t))\,\boldsymbol{\psi}_{\mathrm{s}}^{k}(t) + \boldsymbol{T}_{\mathrm{p}}(\phi_{\mathrm{k}}(t))\,\tfrac{\mathrm{d}}{\mathrm{d}t}\boldsymbol{\psi}_{\mathrm{s}}^{k}(t)\right) \\
&\overset{(39)}{=} R_{\mathrm{s}}\boldsymbol{i}_{\mathrm{s}}^{k}(t) + \underbrace{\boldsymbol{T}_{\mathrm{p}}(\phi_{\mathrm{k}}(t))^{-1}\boldsymbol{T}_{\mathrm{p}}(\phi_{\mathrm{k}}(t))}_{=\boldsymbol{I}_2}\left(\omega_{\mathrm{k}}(t)\boldsymbol{J}\boldsymbol{\psi}_{\mathrm{s}}^{k}(t) + \tfrac{\mathrm{d}}{\mathrm{d}t}\boldsymbol{\psi}_{\mathrm{s}}^{k}(t)\right) \\
&\overset{(80)}{=} R_{\mathrm{s}}\boldsymbol{i}_{\mathrm{s}}^{k}(t) + \omega_{\mathrm{k}}(t)\boldsymbol{J}\left(\boldsymbol{L}_{\mathrm{s}}^{k}\boldsymbol{i}_{\mathrm{s}}^{k}(t) + \boldsymbol{\psi}_{\mathrm{pm}}^{k}(t)\right) + \boldsymbol{L}_{\mathrm{s}}^{k}\tfrac{\mathrm{d}}{\mathrm{d}t}\boldsymbol{i}_{\mathrm{s}}^{k}(t)
\end{aligned} \tag{118}$$

can be derived in flux orientation. For the design of the machine-side control, an understanding of the current dynamics is necessary. Rewriting (118) yields

$$\tfrac{\mathrm{d}}{\mathrm{d}t}\boldsymbol{i}_{s}^{k}(t) = \left(\boldsymbol{L}_{\mathrm{s}}^{k}\right)^{-1}\Big(\boldsymbol{u}_{s}^{k}(t) - R_{\mathrm{s}}\boldsymbol{i}_{s}^{k}(t)\underbrace{-\omega_{\mathrm{k}}(t)\boldsymbol{J}\left(\boldsymbol{L}_{\mathrm{s}}^{k}\boldsymbol{i}_{s}^{k}(t) + \boldsymbol{\psi}_{\mathrm{pm}}^{k}\right)}_{=:\boldsymbol{u}_{\mathrm{s,dist}}^{k}(t)\text{ (disturbance term)}}\Big). \tag{119}$$

Multiplying the result gives the component representation as follows

$$\left.\begin{aligned}
\tfrac{\mathrm{d}}{\mathrm{d}t}i_{\mathrm{s}}^{d}(t) &= \tfrac{1}{L_{\mathrm{s}}^{d}}\Big(u_{s}^{d}(t) - R_{\mathrm{s}}i_{\mathrm{s}}^{d}(t)\overbrace{+\omega_{\mathrm{k}}(t)L_{\mathrm{s}}^{q}i_{\mathrm{s}}^{q}(t)}^{=:u_{\mathrm{s,dist}}^{d}(t)}\Big) \\
\tfrac{\mathrm{d}}{\mathrm{d}t}i_{\mathrm{s}}^{q}(t) &= \tfrac{1}{L_{\mathrm{s}}^{q}}\Big(u_{s}^{q}(t) - R_{\mathrm{s}}i_{\mathrm{s}}^{q}(t)\underbrace{-\omega_{\mathrm{k}}(t)L_{\mathrm{s}}^{d}i_{\mathrm{s}}^{d}(t) - \omega_{\mathrm{k}}(t)\psi_{\mathrm{pm}}}_{=:u_{\mathrm{s,dist}}^{q}(t)}\Big).
\end{aligned}\right\} \tag{120}$$

The angular frequency $\omega_{\mathrm{k}}$ is linked to the mechanical angular frequency $\omega_{\mathrm{m}}$ via the number $n_{\mathrm{p}}$ of pole pairs. It holds that $\omega_{\mathrm{k}}(t) = \omega_{\mathrm{r}}(t) = n_{\mathrm{p}}\,\omega_{\mathrm{m}}(t)$ and, therefore, $\omega_{\mathrm{k}}$ is *state dependent* and *not* time dependent (see (80)). For an abbreviated notation, it is useful to define the time constants

$$T_{\mathrm{s}}^{d} := L_{\mathrm{s}}^{d}/R_{\mathrm{s}}\,(\text{in s}) \quad \text{and} \quad T_{\mathrm{s}}^{q} := L_{\mathrm{s}}^{q}/R_{\mathrm{s}}\,(\text{in s}). \tag{121}$$

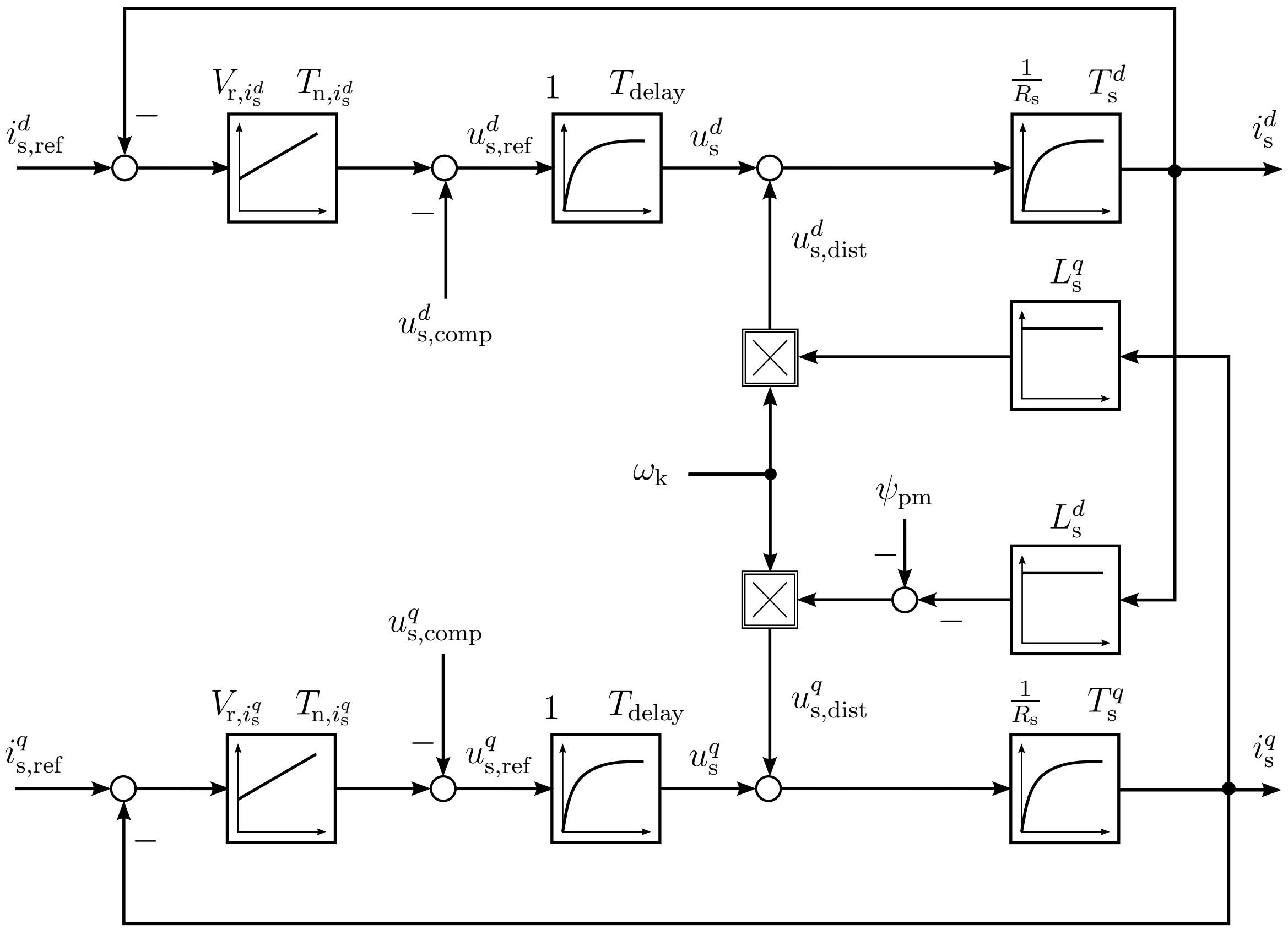

Fig. 33: *Coupled, machine-side current control system in the $k$-reference frame (*flux orientation*) with PI-controllers* (126)*, disturbance compensation* (122)*, approximated converter dynamics* (90) *and generator dynamics* (120)*.*

of the stator in the $d$ and $q$-direction. In Fig. 33, the *nonlinearly* coupled stator current dynamics (120), the approximated converter dynamics (90) and the PI controllers are shown. For isotropic machines, where $L^d_{\rm s} = L^q_{\rm s}$, the time constants in (121) are identical for $d$- and $q$-component.

*3) Compensation of machine-side disturbances:* As in the case of the grid-side current control system (see Section IV-A3), the disturbances $u^d_{\rm s,dist}$ and $u^q_{\rm s,dist}$ in (120) (see Fig. 30) are best compensated for by a *disturbance feedforward network* of the following form

$$\boldsymbol{u}^k_{\rm s,comp}(t) = \begin{pmatrix} u^d_{\rm s,comp}(t) \\ u^q_{\rm s,comp}(t) \end{pmatrix} = \mathscr{L}^{-1}\left\{F_K(s)\right\} \begin{pmatrix} \omega_{\rm k}(t) L^q_{\rm s} i^q_{\rm s}(t) \\ -\omega_{\rm k}(t) L^d_{\rm s} i^d_{\rm s}(t) - \omega_{\rm k}(t)\psi_{\rm pm} \end{pmatrix}. \tag{122}$$

The compensation transfer function $F_K(s)$ is selected as it was in (109). A detailed discussion of the influences of the disturbances (also called cross-coupling) can be found [28, Fig. 16.4]. The cross-coupling is shown in Fig. 33.

*4) Design of the machine-side current controllers:*

**Assumption (A.10)** *For an* ideal *compensation of the machine-side disturbances in* (119) *and* (120)*,*

$$\forall t \geq 0\,{\rm s}: \qquad \boldsymbol{u}^k_{\rm s,comp}(t) = \mathscr{L}^{-1}\left\{1 + s\,T_{\rm delay}\right\} \boldsymbol{u}^k_{\rm s,dist}(t), \tag{123}$$

the influence of the *nonlinear* cross-coupling can be eliminated and the dynamics are decomposed into two decoupled first-order lag systems systems of the form

$$\left.\begin{aligned} \tfrac{\rm d}{\rm dt}\, i^d_{\rm s}(t) &= \tfrac{1}{L^d_{\rm s}}\Big(u^d_{\rm s}(t) - R_{\rm s} i^d_{\rm s}(t)\Big) \quad \circ\!\!-\!\!\bullet \quad \tfrac{i^d_{\rm s}(s)}{u^d_{\rm s}(s)} = \tfrac{\frac{1}{R_{\rm s}}}{1+s\,T^d_{\rm s}} \\ \tfrac{\rm d}{\rm dt}\, i^q_{\rm s}(t) &= \tfrac{1}{L^q_{\rm s}}\Big(u^q_{\rm s}(t) - R_{\rm s} i^q_{\rm s}(t)\Big) \quad \circ\!\!-\!\!\bullet \quad \tfrac{i^q_{\rm s}(s)}{u^q_{\rm s}(s)} = \tfrac{\frac{1}{R_{\rm s}}}{1+s\,T^q_{\rm s}}. \end{aligned}\right\} \tag{124}$$

Taking into account the converter dynamics (90), the following overall system transfer functions

$$\left.\begin{aligned} F_{S,i_{\rm s}^d}(s) &= \frac{i_{\rm s}^d(s)}{u_{\rm s,ref}^d(s)} &= \frac{\frac{1}{R_{\rm s}}}{(1+sT_{\rm s}^d)(1+sT_{\rm delay})} \\ F_{S,i_{\rm s}^q}(s) &= \frac{i_{\rm s}^q(s)}{u_{\rm s,ref}^q(s)} &= \frac{\frac{1}{R_{\rm s}}}{(1+sT_{\rm s}^q)(1+sT_{\rm delay})} \end{aligned}\right\} \text{ with time constants } T_s^{d/q} = \frac{L_s^{d/q}}{R_{\rm s}} \tag{125}$$

are obtained for the controller design. According to the optimization table in [28, p. 81/82], both PI controllers are tuned with the help of the Magnitude Optimum to achieve a fast transient behavior. The following tuning for the $d$ and $q$-current controllers is used:

$$\left.\begin{aligned} F_{C,i_{\rm s}^d}(s) &= V_{{\rm r},i_{\rm s}^d}\frac{1+s\,T_{{\rm n},i_{\rm s}^d}}{s\,T_{{\rm n},i_{\rm s}^d}} \quad \text{with} \quad V_{{\rm r},i_{\rm s}^d} = \frac{T_{\rm s}^d}{2\cdot\frac{1}{R_{\rm s}}\cdot T_{\rm delay}} \quad \text{and} \quad T_{{\rm n},i_{\rm s}^d} = T_{\rm s}^d, \\ F_{C,i_{\rm s}^q}(s) &= V_{{\rm r},i_{\rm s}^q}\frac{1+s\,T_{{\rm n},i_{\rm s}^q}}{s\,T_{{\rm n},i_{\rm s}^q}} \quad \text{with} \quad V_{{\rm r},i_{\rm s}^q} = \frac{T_{\rm s}^q}{2\cdot\frac{1}{R_{\rm s}}\cdot T_{\rm delay}} \quad \text{and} \quad T_{{\rm n},i_{\rm s}^q} = T_{\rm s}^q. \end{aligned}\right\} \tag{126}$$

Clearly, for anisotropic machines with $L_{\rm s}^d \neq L_{\rm s}^q$, the controllers in (126) are parameterized *differently*. Analogous to the derivation of the grid-side current closed-loop dynamics in (114), the machine-side current closed-loop dynamics can be also approximated by the first-order transfer function

$$F_{CL,i_{\rm s}^{d/q}}(s) = \frac{i_{\rm s}^{d/q}(s)}{i_{\rm s,ref}^{d/q}(s)} \approx \frac{1}{1+s\,T_{{\rm app},\boldsymbol{i}_{\rm s}^k}}$$

$$\text{with approximated time constant } T_{{\rm app},\boldsymbol{i}_{\rm s}^k} := 2\,T_{\rm delay} \tag{127}$$

**Remark IV.1.** *As a rule, the effects of actuator saturation (input constraint) due to the limited DC-link voltage* $u_{\rm dc}$ *in the converter must be taken into account. For this purpose, the PI-current controllers* (113) *and* (126) *should be implemented with anti-windup strategies on the machine and grid side, respectively. Windup effects lead to a deteriorated controller performance which should be avoided. A detailed discussion of windup effects and various anti-windup strategies can be found in [28, Sec. 5.6]. There are also limitations due to the assumptions made (cf. [28, Sec. 13.9]). In addition to the controller design in the flux orientation, the machine can also be directly controlled in the stator-fixed reference frame using so called "proportional-resonant controllers" (see [28, Sec. 3.6] or [48]). A detailed description of discrete-time stator control methods can be found in [28, Sec. 14.9].*

*5) Torque generation and approximation of its dynamics:* In (80), the relationship between the stator flux linkage $\boldsymbol{\psi}_{\rm s}^s$, stator currents $\boldsymbol{i}_{\rm s}^s$ (in the $s$-reference frame) and the generator torque $m_{\rm m}$ was established. Now, the torque dynamics in the $k$-coordinate system will be derived for the upcoming speed control problem. Applying the Park transformation (32) to the torque equation in (80) yields

$$\begin{aligned} m_{\rm m}(t) &\overset{(80)}{=} \frac{3}{2}n_{\rm p}\,\big(\underbrace{\boldsymbol{T}_{\rm p}(\phi_{\rm k}(t))\boldsymbol{i}_{\rm s}^k(t)}_{=\boldsymbol{i}_{\rm s}^s(t)}\big)^\top \boldsymbol{J}\,\underbrace{\boldsymbol{T}_{\rm p}(\phi_{\rm k}(t))\boldsymbol{\psi}_{\rm s}^k(t)}_{=\boldsymbol{\psi}_{\rm s}^s(t)} \\ &\overset{(35)}{=} \frac{3}{2}n_{\rm p}\,\boldsymbol{i}_{\rm s}^k(t)^\top\,\underbrace{\boldsymbol{T}_{\rm p}(\phi_{\rm k}(t))^{-1}\boldsymbol{T}_{\rm p}(\phi_{\rm k}(t))}_{=\boldsymbol{I}_2}\,\boldsymbol{J}\boldsymbol{\psi}_{\rm s}^k(t) \\ &\overset{(80)}{=} \frac{3}{2}n_{\rm p}\,\boldsymbol{i}_{\rm s}^k(t)^\top\boldsymbol{J}\Bigg(\underbrace{\begin{bmatrix} L_{\rm s}^d & 0\,\frac{\rm V\,s}{\rm A} \\ 0\,\frac{\rm V\,s}{\rm A} & L_{\rm s}^q \end{bmatrix}}_{=\boldsymbol{L}_{\rm s}^k}\boldsymbol{i}_{\rm s}^k(t) + \underbrace{\begin{pmatrix} \psi_{\rm pm} \\ 0\,{\rm V\,s} \end{pmatrix}}_{=\boldsymbol{\psi}_{\rm pm}^k}\Bigg) \end{aligned} \tag{128}$$

$$\overset{(34)}{=} \underbrace{\frac{3}{2}n_{\rm p}\,\psi_{\rm pm}\,i_{\rm s}^q(t)}_{=:m_{\rm el}(t)\ \text{(PM torque)}} + \underbrace{\frac{3}{2}n_{\rm p}\,(L_{\rm s}^d - L_{\rm s}^q)\,i_{\rm s}^d(t)\,i_{\rm s}^q(t)}_{=:m_{\rm re}(t)\ \text{(reluctance torque)}}. \tag{129}$$

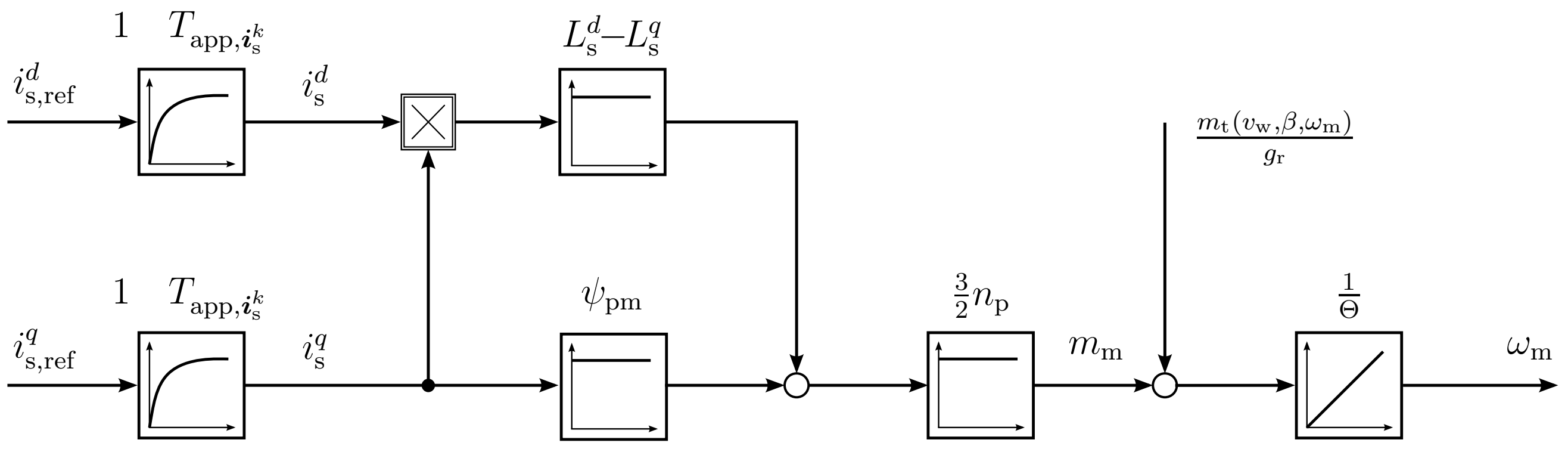

Fig. 34: *Signal flow diagram of the permanent magnet synchronous generator in the flux orientation (k-reference frame) with an approximated current control circuit.*

Here, $n_{\mathrm{p}}$ is the number of poles of the generator, $i^d_{\mathrm{s}}$ (in A) and $i^q_{\mathrm{s}}$ (in A) are the $d$ and $q$-components of the generator stator currents, $L^d_{\mathrm{s}}$ (in $\frac{\mathrm{V\,s}}{\mathrm{A}}$) and $L^q_{\mathrm{s}}$ (in $\frac{\mathrm{V\,s}}{\mathrm{A}}$) are the $d$ and $q$-components of the stator inductance and $\psi_{\mathrm{pm}}$ (in V s) is the flux linkage of the permanent magnet in the rotor. In Fig. 34, the signal flow diagram of the torque generation of the permanent magnet generator is illustrated taking into account the approximated current closed-loop dynamics. For isotropic machines, where $L^d_{\mathrm{s}} = L^q_{\mathrm{s}}$, only the electro-magnetic torque $m_{\mathrm{el}}$ (in N m), and no reluctance torque $m_{\mathrm{re}} = 0\,\mathrm{N\,m}$, is contributing to torque generation (129). For $L^d_{\mathrm{s}} \neq L^q_{\mathrm{s}}$, the reluctance torque $m_{\mathrm{re}}$ can also be exploited (in the sense of "Maximum Torque per Ampere (MTPA)" control, see [28, Sec. 16.7.1] or [49, 50]). In the following, for simplicity, it is assumed

**Assumption (A.11)** *that either an isotropic machine is present, i.e.* $L^d_{\mathrm{s}} = L^q_{\mathrm{s}}$, *or that* $i^d_{\mathrm{s,ref}}(t) = 0\,\mathrm{A}$ *for all* $t \geq 0\,\mathrm{s}$.

and then either $m_{\mathrm{re}}(t) = 0\,\mathrm{N\,m}$ or, because of the (almost) decoupled current control systems, $i^d_{\mathrm{s}}(t) \approx i^d_{\mathrm{s,ref}}(t) = 0\,\mathrm{A}$. In either case $m_{\mathrm{m}}(t) = m_{\mathrm{el}}(t)$ for all $t \geq 0\,\mathrm{s}$. Thus, the simplified dynamic relationship between the current component $i^q_{\mathrm{s,ref}}$ and the generator torque $m_{\mathrm{m}}$ is given by

$$\frac{m_{\mathrm{m}}(s)}{i^q_{\mathrm{s,ref}}(s)} \overset{(129),(A.11)}{=} \frac{3}{2} n_{\mathrm{p}}\, \psi_{\mathrm{pm}} \frac{i^q_{\mathrm{s}}(s)}{i^q_{\mathrm{s,ref}}(s)} \overset{(127)}{=} \frac{3}{2} n_{\mathrm{p}}\, \psi_{\mathrm{pm}} \frac{1}{1 + s\, T_{\mathrm{app},i^q_{\mathrm{s}}}} \approx \frac{3}{2} n_{\mathrm{p}}\, \psi_{\mathrm{pm}}. \tag{130}$$

For wind power plants, the time constant $T_{\mathrm{app},i^q_{\mathrm{s}}} = 2T_{\mathrm{delay}} \ll 1\,\mathrm{s}$ of the current control system is negligibly small with respect to the "time constant" of the mechanical system, such that the approximation in (130) is justified. Thus, for any predefined torque $m_{\mathrm{m,ref}}$ (in N m), the torque feedforward control problem is solved via an appropriate selection of the $q$-reference current and leads to a simple proportional behavior:

$$\forall t \geq 0\,\mathrm{s}: \quad i^q_{\mathrm{s,ref}}(t) = \frac{1}{\frac{3}{2} n_{\mathrm{p}}\, \psi_{\mathrm{pm}}} m_{\mathrm{m,ref}}(t) \quad \overset{(130)}{\Longrightarrow} \quad m_{\mathrm{m}}(t) \approx m_{\mathrm{m,ref}}(t). \tag{131}$$

**Remark IV.2.** *According to* (58), *the stator reactive power is calculated in flux orientation by* $q_{\mathrm{s}}(t) = \frac{3}{2} \boldsymbol{u}^k_{\mathrm{s}}(t)^\top \boldsymbol{J} \boldsymbol{i}^k_{\mathrm{s}}(t) = \frac{3}{2}\big(u^q_{\mathrm{s}}(t) i^d_{\mathrm{s}}(t) - u^d_{\mathrm{s}}(t) i^q_{\mathrm{s}}(t)\big)$ *(in* var*). Note that, for* $i^d_{\mathrm{s}}(t) = 0\,\mathrm{A}$, *the stator reactive power is not zero, since*

$$\forall t \geq 0\,\mathrm{s}: \qquad q_{\mathrm{s}}(t) \overset{i^d_{\mathrm{s}}(t)=0}{=} -\frac{3}{2} u^d_{\mathrm{s}}(t) i^q_{\mathrm{s}}(t). \tag{132}$$

*Depending on the voltage* $u^d_{\mathrm{s}}(t)$ *and current* $i^q_{\mathrm{s}}(t)$, *a large reactive power* $q_{\mathrm{s}}(t)$ *may result. This must be taken into account when the power ratings of the converter are specified (in terms of apparent power).*

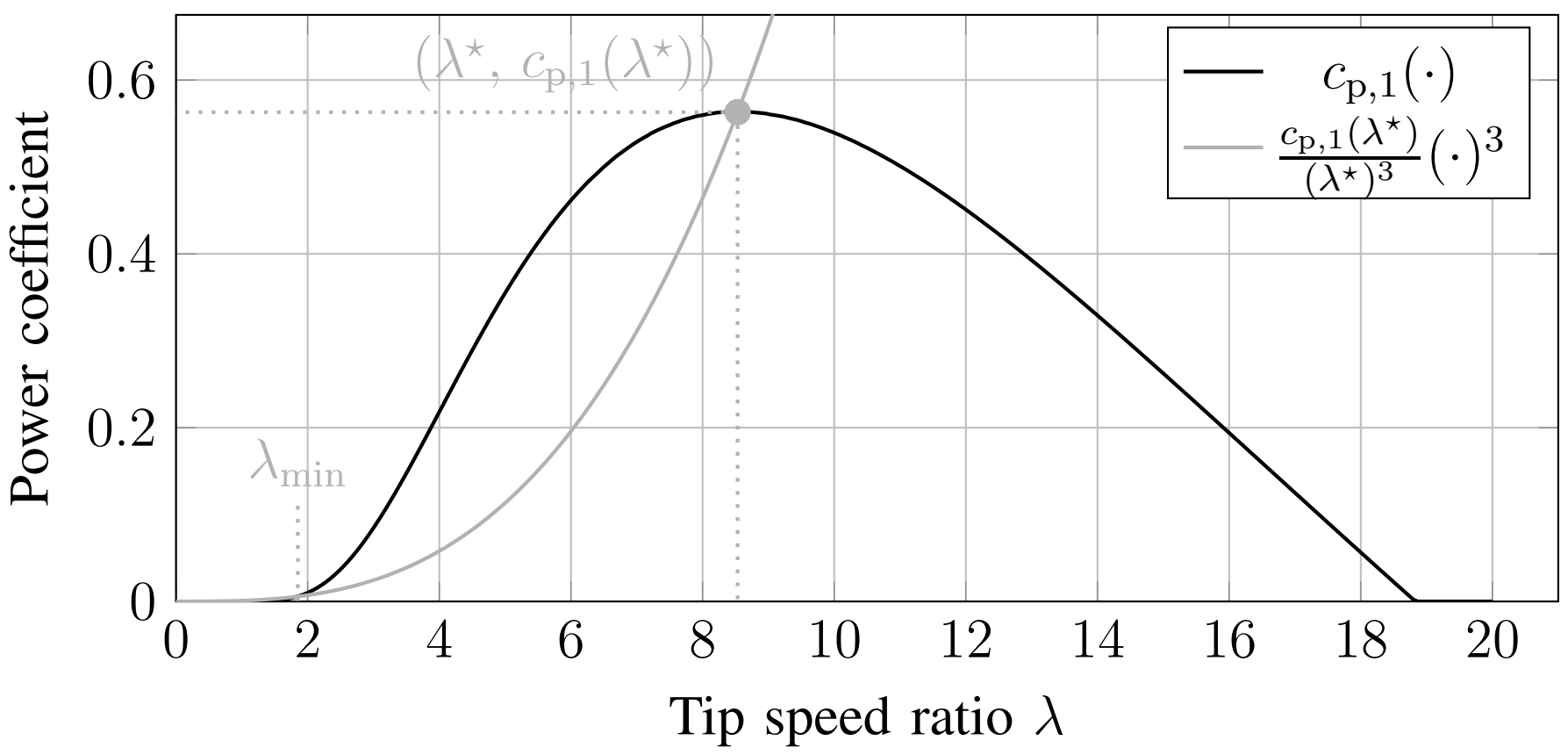

Fig. 35: *Power coefficient* $c_p(0,\cdot) := c_{p,1}(\cdot)$ *of the considered wind turbine system: The goal is to operate the turbine at its optimum tip speed ratio* $\lambda^\star$ *via the speed regulation of the generator.*

## *C. Control of the generator speed (and tip speed ratio)*

The instantaneous or active power which is transferred into the grid must be provided by the turbine via the generator (electric machine). For this reason, the generator must counteract the turbine torque $m_t$ (in $\mathrm{N\,m}$) with the machine torque $m_m$ (in $\mathrm{N\,m}$).

In the operation regimes I and IV (see Fig. 11), the wind turbine is at a standstill (due to too low or too high wind speeds). No power is delivered to the grid and the generator speed is not regulated.

In operation regime III, the generator outputs the rated mechanical power

$$p_{m,mech,nom} := m_{m,nom}\omega_{m,nom} > 0 \ \text{(in W)}$$

to the grid (neglecting losses). The generator is torque-controlled and consistently produces the constant nominal torque $m_{m,nom}$ (in $\mathrm{N\,m}$) at nominal angular velocity $\omega_{m,nom}$ (in $\frac{\mathrm{rad}}{\mathrm{s}}$) (see Section I-D6). The rotational speed is then controlled by the pitch control system by adjusting (reducing) the turbine torque $m_t/g_r$ to equal the machine torque $m_{m,nom}$. Thus, an acceleration or deceleration of the generator away from its nominal operation point is avoided.

In operation range II, the wind turbine operates below its nominal power and the pitch control system is inactive. The aim is to achieve a maximum conversion of the available wind power to electrical power and the following holds

$$\forall\, t \geq 0\,\mathrm{s}: \quad p_m(t) = m_m(t)\omega_m(t) \leq p_{m,nom} \quad \text{and} \quad \beta(t) = 0^\circ. \tag{133}$$

For this purpose, a change of the tip speed ratio and the turbine/generator speed is required. This adjustment is made by controlling the generator speed (or generator angular velocity). The speed control system is superimposed on the closed-loop system of the $q$-component of the stator current. In view of Assumption (A.11), the $d$-component of the stator current is set to zero. For the further discussion of a simplified but nonlinear speed control systems, the following two assumptions are imposed:

**Assumption (A.12)** *The wind turbine will operate in* operation regime II *(see Fig. 11), i.e.,*

$$\forall t \geq 0\,\mathrm{s}: \quad v_{nom} > v_w(t) \geq v_{cut-in} > 0\,\frac{\mathrm{m}}{\mathrm{s}} \ \wedge\ \omega_m(t) \geq \omega_{m,min} > 0\,\frac{\mathrm{rad}}{\mathrm{s}}.$$

**Assumption (A.13)** *The dynamics of the current/torque control systems are sufficiently fast with respect to the mechanics of the wind turbine, and therefore* (131) *holds ideally with*

$$\forall t \geq 0\,\mathrm{s}: \quad m_{m,ref}(t) = m_m(t).$$

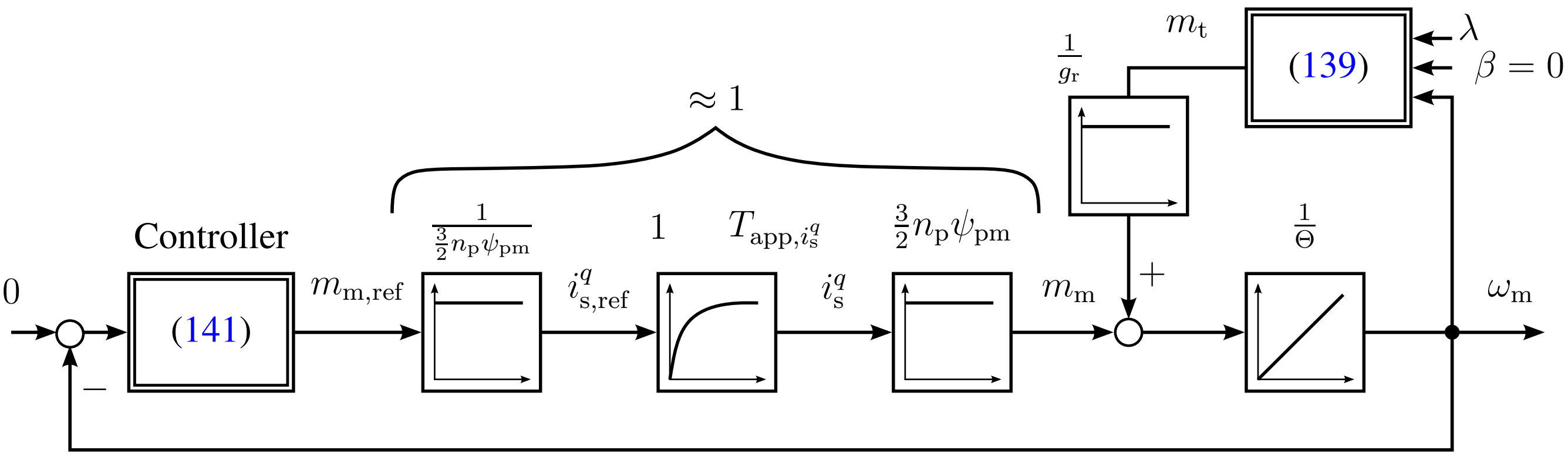

Fig. 36: *Simplified representation of the nonlinear speed control loop.*

*1) Control objective:* In the operation range II (see Fig. 11), the pitch control system is inactive and (133) holds. Thus the turbine power is calculated as

$$p_{\mathrm{t}}(v_{\mathrm{w}},\beta,\lambda) \overset{(133)}{=} p_{\mathrm{t}}(v_{\mathrm{w}},0,\lambda) \overset{(62)}{=} c_{\mathrm{p}}(0,\lambda)\underbrace{\frac{1}{2}\varrho\pi r_{\mathrm{t}}^2 v_{\mathrm{w}}(t)^3}_{=p_{\mathrm{w}}(t)} \overset{(64)}{\leq} c_{\mathrm{p}}(0,\lambda^\star)\, p_{\mathrm{w}}(t).$$

The power yield can then only be influenced by the tip speed ratio $\lambda = \frac{r_{\mathrm{t}}\omega_{\mathrm{t}}}{v_{\mathrm{w}}}$ or, for a known wind speed $v_{\mathrm{w}}$, by the rotational speed $r_{\mathrm{t}}\,\omega_{\mathrm{t}}$. In Fig. 35, the power coefficient $c_{\mathrm{p}}(0,\cdot) := c_{\mathrm{p},1}(\cdot)$ of a $2\,\mathrm{MW}$ wind turbine without a pitch control system is depicted. The goal is to dynamically track the maximum power point $c_{\mathrm{p}}(0,\lambda^\star) := c_{\mathrm{p},1}(\lambda^\star)$ shown in Fig. 35, i.e.,

$$\lambda \to \lambda^\star \qquad \overset{(P_1)-(P_3)\text{ of } c_{\mathrm{p}}(\cdot,\cdot)\text{ (see p. 26)}}{\Longrightarrow} \qquad c_{\mathrm{p}}(0,\lambda) \to c_{\mathrm{p}}(0,\lambda^\star),$$

Ideally, the wind turbine should always operate at the optimum tip speed ratio $\lambda^\star$ and the maximum power point of the power coefficient $c_{\mathrm{p}}(0,\lambda^\star)$. The tip speed ratio

$$\forall\, t \geq 0\,\mathrm{s}: \quad \lambda(t) \overset{(61)}{=} \frac{r_{\mathrm{t}}\omega_{\mathrm{t}}(t)}{v_{\mathrm{w}}(t)} \overset{(75)}{=} \frac{r_{\mathrm{t}}\omega_{\mathrm{m}}(t)}{g_{\mathrm{r}}v_{\mathrm{w}}(t)} \geq \frac{r_{\mathrm{t}}\omega_{M,\mathrm{min}}}{g_{\mathrm{r}}v_{\mathrm{nom}}} > 0 \tag{134}$$

depends on wind speed $v_{\mathrm{w}}$, turbine radius $r_{\mathrm{t}}$, turbine angular velocity $\omega_{\mathrm{t}}$, gear ratio $g_{\mathrm{r}}$, and generator angular velocity $\omega_{\mathrm{m}}$. In light of Assumption (A.12), the tip speed ratio is always positive in operation regime II.

Thus, the task of the generator speed control is to ensure a precise reference tracking capability. Depending on the optimum tip speed ratio $\lambda^\star$ (and therefore the wind velocity $v_{\mathrm{w}}$), the optimum generator angular velocity

$$\forall\, t \geq 0\,\mathrm{s}: \qquad \omega^\star_{\mathrm{m}}(t) := g_{\mathrm{r}}\omega^\star_{\mathrm{t}}(t) := \frac{g_{\mathrm{r}}\lambda^\star}{r_{\mathrm{t}}} v_{\mathrm{w}}(t) \tag{135}$$

must be tracked as good as possible. For optimal power conversion, the designed speed controller should ideally comply with the following control objective:

$$\forall\, t \geq 0\,\mathrm{s}: \quad \omega_{\mathrm{m}}(t) = \omega^\star_{\mathrm{m}}(t) \qquad \text{and} \qquad \lambda(t) = \lambda^\star \tag{136}$$

*2) Rewritten nonlinear dynamics of the mechanical system:* The model of a permanent-magnet synchronous generator was presented in Section III-D. In view of Assumption (A.13), where the current/torque control is considered fast enough, the generator mechanics can be written as follows:

$$\tfrac{\mathrm{d}}{\mathrm{d}t}\,\omega_{\mathrm{m}}(t) \overset{(80)}{=} \frac{1}{\Theta}\left(\frac{m_{\mathrm{t}}(v_{\mathrm{w}},\beta,\omega_{\mathrm{t}})}{g_{\mathrm{r}}} + m_{\mathrm{m}}(t)\right), \qquad \omega_{\mathrm{m}}(0\,\mathrm{s}) = \omega^0_{\mathrm{m}} \overset{(A.12)}{\geq} \omega_{\mathrm{m,min}}$$

$$\overset{(A.13)}{=} \frac{1}{\Theta}\left(\frac{m_t(v_w,\beta,\omega_t)}{g_r} + m_{m,ref}(t)\right). \tag{137}$$

Moreover, Assumption (A.12) ensures, that the generator rotates with a positive initial angular velocity $\omega_m(0\,\mathrm{s}) = \omega_m^0 \geq \omega_{m,min} > 0\,\frac{\mathrm{rad}}{\mathrm{s}}$. By invoking

$$v_w \overset{(134)}{=} \frac{r_t \omega_m}{g_r \lambda}, \tag{138}$$

the turbine torque $m_t(v_w,\beta,\omega_t)$ in (137) can be rewritten as

$$\begin{aligned} m_t(v_w,\beta,\omega_t) &\overset{(69)}{=} \frac{1}{2}\varrho\,\pi\, r_t^2\, v_w^3\, \frac{c_p(v_w,\beta,\omega_t)}{\omega_t} \\ &\overset{(70)}{=} \frac{1}{2}\varrho\,\pi\, r_t^3\, v_w^2\, \frac{c_p(\beta,\lambda)}{\lambda} \\ &\overset{(138)}{=} \frac{1}{2}\frac{\varrho\,\pi\, r_t^5}{g_r^2}\frac{c_p(\beta,\lambda)}{\lambda^3}\omega_m^2 = m_t(\beta,\lambda,\omega_m) \end{aligned} \tag{139}$$

which shows that it depends *nonlinearly* on the state $\omega_m$. Inserting (139) with $\beta = 0^\circ$ into (137) yields the nonlinear dynamics of the machine-side mechanics

$$\frac{\mathrm{d}}{\mathrm{d}t}\omega_m(t) \overset{(137),(139)}{=} \frac{1}{\Theta}\Bigg( \underbrace{\frac{1}{2}\frac{\varrho r_t^5 \pi}{g_r^3}}_{=:c_0>0\,\mathrm{kg\,m^2}} \underbrace{\frac{c_p(0,\lambda(t))}{\lambda(t)^3}}_{\overset{(134)}{>}0} \omega_m(t)^2 + m_{m,ref}(t)\Bigg). \tag{140}$$

Now, a controller with the control output $m_{m,ref}(t)$ (idealized control variable) should be found which stabilizes the nonlinear closed-loop system *and* reaches the control objective (136).

**Remark IV.3.** *The turbine torque* $m_t(\beta,\lambda,\omega_m) = m_t(v_w,\beta,\omega_t) \neq m_t(t)$ *in* (139) *is* not *an external (time-varying) disturbance but a nonlinear function of state* and *time. It depends on pitch angle* $\beta$*, (time-varying) wind speed* $v_w$*, tip speed ratio* $\lambda$*, turbine angular velocity* $\omega_t$ *and generator angular velocity* $\omega_m$ *(system state). A classical tuning of the speed controller (e.g. according to the Symmetrical Optimum), is thus* not *reasonable.*

*3) Nonlinear speed controller:* In [51], the *proportional* but *nonlinear* controller

$$m_{m,ref}(t) = -k_p^\star\,\omega_m(t)^2 \qquad \text{with} \qquad k_p^\star := c_0\,\frac{c_p(0,\lambda^\star)}{(\lambda^\star)^3} > 0\,\mathrm{kg\,m^2} \tag{141}$$

is proposed as speed controller of the wind turbine (140). The closed-loop system of mechanics (140) and controller (141) is shown as a signal flow diagram in Fig. 36. Inserting the nonlinear controller (141) into the mechanics equation (140) of the wind turbine leads to the closed-loop control dynamics

$$\begin{aligned} \frac{\mathrm{d}}{\mathrm{d}t}\omega_m(t) &= \frac{1}{\Theta}\left(c_0\,\frac{c_p(\lambda(t))}{\lambda(t)^3} - k_p^\star\right)\omega_m(t)^2 \\ &\overset{(141)}{=} \frac{c_0}{\Theta}\left(\frac{c_p(\lambda(t))}{\lambda(t)^3} - \frac{c_p(0,\lambda^\star)}{(\lambda^\star)^3}\right)\omega_m(t)^2. \end{aligned} \tag{142}$$

**Remark IV.4.** *It is noteworthy that the nonlinear controller* (141) *works* without *a set-point reference of the form* $\omega_m^\star$ *(recall that* $\omega_m^\star = 0\,\frac{\mathrm{rad}}{\mathrm{s}}$ *in Fig. 36) but still achieves the control objective* (136) *(see [51], [52] and Section IV-C4). Furthermore, note that no positive generator torque can be set by the controller* (141)*, since*

$$\forall t \geq 0\,\mathrm{s}: \qquad m_m(t) \approx m_{m,ref}(t) \overset{(141)}{\leq} 0\,\mathrm{N\,m}.$$

*Thus, an acceleration of the turbine (see* (137)*) is possible only by the turbine toque* $m_t(v_w,\beta,\omega_t)$ *itself.*

*4) Analysis of the closed-loop system and control performance:* In [52], it was shown that, for a *constant* wind velocity (see Assumption A.1 on 6), the following holds:

(i) the closed-loop system (140), (141) has a locally stable equilibrium point in $\omega_{\mathrm{m}}^{\star}$ (or $\lambda^{\star}$) and,
(ii) the equilibrium point $\omega_{\mathrm{m}}^{\star}$ (or $\lambda^{\star}$) is locally attractive, i.e., for all $\omega_{\mathrm{m}}^{0} > \omega_{\mathrm{m,min}}$ (or $\lambda_0 > \lambda_{\min}$), it holds that $\lim_{t\to\infty} \omega_{\mathrm{m}}(t) = \omega_{\mathrm{m}}^{\star}$ (or $\lim_{t\to\infty} \lambda(t) = \lambda^{\star}$).

This result will be discussed in more detail. To do so, three different cases for the initial value of the tip speed ratio

$$\lambda_0 := \lambda(0\,\mathrm{s}) = \frac{r_{\mathrm{t}}\omega_{\mathrm{m}}^{0}}{g_{\mathrm{r}} v_{\mathrm{w}}} \geq \frac{r_{\mathrm{t}}\omega_{\mathrm{m,min}}}{g_{\mathrm{r}} v_{\mathrm{w}}} =: \lambda_{\min} > 0 \qquad (\text{with } v_{\mathrm{w}} = \mathit{const.}) \tag{143}$$

are considered to analyze[16] the behavior of the closed-loop system (140), (141):

**1. Case** $\lambda_0 = \lambda^{\star}$ *(see Fig. 35)*
The wind turbine is already operating in its optimal operation point $\omega_{\mathrm{m}}^{\star}$ (or $\lambda^{\star}$). A change in the tip speed ratio or the generator angular velocity is not required and

$$\forall\, t \geq 0\,\mathrm{s}: \qquad \tfrac{\mathrm{d}}{\mathrm{d}t}\omega_{\mathrm{m}}(t) \overset{(143)}{=} \frac{g_{\mathrm{r}} v_{\mathrm{w}}}{r_{\mathrm{t}}} \tfrac{\mathrm{d}}{\mathrm{d}t}\lambda(t) \overset{!}{=} 0\,\frac{\mathrm{rad}}{\mathrm{s}^2}.$$

should hold further on. The nonlinear closed-loop system (140), (141) shows precisely this behavior, since for $\lambda_0 = \lambda^{\star}$,

$$\forall\, t \geq 0\,\mathrm{s}:\; \tfrac{\mathrm{d}}{\mathrm{d}t}\omega_{\mathrm{m}}(t) = \frac{c_0}{\Theta}\left(\frac{c_{\mathrm{p}}(0,\lambda_0)}{\lambda_0^3} - \frac{c_{\mathrm{p}}(0,\lambda^{\star})}{(\lambda^{\star})^3}\right)\omega_{\mathrm{m}}(t)^2 = 0\,\frac{\mathrm{rad}}{\mathrm{s}^2}.$$

The control objective (136) is (immediately) achieved.

**2. Case** $\lambda_0 \in (\lambda^{\star}, \infty)$ *(right side of $\lambda^{\star}$ in Fig. 35)*
For $\lambda_0 > \lambda^{\star}$ (and a constant wind velocity $v_{\mathrm{w}}$), it follows that $\omega_{\mathrm{m}}^{0} > \omega_{\mathrm{m}}^{\star} \overset{(135)}{=} \frac{g_{\mathrm{r}}\lambda^{\star}}{r_{\mathrm{t}}} v_{\mathrm{w}}$. In order to reach the equilibrium point $\omega_{\mathrm{m}}^{\star}$ and $\lambda^{\star}$, $\frac{\mathrm{d}}{\mathrm{d}t}\,\omega_{\mathrm{m}} < 0\,\frac{\mathrm{rad}}{\mathrm{s}^2}$ or $\frac{\mathrm{d}}{\mathrm{d}t}\,\lambda < 0\,\frac{1}{\mathrm{s}}$ must hold, respectively. Taking into account the following inequalities

$$\forall\, \lambda \in (\lambda^{\star}, \infty): \qquad c_{\mathrm{p}}(\lambda) \leq c_{\mathrm{p}}(\lambda^{\star}) \quad \text{and} \quad \frac{1}{\lambda^3} \leq \frac{1}{(\lambda^{\star})^3}, \tag{144}$$

it can be seen that $\frac{\mathrm{d}}{\mathrm{d}t}\,\omega_{\mathrm{m}} < 0\,\frac{\mathrm{rad}}{\mathrm{s}^2}$ or $\frac{\mathrm{d}}{\mathrm{d}t}\,\lambda < 0\,\frac{1}{\mathrm{s}}$ for the controller parameterization in (141) is satisfied, because

$$\forall \lambda \in (\lambda^{\star}, \infty): \qquad c_0\,\frac{c_{\mathrm{p}}(\lambda)}{\lambda^3} - \underbrace{c_0\,\frac{c_{\mathrm{p}}(\lambda^{\star})}{(\lambda^{\star})^3}}_{k_{\mathrm{p}}^{\star}} \overset{(144)}{<} 0\,\mathrm{kg\,m}^2.$$

Therefore, it follows that $\frac{\mathrm{d}}{\mathrm{d}t}\,|\omega_{\mathrm{m}}| \to 0\,\frac{\mathrm{rad}}{\mathrm{s}^2}$ and $\frac{\mathrm{d}}{\mathrm{d}t}\,\lambda \to 0\,\frac{1}{\mathrm{s}}$, respectively, and eventually the wind turbine remains at the equilibrium point. The control objective (136) is reached asymptotically, i.e. $\omega_{\mathrm{m}} \to \omega_{\mathrm{m}}^{\star}$ and $\lambda \to \lambda^{\star}$.

**3. Case** $\lambda_0 \in [\lambda_{\min}, \lambda^{\star})$ *(left side of $\lambda^{\star}$ in Fig. 35)*
For $\lambda_0 < \lambda^{\star}$ (and constant wind velocity $v_{\mathrm{w}}$), it follows that $\omega_{\mathrm{m}}^{0} < \omega_{\mathrm{m}}^{\star} \overset{(135)}{=} \frac{g_{\mathrm{r}}\lambda^{\star}}{r_{\mathrm{t}}} v_{\mathrm{w}}$. In order to reach the optimum operation point $\omega_{\mathrm{m}}^{\star}$ and $\lambda^{\star}$, it should hold that $\frac{\mathrm{d}}{\mathrm{d}t}\,\omega_{\mathrm{m}} > 0\,\frac{\mathrm{rad}}{\mathrm{s}^2}$ and also $\frac{\mathrm{d}}{\mathrm{d}t}\,\lambda > 0\,\frac{1}{\mathrm{s}}$. For the controller parameterization in (141), the following inequality is satisfied for any given and known power coefficient $c_{\mathrm{p}}(0,\cdot) = c_{\mathrm{p},1}(\cdot)$ (see Fig. 35):

$$\forall \lambda \in (\lambda_{\min},\, \lambda^{\star}): \qquad c_{\mathrm{p}}(\lambda) > \frac{c_{\mathrm{p}}(\lambda^{\star})}{(\lambda^{\star})^3}\lambda^3.$$

[16] A complete mathematical proof can be found in [52].

Hence,

$$\forall \lambda_0 \in [\lambda_{\min},\, \lambda^\star): \qquad c_0 \frac{c_{\mathrm{p}}(\lambda)}{\lambda^3} - \underbrace{c_0 \frac{c_{\mathrm{p}}(\lambda^\star)}{(\lambda^\star)^3}}_{\overset{(141)}{=} k_{\mathrm{p}}^\star} > 0\,\mathrm{kg\,m}^2 \tag{145}$$

from which $\frac{\mathrm{d}}{\mathrm{d}t}\,\omega_{\mathrm{m}} > 0\,\frac{\mathrm{rad}}{\mathrm{s}^2}$ and $\frac{\mathrm{d}}{\mathrm{d}t}\,\lambda > 0\,\frac{1}{\mathrm{s}}$ follow. As soon as $\omega_{\mathrm{m}} \to \omega_{\mathrm{m}}^\star$ and $\lambda \to \lambda^\star$, then $\frac{\mathrm{d}}{\mathrm{d}t}\,|\omega_{\mathrm{m}}| \to 0$ and $\frac{\mathrm{d}}{\mathrm{d}t}\,\lambda \to 0$. As a result, the wind turbine system asymptotically reaches its optimum operation point and the controller achieves the control objective (136) eventually. *Remark:* Note that, for $\lambda < \lambda_{\min}$, the inequality (145) is *not* satisfied (see also Fig. 35).

In conclusion, the nonlinear controller (141) achieves the control objective (136) for all $\lambda_0 \in (\lambda_{\min}, \infty)$ at least asymptotically. For a constant wind velocity and a sufficiently long settling time, it can thus be assumed that the wind turbine will operate at its optimal tip speed ratio $(\lambda^\star, c_{\mathrm{p}}(0, \lambda^\star))$ and will extract the maximally available wind power.

## *D. Control of the DC-link voltage*

The DC-link voltage control system is superimposed on the current control system of the $d$-component of the grid side filter current (see Fig. 37). The main tasks of the DC-link voltage controller are, (i) that both a minimum and a maximum value of the DC-link voltage $u_{\mathrm{dc}}$ are not exceeded, i.e.,

$$\forall t \geq 0\,\mathrm{s}: \qquad 0\,\mathrm{V} < u_{\mathrm{dc,min}} \leq u_{\mathrm{dc}}(t) \leq u_{\mathrm{dc,max}}, \tag{146}$$

and (ii) that the filter voltage $\boldsymbol{u}_{\mathrm{f}}^k$ and the stator voltage $\boldsymbol{u}_{\mathrm{s}}^k$—the control variables for the underlying current closed-loop systems for $\boldsymbol{i}_{\mathrm{f}}^k$ and $\boldsymbol{i}_{\mathrm{s}}^k$—can be adequately defined.

For this purpose, the selection of a *constant* reference $u_{\mathrm{dc,ref}}$ (within the valid voltage range (146)) is beneficial, since, on one hand, a constant DC-link voltage $u_{\mathrm{dc}} \approx u_{\mathrm{dc,ref}} \geq u_{\mathrm{dc,min}}$ simplifies the implementation of the modulation method (e.g. PWM or SVM) and, on the other hand, constant set-point tracking of the DC-link voltage can be realized with a PI controller (as the following discussion will show).

In the following sections, a PI controller is to be designed for controlling the DC-link voltage. It will be shown that the control of the DC-link voltage is a highly nonlinear control problem and the considered system behavior is structurally changing: For a power flow from the DC-link to the grid, the system is minimum phase; whereas, for a power flow from the grid to the DC-link, the system is *non-minimum phase*. This non-minimum phase behavior necessitates a conservative (slow) controller design. The following simplifying assumptions are made for the upcoming discussion:

**Assumption (A.14)** *The dynamics of the grid-side current closed-loop system are approximated as first-order lag system, i.e.* (115) *holds.*

and

**Assumption (A.15)** *The power from the stator of the generator is considered as an external, time-variant, but bounded disturbance, and* not *as a state-dependent variable, i.e.* $p_{\mathrm{s}}(\cdot) \in \mathcal{L}^\infty(I;Y)$[17] *and* $p_{\mathrm{s}}(t) \neq p_{\mathrm{s}}(u_{\mathrm{dc}})$.

*1) Nonlinear model of the DC-link dynamics:* Starting from the dynamics (85) of the DC-link, a performance evaluation can be carried out. The time derivative of the stored energy in the DC-link capacitor $C_{\mathrm{dc}}$ (in $\frac{\mathrm{A\,s}}{\mathrm{V}}$) corresponds to the power consumption $p_{\mathrm{dc}}$ (in W) in the DC-link circuit, i.e.,

$$p_{\mathrm{dc}}(t) := \frac{\mathrm{d}}{\mathrm{d}t}\left(\frac{1}{2} C_{\mathrm{dc}} u_{\mathrm{dc}}(t)^2\right) = \left(\frac{\mathrm{d}}{\mathrm{d}t}\,u_{\mathrm{dc}}(t)\right) C_{\mathrm{dc}} u_{\mathrm{dc}}(t) = -p_{\mathrm{s}}(t) - p_{\mathrm{f}}(t) \tag{147}$$

[17] $\mathcal{L}^\infty_{(\mathrm{loc})}(I;Y)$ denotes the space of measurable, (locally) essentially bounded functions mapping $I \to Y$ equipped with the norm $\|\boldsymbol{f}\|_\infty := \text{ess-sup}_{t\in I}\|\boldsymbol{f}(t)\|$.

where $p_\mathrm{s}(t)$ (in W) is the machine's stator power and $p_\mathrm{f}(t)$ (in W) is the filter power (see also Fig. 20). The filter power can be calculated as follows

$$\begin{aligned} p_\mathrm{f}(t) &= \boldsymbol{u}_\mathrm{f}^{abc}(t)^\top \boldsymbol{i}_\mathrm{f}^{abc}(t) \overset{(55)}{=} \frac{3}{2}\boldsymbol{u}_\mathrm{f}^{k}(t)^\top \boldsymbol{i}_\mathrm{f}^{k}(t) = \frac{3}{2}\boldsymbol{i}_\mathrm{f}^{k}(t)^\top \boldsymbol{u}_\mathrm{f}^{k}(t) \\ &\overset{(105)}{=} \frac{3}{2}\Big[R_\mathrm{f}\|\boldsymbol{i}_\mathrm{f}^{k}(t)\|^2 + \omega_\mathrm{g} L_\mathrm{f} \underbrace{\boldsymbol{i}_\mathrm{f}^{k}(t)^\top \boldsymbol{J}\boldsymbol{i}_\mathrm{f}^{k}(t)}_{\overset{(34)}{=}0\,\mathrm{A}^2} + L_\mathrm{f}\boldsymbol{i}_\mathrm{f}^{k}(t)^\top \tfrac{\mathrm{d}}{\mathrm{d}t}\boldsymbol{i}_\mathrm{f}^{k}(t) + \boldsymbol{i}_\mathrm{f}^{k}(t)^\top \boldsymbol{u}_\mathrm{g}^{k}(t)\Big] \\ &\overset{(97)}{=} \frac{3}{2}R_\mathrm{f}\|\boldsymbol{i}_\mathrm{f}^{k}(t)\|^2 + \frac{3}{2}L_\mathrm{f}\boldsymbol{i}_\mathrm{f}^{k}(t)^\top \tfrac{\mathrm{d}}{\mathrm{d}t}\boldsymbol{i}_\mathrm{f}^{k}(t) + \frac{3}{2}\hat{u}_\mathrm{g} i_\mathrm{f}^{d}(t). \end{aligned} \tag{148}$$

The dynamics of the DC-link circuit (85) can be reformulated as a function of the machine's stator power and the filter power. Taking Assumption (A.15) into consideration, the following *nonlinear* state differential equation is obtained

$$\begin{aligned} \tfrac{\mathrm{d}}{\mathrm{d}t} u_\mathrm{dc}(t) &\overset{(85)}{=} \frac{1}{C_\mathrm{dc} u_\mathrm{dc}(t)}\Big[-\underbrace{u_\mathrm{dc}(t) i_\mathrm{m}(t)}_{=p_\mathrm{s}(t)} - \underbrace{u_\mathrm{dc}(t) i_\mathrm{g}(t)}_{=p_\mathrm{f}(t)}\Big], \quad u_\mathrm{dc}(0\,\mathrm{s}) = u_\mathrm{dc}^0 \\ &\overset{(148)}{=} \frac{1}{C_\mathrm{dc} u_\mathrm{dc}(t)}\Big[-p_\mathrm{s}(t) \\ &\qquad -\frac{3}{2}R_\mathrm{f}\|\boldsymbol{i}_\mathrm{f}^{k}(t)\|^2 - \frac{3}{2}L_\mathrm{f}\boldsymbol{i}_\mathrm{f}^{k}(t)^\top \tfrac{\mathrm{d}}{\mathrm{d}t}\boldsymbol{i}_\mathrm{f}^{k}(t) - \frac{3}{2}\hat{u}_\mathrm{g} i_\mathrm{f}^{d}(t)\Big]. \end{aligned} \tag{149}$$

A positive initial voltage $u_\mathrm{dc}(0\,\mathrm{s}) = u_\mathrm{dc}^0 > 0\,\mathrm{V}$ can be seen as a result of the free-wheeling diodes in the converter (acting like a diode rectifier). The dynamics (149) of the DC-link circuit represent a nonlinear system: Depending on the operation point, the dynamics (149) are either minimum phase or *non*-minimum phase, which makes the DC-link voltage control problem particularly challenging (for a thorough discussion and other control approaches see, e.g., [53–55]).

*2) Control objective:* The goal of the DC-link voltage control system is fast and accurate set-point tracking. The DC-link voltage $u_\mathrm{dc}$ is to be fixed, as close as possible, to the predetermined constant reference set-point $u_\mathrm{dc,ref} \geq u_\mathrm{dc,min} > 0\,\mathrm{V}$. Ideally, the following should hold

$$\forall t \geq 0\,\mathrm{s}: \qquad u_\mathrm{dc}(t) = u_\mathrm{dc,ref}(t) \geq u_\mathrm{dc,min} > 0\,\mathrm{V}. \tag{150}$$

In particular, a good disturbance rejection of the unknown disturbance $p_\mathrm{s}$ should be ensured. If (150) is satisfied, no power flows into the DC-link capacitor, i.e. $p_\mathrm{dc} = 0\,\mathrm{W}$, and thus the total machine power $p_\mathrm{s}$ flows into the direction of the PCC (see Fig. 20). The current $i_\mathrm{f}^d$ is used as control variable. The current $i_\mathrm{f}^q$ will be used later for reactive power control (see (184)). An overview of the DC-link control system with underlying current control system is shown in Fig. 37 as signal flow diagram.

In order to better understand the particular challenge of controlling the nonlinear DC-link dynamics (149), a linearization is performed around an operation point (equilibrium) in the next section. Due to the small-signal approximation, the following results only hold locally near the selected operating point.

*3) Linearization about an operation point:* For the linearization of the nonlinear DC-link dynamics (149), the DC-link voltage $u_\mathrm{dc}$ is defined as the state variable, and the currents $\boldsymbol{i}_\mathrm{f}^k$, the time derivatives of the currents $\frac{\mathrm{d}}{\mathrm{d}t}\boldsymbol{i}_\mathrm{f}^k$, and the machine power $p_\mathrm{s}$ are regarded as inputs. Thus, by defining the state

$$x := u_\mathrm{dc},$$

the input vector

$$\boldsymbol{u} := \big(\underbrace{(u_{11}, u_{12})}_{=:\boldsymbol{u}_1^\top}, \underbrace{(u_{21}, u_{22})}_{=:\boldsymbol{u}_2^\top}, u_3\big)^\top := \begin{pmatrix} \boldsymbol{i}_\mathrm{f}^k \\ \tfrac{\mathrm{d}}{\mathrm{d}t}\boldsymbol{i}_\mathrm{f}^k \\ p_\mathrm{s} \end{pmatrix} \in \mathbb{R}^5,$$

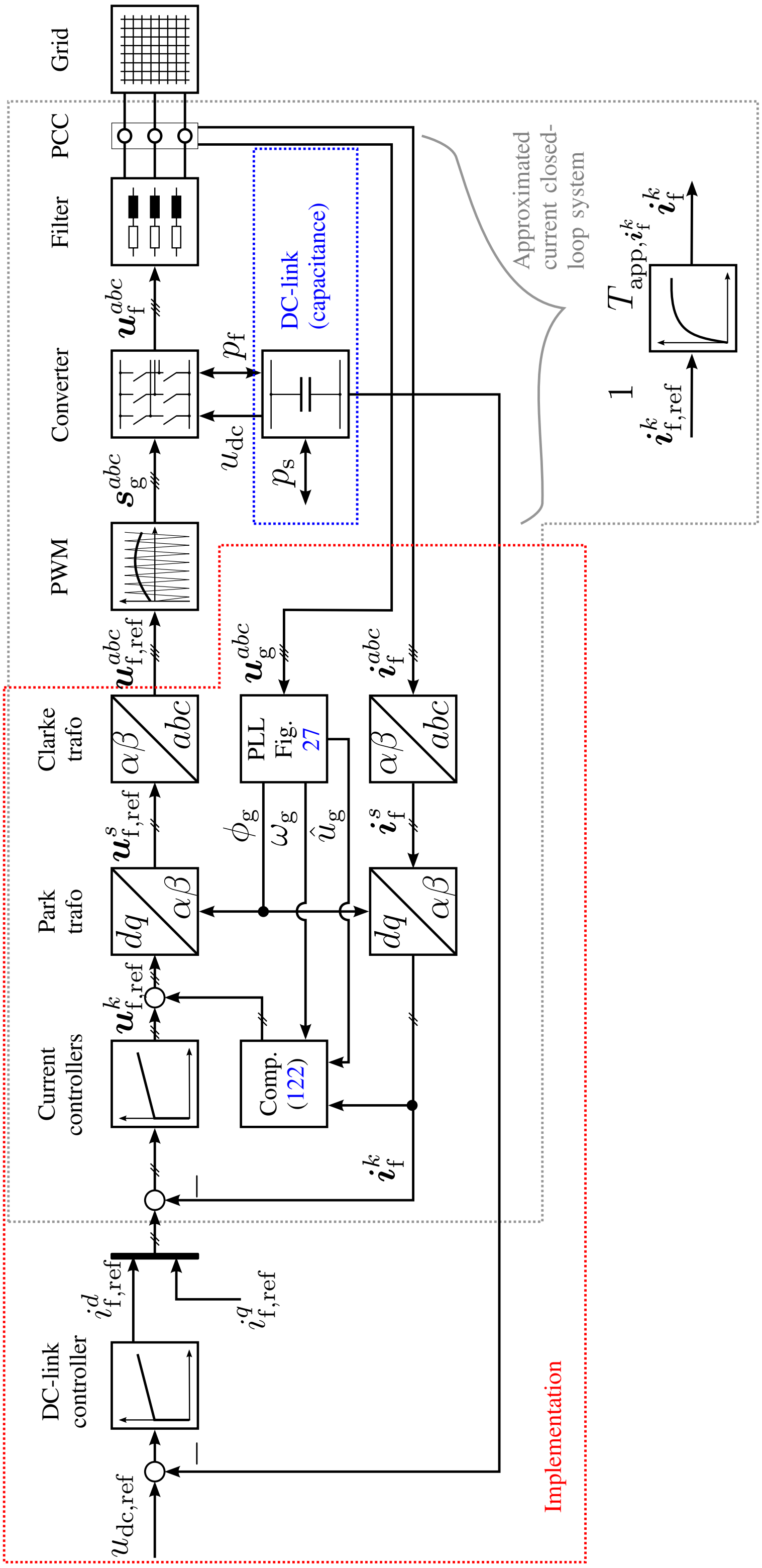

Fig. 37: *Block diagram of the DC-link voltage control system with converter, pulse width modulation (PWM), phase-locked loop (PLL), underlying current controllers and superimposed DC-link controller.*

and the function $f\colon\ \mathbb{R}\times\mathbb{R}^5\to\mathbb{R},\quad (x,\boldsymbol{u})\mapsto$

$$f(x,\boldsymbol{u}) := \frac{1}{C_{\mathrm{dc}}x}\left(-u_3-\frac{3}{2}R_{\mathrm{f}}\|\boldsymbol{u}_1\|^2-\frac{3}{2}L_{\mathrm{f}}\left(\boldsymbol{u}_2\right)^\top\boldsymbol{u}_1-\frac{3}{2}\hat{u}_{\mathrm{g}}u_{11}\right), \tag{151}$$

the DC-link dynamics (149) can be written in the standard form

$$\tfrac{\mathrm{d}}{\mathrm{d}t}\,x = f(x,\boldsymbol{u}), \qquad x(0\,\mathrm{s}) = u_{\mathrm{dc}}^0, \tag{152}$$

and can be linearized around the operation point[18]

$$\begin{pmatrix} x^\star \\ \boldsymbol{u}^\star\end{pmatrix} := \left(\begin{pmatrix} u_{\mathrm{dc}}^\star \\ \boldsymbol{i}_{\mathrm{f}}^{k,\star} \\ \tfrac{\mathrm{d}}{\mathrm{d}t}\,\boldsymbol{i}_{\mathrm{f}}^{k,\star} \\ p_{\mathrm{s}}^\star\end{pmatrix}\right) = \left(\begin{pmatrix} u_{\mathrm{dc}}^\star \\ \boldsymbol{i}_{\mathrm{f}}^{k,\star} \\ (0\,\tfrac{\mathrm{A}}{\mathrm{s}},0\,\tfrac{\mathrm{A}}{\mathrm{s}})^\top \\ p_{\mathrm{s}}^\star\end{pmatrix}\right). \tag{153}$$

At the operating point (153)—an equilibrium of the DC-link dynamics (149) or (152)—the following holds

$$\tfrac{\mathrm{d}}{\mathrm{d}t}\,x^\star = f(x^\star,\boldsymbol{u}^\star) = 0\,\frac{\mathrm{V}}{\mathrm{s}}. \tag{154}$$

For the linearization, the small-signal approximations for the state

$$\widetilde{x}(t) := x(t)-x^\star = \widetilde{u}_{\mathrm{dc}}(t) := u_{\mathrm{dc}}(t)-u_{\mathrm{dc}}^\star \tag{155}$$

and for the input vector

$$\widetilde{\boldsymbol{u}}(t) = \begin{pmatrix} \widetilde{\boldsymbol{i}}_{\mathrm{f}}^k(t) \\ \tfrac{\mathrm{d}}{\mathrm{d}t}\,\widetilde{\boldsymbol{i}}_{\mathrm{f}}^k(t) \\ \widetilde{p}_{\mathrm{s}}(t)\end{pmatrix} := \begin{pmatrix} \boldsymbol{i}_{\mathrm{f}}^k(t)-\boldsymbol{i}_{\mathrm{f}}^{k,\star} \\ \tfrac{\mathrm{d}}{\mathrm{d}t}\,\boldsymbol{i}_{\mathrm{f}}^k(t)-\tfrac{\mathrm{d}}{\mathrm{d}t}\,\boldsymbol{i}_{\mathrm{f}}^{k,\star} \\ p_{\mathrm{s}}(t)-p_{\mathrm{s}}^\star\end{pmatrix} \overset{(153)}{=} \begin{pmatrix} \boldsymbol{i}_{\mathrm{f}}^k(t)-\boldsymbol{i}_{\mathrm{f}}^{k,\star} \\ \tfrac{\mathrm{d}}{\mathrm{d}t}\,\boldsymbol{i}_{\mathrm{f}}^k(t) \\ p_{\mathrm{s}}(t)-p_{\mathrm{s}}^\star\end{pmatrix} \tag{156}$$

are introduced and tagged with $\tilde{\ }$. Thus, the small-signal behavior can be derived from the Taylor series expansion by considering only its first-order terms as follows

$$\begin{aligned}\tfrac{\mathrm{d}}{\mathrm{d}t}\,\widetilde{x} &= \tfrac{\mathrm{d}}{\mathrm{d}t}\,x-\tfrac{\mathrm{d}}{\mathrm{d}t}\,x^\star \\ &\overset{(152),(154)}{\approx} \left.\frac{\partial f(x,\boldsymbol{u})}{\partial x}\right|_{(x^\star,\boldsymbol{u}^\star)}\underbrace{(x-x^\star)}_{=:\widetilde{x}}+\left(\left.\frac{\partial f(x,\boldsymbol{u})}{\partial \boldsymbol{u}}\right|_{(x^\star,\boldsymbol{u}^\star)}\right)^\top\underbrace{(\boldsymbol{u}-\boldsymbol{u}^\star)}_{=:\widetilde{\boldsymbol{u}}}\quad .\end{aligned} \tag{157}$$

Successive calculation of the partial derivatives in (157) and their evaluation at the equilibrium (153) yield

$$\begin{aligned}\left.\tfrac{\partial}{\partial x}\,f(x,\boldsymbol{u})\right|_{(x^\star,\boldsymbol{u}^\star)} &= -\frac{1}{C_{\mathrm{dc}}x^2}\left(-u_3-\frac{3}{2}R_{\mathrm{f}}\|\boldsymbol{u}_1\|^2-\frac{3}{2}L_{\mathrm{f}}\left(\boldsymbol{u}_2\right)^\top\boldsymbol{u}_1-\frac{3}{2}\hat{u}_{\mathrm{g}}u_{11}\right)\Bigg|_{(x^\star,\boldsymbol{u}^\star)} \\ &= -\frac{1}{x^\star}f(x^\star,\boldsymbol{u}^\star) \overset{(154)}{=} 0\,\frac{1}{\mathrm{s}}\end{aligned} \tag{158}$$

and

$$\left.\tfrac{\partial}{\partial \boldsymbol{u}}\,f(x,\boldsymbol{u})\right|_{(x^\star,\boldsymbol{u}^\star)} = \begin{pmatrix} \tfrac{\partial}{\partial \boldsymbol{u}_1}\,f(x,\boldsymbol{u})|_{(x^\star,\boldsymbol{u}^\star)} \\ \tfrac{\partial}{\partial \boldsymbol{u}_2}\,f(x,\boldsymbol{u})|_{(x^\star,\boldsymbol{u}^\star)} \\ \tfrac{\partial}{\partial u_3}\,f(x,\boldsymbol{u})|_{(x^\star,\boldsymbol{u}^\star)}\end{pmatrix}$$

[18]For $n\in\mathbb{N}$. The following zero vector is defined: $\mathbf{0}_n := (\underbrace{0,\ldots,0}_{n-\text{times}})^\top\in\mathbb{R}^n$.

$$\overset{(151)}{=} \begin{pmatrix} \left[\frac{1}{C_{\mathrm{dc}}x}\left(-\frac{3}{2}(2R_{\mathrm{f}}\boldsymbol{u}_1) - \frac{3}{2}L_{\mathrm{f}}\boldsymbol{u}_2 - \frac{3}{2}\begin{pmatrix}\hat{u}_{\mathrm{g}}\\ 0\,\mathrm{V}\end{pmatrix}\right)\right]\Big|_{(x^\star,\boldsymbol{u}^\star)} \\ \left[-\frac{1}{C_{\mathrm{dc}}x}\left(\frac{3}{2}L_{\mathrm{f}}\boldsymbol{u}_1\right)\right]\Big|_{(x^\star,\boldsymbol{u}^\star)} \\ \left[-\frac{1}{C_{\mathrm{dc}}x}\right]\Big|_{(x^\star,\boldsymbol{u}^\star)} \end{pmatrix}$$

$$= \frac{1}{C_{\mathrm{dc}}x^\star}\begin{pmatrix} -\frac{3}{2}(2R_{\mathrm{f}}\boldsymbol{u}_1^\star) - \frac{3}{2}L_{\mathrm{f}}\boldsymbol{u}_2^\star - \frac{3}{2}\begin{pmatrix}\hat{u}_{\mathrm{g}}\\ 0\,\mathrm{V}\end{pmatrix} \\ -\frac{3}{2}L_{\mathrm{f}}\boldsymbol{u}_1^\star \\ -1 \end{pmatrix}. \tag{159}$$

Inserting small-signal approximation (156), and partial derivatives (158) and (159) into the nonlinear DC-link system dynamics (157) leads to the linearized model at the equilibrium (153). It is given by

$$\begin{aligned}
\tfrac{\mathrm{d}}{\mathrm{d}t}\widetilde{u}_{\mathrm{dc}}(t) &\overset{(157),(154)}{=} \frac{\partial}{\partial x}f(x,\boldsymbol{u})\Big|_{(x^\star,\boldsymbol{u}^\star)}\widetilde{x}(t) + \left(\frac{\partial}{\partial \boldsymbol{u}}f(x,\boldsymbol{u})\Big|_{(x^\star,\boldsymbol{u}^\star)}\right)^\top \widetilde{\boldsymbol{u}}(t)\\
&\overset{(158),(159)}{=} \frac{1}{C_{\mathrm{dc}}u_{\mathrm{dc}}^\star}\Bigg(-\widetilde{p}_{\mathrm{s}}(t) - \frac{3}{2}\Bigg[2R_{\mathrm{f}}\boldsymbol{i}_{\mathrm{f}}^{k,\star}(t)^\top\widetilde{\boldsymbol{i}}_{\mathrm{f}}^{k}(t)\\
&\qquad + L_{\mathrm{f}}\underbrace{\left(\tfrac{\mathrm{d}}{\mathrm{d}t}\boldsymbol{i}_{\mathrm{f}}^{k,\star}\right)^\top}_{\overset{(153)}{=}(0\,\frac{\mathrm{A}}{\mathrm{s}},0\,\frac{\mathrm{A}}{\mathrm{s}})^\top}\widetilde{\boldsymbol{i}}_{\mathrm{f}}^{k}(t) + L_{\mathrm{f}}(\boldsymbol{i}_{\mathrm{f}}^{k,\star})^\top\tfrac{\mathrm{d}}{\mathrm{d}t}\widetilde{\boldsymbol{i}}_{\mathrm{f}}^{k}(t) + \hat{u}_{\mathrm{g}}\widetilde{i}_{\mathrm{f}}^{d}(t)\Bigg]\Bigg)\\
&= \frac{1}{C_{\mathrm{dc}}u_{\mathrm{dc}}^\star}\Bigg(-\widetilde{p}_{\mathrm{s}}(t) - \frac{3}{2}\Bigg[2R_{\mathrm{f}}(i_{\mathrm{f}}^{d,\star}\widetilde{i}_{\mathrm{f}}^{d}(t) + i_{\mathrm{f}}^{q,\star}\widetilde{i}_{\mathrm{f}}^{q}(t))\\
&\qquad + L_{\mathrm{f}}(\boldsymbol{i}_{\mathrm{f}}^{k,\star})^\top\tfrac{\mathrm{d}}{\mathrm{d}t}\widetilde{\boldsymbol{i}}_{\mathrm{f}}^{k}(t) + \hat{u}_{\mathrm{g}}\widetilde{i}_{\mathrm{f}}^{d}(t)\Bigg]\Bigg)\\
&= -\frac{1}{C_{\mathrm{dc}}u_{\mathrm{dc}}^\star}\Bigg[\widetilde{p}_{\mathrm{s}}(t) + \frac{3}{2}(\hat{u}_{\mathrm{g}} + 2R_{\mathrm{f}}i_{\mathrm{f}}^{d,\star})\widetilde{i}_{\mathrm{f}}^{d}(t) + \frac{3}{2}L_{\mathrm{f}}i_{\mathrm{f}}^{d,\star}\tfrac{\mathrm{d}}{\mathrm{d}t}\widetilde{i}_{\mathrm{f}}^{d}(t)\\
&\qquad + \frac{3}{2}(2R_{\mathrm{f}}i_{\mathrm{f}}^{q,\star})\widetilde{i}_{\mathrm{f}}^{q}(t) + \frac{3}{2}L_{\mathrm{f}}i_{\mathrm{f}}^{q,\star}\tfrac{\mathrm{d}}{\mathrm{d}t}\widetilde{i}_{\mathrm{f}}^{q}(t)\Bigg].
\end{aligned} \tag{160}$$

*4) Behavior and disturbance behavior of the linearized system:* For the following controller design and stability analysis, the transfer functions are derived. For a physically meaningful operation, it is assumed that:

**Assumption (A.16)** *In the equilibrium (153), the voltage drop across the filter resistance $R_{\mathrm{f}}$ is* smaller *than the grid voltage magnitude $\hat{u}_{\mathrm{g}}$, i.e.,*

$$\hat{u}_{\mathrm{g}} + 2R_{\mathrm{f}}i_{\mathrm{f}}^{d,\star} \geq \hat{u}_{\mathrm{g}} - 2R_{\mathrm{f}}|i_{\mathrm{f}}^{d,\star}| > 0\,\mathrm{V}. \tag{161}$$

**Remark IV.5.** *If Assumption (A.16) does not hold, e.g. $\hat{u}_{\mathrm{g}} + 2R_{\mathrm{f}}i_{\mathrm{f}}^{d,\star} = 0\,\mathrm{V}$, then the DC-link can no longer be controlled via $\widetilde{i}_{\mathrm{f}}^{d}$ (its affect in (160) on the system dynamics is zero).*

In order to derive the system dynamics to analyze set-point tracking performance and disturbance rejection capability, (160) is transformed into the Laplace domain. This transformation into the frequency domain is given by

$$s\widetilde{u}_{\mathrm{dc}}(s)\underbrace{-\widetilde{u}_{\mathrm{dc}}(0\,\mathrm{s})}_{u_{\mathrm{dc}}(0)-u_{\mathrm{dc}}^\star=0\,\mathrm{V}} = -\frac{1}{C_{\mathrm{dc}}u_{\mathrm{dc}}^\star}\Bigg[\widetilde{p}_{\mathrm{s}}(s) + \frac{3}{2}(\hat{u}_{\mathrm{g}} + 2R_{\mathrm{f}}i_{\mathrm{f}}^{d,\star})\widetilde{i}_{\mathrm{f}}^{d}(s) + \frac{3}{2}L_{\mathrm{f}}i_{\mathrm{f}}^{d,\star}s\,\widetilde{i}_{\mathrm{f}}^{d}(s) + \frac{3}{2}(2R_{\mathrm{f}}i_{\mathrm{f}}^{q,\star})\widetilde{i}_{\mathrm{f}}^{q}(s) + \frac{3}{2}L_{\mathrm{f}}i_{\mathrm{f}}^{q,\star}s\,\widetilde{i}_{\mathrm{f}}^{q}(s)\Bigg]. \tag{162}$$

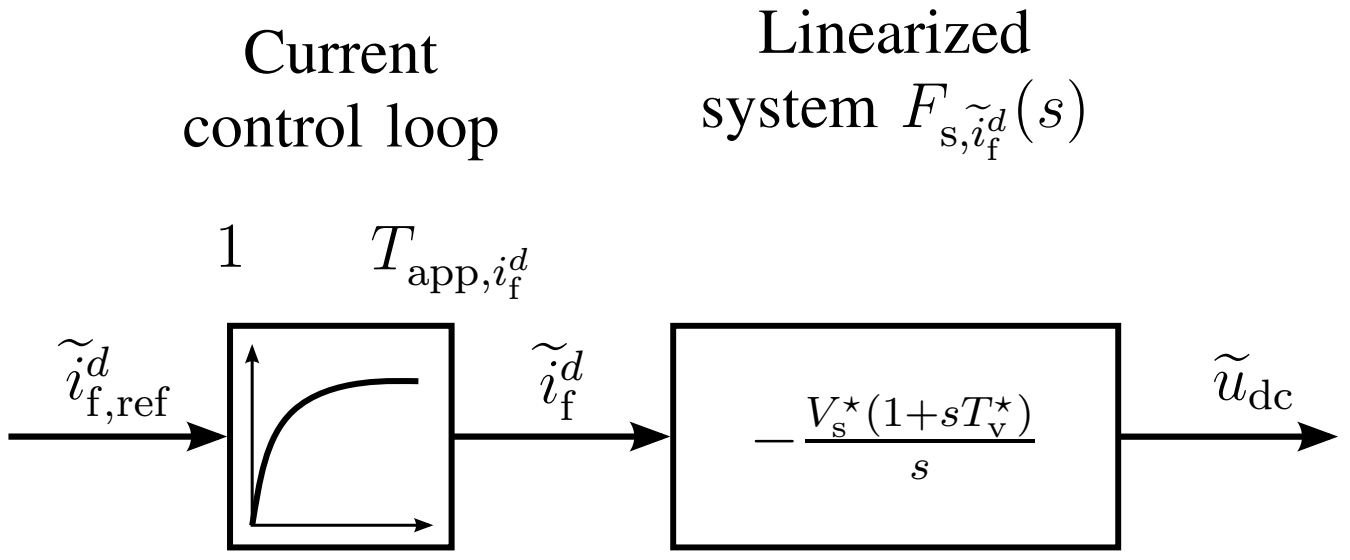

Fig. 38: *Linearization range of the point* (153) *with the appoximated current control circuit dynamics* (115).

By rearranging and neglecting (zeroing) of $\widetilde{i}_{\mathrm{f}}^{q}(s)$ and $\widetilde{p}_{\mathrm{s}}(s)$, the transfer function

$$F_{\mathrm{s},\widetilde{i}_{\mathrm{f}}^{d}}(s) := \frac{\widetilde{u}_{\mathrm{dc}}(s)}{\widetilde{i}_{\mathrm{f}}^{d}(s)} = -\frac{1}{C_{\mathrm{dc}}u_{\mathrm{dc}}^{\star}s}\left(\frac{3}{2}(\hat{u}_{\mathrm{g}} + 2R_{\mathrm{f}}i_{\mathrm{f}}^{d,\star}) + \frac{3}{2}L_{\mathrm{f}}i_{\mathrm{f}}^{d,\star}s\right)$$
$$= -\frac{\frac{3}{2}(\hat{u}_{\mathrm{g}} + 2R_{\mathrm{f}}i_{\mathrm{f}}^{d,\star})}{C_{\mathrm{dc}}u_{\mathrm{dc}}^{\star}s}\left(1 + \frac{L_{\mathrm{f}}i_{\mathrm{f}}^{d,\star}}{\hat{u}_{\mathrm{g}} + 2R_{\mathrm{f}}i_{\mathrm{f}}^{d,\star}}s\right).$$

of the controlled variable can be directly calculated. By defining the *equilibrium dependent* system parameters

$$\left.\boxed{\begin{array}{llll} \text{Gain:} & V_{\mathrm{s}}^{\star}(i_{\mathrm{f}}^{d,\star}) & := \dfrac{\frac{3}{2}(\hat{u}_{\mathrm{g}} + 2R_{\mathrm{f}}i_{\mathrm{f}}^{d,\star})}{C_{\mathrm{dc}}u_{\mathrm{dc}}^{\star}} \overset{(161)}{>} 0\,\frac{\mathrm{V}}{\mathrm{A\,s}} & \text{and} \\ \text{Time constant:} & T_{\mathrm{v}}^{\star}(i_{\mathrm{f}}^{d,\star}) & := \dfrac{L_{\mathrm{f}}i_{\mathrm{f}}^{d,\star}}{\hat{u}_{\mathrm{g}} + 2R_{\mathrm{f}}i_{\mathrm{f}}^{d,\star}}, & \end{array}}\right\} \tag{163}$$

the transfer function can be written in a more compact form

$$F_{\mathrm{s},\widetilde{i}_{\mathrm{f}}^{d}}(s) = \frac{\widetilde{u}_{\mathrm{dc}}(s)}{\widetilde{i}_{\mathrm{f}}^{d}(s)} = -V_{\mathrm{s}}^{\star}\frac{1 + sT_{\mathrm{v}}^{\star}}{s}. \tag{164}$$

The disturbance behavior can also be derived from (162). The disturbance transfer functions are given by

$$\frac{\widetilde{u}_{\mathrm{dc}}(s)}{\widetilde{p}_{\mathrm{s}}(s)} = -\frac{1}{C_{\mathrm{dc}}u_{\mathrm{dc}}^{\star}s} \quad \text{and} \quad \frac{\widetilde{u}_{\mathrm{dc}}(s)}{\widetilde{i}_{\mathrm{f}}^{q}(s)} = -\frac{3i_{\mathrm{f}}^{q,\star}R_{\mathrm{f}}}{C_{\mathrm{dc}}u_{\mathrm{dc}}^{\star}s}\left(1 + s\frac{L_{\mathrm{f}}}{2R_{\mathrm{f}}}\right). \tag{165}$$

The DC-link voltage controller must compensate for theses disturbances (caused by the machine-side power flow $\widetilde{p}_{\mathrm{s}}$ and the filter current $q$-component $\widetilde{i}_{\mathrm{f}}^{q}$) as much as possible.

*5) Analysis of the linearized structure varying system:* For further analysis, the behavior of the linearized DC-link dynamics (164) is to be extended by the (approximated) current control closed-loop dynamics (115) (see Fig. 37). The overall system transfer function is given by

$$\boxed{F_{\mathrm{s},\widetilde{i}_{\mathrm{f,ref}}^{d}}(s) = \frac{\widetilde{u}_{\mathrm{dc}}(s)}{\widetilde{i}_{\mathrm{f,ref}}^{d}(s)} = -V_{\mathrm{s}}^{\star}\frac{1 + sT_{\mathrm{v}}^{\star}}{s(1 + sT_{\mathrm{app},i_{\mathrm{f}}^{d}})}}. \tag{166}$$

Its behavior around the equilibrium (157) with $\left(u_{\mathrm{dc}}^{\star},\ \boldsymbol{i}_{\mathrm{f}}^{k,\star},\ \frac{\mathrm{d}}{\mathrm{d}t}\boldsymbol{i}_{\mathrm{f}}^{k,\star},\ p_{\mathrm{s}}^{\star}\right)^{\top}$ will be analyzed. For this purpose, three types of equilibriums, $i_{\mathrm{f}}^{d,\star} = 0\,\mathrm{A}$, $i_{\mathrm{f}}^{d,\star} > 0\,\mathrm{A}$ and $i_{\mathrm{f}}^{d,\star} < 0\,\mathrm{A}$, are considered. The other equilibrium values remain unchanged. This yields three cases:

**1. Case** $\boxed{i_{\mathrm{f}}^{d,\star} = 0\,\mathrm{A} \overset{(163)}{\Longrightarrow} T_{\mathrm{v}}^{\star} = 0\,\mathrm{s}:}$ Here (166) simplifies to

$$F_{S,\widetilde{i}_{\mathrm{f,ref}}^{d}}(s) = -V_{\mathrm{s}}^{\star}(0\,\mathrm{A})\frac{1}{s(1 + sT_{\mathrm{app},i_{\mathrm{f}}^{d}})},$$

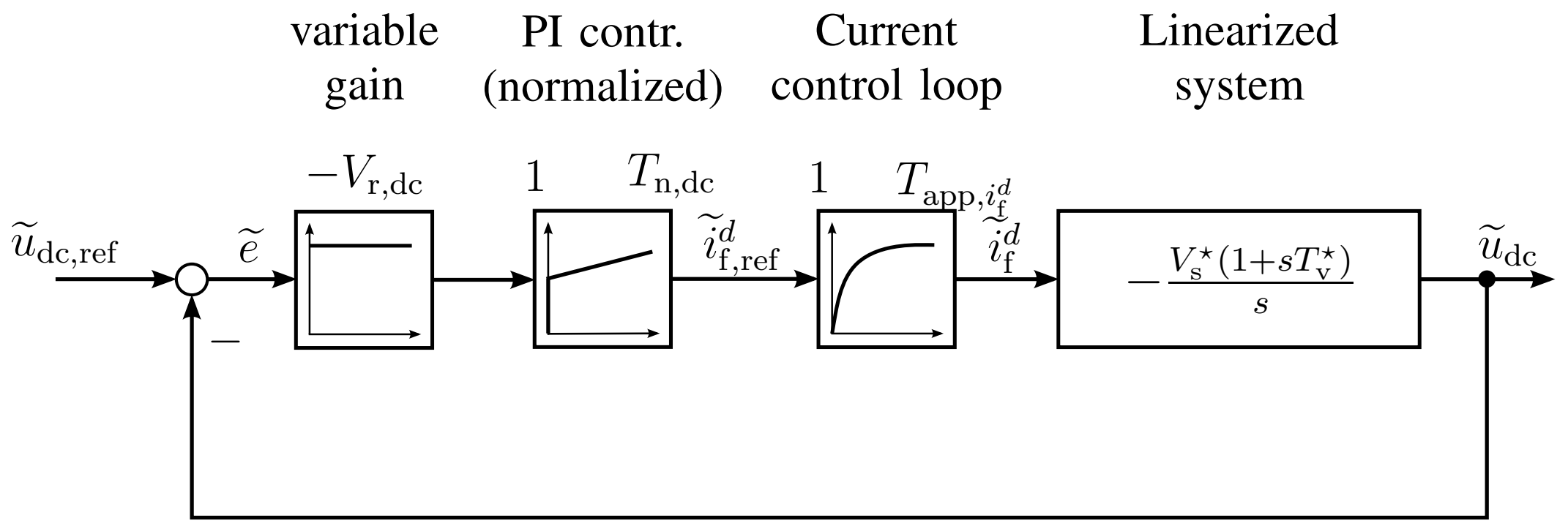

Fig. 39: *Control circuit for the generation of the root locus curve (RLC).*

which is a first-order lag system with additional integrator. For this case, a PI controller tuned according to the Symmetrical Optimum (see [28, p. 81/82]) could be used. *Remark:* A tuning according to the Symmetrical Optimum should not be applied without care for the following two cases. There is a risk of instability of the closed-loop system.

**2. Case** $\boxed{i_{\mathrm{f}}^{d,\star} > 0\,\mathrm{A} \overset{(163)}{\Longrightarrow} T_{\mathrm{v}}^\star > 0\,\mathrm{s}:}$ The transfer function (166) does not change. It is a first-order lag system with additional integrator and a derivative term in the numerator. Its numerator polynomial has a zero in the *left* (stable) complex half-plane. Therefore, the system is *minimum phase*. An investigation of this system using the root locus (see Section IV-D6 and Fig. 40(b)) shows *no* stability problems.

**3. Case** $\boxed{i_{\mathrm{f}}^{d,\star} < 0\,\mathrm{A} \overset{(163)}{\Longrightarrow} T_{\mathrm{v}}^\star < 0\,\mathrm{s}:}$ In this case, the linearized DC-link dynamics (166) can be rewritten as

$$F_{S,\widetilde{i}_{\mathrm{f,ref}}^d}(s) = -V_{\mathrm{s}}^\star \cdot \frac{1 - s|T_{\mathrm{v}}^\star|}{s(1 + sT_{\mathrm{app},i_{\mathrm{f}}^d})}, \tag{167}$$

which is also a first-order lag system with additional integrator and a derivative term in the numerator, but with a zero in the *right* (unstable) complex half-plane. The transfer function (167) is *non-minimum phase*, therefore, too *high* gains lead to an *unstable* closed-loop system (see Section IV-D6 and Fig. 40(c)).

*6) Root locus of the linearized open-loop system:* The difficulty in the control of non-minimum phase systems can be explained by the root locus. The root locus graphically represents the trajectories of the poles of a closed-loop transfer function for increasing values of the controller gain. The closed-loop system , depicted in Fig. 39, will be considered. The linearized system (166) is controlled with a PI controller of the form

$$F_{C,\widetilde{u}_{\mathrm{dc}}}(s) = \frac{\widetilde{i}_{\mathrm{f,ref}}^d(s)}{\widetilde{e}(s)} = -V_{\mathrm{r,dc}} \frac{1 + sT_{\mathrm{n,dc}}}{sT_{\mathrm{n,dc}}} \text{ with } V_{\mathrm{r,dc}} > 0\,\frac{\mathrm{A}}{\mathrm{V}},\; T_{\mathrm{n,dc}} > 0\,\mathrm{s}, \tag{168}$$

having gain $V_{\mathrm{r,dc}}$ and integrator time constant $T_{\mathrm{n,dc}}$ (in s). From the open-loop transfer function

$$\frac{\widetilde{u}_{\mathrm{dc}}(s)}{\widetilde{e}(s)} = V_{\mathrm{r,dc}} \frac{V_{\mathrm{s}}^\star(1 + sT_{\mathrm{n,dc}})(1 + sT_{\mathrm{v}}^\star)}{s^2 T_{\mathrm{n,dc}}(1 + sT_{\mathrm{app},i_{\mathrm{f}}^d})}, \tag{169}$$

the closed-loop transfer function (see Fig. 39)

$$F_{CL,\widetilde{u}_{\mathrm{dc}}}(s) = \frac{\widetilde{u}_{\mathrm{dc}}(s)}{\widetilde{u}_{\mathrm{dc,ref}}(s)} = \frac{V_{\mathrm{r,dc}} \frac{1+sT_{\mathrm{n,dc}}}{sT_{\mathrm{n,dc}}} V_{\mathrm{s}}^\star \frac{1+sT_{\mathrm{v}}^\star}{s(1+sT_{\mathrm{app},i_{\mathrm{f}}^d})}}{1 + V_{\mathrm{r,dc}} \frac{1+sT_{\mathrm{n,dc}}}{sT_{\mathrm{n,dc}}} V_{\mathrm{s}}^\star \frac{1+sT_{\mathrm{v}}^\star}{s(1+sT_{\mathrm{app},i_{\mathrm{f}}^d})}}$$

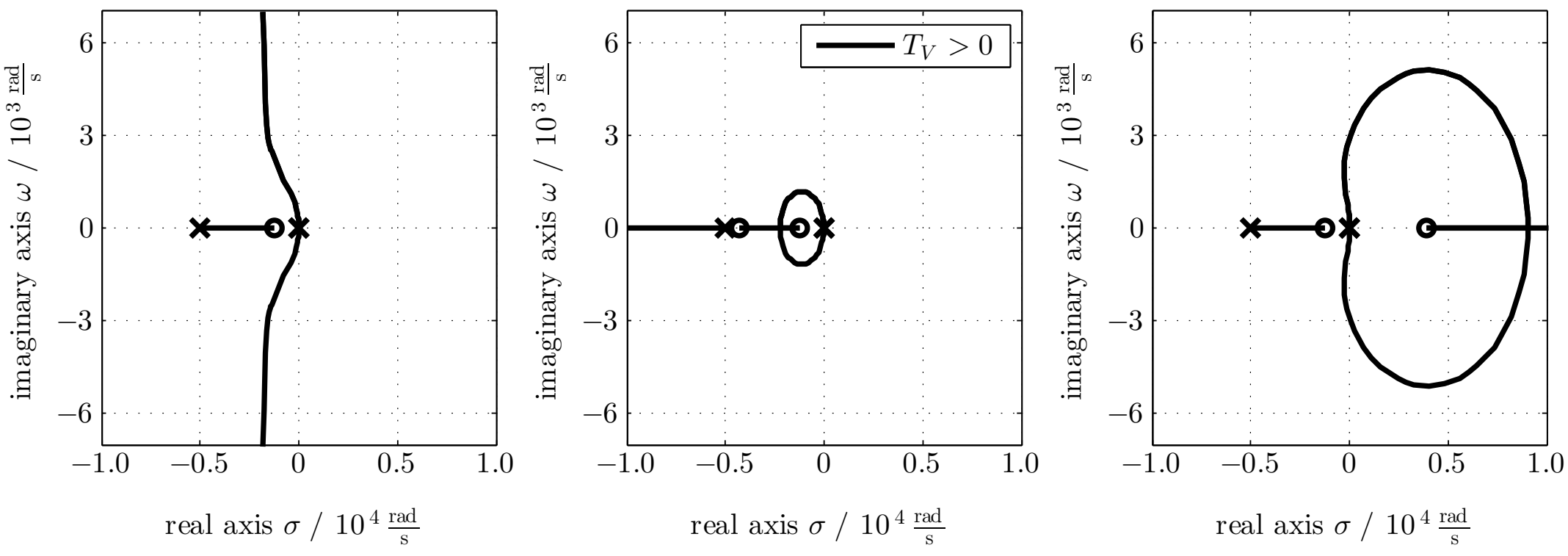

(a) $T_{\mathrm{v}}^{\star} = 0\,\mathrm{s}$. (b) $T_{\mathrm{v}}^{\star} > 0\,\mathrm{s}$. (c) $T_{\mathrm{v}}^{\star} < 0\,\mathrm{s}$.

Fig. 40: *Root loci of the open-loop system* (169) *for the* three *cases: (a)* $T_{\mathrm{v}}^{\star} = 0\,\mathrm{s}$, *(b)* $T_{\mathrm{v}}^{\star} > 0\,\mathrm{s}$ *and (c)* $T_{\mathrm{v}}^{\star} < 0\,\mathrm{s}$. *The locations of the open-loop zeros and poles of* (169) *are shown as* circles *and* crosses, *respectively.*

$$
\begin{aligned}
&= \frac{V_{\mathrm{r,dc}} V_{\mathrm{s}}^{\star}(1 + sT_{\mathrm{n,dc}})(1 + sT_{\mathrm{v}}^{\star})}{(sT_{\mathrm{n,dc}})(s(1 + sT_{\mathrm{app},i_{\mathrm{f}}^d})) + V_{\mathrm{r,dc}} V_{\mathrm{s}}^{\star}(1 + sT_{\mathrm{n,dc}})(1 + sT_{\mathrm{v}}^{\star})} \\
&= \frac{V_{\mathrm{r,dc}} V_{\mathrm{s}}^{\star}(1 + sT_{\mathrm{n,dc}})(1 + sT_{\mathrm{v}}^{\star})}{s^3 \underbrace{T_{\mathrm{n,dc}} T_{\mathrm{app},i_{\mathrm{f}}^d}}_{=:a_3} + s^2 \underbrace{T_{\mathrm{n,dc}}(1 + T_{\mathrm{v}}^{\star} V_{\mathrm{r,dc}} V_{\mathrm{s}}^{\star})}_{=:a_2} + s \underbrace{V_{\mathrm{r,dc}} V_{\mathrm{s}}^{\star}(T_{\mathrm{n,dc}} + T_{\mathrm{v}}^{\star})}_{=:a_1} + \underbrace{V_{\mathrm{r,dc}} V_{\mathrm{s}}^{\star}}_{=:a_0}}.
\end{aligned}
\tag{170}
$$

is obtained. Its denominator and thus the poles (roots of the denominator) of (170) depend on the controller parameters $V_{\mathrm{r,dc}}$ and $T_{\mathrm{n,dc}}$. If the poles of the closed-loop system (170) are plotted for various but increasing gains $V_{\mathrm{r,dc}} \geq 0\,\frac{\mathrm{A}}{\mathrm{V}}$ in the complex plane, the root locus is obtained. The trajectories of the closed-loop poles start, for $V_{\mathrm{r,dc}} = 0\,\frac{\mathrm{A}}{\mathrm{V}}$, from the poles of the open-loop system (169) (see [38, Sec. 2.2]). For $V_{\mathrm{r,dc}} \to \infty\,\frac{\mathrm{A}}{\mathrm{V}}$, the as many closed-loop pole trajectories end in the zeros of the open-loop system (169) as there are zeros (for $T_{\mathrm{v}}^{\star} = 0\,\mathrm{s}$: one zero at $-1/T_{\mathrm{n,dc}}$; for $T_{\mathrm{v}}^{\star} \neq 0\,\mathrm{s}$: *two* zeros at $-1/T_{\mathrm{n,dc}}$ and $-1/T_{\mathrm{v}}^{\star}$). The remaining closed-loop pole trajectories are symmetrical to the real axis of the complex plane. For the three cases described above, the respective root loci are shown in Fig. 40. For $T_{\mathrm{v}}^{\star} < 0\,\mathrm{s}$ (see Fig. 40(c)), the closed-loop poles trajectories enter the *right* complex half-plane for too large controller gains. Therefore, for controller gains larger than a critical limit, the closed-loop system becomes *unstable*. The other two cases with $T_{\mathrm{v}}^{\star} = 0\,\mathrm{s}$ (see Fig. 40(a)) or $T_{\mathrm{v}}^{\star} > 0\,\mathrm{s}$ (see Fig. 40(b)) are not critical, the closed-loop poles remain in the *left* complex plane for each (also just as large) choice of $V_{\mathrm{r,dc}}$.

*7) Hurwitz stability analysis:* The investigation using the root locus plots has shown that an upper limit for the controller gain must not be exceeded for non-minimum phase systems in order to guarantee a stable closed-loop system. For the present case of the DC-link control system (170), an additional condition for the integrator time constant $T_{\mathrm{n,dc}}$ must be considered. For this purpose, the denominator polynomial of the closed-loop system (170) is now checked for stability using the Hurwitz Stability Criterion (see [47, Theorem 3.4.71]). The necessary condition of the Hurwitz Criterion requires that all coefficients must be greater than zero, i.e. $a_3 > 0\,\mathrm{s}^2$, $a_2 > 0\,\mathrm{s}$, $a_1 > 0$, $a_0 > 0\,\frac{1}{\mathrm{s}}$. The necessary condition can be satisfied for the denominator polynomial in (170), if the following implications hold

$$
\left.
\begin{array}{rcl}
T_{\mathrm{n,dc}} \cdot T_{\mathrm{app},i_{\mathrm{f}}^d} \overset{(168),(115)}{>} 0\,\mathrm{s}^2 & \Longrightarrow & a_3 > 0\,\mathrm{s}^2 \\
V_{\mathrm{r,dc}} \cdot V_{\mathrm{s}}^{\star} \overset{(168),(163)}{>} 0\,\frac{1}{\mathrm{s}} & \Longrightarrow & a_0 > 0\,\frac{1}{\mathrm{s}} \\
T_{\mathrm{n,dc}} > |T_{\mathrm{v}}^{\star}| & \Longrightarrow & a_1 > 0 \\
V_{\mathrm{r,dc}} < \frac{1}{|T_{\mathrm{v}}^{\star}| V_{\mathrm{s}}^{\star}} & \Longrightarrow & 1\,\mathrm{s} > -T_{\mathrm{v}}^{\star} V_{\mathrm{r,dc}} V_{\mathrm{s}}^{\star} \Longrightarrow a_2 > 0\,\mathrm{s}.
\end{array}
\right\}
\tag{171}
$$

The first two implications in (171) are trivially satisfied, since $V_{\rm s}^\star > 0\,\frac{\rm V}{\rm As}$ and $T_{{\rm app},i_{\rm f}^d} > 0\,{\rm s}$ hold by physical means, and if $V_{\rm r,dc} > 0\,\frac{\rm A}{\rm V}$ and $T_{\rm n,dc} > 0\,{\rm s}$ are chosen to be positive. The last two implications in (171) apply with a sufficiently large or small choice of $T_{\rm n,dc}$ and $V_{\rm r,dc}$, respectively. Thus, the *necessary* condition is satisfied. However, a stable closed-loop system may *not* be achieved yet. For this, the *sufficient* condition must be checked

$$D_1 = a_2 > 0\,{\rm s} \qquad \text{and} \qquad D_2 = a_2 a_1 - a_3 a_0 > 0\,{\rm s}.$$

$D_1 = a_2 > 0\,{\rm s}$ has already been shown above and holds for $T_{\rm n,dc} > |T_{\rm v}^\star|$ in (171). Expanding $D_2 = a_2 a_1 - a_3 a_0 > 0\,{\rm s}$ and inserting the coefficients $a_0, \dots, a_3$ as in (170) yields

$$T_{\rm n,dc}(1 + T_{\rm v}^\star V_{\rm r,dc} V_{\rm s}^\star) V_{\rm r,dc} V_{\rm s}^\star (T_{\rm n,dc} + T_{\rm v}^\star) - T_{\rm n,dc} T_{{\rm app},i_{\rm f}^d} V_{\rm r,dc} V_{\rm s}^\star > 0\,{\rm s}$$
$$\overset{V_{\rm r,dc}\cdot V_{\rm s}^\star \cdot T_{\rm n,dc} > 0}{\Longrightarrow} \qquad (1 + T_{\rm v}^\star V_{\rm r,dc} V_{\rm s}^\star)(T_{\rm n,dc} + T_{\rm v}^\star) - T_{{\rm app},i_{\rm f}^d} > 0\,{\rm s}.$$

For the **3. case** (“worst-case”), it follows that $0\,{\rm s} > T_{\rm v}^\star = -|T_{\rm v}^\star|$, thus the following inequality is obtained

$$T_{\rm n,dc} > \frac{T_{{\rm app},i_{\rm f}^d}}{1 - |T_{\rm v}^\star| V_{\rm r,dc} V_{\rm s}^\star} + |T_{\rm v}^\star| \quad \Longrightarrow \quad D_2 > 0\,{\rm s},$$

to ensure $D_2 > 0\,{\rm s}$ for all three cases. Concluding, for a stable closed-loop system, the following two conditions

$$\textbf{(C}_1\textbf{)}\ 0\,\frac{\rm V}{\rm A} < V_{\rm r,dc} < \frac{1}{|T_{\rm v}^\star| V_{\rm s}^\star} \qquad \text{and} \qquad \textbf{(C}_2\textbf{)}\ T_{\rm n,dc} > \frac{T_{{\rm app},i_{\rm f}^d}}{1 - V_{\rm r,dc} V_{\rm s}^\star |T_{\rm v}^\star|} + |T_{\rm v}^\star| > 0\,{\rm s}$$

have to be fulfilled by adequate controller parameterization. In Fig. 41 condition ($C_2$) is illustrated. $T_{\rm n,dc}$ must always be within the permissible range above the —— line in Fig. 41, i.e., for larger controller gains $V_{\rm r,dc}$, the integration time constant $T_{\rm n,dc}$ must also be increased. In this way, for *all three* cases $T_{\rm v}^\star = 0\,{\rm s}$, $T_{\rm v}^\star > 0\,{\rm s}$ and $T_{\rm v}^\star < 0\,{\rm s}$, a *locally* stable closed-loop system is guaranteed.
In summary:

| | |
|---|---|
| PI controller: | $F_{C,\tilde{u}_{\rm dc}}(s) = \frac{\tilde{i}^d_{\rm f,ref}(s)}{\tilde{e}(s)} = -V_{\rm r,dc}\frac{1+sT_{\rm n,dc}}{sT_{\rm n,dc}}$ with |
| Controller gain: | $0\,\frac{\rm A}{\rm V} < V_{\rm r,dc} < \frac{1}{\lvert T_{\rm v}^\star\rvert V_{\rm s}^\star}$, |
| Controller time constant: | $T_{\rm n,dc} > \frac{T_{{\rm app},i_{\rm f}^d}}{1-V_{\rm r,dc}V_{\rm s}^\star \lvert T_{\rm v}^\star\rvert} + \lvert T_{\rm v}^\star\rvert > 0\,{\rm s}$, |
| System gain: | $V_{\rm s}^\star(i_{\rm f}^{d,\star}) := \frac{\frac{3}{2}(\hat{u}_{\rm g} + 2R_{\rm f} i_{\rm f}^{d,\star})}{C_{\rm dc} u_{\rm dc}^\star} \overset{(161)}{>} 0\,\frac{\rm V}{\rm As}$ and |
| System time constant: | $T_{\rm v}^\star(i_{\rm f}^{d,\star}) := \frac{L_{\rm f} i_{\rm f}^{d,\star}}{\hat{u}_{\rm g} + 2R_{\rm f} i_{\rm f}^{d,\star}}$, |

## *E. Feedforward control of the power flow (in operation regime II)*

In the previous Sections IV-A– IV-D, the control systems of the filter currents and the stator currents, the generator speed and the DC-link voltage were presented. In this section, the control of the power flow will be explained in more detail.

To this end, starting from the objectives of operation management system, the power flow in wind turbine systems (with permanent-magnet synchronous motors), the power balance at a stationary operating point, as well as the effects of the presented control systems on the power flow are discussed.

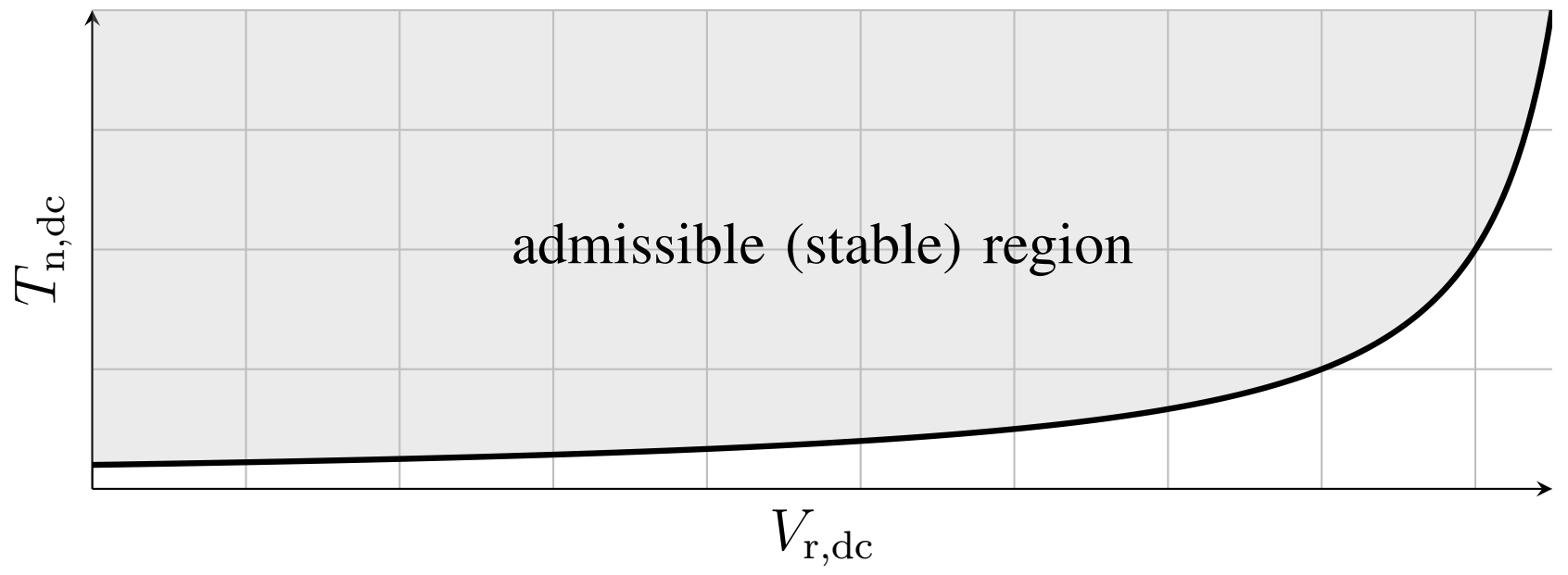

Fig. 41: *Illustration of the Hurwitz criterion condition* (**$C_2$**)*.*

*1) Objectives of the operation management:* The underlying current control systems form the basis for the higher-level control tasks of the wind turbine system: the generator speed regulation for MPPT, the DC-link voltage regulation and the control of the reactive power. The objectives of the operation management are to be discussed, for example, for operation range II of the wind turbine (see Fig. 11). In operation regime II, $\beta_{\mathrm{ref}} = \beta = 0$ holds and the following three *objectives* must be ensured by the operation management (see Fig. 10):

**($O_1$)** Maximum possible power draw from the wind through the turbine;
**($O_2$)** Maximum possible (instantaneous) active power output into the grid, and
**($O_3$)** Providing reactive power on grid operator requests.

The reachability of the objectives ($O_1$)–($O_3$) can be illustrated using the power flow in wind turbine system (e.g., with permanent-magnet synchronous generators) and its resulting power balance in steady-state.

*2) Power flow in wind turbine systems:* Wind turbines can only extract a part of the kinetic energy contained in the wind; only a part of that power can be converted into electrical energy. At the point of common coupling (PCC), the electrical power $p_{\mathrm{pcc}}$ (in W) produced by the wind turbine is fed to the grid. Under the following two assumptions

**Assumption (A.17)** *Losses due to switching in the back-to-back converter are neglected; and*

**Assumption (A.18)** *Losses in the drive trains due to friction are neglected,*

the power flow[19] in a wind turbine with a permanent-magnet synchronous generator, and the incurring losses will be discussed.

The energy source of a wind turbine represents the kinetic energy of the wind. The wind power

$$p_{\mathrm{w}}(t) \overset{(5)}{=} \frac{1}{2}\varrho\pi r_{\mathrm{t}}^2 v_{\mathrm{w}}(t)^3 \tag{172}$$

is partly absorbed by the turbine (in the rotor with its three blades) as turbine power

$$p_{\mathrm{t}}(t) \overset{(68)}{=} m_{\mathrm{t}}(t)\omega_{\mathrm{t}}(t) \overset{(62)}{=} c_{\mathrm{p}}(\lambda,\beta)p_{\mathrm{w}}(t). \tag{173}$$

Due to energy conservation and under Assumption (A.18), the mechanical power of the electrical machine

$$p_{\mathrm{m}}(t) := \omega_{\mathrm{m}}(t)m_{\mathrm{m}}(t) \overset{(78),(A.18)}{=} \underbrace{\Theta\omega_{\mathrm{m}}(t)\tfrac{\mathrm{d}}{\mathrm{d}t}\omega_{\mathrm{m}}(t)}_{\substack{=:p_{\Theta}(t)\\ \text{actual change}\\ \text{of stored}\\ \text{rotational energy}\\ \text{in shaft / inertia}}} - \underbrace{\omega_{\mathrm{t}}(t)m_{\mathrm{t}}(t)}_{\substack{=p_{\mathrm{t}}(t)\\ \text{turbine}\\ \text{power}}} \tag{174}$$

[19]Orientation of the used $k$-reference frames:
– $k$-reference frame of the stator quantities $x_{\mathrm{s}}^k$ with $x \in \{\boldsymbol{u}, \boldsymbol{i}, \boldsymbol{\psi}\}$ in flux orientation.
– $k$-reference frame of the filter quantities $x_{\mathrm{f}}^k$ with $x \in \{\boldsymbol{u}, \boldsymbol{i}, \boldsymbol{\psi}\}$ in grid voltage orientation.

can be derived from the difference between the instantaneous change $p_\Theta$ (in W) of the stored rotational energy (converted to the machine-side shaft) and the turbine power $p_\mathrm{t}$. The generator counteracts the turbine torque $m_\mathrm{t}$ with a *negative* machine torque $m_\mathrm{m}$. For this purpose, the machine-side stator power

$$
\begin{aligned}
p_\mathrm{s}(t) &= \boldsymbol{u}_\mathrm{s}^{abc}(t)^\top \boldsymbol{i}_\mathrm{s}^{abc}(t) \overset{(55)}{=} \frac{3}{2}\boldsymbol{u}_\mathrm{s}^{k}(t)^\top \boldsymbol{i}_\mathrm{s}^{k}(t) = \frac{3}{2}\boldsymbol{i}_\mathrm{s}^{k}(t)^\top \boldsymbol{u}_\mathrm{s}^{k}(t) \\
&\overset{(118)}{=} \frac{3}{2}\Big[R_\mathrm{s}\|\boldsymbol{i}_\mathrm{s}^k(t)\|^2 + \boldsymbol{i}_\mathrm{s}^k(t)^\top \boldsymbol{L}_\mathrm{s}^k \tfrac{\mathrm{d}}{\mathrm{d}t}\boldsymbol{i}_\mathrm{s}^k(t) \\
&\qquad + \underbrace{\omega_\mathrm{k}(t)}_{\overset{(117)}{=} n_\mathrm{p}\,\omega_\mathrm{m}(t)} \boldsymbol{i}_\mathrm{s}^k(t)^\top \big(\boldsymbol{J}\boldsymbol{L}_\mathrm{s}^k \boldsymbol{i}_\mathrm{s}^k(t) + \boldsymbol{\psi}_\mathrm{pm}^k(t)\big)\Big] \\
&= \frac{3}{2}R_\mathrm{s}\|\boldsymbol{i}_\mathrm{s}^k(t)\|^2 + \frac{3}{2}\boldsymbol{i}_\mathrm{s}^k(t)^\top \boldsymbol{L}_\mathrm{s}^k \tfrac{\mathrm{d}}{\mathrm{d}t}\boldsymbol{i}_\mathrm{s}^k(t) + \omega_\mathrm{m}(t)\underbrace{\frac{3}{2}n_\mathrm{p}\,\boldsymbol{i}_\mathrm{s}^k(t)^\top \boldsymbol{J}\big(\boldsymbol{L}_\mathrm{s}^k\boldsymbol{i}_\mathrm{s}^k(t)+\boldsymbol{\psi}_\mathrm{pm}^k\big)}_{\overset{(128)}{=} m_\mathrm{m}(t)} \\
&= \frac{3}{2}R_\mathrm{s}\|\boldsymbol{i}_\mathrm{s}^k(t)\|^2 + \frac{3}{2}\boldsymbol{i}_\mathrm{s}^k(t)^\top \boldsymbol{L}_\mathrm{s}^k \tfrac{\mathrm{d}}{\mathrm{d}t}\boldsymbol{i}_\mathrm{s}^k(t) + \omega_\mathrm{m}(t) m_\mathrm{m}(t) \\
&\overset{(174)}{=} \underbrace{\frac{3}{2}R_\mathrm{s}\|\boldsymbol{i}_\mathrm{s}^k(t)\|^2}_{\substack{=:p_{R_\mathrm{s},\mathrm{loss}}(t) \\ \text{power losses in} \\ \text{stator resistance}}} + \underbrace{\frac{3}{2}\boldsymbol{i}_\mathrm{s}^k(t)^\top \boldsymbol{L}_\mathrm{s}^k \tfrac{\mathrm{d}}{\mathrm{d}t}\boldsymbol{i}_\mathrm{s}^k(t)}_{\substack{=:p_{\boldsymbol{L}_\mathrm{s}^k}(t) \\ \text{actual change} \\ \text{of stored} \\ \text{magnetic energy} \\ \text{in stator inductance}}} + \underbrace{p_\Theta(t) - p_\mathrm{t}(t)}_{\substack{=:p_\mathrm{m}(t) \\ \text{mechanical} \\ \text{machine power}}}
\end{aligned}
\tag{175}
$$

must be applied. In the Back-to-back converter, a power exchange occurs between the filter power $p_\mathrm{f}$ (in W), the DC-link power $p_\mathrm{dc}$ (in W) and the machine-side stator power $p_\mathrm{s}$ (in W) (see Fig. 20). If switching losses are neglected, the following holds

$$
p_\mathrm{dc}(t) \overset{(147)}{=} \underbrace{C_\mathrm{dc} u_\mathrm{dc}(t)\tfrac{\mathrm{d}}{\mathrm{d}t}u_\mathrm{dc}(t)}_{\substack{\text{actual change} \\ \text{of stored} \\ \text{electrical energy} \\ \text{in DC-link capacitance}}} \overset{(A.17)}{=} -p_\mathrm{f}(t) - p_\mathrm{s}(t). \tag{176}
$$

The grid-side power output of the back-to-back converter, i.e. the filter power, is given by

$$
p_\mathrm{f}(t) \overset{(148)}{=} \underbrace{\frac{3}{2}R_\mathrm{f}\|\boldsymbol{i}_\mathrm{f}^k(t)\|^2}_{\substack{=:p_{R_\mathrm{f},\mathrm{loss}}(t) \\ \text{power losses in} \\ \text{filter resistance}}} + \underbrace{\frac{3}{2}L_\mathrm{f}\boldsymbol{i}_\mathrm{f}^k(t)^\top \tfrac{\mathrm{d}}{\mathrm{d}t}\boldsymbol{i}_\mathrm{f}^k(t)}_{\substack{=:p_{L_\mathrm{f}}(t) \\ \text{actual change} \\ \text{of stored} \\ \text{magnetic energy} \\ \text{in filter inductance}}} + \underbrace{\frac{3}{2}\hat{u}_\mathrm{g} i_\mathrm{f}^d(t)}_{\substack{=:p_\mathrm{pcc}(t) \\ \text{induced power} \\ \text{at PCC}}}, \tag{177}
$$

where $\frac{3}{2}\hat{u}_\mathrm{g} i_\mathrm{f}^d$ (in W) corresponds to the power fed in at the PCC, since, according to Fig. 20, the following holds

$$
p_\mathrm{pcc}(t) = \boldsymbol{u}_\mathrm{g}^{abc}(t)^\top \boldsymbol{i}_\mathrm{f}^{abc}(t) \overset{(55)}{=} \frac{3}{2}\boldsymbol{u}_\mathrm{g}^k(t)^\top \boldsymbol{i}_\mathrm{f}^k(t) \overset{(97)}{=} \frac{3}{2}\hat{u}_\mathrm{g} i_\mathrm{f}^d(t). \tag{178}
$$

Summarizing all the results so far, one obtains the total power flow of the wind turbine system at the PCC:

$$
\begin{aligned}
p_\mathrm{pcc}(t) &\overset{(178)}{=} \frac{3}{2}\hat{u}_\mathrm{g} i_\mathrm{f}^d(t) \overset{(177)}{=} p_\mathrm{f}(t) - p_{R_\mathrm{f},\mathrm{loss}}(t) - p_{L_\mathrm{f}}(t) \\
&\overset{(176)}{=} -p_\mathrm{s}(t) - p_\mathrm{dc}(t) - p_{R_\mathrm{f},\mathrm{loss}}(t) - p_{L_\mathrm{f}}(t) \\
&\overset{(175)}{=} p_\mathrm{t}(t) - p_\Theta(t) - p_{R_\mathrm{s},\mathrm{loss}}(t) - p_{\boldsymbol{L}_\mathrm{s}^k}(t) \\
&\qquad - p_\mathrm{dc}(t) - p_{R_\mathrm{f},\mathrm{loss}}(t) - p_{L_\mathrm{f}}(t)
\end{aligned}
$$

$$\stackrel{(173)}{=} c_{\rm p}(\lambda,\beta)p_{\rm w}(t) - p_{\Theta}(t) - p_{R_{\rm s},\rm loss}(t) - p_{\boldsymbol{L}_{\rm s}^k}(t) \\ -p_{\rm dc}(t) - p_{R_{\rm f},\rm loss}(t) - p_{L_{\rm f}}(t). \tag{179}$$

The power $p_{\rm pcc}$ fed in at the PCC is therefore composed of the fraction $p_{\rm t} = c_{\rm p}p_{\rm w}$ of the wind power $p_{\rm w}$ taken from the turbine minus the change $p_{\Theta} = \Theta\omega_{\rm m}\frac{\rm d}{{\rm d}t}\omega_{\rm m}$ of the stored rotational energy, the power dissipated $p_{R_{\rm s},\rm loss} = \frac{3}{2}R_{\rm s}\|\boldsymbol{i}_{\rm s}^k\|^2$ at the stator resistance $R_{\rm s}$, the magnetization power $p_{\boldsymbol{L}_{\rm s}^k} = \frac{3}{2}(\boldsymbol{i}_{\rm s}^k)^\top\boldsymbol{L}_{\rm s}^k\frac{\rm d}{{\rm d}t}\boldsymbol{i}_{\rm s}^k$ of the stator current inductances $\boldsymbol{L}_{\rm s}^k$, the change $p_{\rm dc} = C_{\rm dc}u_{\rm dc}\frac{\rm d}{{\rm d}t}u_{\rm dc}$ of the stored electrical energy in the DC-link capacitor, the power dissipation $p_{R_{\rm f},\rm loss} = \frac{3}{2}R_{\rm f}\|\boldsymbol{i}_{\rm f}^k\|^2$ at the filter resistance $R_{\rm f}$ and the magnetization power $p_{L_{\rm f}} = \frac{3}{2}L_{\rm f}(\boldsymbol{i}_{\rm f}^k)^\top\frac{\rm d}{{\rm d}t}\boldsymbol{i}_{\rm f}^k$ of the filter inductance $L_{\rm f}$.

Figure 42 shows the power flow (179) of the wind turbine system in the form of a power flow diagram. Overall, the modeled wind power plant has *four* reservoirs (or energy storages): the turbine/machine shaft, the stator inductance, the DC-link capacitance and the filter inductance. These can store energy and return it to the system without dissipating energy (bi-directional power/energy flow without losses). On average, the storages have a power consumption or power output of zero. Losses occur in the wind turbine at the stator and filter resistances. Since the turbine can extract only a part of the wind power $p_{\rm w}$ from the wind, the part $p_{\rm t,loss} = (1-c_{\rm p}(\beta,\lambda))p_{\rm w}$ is "lost" (and can be counted as loss). Because of the simplifying Assumptions (A.17) and (A.18), the back-to-back converters and the drive train do not cause any losses, i.e. $p_{\rm s,loss} = p_{\rm f,loss} = 0\,\rm W$.

*3) Steady-state power balance:* In order to determine the reachability of the objectives ($\rm O_1$)–($\rm O_3$) stated in Sect. IV-E1, a simplified consideration of the power flow can be carried out in steady-state operation. In steady state[20], the following holds

$$\tfrac{\rm d}{{\rm d}t}\boldsymbol{i}_{\rm s}^k = \tfrac{\rm d}{{\rm d}t}\boldsymbol{i}_{\rm f}^k = (0\,\frac{\rm A}{\rm s}, 0\,\frac{\rm A}{\rm s})^\top,\ \tfrac{\rm d}{{\rm d}t}\omega_{\rm m} = 0\,\frac{\rm rad}{\rm s} \text{ and } \tfrac{\rm d}{{\rm d}t}u_{\rm dc} = 0\,\frac{\rm V}{\rm s}, \tag{180}$$

and *no* energy is stored in or released from the energy storages (now power flow into or out of the reservoir), since

$$\left.\begin{array}{lcl} p_{\Theta} & \stackrel{(174)}{=} & \Theta\omega_{\rm m}\frac{\rm d}{{\rm d}t}\omega_{\rm m} \\ p_{\boldsymbol{L}_{\rm s}^k} & \stackrel{(175)}{=} & \frac{3}{2}(\boldsymbol{i}_{\rm s}^k)^\top\boldsymbol{L}_{\rm s}^k\frac{\rm d}{{\rm d}t}\boldsymbol{i}_{\rm s}^k \\ p_{\rm dc} & \stackrel{(176)}{=} & u_{\rm dc}C_{\rm dc}\frac{\rm d}{{\rm d}t}u_{\rm dc} \\ p_{L_{\rm f}} & \stackrel{(177)}{=} & \frac{3}{2}L_{\rm f}(\boldsymbol{i}_{\rm f}^k)^\top\frac{\rm d}{{\rm d}t}\boldsymbol{i}_{\rm f}^k \end{array}\right\} \stackrel{(180)}{=} 0\,\rm W. \tag{181}$$

Thus, the power flow (179) is reduced to the steady-state *power balance*:

$$p_{\rm pcc} \stackrel{(179),(181)}{=} \underbrace{c_{\rm p}(\lambda,\beta)\,\frac{1}{2}\varrho\pi r_{\rm t}^2 v_{\rm w}^3}_{\stackrel{(18)}{=}p_{\rm t}} - \underbrace{\frac{3}{2}R_{\rm s}\|\boldsymbol{i}_{\rm s}^k\|^2}_{\stackrel{(175)}{=}p_{R_{\rm s},\rm loss}} - \underbrace{\frac{3}{2}R_{\rm f}\|\boldsymbol{i}_{\rm f}^k\|^2}_{\stackrel{(177)}{=}p_{R_{\rm f},\rm loss}}. \tag{182}$$

For a physically meaningful operation

$$c_{\rm p}(\lambda,\beta)\frac{1}{2}\varrho\pi r_{\rm t}^2 v_{\rm w}^3 - \frac{3}{2}R_{\rm s}\|\boldsymbol{i}_{\rm s}^k\|^2 - \frac{3}{2}R_{\rm f}\|\boldsymbol{i}_{\rm f}^k\|^2 > 0\,\rm W$$

must hold in the long term, otherwise the losses would surpass the converted energy and no electricity would be produced. Using the power balance (182), it is possible to derive conditions (at least in steady state), which state what is required to achieve objectives ($\rm O_1$) and ($\rm O_2$) in operation regime II:

- The wind turbine must operate at its optimum operation point $(0,\lambda^\star)$ in order to extract the maximally possible power from the wind (=$\rm O_1$). This results in the maximum power coefficient at $c_{\rm p}^\star := c_{\rm p}(0,\lambda^\star)$ and the turbine power $p_{\rm t} = c_{\rm p}^\star p_{\rm w}$ becomes maximum.

[20] Due to the switching behavior of the inverter, a steady state is never reached.

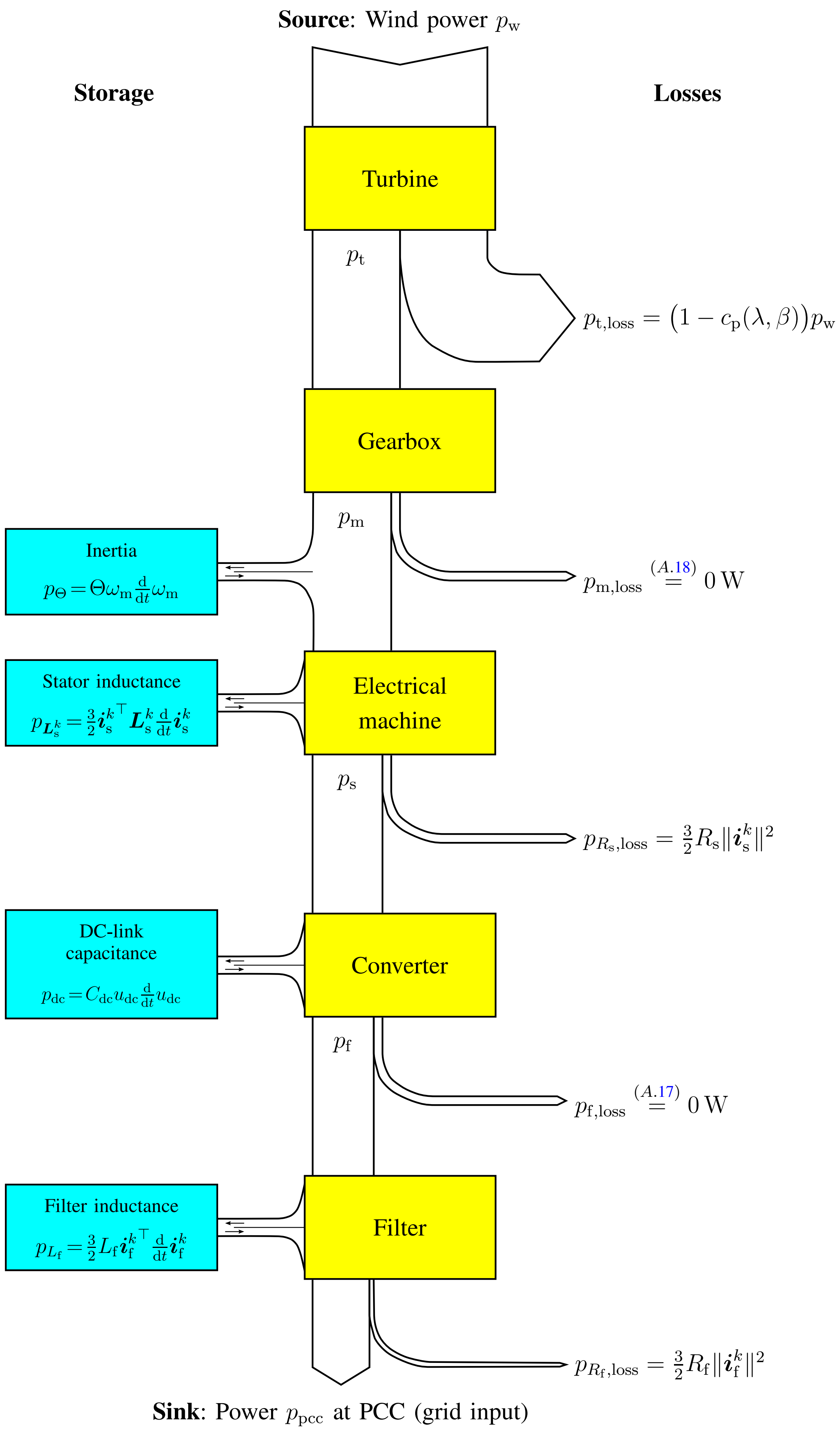

Fig. 42: *Qualitative power flow diagram of a wind turbine with PMSG.*

- In order to feed the greatest possible instantaneous (active) power into the grid (=O$_2$), the losses of the overall wind turbine system have to be minimized. In the simplified case considered in (182), the current amplitudes must be as small as possible.

On the basis of the power balance, there is no insight into the reachability of the objective (O$_3$). To this end, the impact and the remaining degrees of freedom of the presented control system must be discussed.

*4) Active & reactive power at the PCC:* In the case of three-phase systems, in addition to the instantaneous active power, it is also important to consider reactive power. In wind turbine systems, active and reactive power at the PCC can be expressed, under Assumptions (A.6), (A.7) and (A.8), as follows (recall Section II-B3):

$$\left.\begin{aligned} P_{\text{pcc}}(t) &\overset{(57),(A.6)-(A.8)}{=} \tfrac{3}{2}\boldsymbol{u}_{\text{g}}^{k}(t)^{\top}\boldsymbol{i}_{\text{f}}^{k}(t) \overset{(97)}{=} \tfrac{3}{2}\hat{u}_{\text{g}} i_{\text{f}}^{d}(t) \text{ (in W) and} \\ Q_{\text{pcc}}(t) &\overset{(58),(A.6)-(A.8)}{=} \tfrac{3}{2}\boldsymbol{u}_{\text{g}}^{k}(t)^{\top}\boldsymbol{J}\boldsymbol{i}_{F}^{k}(t) \overset{(97)}{=} -\tfrac{3}{2}\hat{u}_{\text{g}} i_{\text{f}}^{q}(t) \text{ (in var).} \end{aligned}\right\} \tag{183}$$

**Active power:** Because of (178) and (183) in the *steady state case*, the active power $P_{\text{pcc}}$ corresponds to the instantaneous power $p_{\text{pcc}}$ at the PCC. A maximization of the instantaneous power $p_{\text{pcc}}$ also maximizes the active power $P_{\text{pcc}}$ (at least in steady state).

Furthermore, according to (183), the active power output $P_{\text{pcc}}$ to the grid is directly proportional to the $d$-component $i_{\text{f}}^{d}$ of the filter current. The current component $i_{\text{f}}^{d}$ was used in Section IV-D to control the DC-link voltage. This means that $P_{\text{pcc}}$ can not be directly controlled. The following Section IV-E5 shows that $p_{\text{pcc}}$ and thus $P_{\text{pcc}}$ can only be indirectly influenced by the DC-link voltage control and the generator speed control.

**Reactive power:** The reactive power output $Q_{\text{pcc}}$ to the grid is proportional to the $q$-component $i_{\text{f}}^{q}$ of the filter current according to (183). The reactive power $Q_{\text{pcc}}$ can thus be controlled via a corresponding manipulation of the current component $i_{\text{f}}^{q}$. For a given (possibly time-varying) reactive power reference $q_{\text{pcc,ref}}\colon \mathbb{R}_{>0} \to \mathbb{R}$, simply the current reference

$$\forall t \geq 0\,\text{s}: \qquad i_{\text{f,ref}}^{q}(t) = -\frac{2}{3\hat{u}_{\text{g}}} Q_{\text{pcc,ref}}(t) \tag{184}$$

of the $q$-current component has to be adapted. The fast dynamics of the subordinate current control system of the $q$-component ensure that the reference (184) can be tracked quickly and, thus, provided that grid voltage magnitude $\hat{u}_{\text{g}}$ is known, that $Q_{\text{pcc}}(t) = Q_{\text{pcc,ref}}(t)$ holds for (almost) all $t \geq 0$.

*5) Impact of operation management and control system on the power flow:* To achieve the *three* objective (O$_1$), (O$_2$) and (O$_3$) with the *four* controlled current components $i_{\text{f}}^{d}$, $i_{\text{f}}^{q}$, $i_{\text{s}}^{d}$ and $i_{\text{s}}^{q}$ (see Section IV-A and IV-B), there are, in principle, *four* degrees of freedom to influence the power flow in the wind turbine system. The presented current control systems (see Section IV-A–IV-D) allow for:

- Feedforward control of the machine torque by the stator currents $i_{\text{s}}^{d}$ and $i_{\text{s}}^{q}$ (e.g. $i_{\text{s,ref}}^{d} = 0\,\text{A}$ and $i_{\text{s,ref}}^{q} = \frac{1}{\frac{3}{2}n_{\text{p}}\,\psi_{\text{pm}}} m_{\text{m,ref}}$ for isotropic machines, see Assumption (A. 11) or Eq. (131)) and
- Control of the DC-link voltage and feedforward control of the reactive power flow at the PCC by the filter currents $i_{\text{f}}^{d}$ and $i_{\text{f}}^{q}$ (e.g. $i_{\text{f,ref}}^{d}$ for DC-link voltage stabilization and $i_{\text{f,ref}}^{q}$ as in (184) for reactive power control).

Thus, there is no possibility to achieve direct control of the active power output to the grid. Objective (O$_2$) is nevertheless achieved, though indirectly, by the control of the DC-link voltage. The achievement of the objectives (O$_1$), (O$_2$) and (O$_3$) by the operation management is now explained individually for each control objective. From the Assumptions (A.1)–(A.18), the following can be inferred:

**Objective (O$_1$)**: In operation regime II, the nonlinear speed controller (141), superimposed on the stator current controllers (126), reaches its control objective (136) (at least asymptotically): For a constant pitch angle $\beta_{\text{ref}} = \beta = 0°$, the generator angular velocity $\omega_{\text{m}}$ and the tip speed ratio $\lambda$ asymptotically approach the optimum angular velocity $\omega_{\mathrm{m}}^{\star}$ and the optimum tip speed ratio $\lambda^{\star}$, respectively. This maximizes the power coefficient and the turbine output, since $c_{\mathrm{p}}(\beta, \lambda) \to c_{\mathrm{p}}(0, \lambda^{\star}) =: c_{\mathrm{p}}^{\star}$ and thus $p_{\mathrm{t}} \to c_{\mathrm{p}}^{\star} p_{\mathrm{w}}$. Hence, objective ($O_1$) is achieved by the "Maximum Power Point Tracking" of the speed controller (for constant wind speeds).

**Objective ($\mathbf{O_2}$)**: As already noted, there is no direct influence of the current control system on the active output power into the grid. The speed control ensures maximization of the turbine power (ideally: $p_{\mathrm{t}} = c_{\mathrm{p}}^{\star} p_{\mathrm{w}}$). The DC-link voltage control (168) guarantees (at least locally) a stable set-point tracking of the DC-link voltage $u_{\mathrm{dc}} = u_{\mathrm{dc,ref}} > 0\,\mathrm{V}$; which means, on average, *no* power is exchanged with the DC-link capacitor, i.e. $p_{\mathrm{f}} \overset{(176)}{=} -p_{\mathrm{s}}$. The power balance (182) thus ensures during steady-state operation of the wind turbine system, that the maximum possible instantaneous power is also output into the grid if resistance losses are minimized and turbine power is maximized. Objective ($O_2$) is thus indirectly achieved.

**Objective ($\mathbf{O_3}$)**: The desired reactive power output to the grid is calculated according to (183) and, for a given reactive power reference $Q_{\mathrm{pcc,ref}} \colon \mathbb{R}_{>0} \to \mathbb{R}$, the underlying grid-side current controller (113) of the $q$-component of the filter current $i_{\mathrm{f}}^{q}$ guarantees that the desired reactive power (184) is fed into the grid.

**Remark IV.6.** *In view of Assumption (A.11), no reluctance torque was considered, i.e.* $L_{\mathrm{s}}^{d} = L_{\mathrm{s}}^{q}$ *in* (129) *(see Sec. IV-B5). For such isotropic machines, where* $L_{\mathrm{s}}^{d} = L_{\mathrm{s}}^{q}$ *in* (129)*, the* $d$*-component of the stator current* $i_{\mathrm{s}}^{d}$ *is freely selectable and could, for example, be used to reduce the reactive power in the machine. However,* $i_{\mathrm{s}}^{d} \neq 0\,\mathrm{A}$ *inevitably leads to increased losses in the stator resistance (see the power balance in* (182)*). Thus, the benefit of* $i_{\mathrm{s}}^{d} \neq 0\,\mathrm{A}$ *should be weighed against the accompanying increase in copper losses.*

*In contrast, for anisotropic machines, where* $L_{\mathrm{s}}^{d} \neq L_{\mathrm{s}}^{q}$ *in* (129)*, the reluctance torque and the* $d$*-current* $i_{\mathrm{s}}^{d}$ *can be used to increase the machine torque* and *to reduce the copper losses (for example, by using a "Maximum Torque per Ampere" algorithm [49, 50]). With regard to the objectives ($O_1$), ($O_2$) and ($O_3$), this corresponds to a physically motivated relaxation of Assumption (A.11) to exploit the reluctance torque additionally.*

### *F. Further control concepts for wind turbines*

In this chapter, one of several *possible* control system variants for wind turbine systems was presented. For wind turbines (with e.g. isotropic permanent-magnet synchronous generators, i.e. $L_{\mathrm{s}}^{d} = L_{\mathrm{s}}^{q}$, see Assumption (A.11)) and back-to-back converters, there are a total of four variants available.

- **Maximum Torque per Ampere control [56] (see Fig. 43)**
  This control method corresponds to the presented variant in this chapter.
  *Machine-side control*: The machine speed is controlled via $i_{\mathrm{s}}^{q}$ while $i_{\mathrm{s}}^{d} = i_{\mathrm{s,ref}}^{d} = 0\,\mathrm{A}$. Then, for isotropic machines, the maximum torque can be generated and the electric machine can be dimensioned smaller.
  *Grid-side control*: The DC-link voltage is controlled via $i_{\mathrm{f}}^{d}$ to track a constant set-point value. The control of the reactive power, exchanged with the grid, is achieved via $i_{\mathrm{f}}^{q}$.
- **Unity power factor control of generator [57]**
  *Machine-side control*: The machine speed is also controlled via $i_{\mathrm{s}}^{q}$, but $i_{\mathrm{s}}^{d}$ is set such that the reactive power exchanged with the machine-side converter becomes zero, i.e., the following should hold

$$\forall t \geq 0\,\mathrm{s}: \qquad Q_{\mathrm{s}}(t) \overset{(58)}{=} \frac{3}{2}\big(u_{\mathrm{s}}^{q}(t) i_{\mathrm{s}}^{d}(t) - u_{\mathrm{s}}^{d}(t) i_{\mathrm{s}}^{q}(t)\big) \overset{!}{=} 0\,\mathrm{var}.$$

  Thus, the machine-side converter can be dimensioned smaller, since on the machine side, the apparent power corresponds to the active power (Remark IV.6 must be observed).
  *Grid-side control*: the DC-link voltage is controlled to a constant value via $i_{\mathrm{f}}^{d}$. The control of the reactive power exchange with the grid is again achieved via $i_{\mathrm{f}}^{q}$.

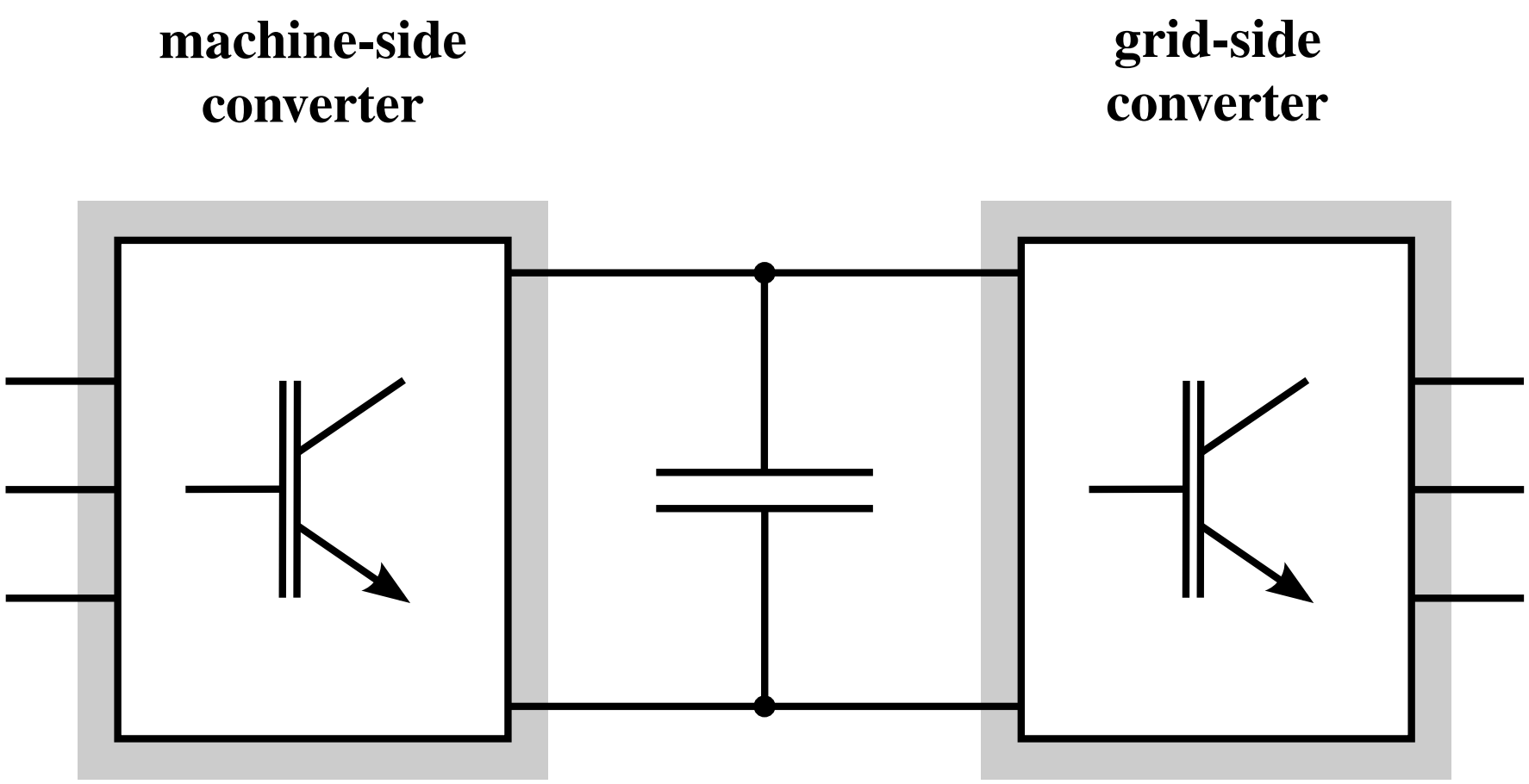

Fig. 43: *Control tasks of the "Maximum Torque per Ampere" control variant.*

- **Constant stator voltage control [58, p. 99–108]**
  *Machine-side control*: The active power output of the electric machine is controlled via the stator current $i_s^\alpha$. In this case, the stator voltage component $u_s^\alpha$ is adjusted via $i_s^\beta$ (as a function of $i_s^\alpha$) such that $|u_s^\alpha|$ corresponds to the nominal voltage of the machine.
  *Grid-side control*: The DC-link voltage is controlled to a constant value via $i_f^d$. The control of the reactive power to be fed into the grid takes place again via $i_f^q$ instead.
- **P-Q control by means of the grid-side converter [59]**
  *Machine-side control*: If a lossless converter is assumed, $p_{dc} = -p_s - p_f$ holds. Thus, with good DC-link voltage regulation (i.e. $\frac{d}{dt}u_{dc} = 0\,\frac{\mathrm{V}}{\mathrm{s}} \Rightarrow p_{dc} = 0\,\mathrm{W}$), the control of the active power and the control of the DC-link voltage can be interchanged. Then, on the machine side, the stator voltage component $u_s^\alpha$ and the DC-link circuit voltage $u_{dc}$ are controlled.
  *Grid-side control*: The active power output to the grid is controlled via $i_f^d$. The reactive power control is again achieved via $i_f^q$.

Finally, the "Maximum Torque per Ampere" control variant, presented in this chapter, and its underlying control tasks are illustrated in Fig. 43.

## V. Simulation of the overall and controlled wind turbine system

To illustrate and validate the models and control systems presented in Sections III & IV, simulation results are presented and discussed for a complete wind turbine system with permanent-magnet synchronous machine.

### A. Implementation

The complete wind turbine (see Section III) and its control (see Section IV) were implemented and simulated using Matlab/Simulink® from MathWorks®.

*1) Wind turbine:* The wind turbine system was implemented in Matlab/Simulink® according to the modeling equations in Section III:

- Turbine power coefficient (66) and turbine torque (68);
- Gearbox with turbine speed conversion and turbine torque according to (75), and with machine-side total inertia (77) of the wind turbine;

- Permanent-magnet synchronous generator (80) with stator circuit, flux linkage, mechanical system and machine torque;
- Back-to-back converter with DC-link (85), line-to-line voltages (86) (or (92)) and DC-link currents (87) as functions of the switching vectors. The implementation of the modulation method corresponds to the "regular-sampled" PWM with "symmetric sampling" (see [44, Sec. 8.4.12]);
- Grid filter (83) (or (84)) with filter resistance, filter inductance and ideal (balanced) grid voltages (82).

*2) Control:* The control systems of the wind turbine system were designed according to the equations of Section IV and implemented in Matlab/Simulink®:

- Grid-side current control in *grid voltage orientation* (97) with phase-locked loop $PI$-controllers (102), disturbance compensation (108) and $PI$-current controller (113) (designed according to the Magnitude Optimum);
- Machine-side current control in *permanent-magnet flux orientation* (116) with disturbance compensation (122), $PI$-current controllers (126) (designed according to the Magnitude Optimum);
- Nonlinear speed controller (141) for the machine/turbine;
- DC-link voltage $PI$-controller (168) which is locally stable and designed according to the Hurwitz conditions ($C_1$) and ($C_2$) from Section IV-D7;
- Feedforward control of the reactive power (184) by means of a corresponding reference generation for the $q$-component of the filter current.

*3) Wind data used:* For the simulation, $600\,\mathrm{s}$ of real wind data from the research platform FINO1 were used. The wind data was recorded on the FINO1 measuring platform (coordinates: 54° 00' 53.5" N, 06° 35' 15.5" E) on the 24th November 2012 between 11:40–11:50 am with a resolution of $10\,\mathrm{Hz}$. In Fig. 44 (top), the measured wind speed profile $v_{\mathrm{w}}(\cdot)$ is shown. The average wind speed $\bar{v}_W$ is approximately $5.5\,\frac{\mathrm{m}}{\mathrm{s}}$, which corresponds to operation regime II (see Fig. 11).

**Acknowledgement.** *For the provided wind data profile, the authors would like to thank the contributing organizations of the FINO-Project: Bundesministerium für Umwelt, Naturschutz, Bau und Reaktorsicherheit (BMUB), Projektträger Jülich (PTJ)/Forschungszentrum Jülich GmbH, Bundesamt für Seeschifffahrt und Hydrographie (BSH) and DEWI GmbH.*

*4) Simulation, model and control parameters:* The implementation of the model of the overall and controller wind turbine system used the numerical fixed-step ODE-Solver `Runge-Kutta (ode4)` and had a time step of $h = 4\,\mu\mathrm{s}$. The model parameters are taken from a wind turbine with a permanent magnet synchronous generator and a nominal output power of $2\,\mathrm{MW}$ [58, p. 124,199,204] and [60]. The controller parameters are determined according to the corresponding derivations in Section IV-C. For controller parameterization and implementation, it was assumed that all model parameters are exactly known. The simulation, model and controller parameters are summarized in Tab. V.

### *B. Discussion of the simulation results*

The simulation results are shown in Fig. 44, 45 and 46. The individual figures are discussed below.

*Wind, power harvesting and speed control:* In Fig. 44, the wind speed $v_{\mathrm{w}}$ impinging on the turbine and its mean value $\bar{v}_{\mathrm{w}}$ (averaged over the simulation duration of $600\,\mathrm{s}$) are depicted in the uppermost subplot. In the second subplot (from above), the wind power $p_{\mathrm{w}}$, turbine power $p_{\mathrm{t}}$ and instantaneous (active) power $p_{\mathrm{pcc}}$ at the PCC are shown. The third subplot shows the optimal tip speed ratio $\lambda^{\star}$ and the evolution of the tip speed ratio $\lambda$ of the turbine. In the fourth subplot, the Betz limit $c_{\mathrm{p,Betz}}$, the maximum power coefficient $c_{\mathrm{p,1}}(\lambda^{\star})$ and the curve of the power coefficient $c_{\mathrm{p,1}}$ of the wind turbine are plotted. The lowest subplot shows the machine angular velocity $\omega_{\mathrm{m}}$ and the optimal machine angular velocity $\omega_{\mathrm{m}}^{\star}$ (see Section IV-C).

Despite small changes in the wind speed (in the range of $4 - 6\,\frac{\mathrm{m}}{\mathrm{s}}$), there are strong fluctuations in the wind power $p_{\mathrm{w}}$ because of its cubic relationship to wind speed (see (5)). As expected, the turbine power $p_{\mathrm{t}}$ is always lower than the wind power $p_{\mathrm{w}}$, whereas the instantaneous (active) power $p_{\mathrm{pcc}} = P_{\mathrm{pcc}}$ (which corresponds to the active power $P_{\mathrm{pcc}}$ for a constant grid frequency, recall (54)) which is fed into the grid, can temporarily exceed the turbine power $p_\mathrm{t}$ (due to the energy storages present within the system, see Fig. 42). On average, the feed-in active power $p_\mathrm{pcc}$ must be smaller than the turbine power $p_\mathrm{t}$ due to the losses occurring within the turbine.

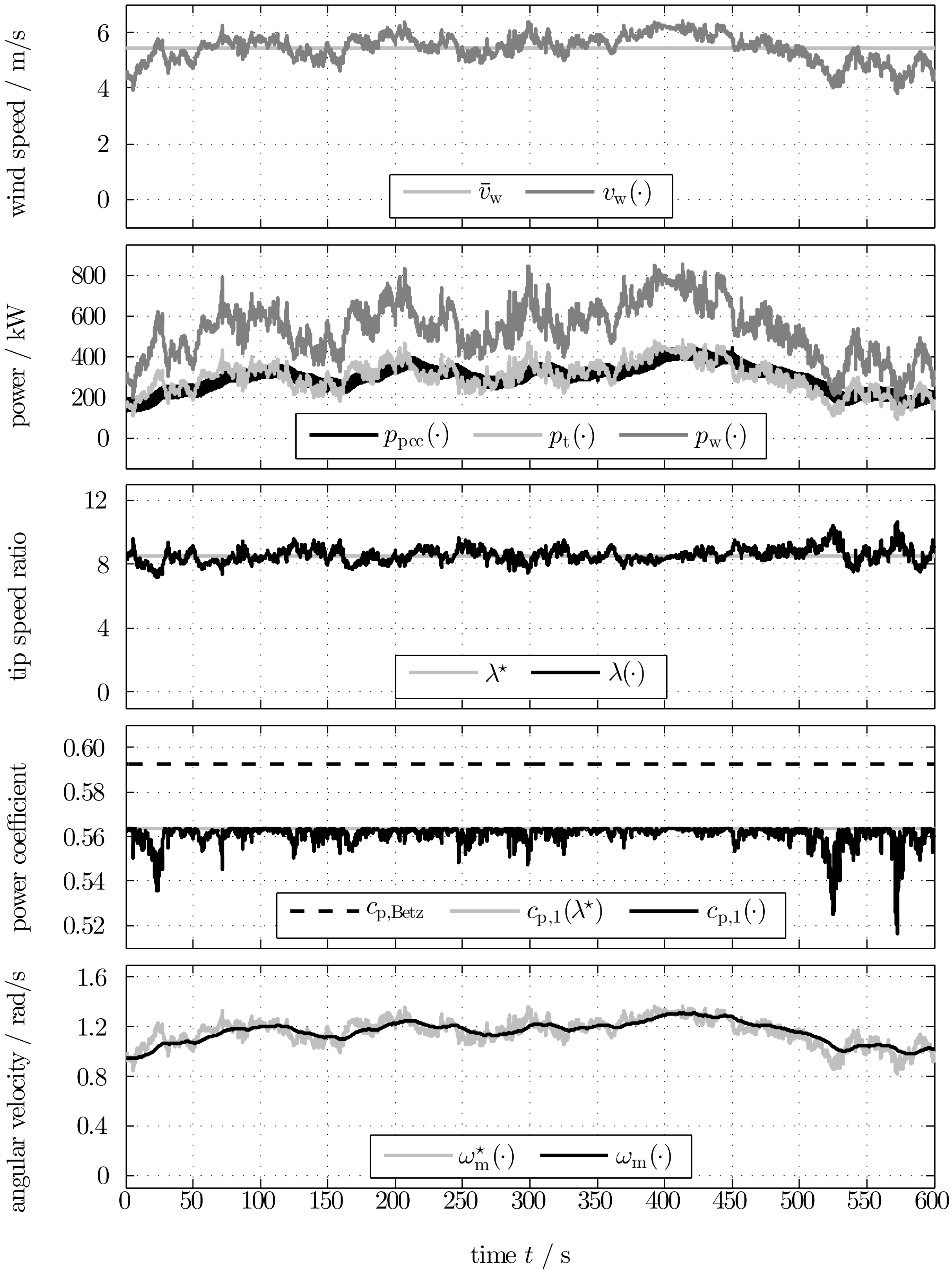

Fig. 44: *Time plots of the overall system simulation (from top to bottom): Wind speed $v_w(\cdot)$ and its mean $\bar{v}_W$; instantaneous (active) power $p_{pcc}(\cdot)$ at PCC, turbine power $p_t(\cdot)$ and wind power $p_w(\cdot)$; tip speed ratio $\lambda(\cdot)$ and its optimum value $\lambda^\star$; maximum power coefficient $c_{p,1}(\cdot)$ and its maximum $c_{p,1}(\lambda^\star)$ and Betz limit $c_{p,Betz}$; machine angular velocity $\omega_m(\cdot)$ and its optimal value $\omega_m^\star(\cdot)$*

TABLE V: *Simulation, model and controller parameters for implementation and simulation.*

| **Description** | **Symbol** | **Value (with unit)** |
|---|---|---|
| *Implementation* | | |
| ODE solver (fixed-step) | | `Runge-Kutta (ode4)` |
| Sampling time | $h$ | $4\,\mu\mathrm{s}$ |
| *Turbine & gear (direct drive)* | | |
| Air density | $\varrho$ | $1.293\,\frac{\mathrm{kg}}{\mathrm{m}^3}$ |
| Turbine radius | $r_\mathrm{t}$ | $40\,\mathrm{m}$ |
| Turbine inertia | $\Theta_\mathrm{t}$ | $8.6 \cdot 10^6\,\mathrm{kg\,m}^2$ |
| Power coefficient | $c_{\mathrm{p},1}(\cdot)$ | as in (66) |
| Gear ratio | $g_\mathrm{r}$ | 1 |
| *Permanent-magnet synchronous generator (isotropic)* | | |
| Number of pole pairs | $n_\mathrm{p}$ | 48 |
| Stator resistance | $R_\mathrm{s}$ | $0.01\,\Omega$ |
| Stator inductance(s) | $L_\mathrm{s}^d = L_\mathrm{s}^q$ | $3.0\,\mathrm{mH}$ |
| flux linkage of permanent magnet | $\psi_\mathrm{pm}$ | $12.9\,\mathrm{V\,s}$ |
| Generator inertia | $\Theta_\mathrm{m}$ | $1.3 \cdot 10^6\,\mathrm{kg\,m}^2$ |
| *Back-to-back converter* | | |
| DC-link capacitance | $C_\mathrm{dc}$ | $2.4\,\mathrm{mF}$ |
| Switching frequency | $f_\mathrm{sw}$ | $2.5\,\mathrm{kHz}$ |
| Delay | $T_\mathrm{delay} = \frac{1}{f_\mathrm{sw}}$ | $0.4\,\mathrm{ms}$ |
| *Filter & grid voltage* | | |
| Filter resistance | $R_\mathrm{f}$ | $0.1\,\Omega$ |
| Filter inductance | $L_\mathrm{f}$ | $24\,\mathrm{mH}$ |
| Grid angular frequency | $\omega_\mathrm{g} = 2\pi f_\mathrm{g}$ | $100\pi\,\frac{\mathrm{rad}}{\mathrm{s}}$ |
| Grid voltage amplitude | $\hat{u}_\mathrm{g}$ | $2.7\,\mathrm{kV}$ |
| Grid voltage initial angle | $\alpha_0$ | $0\,\mathrm{rad}$ |
| *Controller parameters* | | |
| PI current controller (113) | $V_{\mathrm{r},\boldsymbol{i}_\mathrm{f}^k}$ | $30\,\Omega$ |
| (grid-side) | $T_{\mathrm{n},\boldsymbol{i}_\mathrm{f}^k}$ | $0.24\,\mathrm{s}$ |
| PI current controller (126) | $V_{\mathrm{r},\boldsymbol{i}_\mathrm{s}^k}$ | $3.75\,\Omega$ |
| (machine-side) | $T_{\mathrm{n},\boldsymbol{i}_\mathrm{s}^k}$ | $0.3\,\mathrm{s}$ |
| Speed controller (141) | $k_\mathrm{p}^\star$ | $188.73\,\frac{\mathrm{kN\,m}}{\mathrm{s}^2}$ |
| DC-link voltage | $V_\mathrm{r,dc}$ | $1.44\,\frac{\mathrm{A}}{\mathrm{V}}$ |
| PI controller (168) | $T_\mathrm{n,dc}$ | $18.9\,\mathrm{ms}$ |
| Phased-locked loop | $V_\mathrm{r,pll}$ | $20\,000\,\frac{1}{\mathrm{s}}$ |
| PI controller (102) | $T_\mathrm{n,pll}$ | $0.2\,\mathrm{ms}$ |

The generator speed control achieves an acceptable reference tracking performance. On average, the machine angular velocity $\omega_\mathrm{m}$ follows the optimum angular velocity $\omega_\mathrm{m}^\star$. Thus, the tip speed ratio $\lambda$ remains close to its optimum value $\lambda^\star$.

However, due to the deviation between $\lambda$ and $\lambda^\star$, the wind turbine is *not* able to always operate at its optimum: The power coefficient $c_{\mathrm{p},1}$ sometimes stays (far) below its maximum value $c_{\mathrm{p},1}(\lambda^\star)$. Thus, a maximum energy yield can not be assumed in general. On one hand, this sub-optimum operation can be attributed to the (very) high total inertia of the wind turbine and, on the other hand, to the speed control implemented (the speed control can *not* support the motor during acceleration of the turbine, see Remark IV.4).

*Turbine power and torque/current control:* Figure 45 shows the turbine power $p_\mathrm{t}$, the motor torque $m_\mathrm{m}$ and its reference $m_\mathrm{m,ref}$ as well as the $q$-component $i_\mathrm{s}^q$ of the stator current and its stator current reference

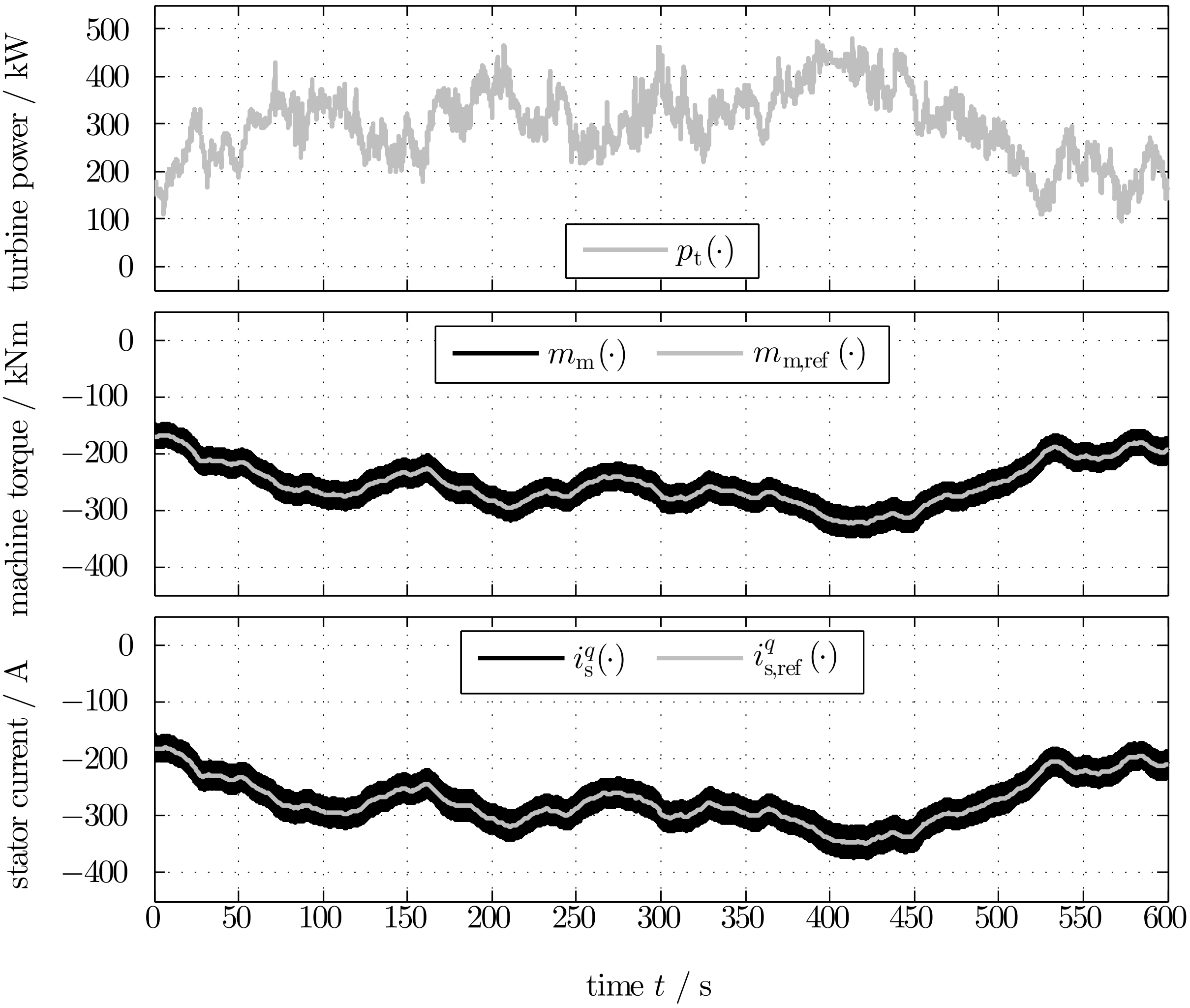

Fig. 45: *Time plots of the overall system simulation (from top to bottom): Turbine power $p_t(\cdot)$; machine torque $m_m(\cdot)$ and its reference $m_{m,ref}(\cdot)$; q-component $i_s^q(\cdot)$ of the stator current (torque generation) and its reference $i_{s,ref}^q(\cdot)$.*

$i_{s,ref}^q$.

In order to be able to withdraw power from the turbine, or the wind, respectively, a (negative) machine torque $m_m$ must be generated. The simulation shows that the implemented current control of the torque-causing $q$-component of the stator current has a fast dynamic response. On average, the current component $i_s^q$ and also the machine torque $m_m$ (see (131)) follow their references $i_{s,ref}^q$ and $m_{m,ref}$ respectively, (nearly) instantaneously. This observation underpins the assumption of a sufficiently fast torque generation in the electric machine (see Assumption (A.13)).

*Reactive power control and DC-link voltage control:* In the top subplot of Fig. 46, the reactive power $Q_{pcc}$ as well as its reference $Q_{pcc,ref}$ at the PCC are shown. The second subplot shows the corresponding $q$-component $i_f^q$ of the filter current and its reference $i_{f,ref}^q$. The active power $p_{pcc}$ at the PCC is shown in the middle subplot. The fourth subplot shows the DC-link voltage $u_{dc}$ with its reference $u_{dc,ref}$. The current component $i_f^d$ and its reference $i_{f,ref}^d$ are shown in the lowest subplot.

The reactive power $Q_{pcc} = -\frac{3}{2}\hat{u}_g i_f^q$ (see (183)) follows its reference $Q_{pcc,ref}$ almost instantaneously. This is achieved by the fast control of the $q$-current component: The filter current component $i_f^q$ follows its reference $i_{f,ref}^q = -2Q_{pcc,ref}/(3\hat{u}_g)$ almost without delay (see (184)).

The negative reactive power jump at $150\,\mathrm{s}$ (with negative phase angle $\varphi = \arctan(Q_{pcc}/P_{pcc}) =$

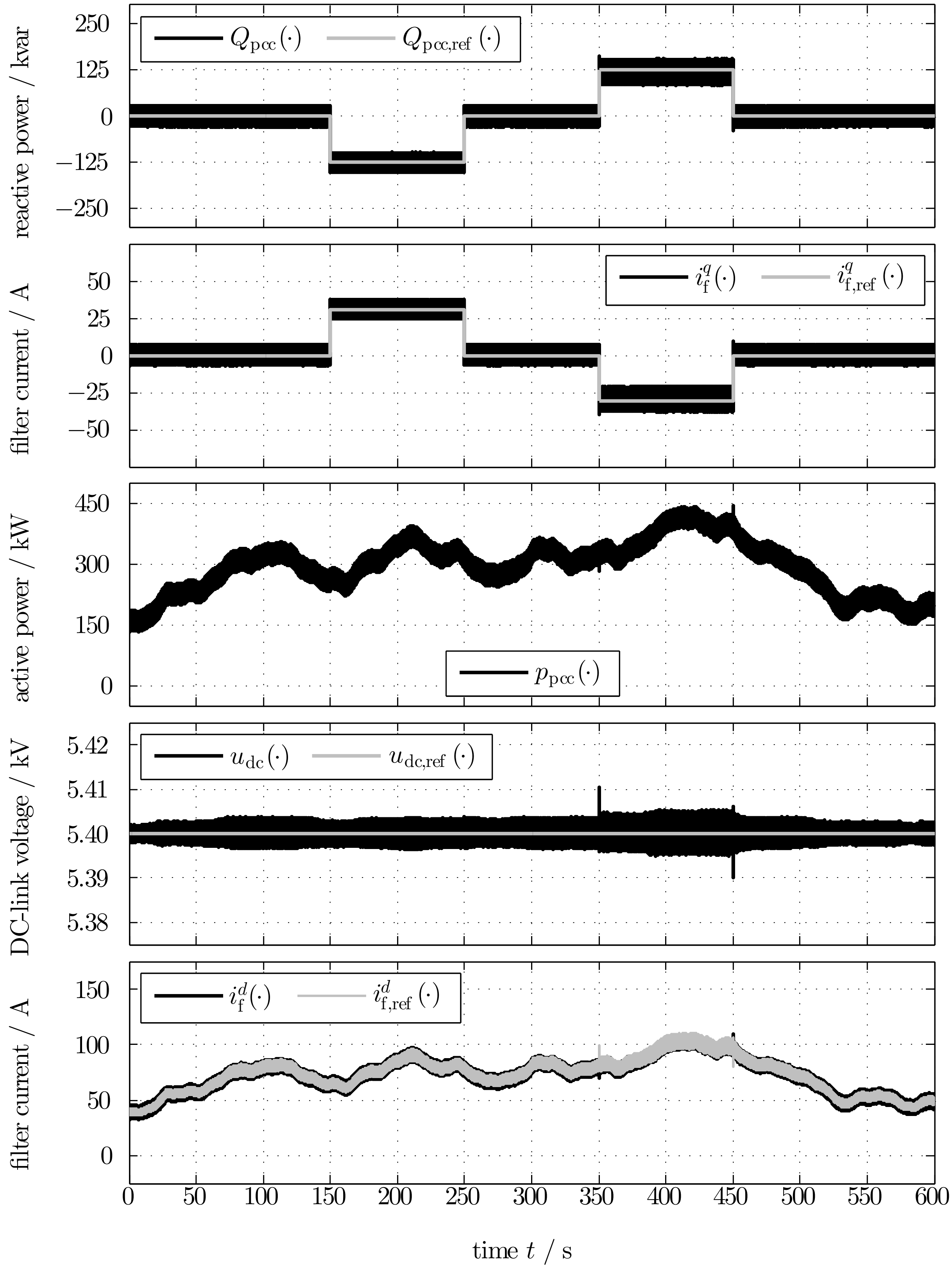

Fig. 46: *Time plots of the overall system simulation (from top to bottom): Reactive power* $Q_{\mathrm{pcc}}(\cdot)$ *and its reference* $Q_{\mathrm{pcc,ref}}(\cdot)$*; q-component* $i_{\mathrm{f}}^{q}(\cdot)$ *of the filter current and its reference* $i_{\mathrm{f,ref}}^{q}(\cdot)$ *(reactive power control); instantaneous (active) power* $p_{\mathrm{pcc}}(\cdot)$*; DC-link voltage* $u_{\mathrm{dc}}(\cdot)$ *and its reference* $u_{\mathrm{dc,ref}}(\cdot)$*; d-component* $i_{\mathrm{f}}^{d}(\cdot)$ *of the filter current and its reference* $i_{\mathrm{f,ref}}^{d}(\cdot)$ *(DC-link control).*

$\arctan\left(-i_{\mathrm{f}}^{q}/i_{\mathrm{f}}^{d}\right) \approx -25°$) and the positive reactive power jump at $350\,\mathrm{s}$ (with positive phase angle $\varphi = \arctan\left(Q_{\mathrm{pcc}}/P_{\mathrm{pcc}}\right) = \arctan\left(-i_{\mathrm{f}}^{q}/i_{\mathrm{f}}^{d}\right) \approx 25°$) requires a positive and negative $q$-current in the filter, respectively. During the intervals $[150,\, 250]\mathrm{s}$ and $[350,\, 450]\mathrm{s}$ the grid-side electrical network (converter and filter) is loaded with an increased apparent power $S_{\mathrm{pcc}} = \sqrt{P_{\mathrm{pcc}}^2 + Q_{\mathrm{pcc}}^2}$.

The DC-link voltage $u_{\mathrm{dc}}$ is controlled via the current component $i_{\mathrm{f}}^{d}$ (see Section IV-D). It can be seen in the plots that the DC-link voltage control-loop is stable and reliably operates at the set-point $u_{\mathrm{dc,ref}} = 5\,400\,\mathrm{V}$. Solely in the case of the third and fourth change of the reactive power, the DC-link voltage $u_{\mathrm{dc}}$ deviates by about $\pm 0.2\,\%$ from its reference. These small peaks also lead to peaks in the $d$-component of the current reference $i_{\mathrm{f,ref}}^{d}$ and of the actual current $i_{\mathrm{f}}^{d}$. Due to the relationship $p_{\mathrm{pcc}} = \frac{3}{2}\hat{u}_{\mathrm{g}} i_{\mathrm{f}}^{d}$ (see (178)) also sudden changes in the instantaneous (active) power are visible.